%% file: arxiv.tex
\documentclass{article}
\usepackage[preprint]{neurips_2021}

\usepackage{graphicx}
\usepackage{dirtytalk}
\usepackage{tabularx}
\usepackage{amsmath}
\usepackage{multirow}

\usepackage[utf8]{inputenc} % allow utf-8 input
\usepackage[T1]{fontenc}    % use 8-bit T1 fonts
\usepackage[hyphens]{url}            % simple URL typesetting
\usepackage[colorlinks=false,hidelinks]{hyperref}
\usepackage{booktabs}       % professional-quality tables
\usepackage{amsfonts}       % blackboard math symbols
\usepackage{nicefrac}       % compact symbols for 1/2, etc.
\usepackage{microtype}      % microtypography
\usepackage{xcolor}         % colors

\setcitestyle{numbers,square}

\title{A Survey of JSON-compatible Binary Serialization Specifications}

\author{
  Juan Cruz~Viotti\thanks{\url{https://www.jviotti.com}} \\
  Department of Computer Science \\
  University of Oxford \\
  Oxford, GB OX1 3QD \\
  \texttt{juancruz.viotti@kellogg.ox.ac.uk} \\
  \and
  Mital~Kinderkhedia \\
  Department of Computer Science \\
  University of Oxford \\
  Oxford, GB OX1 3QD \\
  \texttt{mital.kinderkhedia@cs.ox.ac.uk} \\
}

\newcommand{\paperauthors}{authors }
\newcommand{\We}{We }
\newcommand{\we}{we }
\newcommand{\our}{our }
\newcommand{\Our}{Our }

\begin{document}

\maketitle

\begin{abstract}
\input{sections/abstract.tex}
\end{abstract}

\section{Introduction}
\input{sections/introduction.tex}

\subsection{Schema-less as a Subset of Schema-driven}
\input{sections/schema-less-subset-schema-driven.tex}

\subsection{Related Work}
\input{sections/related-work.tex}

\subsection{Contributions}
\input{sections/contributions.tex}
\input{sections/patches.tex}

\subsection{Paper Organisation}
\input{sections/organisation.tex}

\clearpage
\section{JSON}
\input{sections/json.tex}

\section{Methodology}
\input{sections/methodology.tex}

\section{Encoding Processes}
\label{sec:background}
\subsection{Little Endian Base 128 Encoding}
\input{sections/varints.tex}
\subsection{ZigZag Integer Encoding}
\input{sections/zigzag-encoding.tex}

\clearpage
\section{Schema-driven Specifications}
\label{sec:survey-schema-driven}

\subsection{ASN.1}
\input{sections/formats/asn1.tex}

\clearpage
\subsection{Apache Avro}
\input{sections/formats/avro.tex}

\clearpage
\subsection{Microsoft Bond}
\input{sections/formats/bond.tex}

\clearpage
\subsection{Cap'n Proto}
\input{sections/formats/capnproto.tex}

\clearpage
\subsection{FlatBuffers}
\input{sections/formats/flatbuffers.tex}

\clearpage
\subsection{Protocol Buffers}
\input{sections/formats/protocolbuffers.tex}

\clearpage
\subsection{Apache Thrift}
\input{sections/formats/thrift.tex}

\clearpage
\section{Schema-less Specifications}
\label{sec:survey-schema-less}
\input{sections/schema-less-intro.tex}

\subsection{BSON}
\input{sections/formats/bson.tex}

\clearpage
\subsection{CBOR}
\input{sections/formats/cbor.tex}

\clearpage
\subsection{FlexBuffers}
\input{sections/formats/flexbuffers.tex}

\clearpage
\subsection{MessagePack}
\input{sections/formats/messagepack.tex}

\clearpage
\subsection{Smile}
\input{sections/formats/smile.tex}

\clearpage
\subsection{UBJSON}
\input{sections/formats/ubjson.tex}

\section{Schema Evolution}
\input{sections/evolution.tex}
\input{sections/evolution-theory.tex}

\subsection{Deploying Schema Transformations}
\input{sections/deploying-schema-transformations.tex}

\subsection{Compatibility Analysis}
\input{sections/evolution-analysis.tex}

\section{Conclusions}
\label{sec:conclusions}

\subsection{Use Cases}
\input{sections/conclusions/use-cases.tex}
\subsection{Sequential and Pointer-based Serialization Specifications}
\input{sections/conclusions/sequential-vs-pointer.tex}
\subsection{Types of Schema Compatibility Resolution}
\input{sections/conclusions/compatibility-resolution.tex}
\subsection{Similarities}
\input{sections/conclusions/similarities.tex}

\section{Reproducibility}
\input{sections/reproducibility.tex}

\section{Future Work}
\input{sections/future-work.tex}

\section*{Acknowledgments}

In no particular order, many thanks to Wouter van Oortmerssen, Alessandro
Triglia, and Christopher Warrington for helpful discussions. Many thanks to OSS
Nokalva \footnote{\url{https://www.ossnokalva.com}} for offering access to and
help for using their proprietary ASN-1Step ASN.1 \cite{asn1} implementation and
IDE.

\clearpage
\bibliographystyle{ACM-Reference-Format}
\bibliography{arxiv}

\end{document}

%% file: sections/abstract.tex
In this paper, \we present the recent advances that highlight the
characteristics of JSON-compatible binary serialization specifications.  \We
motivate the discussion by covering the history and evolution of binary
serialization specifications across the years starting from 1960s to 2000s and
onwards. \We analyze the use cases of the most popular serialization
specifications across the industries.  Drawing on the schema-driven (ASN.1,
Apache Avro, Microsoft Bond, Cap'n Proto, FlatBuffers, Protocol Buffers, and
Apache Thrift) and schema-less (BSON, CBOR, FlexBuffers, MessagePack, Smile,
and UBJSON) JSON-compatible binary serialization specifications, \we compare
and contrast their inner workings through \our analysis.  \We explore a set of
non-standardized binary integer encoding techniques (ZigZag integer encoding
and Little Endian Base 128 variable-length integer encoding) that are essential
to understand the various JSON-compatible binary serialization specifications.
\We systematically discuss the history, the characteristics, and the
serialization processes of the selection of schema-driven and schema-less
binary serialization specifications and \we identify the challenges associated
with schema evolution in the context of binary serialization.  Through
reflective exercise, \we explain \our observations of the selection of
JSON-compatible binary serialization specifications. This paper aims to guide
the reader to make informed decisions on the choice between schema-driven or
schema-less JSON-compatible binary serialization specifications.

%% file: sections/introduction.tex
%%%%%%%%%%%%%%%%%%%%%%%%%%%%%%%%%%%%%%%%%%%%%%%%%%%%%%%%%%%%%%%%%%%%%
% DATA SERIALIZATION AND DATA SERIALIZATION FORMATS
%%%%%%%%%%%%%%%%%%%%%%%%%%%%%%%%%%%%%%%%%%%%%%%%%%%%%%%%%%%%%%%%%%%%%

\subsection{Serialization and Deserialization}
\input{sections/serialization.tex}

\subsection{History and Evolution of Serialization Specifications}
\input{sections/history.tex}

%%%%%%%%%%%%%%%%%%%%%%%%%%%%%%%%%%%%%%%%%%%%%%%%%%%%%%%%%%%%%%%%%%%%%
% TEXTUAL AND BINARY SERIALIZATION
%%%%%%%%%%%%%%%%%%%%%%%%%%%%%%%%%%%%%%%%%%%%%%%%%%%%%%%%%%%%%%%%%%%%%

\clearpage
\subsection{Textual and Binary Serialization Specifications}
\input{sections/textual-binary.tex}

%%%%%%%%%%%%%%%%%%%%%%%%%%%%%%%%%%%%%%%%%%%%%%%%%%%%%%%%%%%%%%%%%%%%%
% SCHEMA-LESS AND SCHEMA-DRIVEN SERIALIZATION
%%%%%%%%%%%%%%%%%%%%%%%%%%%%%%%%%%%%%%%%%%%%%%%%%%%%%%%%%%%%%%%%%%%%%

\subsection{Schema-less and Schema-driven Serialization Specifications}
\input{sections/schema-less-driven.tex}

%% file: sections/serialization.tex
\emph{Serialization} is the process of translating a data structure into a
\emph{bit-string} (a sequence of bits) for storage or transmission purposes.
The original data structure can be reconstructed from the bit-string using a
process called \emph{deserialization}. Serialization specifications define the
bidirectional translation between data structures and bit-strings.
Serialization specifications support persistence and the exchange of
information in a machine-and-language-independent manner.

\begin{figure}[hb!]
  \frame{\includegraphics[width=\linewidth]{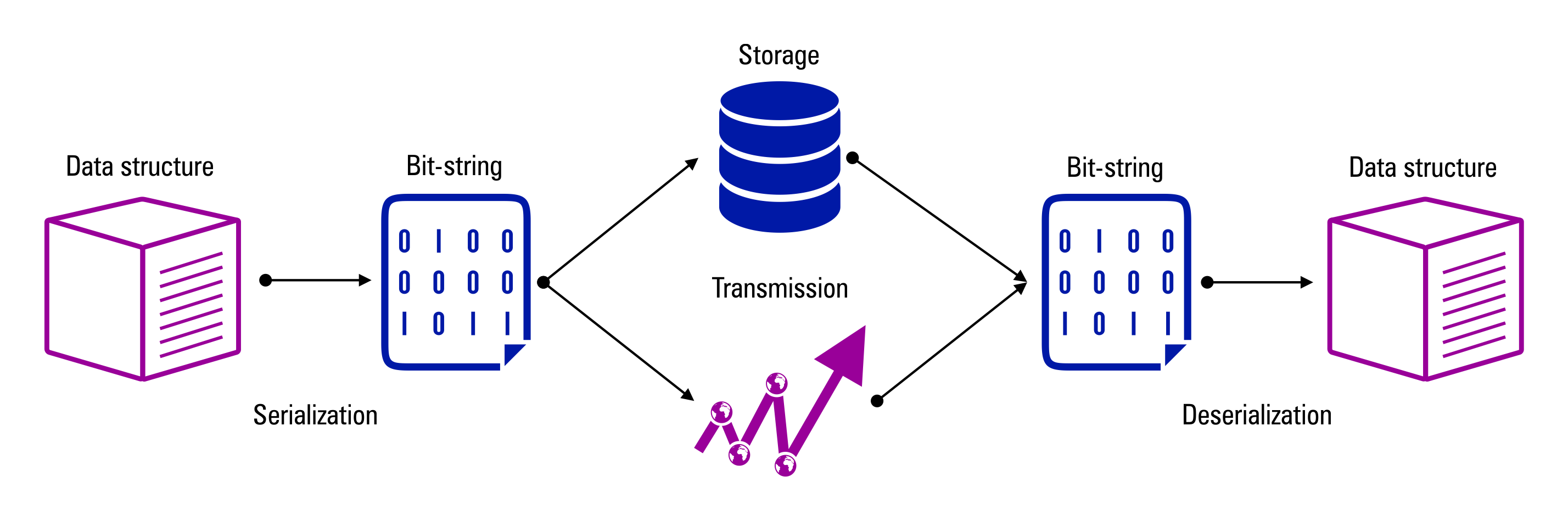}}
  \caption{The process of translating a data structure to a bit-string is
  called \emph{serialization}. The process of translating a bit-string back to
its original data structure is called \emph{deserialization}.}
\label{fig:data-serialization} \end{figure}

Serialization specifications are categorized based on how the information is
represented as a bit-string i.e. \textit{textual} or \textit{binary} and
whether the serialization and deserialization processes require a formal
description of the data structure i.e. \textit{schema-driven} or
\textit{schema-less}. Before \we go into a detailed discussion about textual
and binary serialization specifications, \we motivate by discussing the history
and evolution of serialization specifications.

%% file: sections/history.tex
\begin{figure}[hb!]
\frame{\includegraphics[width=\textwidth]{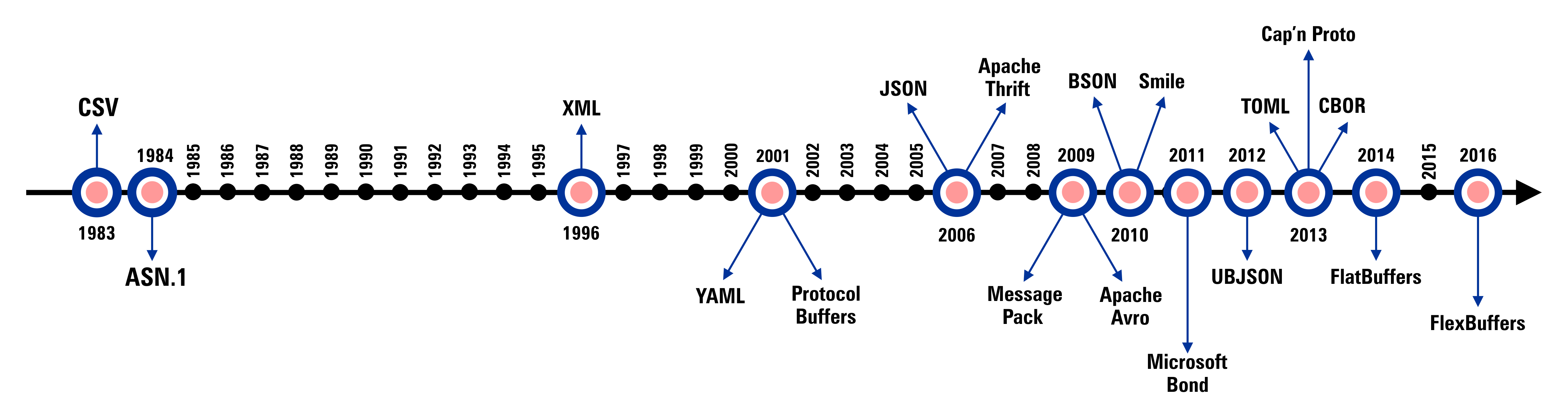}} \caption{A
timeline showcasing some of the most popular serialization specifications since
the early 1980s.} \end{figure}

\textbf{1960s}. In 1969, IBM developed \emph{GML} (Generalized Markup Language)
\footnote{\url{http://www.sgmlsource.com/history/roots.htm}}, a markup language
and schema-less textual serialization specification to define meaning behind
textual documents. Decades later, XML \cite{Paoli:06:EML} was inspired by GML.

\textbf{1970s}. In 1972, IBM OS/360 introduced a general-purpose schema-less
serialization specification as part of their FORTRAN suite
\cite{ibm-list-directed-io}. The IBM FORTRAN manuals referred to the
serialization specification as \emph{List-Directed Input/Output}. It consisted
of comma-separated or space-separated values that now resemble the popular CSV
\cite{RFC4180} schema-less textual serialization specification.

\textbf{1980s}. Microsoft invented the \emph{INI} general purpose schema-less
textual serialization specification
\footnote{\url{https://docs.microsoft.com/en-us/previous-versions/windows/it-pro/windows-server-2008-R2-and-2008/cc731332(v=ws.11)?redirectedfrom=MSDN}}
as part of their MS-DOS operating system in the early 1980s (the exact year is
unclear). In 2021, the Microsoft Windows operating system continues to make use
of the INI specification. INI also inspired the syntax of configuration file
formats in popular software products such as \emph{git}
\footnote{\url{https://git-scm.com}} and \emph{PHP}
\footnote{\url{https://www.php.net}}. In 1983, the Osborne Executive portable
computer Reference Guide \cite{osborne-executive-reference-guide} used the term
\emph{CSV} to refer to files containing comma-separated rows of data. In 1984,
the \emph{International Telecommunication Union}
\footnote{\url{https://www.itu.int/ITU-T/recommendations/index.aspx?ser=X}}
specified the \emph{ASN.1} schema-driven binary serialization specification as
part of the \cite{asn1-old} standard. The ASN.1 serialization specification
became a standalone standard in 1988.  In 1986, the \emph{SGML} (Standard
Generalized Markup Language), a descendant of IBM GML to define custom markup
languages, was proposed as an ISO standard \cite{ISO16387}. In the late 1980s,
NeXT introduced the schema-less textual ASCII-based \cite{STD80} \emph{Property
List} serialization format
\footnote{\url{https://developer.apple.com/library/archive/documentation/Cocoa/Conceptual/PropertyLists/OldStylePlists/OldStylePLists.html}}
which is now popular in Apple's operating systems.

\textbf{1990s}. In 1996, the Netscape web browser used stringified
representations of JavaScript data structures for data interchange
\cite{the-json-saga}. This serialization approach would be standardized as JSON
\cite{RFC8259} a decade later. Also, the SGML \cite{ISO16387} language inspired
the first working draft of a general-purpose schema-less textual serialization
specification named \emph{XML} (Extensible Markup Language) \cite{xml-1996}. In
1997, The Java programming language JDK 1.1 defined a \emph{Serializable}
interface
\footnote{\url{https://docs.oracle.com/javase/7/docs/api/java/io/Serializable.html}}
that provided binary, versioned and streamable serialization of Java classes
and their corresponding state. This serialization specification is referred to
as \emph{Java Object Serialization}
\footnote{\url{https://docs.oracle.com/javase/8/docs/platform/serialization/spec/serialTOC.html}}
and it is still in use. In 1998, \cite{776421} further improved this object
serialization technique. In 1999, XML became a W3C (World Wide Web Consortium)
\footnote{\url{https://www.w3.org}} recommendation \cite{xml-1996}.

\textbf{2000s}. In 2000, Apple (who acquired NeXT in 1997) introduced a binary
encoding for the \emph{Property List} serialization specification
\footnote{\url{https://opensource.apple.com/source/CF/CF-1153.18/CFBinaryPList.c.auto.html}}.
A year later, Apple replaced the ASCII-based \cite{STD80} original
\emph{Property List} encoding with an XML-based encoding
\footnote{\url{https://web.archive.org/web/20140219093104/http://www.appleexaminer.com/MacsAndOS/Analysis/PLIST/PLIST.html}}.
In 2001, Google developed an internal schema-driven binary serialization
specification and RPC protocol named \emph{Protocol Buffers}
\cite{protocolbuffers}. In the same year, the first draft of the schema-less
textual \emph{YAML} \cite{yaml-spec} serialization specification was published
as a human-friendly alternative to XML \cite{Paoli:06:EML}. Refer to
\cite{eriksson2011comparison} for a detailed discussing of the differences
between YAML and JSON. The widely-used \emph{CSV} \cite{RFC4180} schema-less
textual serialization specification was standardized in 2005. The first draft
of the \emph{JSON} schema-less textual serialization specification was
published in 2006 \cite{json-draft-00}. In the same year, Facebook developed an
open-source schema-driven binary serialization specification and RPC protocol
similar to Protocol Buffers \cite{protocolbuffers} named \emph{Apache Thrift}
\cite{slee2007thrift}. In 2008, Google open-sourced \emph{Protocol Buffers}
\cite{protocolbuffers}. In 2009, the \emph{MessagePack} \cite{messagepack}
schema-less binary serialization specification was introduced by Sadayuki
Furuhashi \footnote{\url{https://github.com/frsyuki}}.  Two other binary
serialization specifications were released in 2009: The Apache Hadoop
\footnote{\url{https://hadoop.apache.org}} framework introduced the
\emph{Apache Avro} \cite{avro} schema-driven serialization specification.  The
MongoDB database \footnote{\url{https://www.mongodb.com}} introduced a
schema-less serialization alternative to JSON \cite{RFC8259} named \emph{BSON}
(Binary JSON) \cite{bson}.

\textbf{Advances since 2010}. Two new schema-less binary serialization
specification alternatives to JSON \cite{RFC8259} were conceived in 2010 and
2012: \emph{Smile} \cite{smile} and \emph{UBJSON} \cite{ubjson}, respectively.
In 2011, Microsoft developed \emph{Bond} \cite{microsoft-bond}, a schema-driven
binary serialization specification and RPC protocol inspired by Protocol
Buffers \cite{protocolbuffers} and Apache Thrift \cite{slee2007thrift}. In
2013, the lessons learned from Protocol Buffers \cite{protocolbuffers} inspired
one of its original authors to create an open-source schema-driven binary
serialization specification and RPC protocol named \emph{Cap'n Proto}
\cite{capnproto}.  Two schema-less serialization specifications were created on
2013: a textual serialization specification inspired by INI named \emph{TOML}
\cite{toml-spec} and a binary serialization specification designed for the
Internet of Things named \emph{CBOR} \cite{RFC7049}. In 2014, Google released
\emph{FlatBuffers} \cite{flatbuffers}, a schema-driven binary serialization
specification that was later found to share some similarities to Cap'n Proto
\cite{capnproto}. In 2015, Microsoft open-sourced \emph{Bond}
\cite{microsoft-bond}. In 2016, Google introduced \emph{FlexBuffers}
\cite{flexbuffers}, a schema-less variant of FlatBuffers \cite{flatbuffers}.

\begin{figure}[hb!]
\frame{\includegraphics[width=\linewidth]{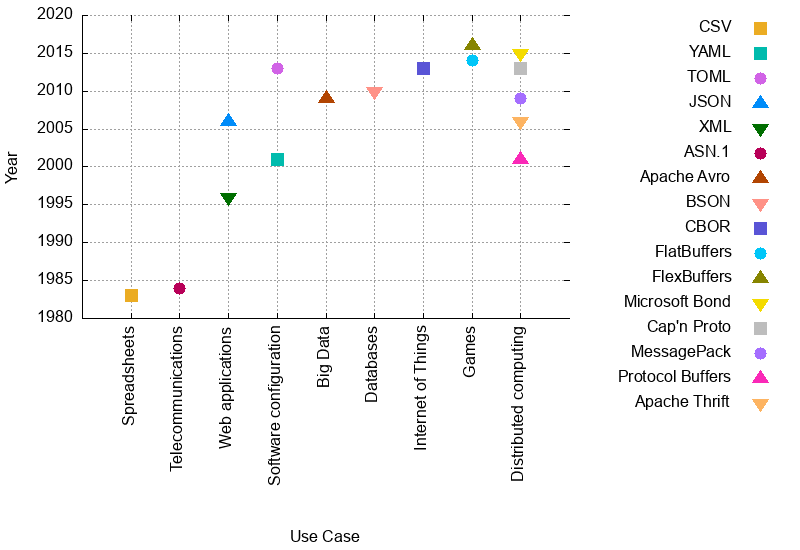}} \caption{The
most popular serialization specifications by their use case.}
\end{figure}

\begin{table}[h!]
\caption{A non-exhaustive list of companies that publicly state that they are using the binary serialization specifications discussed in this paper.}
\label{table:format-companies}
\begin{tabularx}{\linewidth}{l|X}
  \toprule
  \textbf{Serialization Specification} & \textbf{Companies} \\
  \midrule
  ASN.1 & Broadcom, Cisco, Ericsson, Hewlett-Packard, Huawei, IBM, LG Electronics, Mitsubishi, Motorola, NASA, Panasonic, Samsung, Siemens \footnotemark \\ \hline
  Apache Avro & Bol, Cloudera, Confluent, Florida Blue, Imply, LinkedIn, Nuxeo, Spotify, Optus, Twitter \footnotemark \\ \hline
  Microsoft Bond & Microsoft \\ \hline
  Cap'n Proto & Sandstorm, Cloudflare \footnotemark \\ \hline
  FlatBuffers / FlexBuffers & Apple, Electronic Arts, Facebook, Google, Grafana, JetBrains, Netflix, Nintendo, NPM, NVidia, Tesla \footnotemark, \footnotemark \\ \hline
  Protocol Buffers & Alibaba, Amazon, Baidu, Bloomberg, Cisco, Confluent, Datadog, Dropbox, EACOMM, Facebook, Google, Huawei, Intel, Lyft, Microsoft, Mozilla, Netflix, NVidia, PayPal, Sony, Spotify, Twitch, Uber, Unity, Yahoo, Yandex \footnotemark, \footnotemark \\ \hline
  Apache Thrift & Facebook, Cloudera, Evernote, Mendeley, last.fm, OpenX, Pinterest, Quora, RapLeaf, reCaptha, Siemens, Uber \footnotemark \\ \hline
  BSON & MongoDB \\ \hline
  CBOR & Intel \footnotemark, Microsoft \footnotemark, Outfox \footnotemark, Pelion \footnotemark \\ \hline
  MessagePack & Amazon, Atlassian, CODESYS, Datadog, Deliveroo, GitHub, Google, GoSquared, LinkedIn, Microsoft, Mozilla, NASA, National Public Radio, NPM, Pinterest, Sentry, Shopify, Treasure Data, Yelp \footnotemark, \footnotemark, \footnotemark, \footnotemark \\ \hline
  Smile & Ning, Elastic \footnotemark \\ \hline
  UBJSON & Teradata \footnotemark, Wolfram \footnotemark \\
  \bottomrule
\end{tabularx}
\end{table}

\footnotetext[\numexpr\thefootnote-18]{\url{https://www.oss.com/company/customers.html}}
\footnotetext[\numexpr\thefootnote-17]{\url{https://lists.apache.org/thread.html/rc11fcfbc294bb064c6e59167f21b38f3eb6d14e09b9af60970b237d6\%40\%3Cuser.avro.apache.org\%3E}}
\footnotetext[\numexpr\thefootnote-16]{\url{https://www.linkedin.com/in/kenton-varda-5b96a2a4/}}
\footnotetext[\numexpr\thefootnote-15]{\url{https://github.com/google/flatbuffers/issues/6424}}
\footnotetext[\numexpr\thefootnote-14]{\url{https://github.com/google/flatbuffers/network/dependents}}
\footnotetext[\numexpr\thefootnote-13]{\url{https://groups.google.com/g/protobuf/c/tJVbWK3y\_TA/m/vpOiSFfqAQAJ}}
\footnotetext[\numexpr\thefootnote-12]{\url{https://github.com/protocolbuffers/protobuf/network/dependents}}
\footnotetext[\numexpr\thefootnote-11]{\url{https://thrift.apache.org/about\#powered-by-apache-thrift}}
\footnotetext[\numexpr\thefootnote-10]{\url{https://www.npmjs.com/package/tinycbor}}
\footnotetext[\numexpr\thefootnote-9]{\url{https://github.com/OneNoteDev/GoldFish}}
\footnotetext[\numexpr\thefootnote-8]{\url{https://github.com/outfoxx/PotentCodables}}
\footnotetext[\numexpr\thefootnote-7]{\url{https://github.com/PelionIoT/cbor-sync}}
\footnotetext[\numexpr\thefootnote-6]{\url{https://github.com/msgpack/msgpack-node/network/dependents}}
\footnotetext[\numexpr\thefootnote-5]{\url{https://github.com/msgpack/msgpack-ruby/network/dependents}}
\footnotetext[\numexpr\thefootnote-4]{\url{https://github.com/msgpack/msgpack-javascript/network/dependents}}
\footnotetext[\numexpr\thefootnote-3]{\url{https://github.com/msgpack/msgpack/issues/295}}
\footnotetext[\numexpr\thefootnote-2]{\url{https://github.com/FasterXML/smile-format-specification/issues/15}}
\footnotetext[\numexpr\thefootnote-1]{\url{https://docs.teradata.com/reader/C8cVEJ54PO4~YXWXeXGvsA/b9kd0QOTMB3uZp9z5QT2aw}}
\footnotetext[\numexpr\thefootnote]{\url{https://reference.wolfram.com/language/ref/format/UBJSON.html}}

%% file: sections/textual-binary.tex
A serialization specification is \emph{textual} if the bit-strings it produces
correspond to sequences of characters in a text encoding such as ASCII
\cite{STD80}, EBCDIC/CCSID 037 \cite{cpgid00037}, or UTF-8
\cite{UnicodeStandard}, otherwise the serialization specification is
\emph{binary}.

We can think of a textual serialization specification as a set of conventions
within the boundaries of a text encoding such as UTF-8 \cite{UnicodeStandard}.
The availability and diversity of computer tools to operate on popular text
encodings makes textual serialization specifications to be perceived as
human-friendly. In comparison, binary serialization specifications are not
constrained by a text encoding. This flexibility typically results in efficient
representation of data at the expense of requiring accompanying documentation
and specialized tools.

\begin{figure}[hb!]
  \frame{\includegraphics[width=\linewidth]{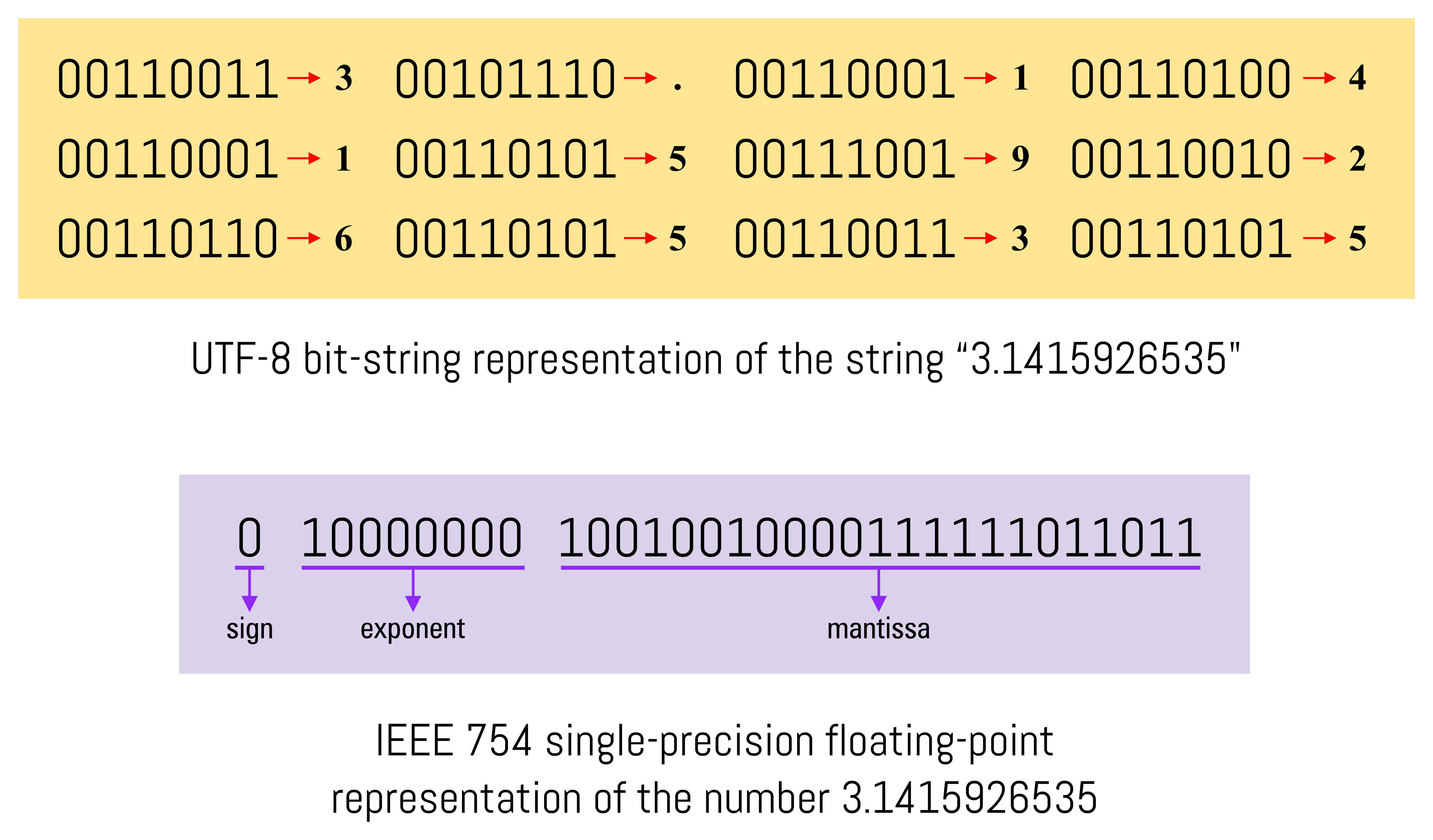}}
  \caption{Textual and binary representations of the decimal number
  $3.1415926535$. The textual representation encodes the decimal number as a
  96-bits sequence of numeric characters ("3" followed by ".", followed by "1",
  and so forth) that we can easily inspect and understand using a text editor.
  On the other hand, the binary representation encodes the decimal number in
  terms of its sign, exponent, and mantissa. The resulting bit-string is only
32 bits long - three times smaller than the textual representation. However, we
are unable to understand it using generally-available text-based tools.}
\label{fig:textual-binary-pi} \end{figure}

%% file: sections/schema-less-driven.tex
\label{sec:schema-less-driven}

% What is a schema?

A \emph{schema} is a formal definition of a data structure.  For example, a
schema may describe a data structure as consisting of two Big Endian IEEE 754
single-precision floating-point numbers \cite{8766229}.  A serialization
specification is \emph{schema-less} if it produces bit-strings that can be
deserialized without prior knowledge of its structure and \emph{schema-driven}
otherwise.

% Advantages and disadvantages

Implementations of schema-less serialization specifications embed the
information provided by a schema into the resulting bit-strings to produce
bit-strings that are standalone with respect to deserialization. In comparison
to schema-driven serialization specifications, schema-less serialization
specifications are perceived as easier to use because receivers can deserialize
any bit-string produced by the implementation and not only the ones the
receivers know about in advance. Alternatively, schema-driven specification
implementations can avoid encoding certain structural information into the
bit-strings they produce. This typically results in compact space-efficient
bit-strings.  For this reason, network-efficient systems tend to adopt
schema-driven serialization specifications \cite{7765670}. Schema-driven
serialization specifications are typically concerned with space efficiency and
therefore tend to be binary.  However, \cite{carrera-rosales} propose a textual
JSON-compatible schema-driven serialization specification. In the case of
communication links with large bandwidths or small datasets, the gains are
negligible but considering slow communication links or large datasets which
could be terabytes in size, the choice of serialization specification could
have a big impact.

Writing and maintaining schema definitions is a core part of using
schema-driven serialization specifications.  Most schema-driven serialization
specifications implement a custom schema language that is not usable by any
other schema-driven serialization specification. A schema-driven serialization
specification may use a low-level or a high-level schema definition language.
Low-level schema definition languages such as \emph{PADS}
\cite{10.1145/1064978.1065046}, \emph{BinX} \cite{1261708}, and \emph{Data
Format Description Language} (DFDL) \cite{1261708} are concerned with
describing the contents of bit-strings while high-level schema definition
languages such as \emph{ASN.1} \cite{asn1} and \emph{JSON Schema}
\cite{jsonschema-core} abstractly describe data structures and depend on the
serialization specification implementation to provide the corresponding
encoding rules.

Often, schemas are auto-generated from formal definitions such as database
tables \footnote{\url{https://github.com/SpringTree/pg-tables-to-jsonschema}},
other schema languages
\footnote{\url{https://github.com/okdistribute/jsonschema-protobuf}}, or
inferred from the data \cite{8424731} \cite{mci/Klettke2015}
\cite{baazizi2019parametric} \cite{CANOVASIZQUIERDO201652}
\cite{10.1145/2187980.2188227} \cite{10.1007/978-3-319-61482-3_16}
\cite{10.1007/978-3-642-39200-9_8} \cite{svoboda2020json}
\cite{10.1145/3122831.3122837}. There are also examples of domain-specific
schema-driven serialization specifications where the schema definition is
implicitly embedded into the serialization and deserialization implementation
routines, such as the SOL binary representation for sensor measurements
\cite{7794864}.

\begin{figure}[hb!]
  \frame{\includegraphics[width=\linewidth]{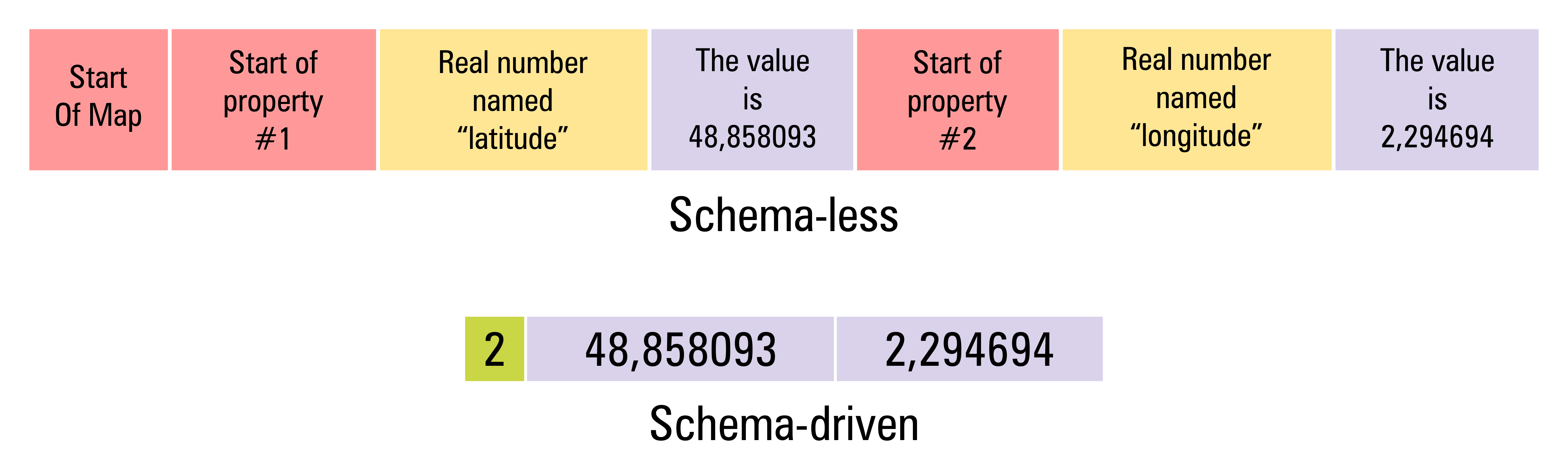}}
  \caption{An associative array (also known as a map) that consists of two
  decimal number properties, "latitude" and "longitude", serialized with
  fictitious schema-less and schema-driven representations. The schema-less
  representation (top) is self-descriptive and each property is self-delimited.
  Alternatively, schema-driven representations (bottom) omit most
  self-descriptive information except for the length of the associative array
as an integer prefix. A reader cannot understand how the schema-driven
representation translates to the original data structure without additional
information such as a schema definition. } \label{fig:schema-driven-less}
\end{figure}

%% file: sections/schema-less-subset-schema-driven.tex
\label{sec:schema-less-subset-schema-driven}

Schema-driven serialization specifications avoid embedding structural
information into the resulting bit-strings for space-efficiency purposes.  If a
schema fails to capture essential structural information then the serialization
specification has to embed that information into the resulting bit-strings. We
can reason about schema-less serialization specifications as schema-driven
specifications where the schema is generic enough that it describes any
bit-string and as a consequence carries no substantial information about any
particular instance. For example, a schema that defines bit-strings as
sequences of bits can describe any bit-string while providing no useful
information for understanding the semantics of such bit-strings.

The amount of information included in a schema can be thought as being
\emph{inversely proportional} to the information that needs to be encoded in a
bit-string described by such schema. However, schema-driven serialization
specifications may still embed redundant information into the bit-strings with
respect to the schema for runtime-efficiency, compatibility or error tolerance.
We can rank schema-driven serialization specifications based on the extent of
information that is necessary to include in the bit-strings.

\begin{figure}[hb!]
  \frame{\includegraphics[width=\linewidth]{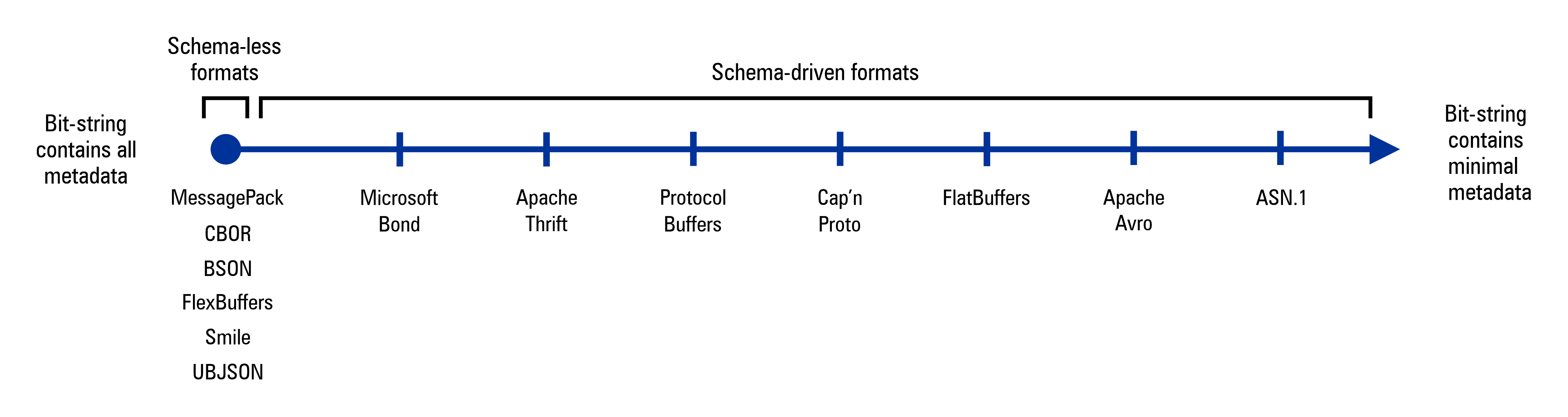}}
  \caption{We can compare schema-less and schema-driven serialization
  specifications based on how much information their resulting bit-strings
  contain.  Schema-less specifications are equivalent in that they all consist
  of implicit schemas that convey no information. However, not all schema
  languages supported by schema-driven specifications can describe the same
  amount of information. For this reason, some schema-driven specifications
  need to encode more structural information into their resulting bit-strings
  than others, placing them on the left hand side of the line.  Schema-driven
  specifications that enable meticulously defined schemas are placed on the
  right hand side of the line.} \label{fig:schema-less-driven-slider}
\end{figure}

%% file: sections/related-work.tex
\autoref{table:related-work} and \autoref{table:related-work-1} list existing
literature that discusses different sets of serialization specifications, both
textual and binary, schema-less and schema-driven. However, many of these
publications discuss serialization specifications that are either not
JSON-compatible, cannot be considered general-purpose serialization
specifications, or are out of date. For example, Java Object Serialization, as
its name implies, is only concerned with serialization of object instances in
the Java programming language. The first Protocol Buffers
\cite{protocolbuffers} version 3 release was published on GitHub in 2014
\footnote{\url{https://github.com/protocolbuffers/protobuf/releases/tag/v3.0.0-alpha-1}},
yet there are several publications listed in \autoref{table:related-work} and
\autoref{table:related-work-1} released before that year discussing the
now-obsolete Protocol Buffers version 2 \cite{microblogging-protobuf}
\cite{10.1145/2016716.2016718} \cite{MaedaK2012Peoo}
\cite{SumarayAudie2012Acod} \cite{VanuraJ.2018PeoJ} \cite{6784954}. As another
example, \cite{6769329} discusses an ASN.1 \cite{asn1} encoding (LWER) that has
been abandoned in the 1990s \cite{asn1-lwer}.

\ifx\thesis\undefined

Furthermore, many of the publications listed in \autoref{table:related-work}
and \autoref{table:related-work-1} are concerned with benchmarking. They tend
to describe the respective specifications in a high-level manner and do not get
into the encoding details of non-trivial examples, if any.

\fi

\begin{table}[hb!]
  \caption{A list of publications that discuss binary serialization specifications. This table is continued in \autoref{table:related-work-1}.}
\label{table:related-work}
\begin{tabularx}{\linewidth}{X|l|p{2cm}|X}
  \toprule
  \textbf{Publication} & \textbf{Year} & \textbf{Context} & \textbf{Discussed Serialization Specifications} \\
  \midrule
  {\small An overview of ASN.1} \cite{NEUFELD1992393} & {\small 1992} & {\small Networking} & {\small ASN.1 BER \cite{asn1-ber-cer-der}, ASN.1 PER \cite{asn1-per}} \\ \hline
  {\small Efficient encoding rules for ASN.1-based protocols} \cite{6769329} & {\small 1994} & {\small Networking} & {\small ASN.1 BER \cite{asn1-ber-cer-der}, ASN.1 CER \cite{asn1-ber-cer-der}, ASN.1 DER \cite{asn1-ber-cer-der}, ASN.1 LWER \cite{asn1-lwer}, ASN.1 PER \cite{asn1-per}} \\ \hline
  {\small Evaluation of Protocol Buffers as Data Serialization Format for Microblogging Communication} \cite{microblogging-protobuf} & {\small 2011} & {\small Microblogging} & {\small JSON \cite{ECMA-404}, Protocol Buffers \cite{protocolbuffers}} \\ \hline
  {\small Impacts of Data Interchange Formats on Energy Consumption and Performance in Smartphones} \cite{10.1145/2016716.2016718} & {\small 2011} & {\small Mobile} & {\small JSON \cite{ECMA-404}, Protocol Buffers \cite{protocolbuffers}, XML \cite{Paoli:06:EML}} \\ \hline
  {\small Performance evaluation of object serialization libraries in XML, JSON and binary formats} \cite{MaedaK2012Peoo} & {\small 2012} & {\small Java Object Serialization} & {\small Apache Avro \cite{avro}, Apache Thrift \cite{slee2007thrift}, Java Object Serialization \footnotemark, JSON \cite{ECMA-404}, Protocol Buffers \cite{protocolbuffers}, XML \cite{Paoli:06:EML}} \\ \hline
  {\small A comparison of data serialization formats for optimal efficiency on a mobile platform} \cite{SumarayAudie2012Acod} & {\small 2012} & {\small Mobile} & {\small Apache Thrift \cite{slee2007thrift}, JSON \cite{ECMA-404}, Protocol Buffers \cite{protocolbuffers}, XML \cite{Paoli:06:EML}} \\ \hline
  {\small Performance evaluation of Java, JavaScript and PHP serialization libraries for XML, JSON and binary formats} \cite{VanuraJ.2018PeoJ} & {\small 2012} & {\small Web Services} & {\small Apache Avro \cite{avro}, Java Object Serialization \footnotemark, JSON \cite{ECMA-404}, MessagePack \cite{messagepack}, Protocol Buffers \cite{protocolbuffers}, XML \cite{Paoli:06:EML}} \\ \hline
  {\small Google protocol buffers research and application in online game} \cite{6784954} & {\small 2013} & {\small Gaming} & {\small Protocol Buffers \cite{protocolbuffers}} \\ \hline
  {\small Integrating a System for Symbol Programming of Real Processes with a Cloud Service} \cite{Kyurkchiev-msgpack-json} & {\small 2015} & {\small Web Services} & {\small JSON \cite{ECMA-404}, MessagePack \cite{messagepack}, XML \cite{Paoli:06:EML}, YAML \cite{yaml-spec}} \\

  \bottomrule
\end{tabularx}
\end{table}

\begin{table}[hb!]
\caption{Continuation of \autoref{table:related-work}.}
\label{table:related-work-1}
\begin{tabularx}{\linewidth}{X|l|p{2cm}|X}
  \toprule
  \textbf{Publication} & \textbf{Year} & \textbf{Context} & \textbf{Discussed Serialization Specifications} \\
  \midrule

  {\small Serialization and deserialization of complex data structures, and applications in high performance computing} \cite{zaluzhnyi2016serialization} & {\small 2016} & {\small Scientific Computing} & {\small Apache Avro \cite{avro}, Boost::serialize \footnotemark, Cap'n Proto \cite{capnproto}, OpenFPM Packer/Unpacker \cite{incardona2019openfpm}, Protocol Buffers \cite{protocolbuffers}} \\ \hline
  {\small Smart grid serialization comparison: Comparison of serialization for distributed control in the context of the internet of things} \cite{petersen2017smart} & {\small 2017} & {\small Internet of Things} & {\small Apache Avro \cite{avro}, BSON \cite{bson}, CBOR \cite{RFC7049}, FST \footnotemark, Hessian \footnotemark, Java Object Serialization \footnotemark, Kryo \footnotemark, MessagePack \cite{messagepack}, Protocol Buffers \cite{protocolbuffers}, ProtoStuff \footnotemark, Smile \cite{smile}, XML \cite{Paoli:06:EML}, YAML \cite{yaml-spec}} \\ \hline
  {\small Binary Representation of Device Descriptions: CBOR versus RDF HDT} \cite{sahlmann2018binary} & {\small 2018} & {\small Internet of Things} & {\small CBOR \cite{RFC7049}, JSON \cite{ECMA-404}, RFD HDT \footnotemark} \\ \hline
  {\small Evaluating serialization for a publish-subscribe based middleware for MPSoCs} \cite{hamerski2018evaluating} & {\small 2018} & {\small Embedded Development} & {\small FlatBuffers \cite{flatbuffers}, MessagePack \cite{messagepack}, Protocol Buffers \cite{protocolbuffers}, YAS \footnotemark} \\ \hline
  {\small Analytical Assessment of Binary Data Serialization Techniques in IoT Context} \cite{10589/150617} & {\small 2019} & {\small Internet of Things} & {\small BSON \cite{bson}, FlatBuffers \cite{flatbuffers}, MessagePack \cite{messagepack}, Protocol Buffers \cite{protocolbuffers}} \\ \hline
  {\small FlatBuffers Implementation on MQTT Publish/Subscribe Communication as Data Delivery Format} \cite{8977050} & {\small 2019} & {\small Internet of Things} & {\small CSV \cite{RFC4180}, FlatBuffers \cite{flatbuffers}, JSON \cite{ECMA-404}, XML \cite{Paoli:06:EML}} \\ \hline
  {\small Enabling Model-Driven Software Development Tools for the Internet of Things} \cite{8876986} & {\small 2019} & {\small Internet of Things} & {\small Apache Avro \cite{avro}, Apache Thrift \cite{slee2007thrift}, FlatBuffers \cite{flatbuffers}, JSON \cite{ECMA-404}, Protocol Buffers \cite{protocolbuffers}} \\ \hline
  {\small Performance Comparison of Messaging Protocols and Serialization Formats for Digital Twins in IoV} \cite{9142787} & {\small 2020} & {\small Automobile} & {\small Cap'n Proto \cite{capnproto}, FlatBuffers \cite{flatbuffers}, Protocol Buffers \cite{protocolbuffers}} \\ \hline
  {\small Performance Analysis and Optimization of Serialization Techniques for Deep Neural Networks} \cite{10.1007/978-981-15-8697-2_23} & {\small 2020} & {\small Machine Learning} & {\small FlatBuffers \cite{flatbuffers}, Protocol Buffers \cite{protocolbuffers}} \\
  \bottomrule
\end{tabularx}
\end{table}

\footnotetext[\numexpr\thefootnote-9]{\url{https://docs.oracle.com/javase/8/docs/platform/serialization/spec/serialTOC.html}}
\footnotetext[\numexpr\thefootnote-8]{\url{https://docs.oracle.com/javase/8/docs/platform/serialization/spec/serialTOC.html}}
\footnotetext[\numexpr\thefootnote-7]{\url{https://www.boost.org/doc/libs/1_55_0/libs/serialization/doc/index.html}}
\footnotetext[\numexpr\thefootnote-6]{\url{https://github.com/RuedigerMoeller/fast-serialization}}
\footnotetext[\numexpr\thefootnote-5]{\url{http://hessian.caucho.com/doc/hessian-serialization.html}}
\footnotetext[\numexpr\thefootnote-4]{\url{https://docs.oracle.com/javase/8/docs/platform/serialization/spec/serialTOC.html}}
\footnotetext[\numexpr\thefootnote-3]{\url{https://github.com/EsotericSoftware/kryo}}
\footnotetext[\numexpr\thefootnote-2]{\url{https://github.com/protostuff/protostuff}}
\footnotetext[\numexpr\thefootnote-1]{\url{https://www.rdfhdt.org}}
\footnotetext[\numexpr\thefootnote]{\url{https://github.com/niXman/yas}}

%% file: sections/contributions.tex
To the best of \our knowledge, there exists gaps in the current literature
resulting in a lack of discussion on a wide range of JSON-compatible binary
serialization specifications in depth. \We aim to fill that gap by providing a
detailed comparative analysis of the most popular JSON-compatible binary
serialization specifications.  Through the process of conducting a literature
review, \we identified and resolved 13 issues with the documentation and
specifications of the Apache Avro \cite{avro}, Apache Thrift
\cite{slee2007thrift}, FlatBuffers \cite{flatbuffers}, FlexBuffers
\cite{flexbuffers}, Microsoft Bond \cite{microsoft-bond}, and Smile
\cite{smile} open-source binary serialization specifications. \Our fixes, the
corresponding patches and links are listed in \autoref{table:commits-1} and
\autoref{table:commits-2}.

%% file: sections/patches.tex
\begin{table}[ht!]
\caption{A list of open-source contributions made by the \paperauthors in the process of writing this survey paper. This table is continued in \autoref{table:commits-2}.}
\label{table:commits-1}
\begin{tabularx}{\linewidth}{p{2cm}|p{3cm}|l|X}
  \toprule
  \textbf{Specification} & \textbf{Repository} & \textbf{Commit} & \textbf{Description} \\
  \midrule

  Apache Avro \cite{avro} & \url{https://github.com/apache/avro} & \href{https://github.com/apache/avro/commit/afe8fa1adfbed7971c077338cd9441b14503507a}{\texttt{afe8fa1}} & Improve the encoding specification to clarify that records encode fields even if they equal their explicitly-set defaults and that the \texttt{default} keyword is only used for schema evolution purposes \\ \hline
  Apache Thrift \cite{slee2007thrift} & \url{https://github.com/apache/thrift} & \href{https://github.com/apache/thrift/commit/2e7f39f6b69d98fccba714266f3fa92bbce934cd}{\texttt{2e7f39f}} & Improve the Compact Protocol specification to clarify the Little Endian Base 128 (LEB128) variable-length integer encoding procedure and include a serialization example \\ \hline
  Apache Thrift \cite{slee2007thrift} & \url{https://github.com/apache/thrift} & \href{https://github.com/apache/thrift/commit/47b3d3b148c5181c02f4f871444fe93ad4ec65f2}{\texttt{47b3d3b}} & Improve the Compact Protocol specification to clarify that strings are not delimited with the \emph{NUL} ASCII \cite{STD80} character \\ \hline
  FlatBuffers \cite{flatbuffers} & \url{https://github.com/google/flatbuffers} & \href{https://github.com/google/flatbuffers/commit/4aff1198dd72b4c79652d0c42fbb1dfa2e21afa9}{\texttt{4aff119}} & Extend the binary encoding specification to document how union types are encoded \\ \hline
  FlatBuffers \cite{flatbuffers} & \url{https://github.com/google/flatbuffers} & \href{https://github.com/google/flatbuffers/commit/7b1ee31d808382fb8ad6213373c7213634601265}{\texttt{7b1ee31}} & Improve the documentation to clarify that the schema language does not permit unions of scalar types but that the C++ \cite{ISO50372} implementation has experimental support for unions of structs and strings \\ \hline
  FlatBuffers \cite{flatbuffers} & \url{https://github.com/google/flatbuffers} & \href{https://github.com/google/flatbuffers/commit/52e2177069750c05adeb24f76c3a9d538c5d33b7}{\texttt{52e2177}} & Remove from the documentation an outdated claim that Protocol Buffers \cite{protocolbuffers} does not support union types \\ \hline
  FlatBuffers \cite{flatbuffers} & \url{https://github.com/google/flatbuffers} & \href{https://github.com/google/flatbuffers/commit/796ed68faf434cac90f094a5fdf47137ba74e5a2}{\texttt{796ed68}} & Improve the FlatBuffers \cite{flatbuffers} and FlexBuffers \cite{flexbuffers} encoding specification to clarify that neither specifications deduplicate vector elements but that vectors may include more than one offset pointing to the same value \\

  \bottomrule
\end{tabularx}
\end{table}

\begin{table}[ht!]
\caption{Continuation of \autoref{table:commits-1}.}
\label{table:commits-2}
\begin{tabularx}{\linewidth}{p{2cm}|p{3cm}|l|X}
  \toprule
  \textbf{Specification} & \textbf{Repository} & \textbf{Commit} & \textbf{Description} \\
  \midrule

  Microsoft Bond \cite{microsoft-bond} & \url{https://github.com/microsoft/bond} & \href{https://github.com/microsoft/bond/commit/4acf83bd088ec97de8e04510481710b2e51d613f}{\texttt{4acf83b}} & Improve the documentation to explain how to enable the Compact Binary version 2 encoding in the C++ \cite{ISO50372} implementation \\ \hline
  Microsoft Bond \cite{microsoft-bond} & \url{https://github.com/microsoft/bond} & \href{https://github.com/microsoft/bond/commit/0012d9942f754e5cb74c70a7ab8d14f3fc367690}{\texttt{0012d99}} & Clarify the Compact Binary encoding specification to clarify that ID bits are encoded as Big Endian unsigned integers, that signed 8-bit integers use Two's Complement \cite{twos-complement}, formalize the concept of variable-length integers as Little Endian Base 128 (LEB128) encoding, clarify that real numbers are encoded as IEEE 764 floating-point numbers \cite{8766229}, and that enumeration constants are encoded as signed 32-bit integers \\ \hline
  Smile \cite{smile} & \url{https://github.com/FasterXML/smile-format-specification} & \href{https://github.com/FasterXML/smile-format-specification/commit/ac82c6b9aa6f2955513cd669b23c0c39c83055d3}{\texttt{ac82c6b}} & Fix the fact that the specification refers to ASCII \cite{STD80} string of 33 to 64 bytes and Unicode \cite{UnicodeStandard} strings of 34 to 65 bytes using two different names \\ \hline
  Smile \cite{smile} & \url{https://github.com/FasterXML/smile-format-specification} & \href{https://github.com/FasterXML/smile-format-specification/commit/7a53b0a460aaaff57ed99e41ccaef0db32fbb6ef}{\texttt{7a53b0a}} & Improve the specification by adding an example of how real numbers are represented using the custom 7-bit variant of IEEE 764 floating-point number encoding \cite{8766229} \\ \hline
  Smile \cite{smile} & \url{https://github.com/FasterXML/smile-format-specification} & \href{https://github.com/FasterXML/smile-format-specification/commit/95133dd7da41980997847f6fcaa9aac79f56e3b5}{\texttt{95133dd}} & Improve the specification to clarify how the byte-length prefixes from the \emph{Tiny Unicode} and \emph{Small Unicode} string encodings are computed differently compared to their ASCII \cite{STD80} counterparts \\ \hline
  Smile \cite{smile} & \url{https://github.com/FasterXML/smile-format-specification} & \href{https://github.com/FasterXML/smile-format-specification/commit/c56793fb20348ee5820a33ddac44dd144b56102d}{\texttt{c56793f}} & Clarify that the encoding attempts to reserve the \texttt{\textbf{0xff}} byte for message framing purposes but that reserving such byte is no longer a requirement to make the format suitable for use with WebSockets \\

  \bottomrule
\end{tabularx}
\end{table}

%% file: sections/organisation.tex
In \autoref{sec:json}, \we introduce the JSON serialization specification and
discuss its role and limitations.  In \autoref{sec:methodology}, \we list the
binary serialization specifications that are discussed in depth in this paper
and \our approach to analyzing them.  In \autoref{sec:background}, \we explore
a set of non-standardized binary integer encoding techniques that are essential
for understanding the inner workings of the various binary serialization
specifications.  In \autoref{sec:survey-schema-driven} and
\autoref{sec:survey-schema-less}, \we systematically study the history, the
characteristics, and the serialization processes of the selection of
schema-driven and schema-less binary serialization specifications. In
\autoref{sec:evolution}, \we introduce the challenges associated with schema
evolution. Further, \we discuss the schema evolution features that the
selection of schema-driven serialization specifications provide. In
\autoref{sec:conclusions}, \we reflect over what \we learned in the process of
closely inspecting the selection of JSON-compatible binary serialization
specifications. In \autoref{sec:reproducibility}, \we provide instructions for
the reader to recreate the bit-strings analyzed in the survey.  Finally, in
\autoref{sec:future-work}, \we discuss a set of next-steps to continue and
broaden the understanding of JSON-compatible binary serialization
specifications.

%% file: sections/json.tex
\label{sec:json}

%%%%%%%%%%%%%%%%%%%%%%%%%%%%%%%%%%%%%%%%%%%%%%%%%
% HISTORY
%%%%%%%%%%%%%%%%%%%%%%%%%%%%%%%%%%%%%%%%%%%%%%%%%

\subsection{History and Characteristics}

\input{sections/json/history.tex}
\input{sections/json/characteristics.tex}

%%%%%%%%%%%%%%%%%%%%%%%%%%%%%%%%%%%%%%%%%%%%%%%%%%
% RELEVANCE OF JSON TODAY
%%%%%%%%%%%%%%%%%%%%%%%%%%%%%%%%%%%%%%%%%%%%%%%%%%

\subsection{Relevance of JSON}
\input{sections/json/relevance.tex}

%%%%%%%%%%%%%%%%%%%%%%%%%%%%%%%%%%%%%%%%%%%%%%%%%
% SHORTCOMINGS
%%%%%%%%%%%%%%%%%%%%%%%%%%%%%%%%%%%%%%%%%%%%%%%%%

\subsection{Shortcomings}
\input{sections/json/shortcomings.tex}

%% file: sections/json/history.tex
JSON (JavaScript Object Notation) is a standard \emph{schema-less} and
\emph{textual} serialization specification \emph{inspired} by a subset
\footnote{\url{http://timelessrepo.com/json-isnt-a-javascript-subset}} of the
JavaScript programming language \cite{ECMA-262}. Douglas Crockford
\footnote{\url{https://www.crockford.com/}}, currently a Distinguished
Architect at PayPal, described the JSON serialization specification online
\footnote{\url{https://www.json.org}} in 2002 and published the first draft of
the JSON serialization specification in 2006 \cite{json-draft-00}.  Douglas
Crockford claims he \emph{discovered} and \emph{named} JSON, whilst Netscape
was already using an unspecified variant as an interchange format as early as
in 1996 in their web browser \cite{the-json-saga}.

\begin{figure}[hb!]
\frame{\includegraphics[width=\linewidth]{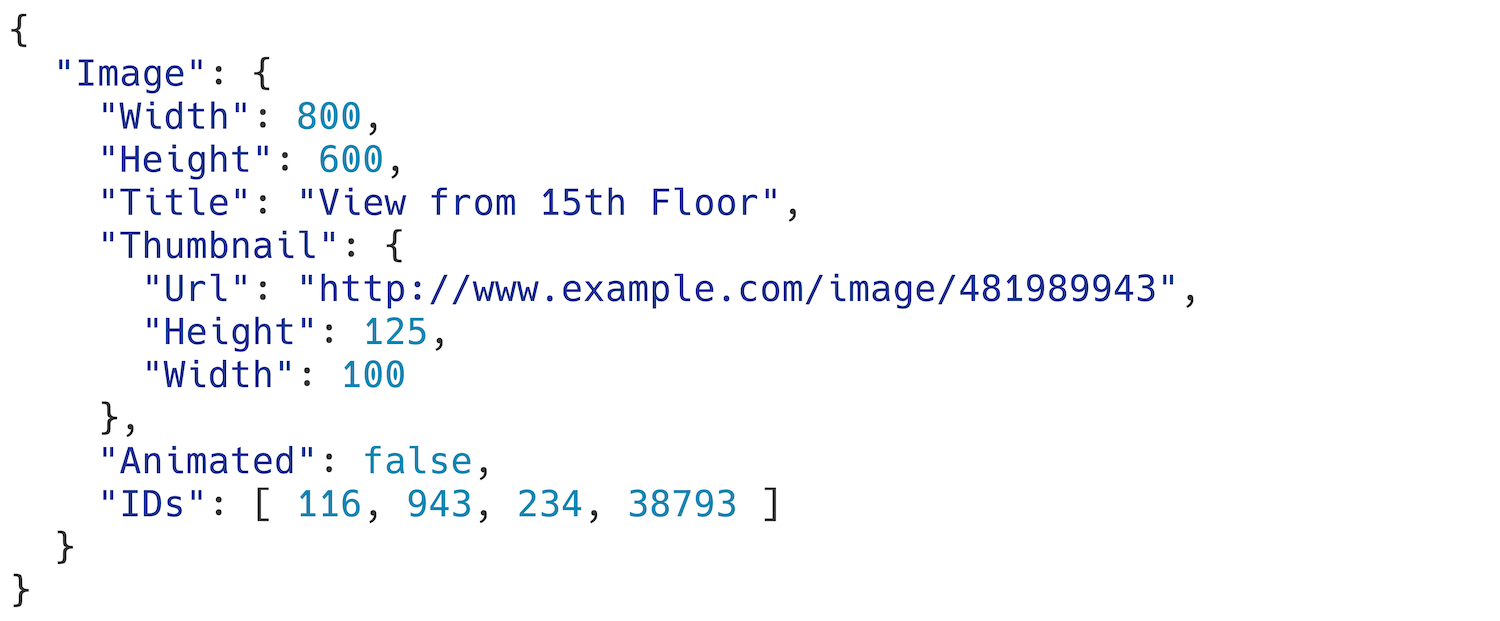}}
\caption{A JSON document example taken from the official specification \cite{RFC8259}.}
\label{lst:json-object-example} \end{figure}

%% file: sections/json/characteristics.tex
\begin{figure}[hb!]
  \frame{\includegraphics[width=\linewidth]{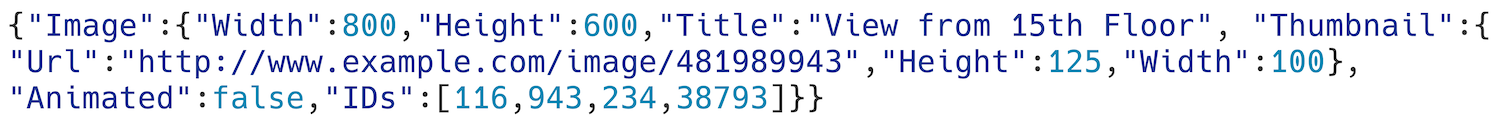}}
  \caption{A \emph{minified} and semantically-equivalent version of the JSON
  document from \autoref{lst:json-object-example}.}
  \label{lst:minified-json-object-example} \end{figure}

JSON is a human-readable open standard specification that consists of
structures built on key-value pairs or list of ordered items. The data types it
supports are objects, arrays, numbers, strings, \texttt{null}, and boolean
constants \texttt{true} and \texttt{false}. A data structure encoded using JSON
is referred to as a \emph{JSON document}.  \cite{ECMA-404} states that JSON
documents are serialized as either UTF-8, UTF-16, or UTF-32 strings
\cite{UnicodeStandard}.  However, \cite{RFC8259} mandate the use of UTF-8 for
interoperability purposes.  The serialization process involves recursively
converting keys and values to strings and optionally removing meaningless new
line, tab, and white space characters (a process known as \emph{minification})
as shown in \autoref{lst:minified-json-object-example}.

%% file: sections/json/relevance.tex
JSON \cite{ECMA-404} is popular interchange specification in the context of
cloud computing. In 2019, \cite{10.1145/3355369.3355594} found that JSON
documents constitute a growing majority of the HTTP \cite{RFC2616} responses
served by Akamai, a leading Content Delivery Network (CDN) that serves around 3
trillion HTTP requests daily.  Gartner
\footnote{\url{https://www.gartner.com}}, a business insight research and
advisory firm, forecasts that the cloud services market will grow 17\% in 2020
to total \$266.4 billion and that SaaS will remain the largest market segment
\footnote{\url{https://www.gartner.com/en/newsroom/press-releases/2019-11-13-gartner-forecasts-worldwide-public-cloud-revenue-to-grow-17-percent-in-2020}}.
SaaS systems typically provide \emph{application programming interfaces} (APIs)
and JSON was found to be the most common request and response format for APIs
\footnote{\url{https://www.programmableweb.com/news/json-clearly-king-api-data-formats-2020/research/2020/04/03}}.
According to their study, JSON was used more than XML \cite{Paoli:06:EML}. JSON
popularity over XML can be attributed to the fact that in comparison to XML,
JSON results in smaller bit-strings and in runtime and memory efficient
serialization and deserialization implementations
\cite{nurseitov2009comparison}.

There is an on-going interest in JSON within the research community.
\cite{10.1145/3034786.3056120} describe the first formal framework for JSON
documents and introduce a query language for JSON documents named \emph{JSON
Navigational Logic} (JNL). There is a growing number of publications that
discuss JSON in the context of APIs \cite{10.1145/2187980.2188227}
\cite{10.1007/978-3-319-61482-3_16} \cite{espinoza2020mapping} and technologies
that emerged from the JSON ecosystem such as the JSON Schema specification
\cite{PezoaF.2016FoJs} \cite{10.1007/978-3-030-34146-6_9} \cite{habib2019type}
\cite{fruth2020challenges}. Apart from cloud computing, JSON is relevant in
areas such as databases \cite{chillon2020deimos} \cite{mci/Klettke2015}
\cite{10.1145/3299815.3314467} \cite{8424731} \cite{10.1145/3428757.3429103},
big data \cite{json-in-big-data}, analytics \cite{10.14778/3236187.3236207}
\cite{10.14778/3115404.3115416}, mobile applications
\cite{SumarayAudie2012Acod} \cite{10.1145/2016716.2016718}, 3D modelling
\cite{LedouxH.2019Caca}, IoT \cite{6861361} \cite{NordahlMattias2015AlDI},
biomedical research \cite{Izzo2016}, and configuration files, for example
\footnote{\url{https://www.typescriptlang.org/docs/handbook/tsconfig-json.html}}.
\cite{10.1145/3299869.3314032} presents a high-level overview of the JSON
ecosystem including a survey of popular schema languages and implementations,
schema extraction technologies and novel parsing tools.

%% file: sections/json/shortcomings.tex
Despite its popularity, JSON is neither a runtime-efficient nor a
space-efficient serialization specification.

\textbf{Runtime-efficiency.} Serialization and deserialization often become a
bottleneck in data-intensive, battery-powered, and resource-constrained
systems.  \cite{10.14778/3236187.3236207} state that big data applications may
spend up to 90\% of their execution time deserializing JSON documents, given
that deserialization of textual specifications such as JSON is typically
expensive using traditional state-machine-based parsing algorithms.
\cite{10.1145/2016716.2016718} and \cite{SumarayAudie2012Acod} highlight the
impact of serialization and deserialization speed on mobile battery consumption
and resource-constrained mobile platforms. As a solution,
\cite{10.14778/3137765.3137782} propose a promising JSON encoder and decoder
that infers JSON usage patterns at runtime and self-optimizes by generating
encoding and decoding machine code on the fly.  Additionally,
\cite{LangdaleGeoff2019PgoJ} propose a novel approach to efficiently parse JSON
document by relying on SIMD processor instructions.
\cite{10.14778/3115404.3115416} claim that applications parse entire JSON
documents but typically only make use of certain fields. As a suggestion for
optimization, they propose a lazy JSON parser that infers schemas for JSON
documents at runtime and uses those schemas to speculate on the position of the
fields that an application requires in order to avoid deserializing unnecessary
areas of the JSON documents.

\textbf{Space-efficiency.}
\ifx\thesis\undefined
Cloud services are typically accessed over the
internet. As a result, these types of software systems are particularly
sensitive to substandard network performance. For example, in 2007, Akamai, a
global content delivery network (CDN), found out that \say{\textit{a
100-millisecond delay in website load time can hurt conversion rates by 7
percent}} and that \say{\textit{a two-second delay in web page load time
increases bounce rates by 103 percent}}
\footnote{\url{https://www.akamai.com/uk/en/about/news/press/2017-press/akamai-releases-spring-2017-state-of-online-retail-performance-report.jsp}}.
Cloud services frequently run on top of \emph{infrastructure as a service}
(IaaS) or \emph{platform as a service} (PaaS) providers such as Amazon Web
Services (AWS) \footnote{\url{https://aws.amazon.com}}. These providers
typically operate on a pay-as-you-go model where they charge per resource
utilization.  Therefore, transmitting data over the network directly translates
to operational expenses.
\fi
In comparison to JSON, \cite{7765670} found that
using a custom binary serialization specification reduced the overall network
traffic by up to 94\%. \cite{BittlSebastian2014Pcoe} conclude that JSON is not
an appropriate specification for bandwidth-constrained communication systems
citing the size of the documents as the main reason.
\cite{queirs:OASIcs:2014:4562} states that network traffic is one of the two
biggest causes for battery consumption on mobile devices and therefore a
space-efficient serialization specification could have a positive impact on
energy savings.

There are several JSON-based specifications that highlight a need for compact
JSON encodings:

\begin{itemize}

\item JSON Patch \cite{RFC6902} is a specification for expressing a sequence of
  operations on JSON documents. \cite{10.1007/978-3-319-46295-0_27} describe an
    algorithm called JDR to compute JSON Patch differences between two JSON
    documents optimized to keep the number of JSON Patch operations to a
    minimum for space-efficiency reasons.

\item CityGML is an XML-based specification to represent 3D models of cities
  and landscapes.  \cite{LedouxH.2019Caca} introduce a JSON-based alternative
    to CityGML called CityJSON citing that CityGML documents are large enough
    that makes it difficult or even impossible to transmit and process them on
    the web. In comparison to CityGML, CityJSON results in smaller document
    size.  However, the authors are looking for a binary JSON encoding to
    compress the CityJSON documents even further.  Additionally,
    \cite{4aad07f4-8f64-46b1-aad3-3d4abe36c5bf} explores how CityJSON documents
    can be compressed further using binary serialization specifications such as
    BSON \cite{bson} and CBOR \cite{RFC7049}.

\item In the context of bioinformatics, \emph{mmJSON} is a popular
  serialization specification used to encode representations of macromolecular
    structures.  \cite{10.1371/journal.pcbi.1005575} introduce \emph{MMTF}, a
    binary serialization specification to encode macromolecular structures
    based on MessagePack \cite{messagepack} to address the space-efficiency and
    runtime-efficiency concerns of using \emph{mmJSON} to perform web-based
    structure visualization. In particular, using \emph{mmJSON}, even after
    applying GZIP \cite{RFC1952} compression, results in large macromolecular
    structures that are slow to download. Due to their size, the largest
    macromolecular structures results in deserialization memory requirements
    that exceed the amount of memory typically available in web browsers.

\end{itemize}

%% file: sections/methodology.tex
\label{sec:methodology}

%%%%%%%%%%%%%%%%%%%%%%%%%%%%%%%%%%%%%%%%%%%%%%%%%
% WHY
%%%%%%%%%%%%%%%%%%%%%%%%%%%%%%%%%%%%%%%%%%%%%%%%%

\Our approach to extend the body of literature through meticulous study of
JSON-compatible binary serialization specifications is based on the following
methodology.  While several serialization specifications have characteristics
outside of the context of JSON \cite{RFC8259}, the scope of this study is
limited to those characteristics that are relevant for encoding JSON documents.

%%%%%%%%%%%%%%%%%%%%%%%%%%%%%%%%%%%%%%%%%%%%%%%%%
% HOW
%%%%%%%%%%%%%%%%%%%%%%%%%%%%%%%%%%%%%%%%%%%%%%%%%

\begin{enumerate}

  \item \textbf{Identify JSON-compatible Binary Serialization Specifications.}
    Research and select a set of schema-driven and schema-less JSON-compatible
    binary serialization specifications.

  \item \textbf{Craft a JSON Test Document.} Design a sufficiently complex yet
    succinct JSON document in an attempt to highlight the challenges of
    encoding JSON data. This JSON document is referred to as the \emph{input
    data}.

  \item \textbf{Write Schemas for the Schema-driven Serialization
    Specifications.} Present schemas that describe the \emph{input data} for
    each of the selected schema-driven serialization specifications. The
    schemas are designed to produce space-efficient results given the features
    documented by the corresponding specifications.

  \item \textbf{Serialize the JSON Test Document.} Serialize the \emph{input
    data} using each of the selected binary serialization specifications.

  \item \textbf{Analyze the Bit-strings Produced by Each Serialization
    Specification.} Study the resulting bit-strings and present annotated
    hexadecimal diagrams that guide the reader in understanding the inner
    workings of each binary serialization specification.

  \item \textbf{Discuss the Characteristics of Each Serialization
    Specification.} For each selected binary serialization specification,
    discuss the characteristics, advantages and optimizations that are relevant
    in the context of serializing JSON \cite{ECMA-404} documents.

\end{enumerate}

%%%%%%%%%%%%%%%%%%%%%%%%%%%%%%%%%%%%%%%%%%%%%%%%%
% WHICH
%%%%%%%%%%%%%%%%%%%%%%%%%%%%%%%%%%%%%%%%%%%%%%%%%
\subsection{Serialization Specifications}
\label{sec:serialization-format-selection}

\We selected a set of general-purpose schema-driven and schema-less
serialization specifications that are popular within the open-source community.
Some of the selected schema-driven serialization specifications support more
than one type of encoding. In these cases, \we chose the most space-efficient
encoding. The implementations used in this study are freely available under
open-source licenses with the exception of ASN.1 \cite{asn1}, for which a
proprietary implementation is used. The choice of JSON-compatible serialization
specifications, the selected encodings and the respective implementations are
documented in \autoref{table:versions-schema-driven} and
\autoref{table:versions-schema-less}.

\begin{table}[hb!]
\caption{The schema-driven binary serialization specifications, encodings and implementations discussed in this study.}
\label{table:versions-schema-driven}
\begin{tabularx}{\linewidth}{l|X|X}
\toprule
\textbf{Specification} & \textbf{Implementation} & \textbf{Encoding} \\
\midrule
  ASN.1 & OSS ASN-1Step Version 10.0.1 (proprietary) & PER Unaligned \cite{asn1-per} \\ \hline
Apache Avro & Python \texttt{avro} (pip) 1.10.0 & Binary Encoding \footnotemark with no framing \\ \hline
Microsoft Bond & C++ library 9.0.3 & Compact Binary v1 \footnotemark \\ \hline
Cap'n Proto & \texttt{capnp} command-line tool 0.8.0 & Packed Encoding \footnotemark \\ \hline
  FlatBuffers & \texttt{flatc} command-line tool 1.12.0 & Binary Wire Format \footnotemark \\ \hline
Protocol Buffers & Python \texttt{protobuf} (pip) 3.13.0 & Binary Wire Format \footnotemark \\ \hline
Apache Thrift & Python \texttt{thrift} (pip) 0.13.0 & Compact Protocol \footnotemark \\
\bottomrule
\end{tabularx}
\end{table}

\footnotetext[\numexpr\thefootnote-5]{\url{https://avro.apache.org/docs/current/spec.html\#binary\_encoding}}
\footnotetext[\numexpr\thefootnote-4]{\url{https://microsoft.github.io/bond/reference/cpp/compact\_\_binary\_8h\_source.html}}
\footnotetext[\numexpr\thefootnote-3]{\url{https://capnproto.org/encoding.html\#packing}}
\footnotetext[\numexpr\thefootnote-2]{\url{https://google.github.io/flatbuffers/flatbuffers\_internals.html}}
\footnotetext[\numexpr\thefootnote-1]{\url{https://developers.google.com/protocol-buffers/docs/encoding}}
\footnotetext[\numexpr\thefootnote]{\url{https://github.com/apache/thrift/blob/master/doc/specs/thrift-compact-protocol.md}}

\begin{table}[hb!]
\caption{The schema-less binary serialization specifications, encodings and implementations discussed in this study.}
\label{table:versions-schema-less}
\begin{tabularx}{\linewidth}{l|X}
\toprule
\textbf{Specification} & \textbf{Implementation} \\
\midrule
BSON & Python \texttt{bson} (pip) 0.5.10 \\ \hline
CBOR & Python \texttt{cbor2} (pip) 5.1.2 \\ \hline
FlexBuffers & \texttt{flatc} command-line tool 1.12.0 \\ \hline
MessagePack & \texttt{json2msgpack} command-line tool 0.6 with \texttt{MPack} 0.9dev \\ \hline
Smile & Python \texttt{pysmile} (pip) 0.2 \\ \hline
UBJSON & Python \texttt{py-ubjson} (pip) 0.16.1 \\
\bottomrule
\end{tabularx}
\end{table}

\begin{figure}[ht!]
  \frame{\includegraphics[width=\linewidth]{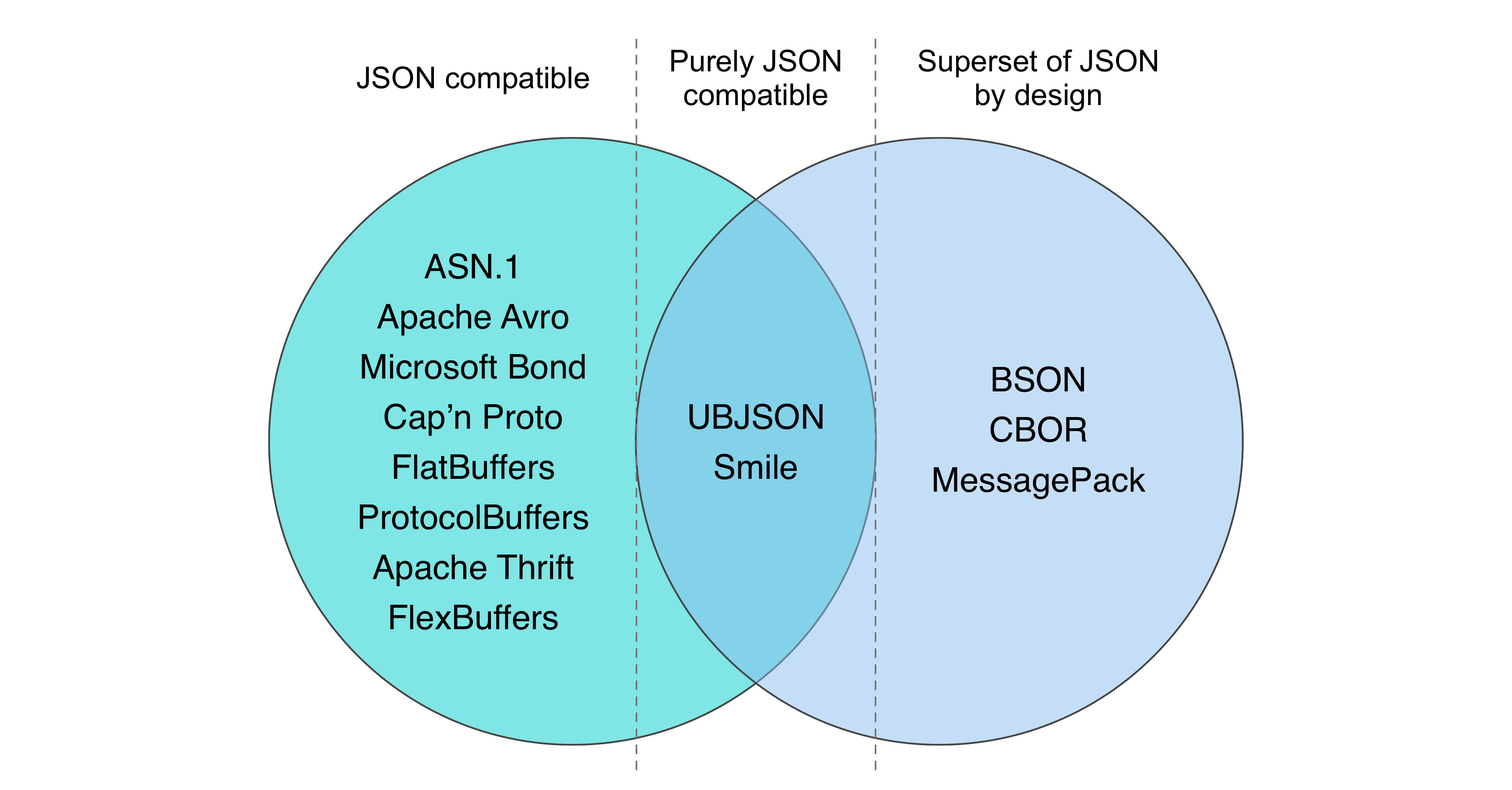}} \caption{The
  binary serialization specifications discussed in this study divided by
  whether they are purely JSON compatible (center), whether they consider JSON
compatibility but are a superset of JSON (right), or whether \we found them to
be JSON compatible but that's not one of their design goals (left).}
\label{fig:venn-json} \end{figure}

As part of this study, \we chose not to discuss serialization specifications
that could not encode the \emph{input data} JSON document from
\autoref{lst:json-object-test} without significant changes, such as
\emph{Simple Binary Encoding} (SBE)
\footnote{\url{https://github.com/real-logic/simple-binary-encoding}} and
Apple's \emph{Binary Property List} (BPList)
\footnote{\url{https://opensource.apple.com/source/CF/CF-550/CFBinaryPList.c}}
or that could not be considered general-purpose serialization specifications,
such as \emph{Java Object Serialization}
\footnote{\url{https://docs.oracle.com/javase/8/docs/platform/serialization/spec/serialTOC.html}}
and \emph{YAS} \footnote{\url{https://github.com/niXman/yas}}.  \We also chose
not to discuss serialization specifications that remain unused in the industry
at present, such as \emph{PalCom Object Notation} (PON)
\cite{NordahlMattias2015AlDI} and as a consequence lack a well-documented and
usable implementation, such as \emph{SJSON} \cite{AnjosEdman2016SAsr} and the
\emph{JSON-B}, \emph{JSON-C} and \emph{JSON-D} family of schema-less
serialization specifications \cite{draft-hallambaker-jsonbcd-11}.

\ifx\thesis\undefined
\begin{table}[hb!]
\caption{Each serialization specification report will include a summary reference table following this layout.}
\label{table:format-legend}
\begin{tabularx}{\linewidth}{l|X}
\toprule
\textbf{Website} & The website address of the serialization specification project \\ \hline
\textbf{Company / Individual} & The company or individual behind the serialization specification \\ \hline
\textbf{Year} & The year the serialization specification was conceived \\ \hline
\textbf{Specification} & A link or pointer to official specification of the serialization specification \\ \hline
\textbf{License} & The software license used to release the specification and implementations  \\ \hline
\textbf{Schema Language} & The schema language used by the serialization specification, if schema-driven \\ \hline
\textbf{Layout} & The overall structure of the bit-strings created by the serialization specification \\ \hline
\textbf{Languages} & The programming languages featuring official and/or third-party implementations of the serialization specification  \\ \hline
\textbf{Types} & A brief description of the data types supported by the serialization specification \\
\bottomrule
\end{tabularx}
\end{table}
\fi

%%%%%%%%%%%%%%%%%%%%%%%%%%%%%%%%%%%%%%%%%%%%%%%%%
% TEST DOCUMENT
%%%%%%%%%%%%%%%%%%%%%%%%%%%%%%%%%%%%%%%%%%%%%%%%%

\subsection{Input Data}
\label{sec:survey-methodology-input-data}

\begin{figure}[ht!]
  \frame{\includegraphics[width=\linewidth]{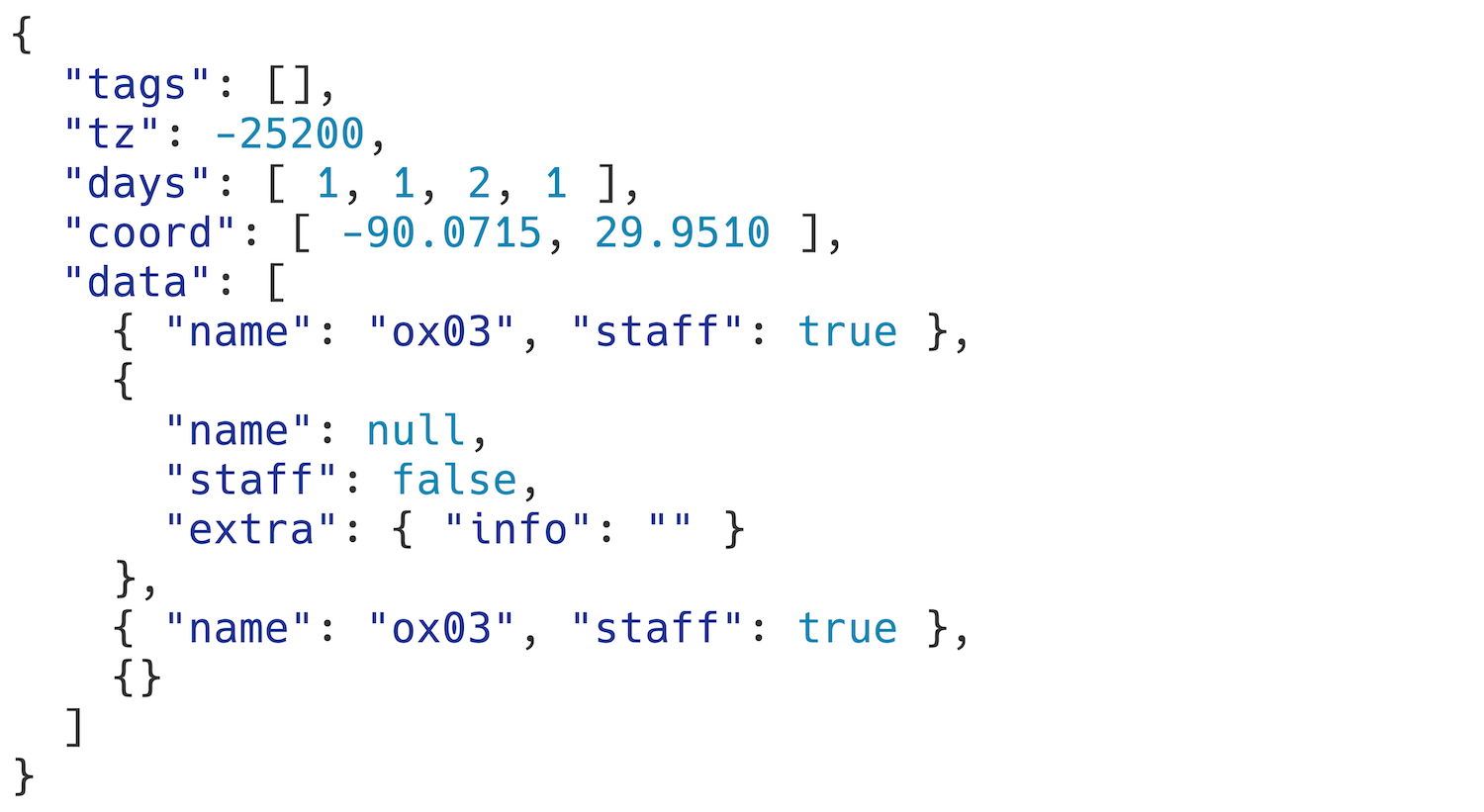}}
  \caption{The JSON test document that will be used as a basis for analyzing
  various binary serialization specifications.} \label{lst:json-object-test}
\end{figure}

\We designed a test JSON \cite{ECMA-404} document that is used to showcase the
challenges of serializing JSON \cite{ECMA-404} data and attempt to highlight
the interesting aspects of each selected serialization specification. The JSON
document \we created, presented in \autoref{lst:json-object-test}, has the
following characteristics:

\begin{itemize}

\item It contains an empty array, an empty object, and an empty string.
\item It contains nested objects and homogeneous and heterogeneous arrays.
\item It contains an array of scalars with and without duplicate values.
\item It contains an array of composite values with duplicate values.
\item It contains a set and an unset nullable string.
\item It contains positive and negative integers and floating-point numbers.
\item It contains true and false boolean values.

\end{itemize}

The \emph{input data} is not representative of every JSON document that the
reader may encounter in practice. \We hope that the characteristics of the input
data and output demonstrate how serialization specifications perform with
respect to various JSON documents.

%%%%%%%%%%%%%%%%%%%%%%%%%%%%%%%%%%%%%%%%%%%%%%%%%
% WHERE
%%%%%%%%%%%%%%%%%%%%%%%%%%%%%%%%%%%%%%%%%%%%%%%%%
\ifx\thesis\undefined
\subsection{System Specification}

The implementations of the serialization specifications were executed on a
MacBook Pro 13" Quad-Core Intel Core i7 2.7 GHz with 4 cores and 16 GB of
memory (model identifier \texttt{MacBookPro15,2}) running macOS Catalina
10.15.7, Xcode 11.3 (11C29), clang 1100.0.33.16, and Python 3.7.3.
\fi

%% file: sections/varints.tex
\label{sec:varints}

\emph{Little Endian Base 128} (LEB128) variable-length integer encoding stores
arbitrary large integers into a small variable amount of bytes.  Decoders do
not need to know how large the integer is in advance. The LEB128 encoding was
introduced by the DWARF \cite{dwarf-leb128} debug data format and the concept
of variable-length integers in the context of serialization was popularized by
the Protocol Buffers \cite{protocolbuffers} schema-driven serialization
specification.

The encoding process consists of:

\begin{itemize}

\item Obtaining the Big Endian binary representation of the input integer.

\item Padding the bit-string to make it a multiple of 7-bits.

\item Splitting the bit-string into 7-bits groups, prefixing the
  most-significant 7-bit group with a zero-bit.

\item Prefixing the remaining groups with set bits.

\item Storing the result as Little Endian.

\end{itemize}

The decoding process is the inverse of the encoding process. The decoder knows
when to stop parsing the variable-length integer bytes through the
most-significant bit in each group: the most-significant bit in the last byte
is equal to 0.

\subsubsection{Unsigned Integer}
\label{sec:varints-unsigned-example}

\begin{figure}[ht!]
\frame{\includegraphics[width=\linewidth]{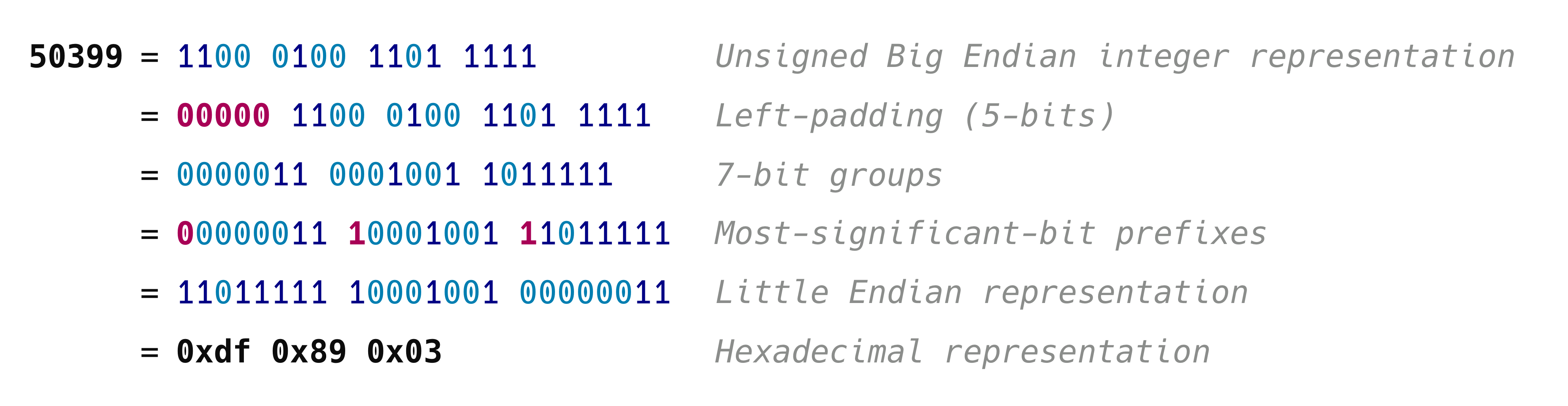}}
\caption{Little Endian Base 128 (LEB128) encoding of the unsigned 32-bit
integer $50399$.} \label{fig:varints-unsigned-example} \end{figure}

In this example, \we consider the unsigned integer $50399$ in
\autoref{fig:varints-unsigned-example}.

The Big Endian unsigned integer representation of the number $50399$ is 16-bits
wide. 16 is not a multiple of 7, therefore \we left-pad the bit-string with
five zeroes to make it 21-bit wide. Then, \we split the bit-string into 7-bit
groups.  Next, \we prefix the most-significant group with \texttt{\textbf{0}}
and the rest of the groups with \texttt{\textbf{1}}. The Little Endian
representation of the bit-string equals \texttt{\textbf{0xdf 0x89 0x03}} which
is the Little Endian Base 128 (LEB128) encoding of the unsigned integer $50399$
as shown in \autoref{fig:varints-unsigned-example}.

\subsubsection{Signed Integer Using Two's Complement}
\label{sec:varints-signed-example}

The canonical approach to encoding signed integers using Little Endian Base 128
(LEB128) encoding is to apply Two's Complement \cite{twos-complement} and
proceed as if the integer is unsigned.

\begin{figure}[ht!]
\frame{\includegraphics[width=\linewidth]{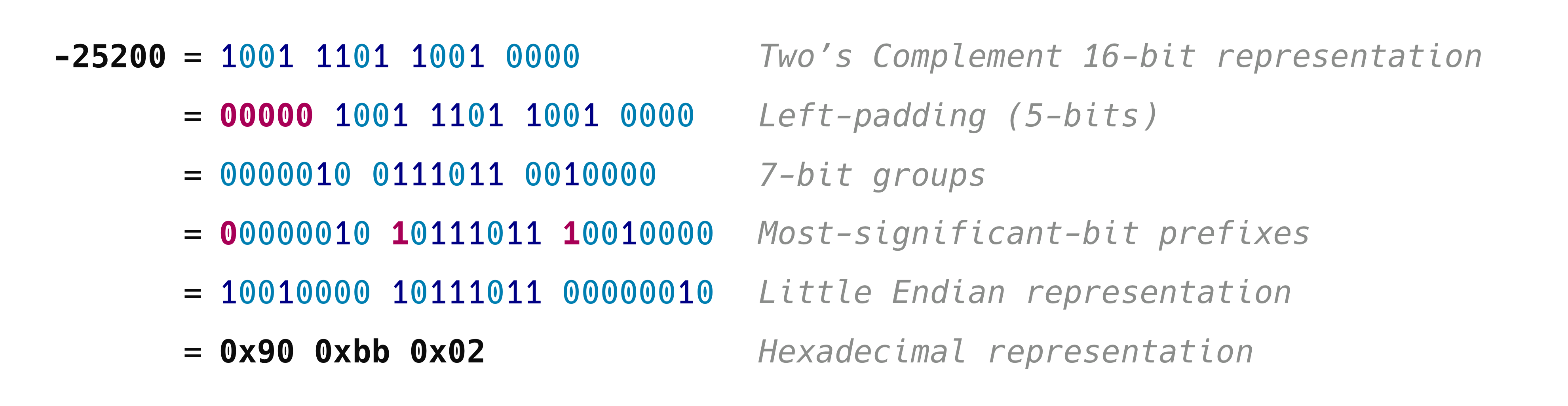}}
\caption{Little Endian Base 128 (LEB128) encoding of the signed 32-bit integer
$-25200$.} \label{fig:varints-signed-example} \end{figure}

For example, consider the signed integer $-25200$ as shown in
\autoref{fig:varints-signed-example}.

%% file: sections/zigzag-encoding.tex
\label{sec:zigzag-encoding}

% https://developers.google.com/protocol-buffers/docs/encoding#signed_integers
% https://gist.github.com/mfuerstenau/ba870a29e16536fdbaba
% https://en.wikibooks.org/wiki/Data_Compression/Coding#ZigZag_encoding
% https://stackoverflow.com/questions/4533076/google-protocol-buffers-zigzag-encoding

ZigZag integer encoding, pioneered by Protocol Buffers \cite{protocolbuffers},
is an approach to map signed integers to unsigned integers. The goal of ZigZag
integer encoding is that the absolute values of the resulting unsigned integers
are relatively close to the absolute values of the original negative integers.
In comparison, Two's Complement \cite{twos-complement} converts negative
integers into large unsigned integers. The encoding works in a \emph{zig-zag}
fashion between positive and negative integers. ZigZag integer encoding does
not affect the range of signed integer values that can be encoded in a given
number of bits. \emph{Little Endian Base 128} (LEB128) (\ref{sec:varints})
variable-length integer encoding encodes unsigned integers using a number of
bytes that are proportial to the absolute value of the unsigned integer.
Therefore, ZigZag integer encoding results in space-efficiency when encoding
negative integers in combination with Little Endian Base 128 (LEB128) encoding.

\begin{figure}[ht!]
\frame{\includegraphics[width=\linewidth]{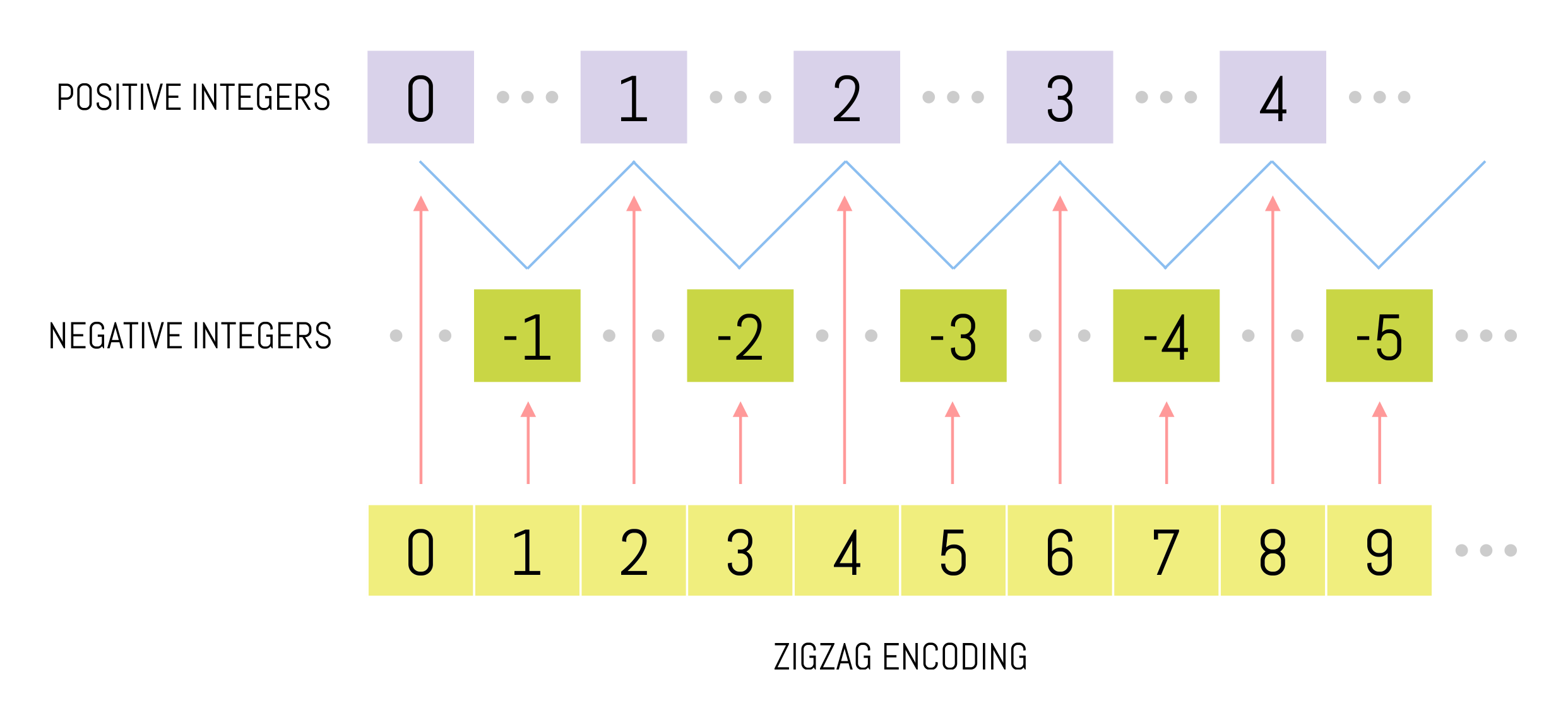}}
\caption{ZigZag integer encoding "zig-zags" between positive and negative
integers.} \label{fig:zigzag-encoding} \end{figure}

\textbf{ZigZag Encoding}:

\begin{itemize}

\item Encode the signed integer using Two's Complement
  \cite{twos-complement}.

\item Left-shift the bit-string by one position.

\item Right-shift the bit-string by the number of desired bits minus one.

\item Calculate the exclusive disjunction (XOR) between both bit-shifted
  bit-strings.

\end{itemize}

\textbf{ZigZag Decoding}:

\begin{itemize}

\item Right-shift the unsigned integer bit-string by one position.

\item Calculate the bitwise conjunction (AND) between the bit-string and $1$.

\item Calculate the Two's Complement \cite{twos-complement} of the result
  of negating the integer represented by the bit-string.

\item Calculate the exclusive disjunction (XOR) between both numbers.

\end{itemize}

\textbf{Alternative definition of ZigZag integer encoding} is that positive
integers are multiplied by two. Negative integers are equal to their absolute
values multiplied by two, minus one.  ZigZag integer decoding is the inverse of
the encoding process.

\begin{figure}[ht!]
  \frame{\includegraphics[width=\linewidth]{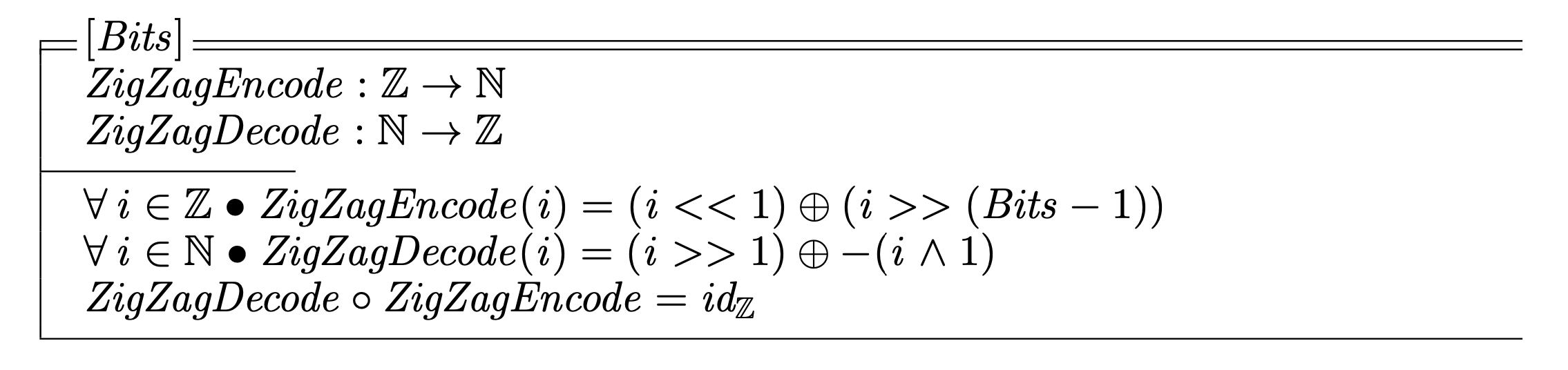}} \caption{
    A formal definition of ZigZag integer encoding using Z notation
    \cite{ISO13568-2002}. \We capture the relationship between the encoding
    and decoding functions by stating that the functional composition of
\texttt{\textbf{ZigZagDecode}} and \texttt{\textbf{ZigZagEncode}} equals the
identity function.  Decoding an encoded integer equals the original integer.}
\label{fig:zigzag-spec} \end{figure}

% RENDERED FROM:
% \begin{gendef}[Bits]
% ZigZagEncode : \mathbb{Z} \fun \nat \\
% ZigZagDecode : \nat \fun \mathbb{Z}
% \where
% \forall i \in \mathbb{Z} @ ZigZagEncode(i) = (i << 1) \oplus (i >> (Bits - 1)) \\
% \forall i \in \nat @ ZigZagDecode(i) = (i >> 1) \oplus -(i \land 1) \\
% ZigZagDecode \circ ZigZagEncode = id_{\mathbb{Z}}
% \end{gendef}

\subsubsection{ZigZag Encoding}
\label{sec:zigzag-encoding-example}

The formal definition from \autoref{fig:zigzag-spec} states that the ZigZag
integer encoding of the 32-bit signed negative integer $-25200$ is equal to
$(-25200 << 1) \thinspace \oplus \thinspace (-25200 >> 31)$.

\begin{figure}[ht!]
\frame{\includegraphics[width=\linewidth]{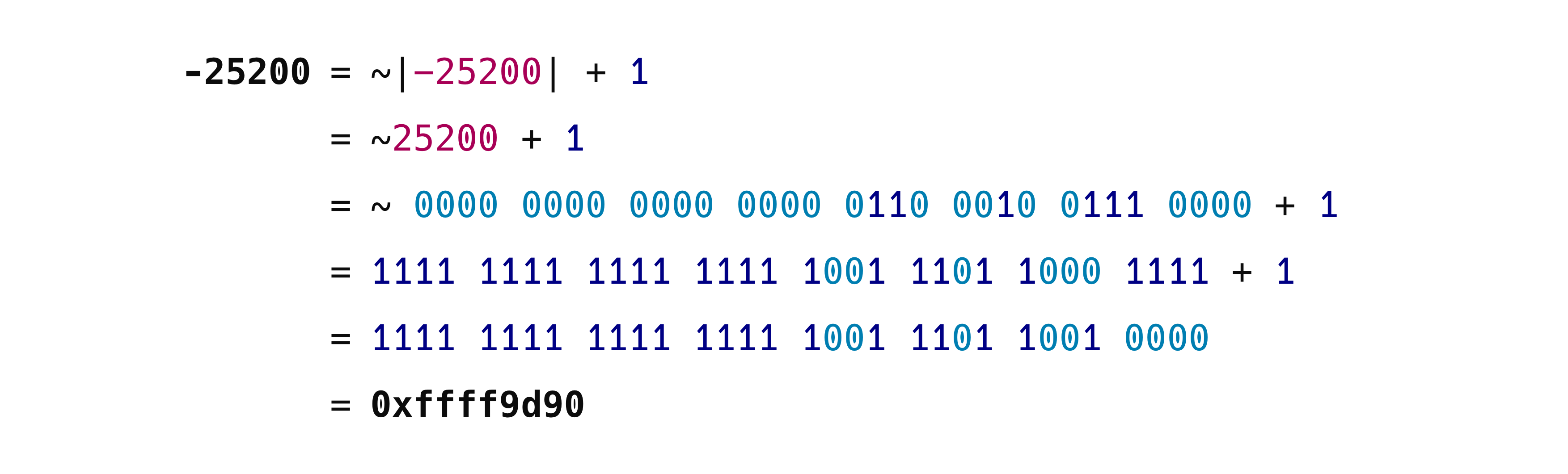}}
\caption{The Two's Complement \cite{twos-complement} of the 32-bit signed
  integer $-25200$ is \texttt{\textbf{0xffff9d90}}.}
\label{fig:zigzag-encoding-example-twos-complement} \end{figure}

First, we calculate the Two's Complement \cite{twos-complement} of the
32-bit signed integer $-25200$ which involves calculating the unsigned integer
representation of its absolute value, calculating its binary complement, and
adding 1 to the result as shown in
\autoref{fig:zigzag-encoding-example-twos-complement}.

\begin{figure}[ht!]
  \frame{\includegraphics[width=\linewidth]{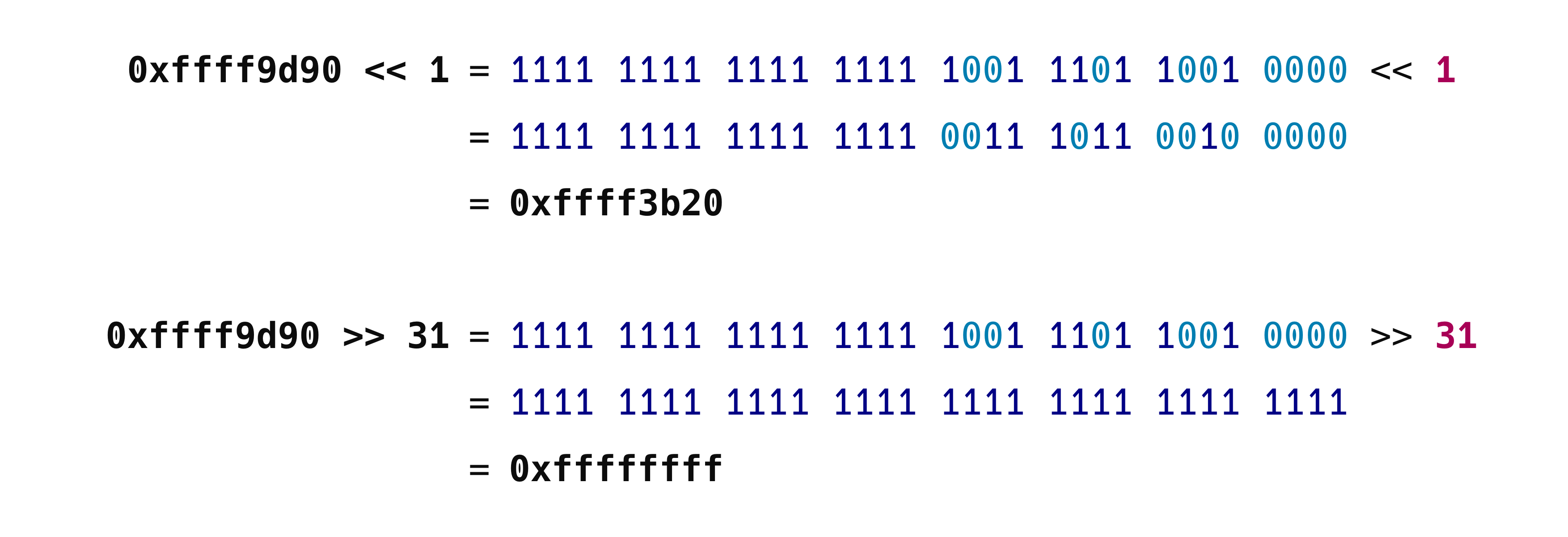}}
  \caption{Bit-shifting the Two's Complement \cite{twos-complement} of the
  32-bit signed integer $-25200$ as per the formal definition of ZigZag integer
  encoding from \autoref{fig:zigzag-spec}.}
\label{fig:zigzag-encoding-example-bit-shifts} \end{figure}

Then, we left-shift \texttt{\textbf{0xffff9d90}} by one position and
right-shift \texttt{\textbf{0xffff9d90}} by 31 positions (32-bits minus 1) as
shown in \autoref{fig:zigzag-encoding-example-bit-shifts}.

\begin{figure}[ht!]
  \frame{\includegraphics[width=\linewidth]{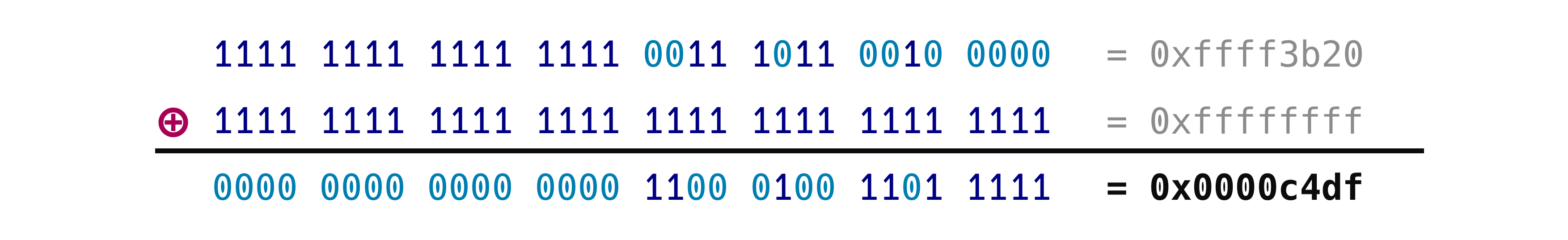}}
  \caption{The exclusive disjunction (XOR) between \texttt{\textbf{0xffff3b20}}
  and \texttt{\textbf{0xffffffff}} results in \texttt{\textbf{0x0000c4df}}.}
\label{fig:zigzag-encoding-example-xor} \end{figure}

Finally, we calculate the exclusive disjunction (XOR) between the two
hexadecimal strings obtained before as shown in
\autoref{fig:zigzag-encoding-example-xor}.

Therefore, the ZigZag integer encoding of the signed 32-bit integer $-25200$ is
$\texttt{\textbf{0x0000c4df}}$, which is the 32-bit unsigned integer $50399$.

\subsubsection{ZigZag Decoding}
\label{sec:zigzag-decoding-example}

The formal definition from \autoref{fig:zigzag-spec} states that the ZigZag
integer decoding of the bit-string \texttt{\textbf{0x0000c4df}} we obtained in
\autoref{sec:zigzag-encoding-example} is equal to $(\texttt{0x0000c4df} >> 1)
\oplus -(\texttt{0x0000c4df} \land 1)$.

\begin{figure}[ht!]
  \frame{\includegraphics[width=\linewidth]{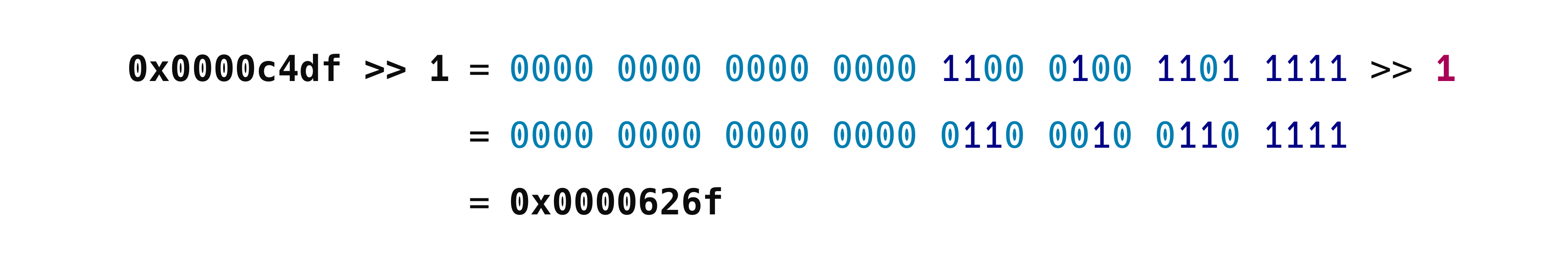}}
  \caption{Righ-shifting the ZigZag integer encoding of the 32-bit signed
  integer $-25200$ by one position results in \texttt{\textbf{0x0000626f}}.}
\label{fig:zigzag-decoding-example-bit-shifts} \end{figure}

First, we right-shift \texttt{\textbf{0x0000c4df}} by one position as shown in
\autoref{fig:zigzag-decoding-example-bit-shifts}.

\begin{figure}[ht!]
  \frame{\includegraphics[width=\linewidth]{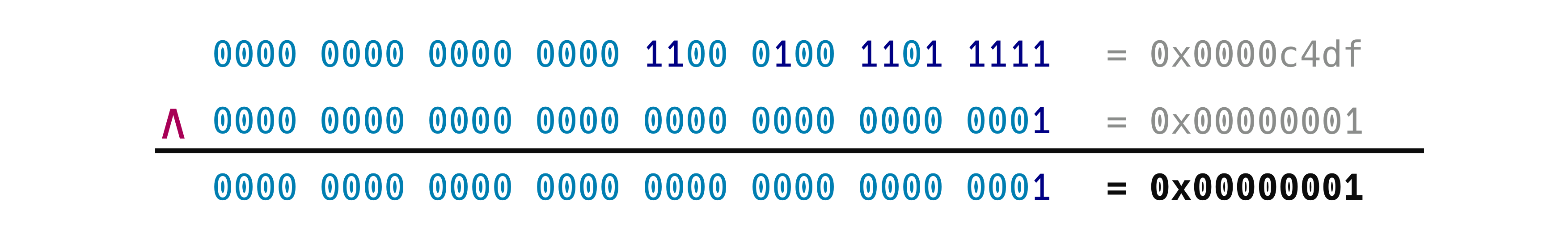}}
  \caption{The bitwise conjunction (AND) between \texttt{\textbf{0x0000c4df}}
  and \texttt{\textbf{0x00000001}} results in \texttt{\textbf{0x00000001}}.}
\label{fig:zigzag-decoding-example-and} \end{figure}

Then, we calculate the bitwise conjunction (AND) between
\texttt{\textbf{0x0000c4df}} and \texttt{\textbf{0x00000001}} as shown in
\autoref{fig:zigzag-decoding-example-and}.

\begin{figure}[ht!]
  \frame{\includegraphics[width=\linewidth]{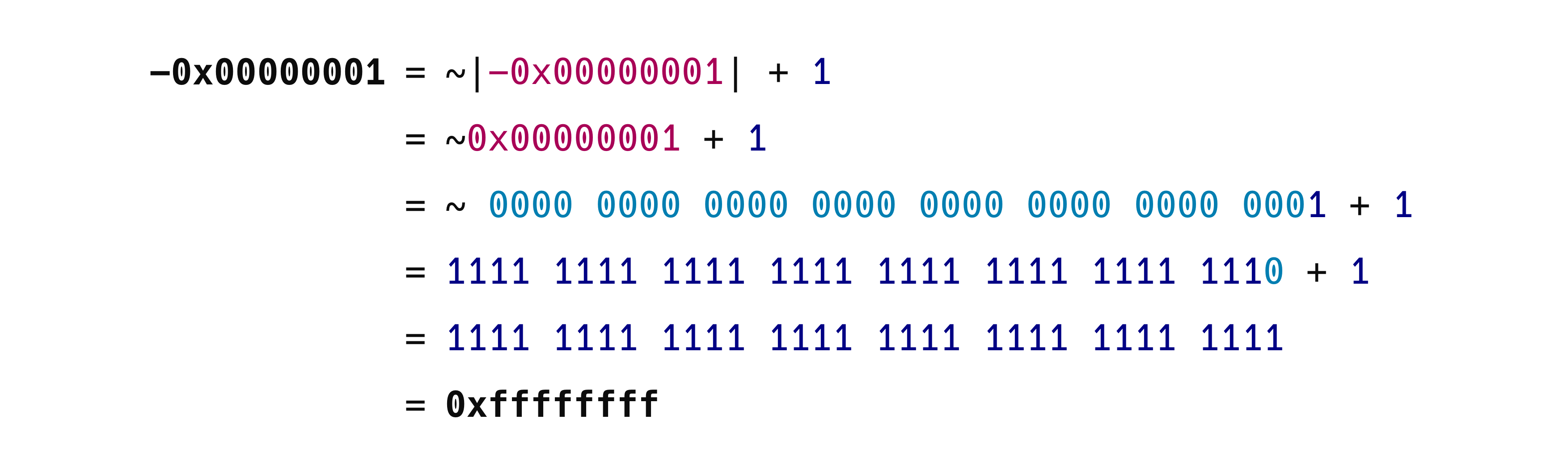}}
  \caption{The Two's Complement \cite{twos-complement} of
  \texttt{\textbf{-0x00000001}} is \texttt{\textbf{0xffffffff}}.}
\label{fig:zigzag-decoding-example-twos-complement} \end{figure}

Next, we calculate the Two's Complement \cite{twos-complement} of
\texttt{\textbf{-0x00000001}} as shown in
\autoref{fig:zigzag-decoding-example-twos-complement}.

\begin{figure}[ht!]
  \frame{\includegraphics[width=\linewidth]{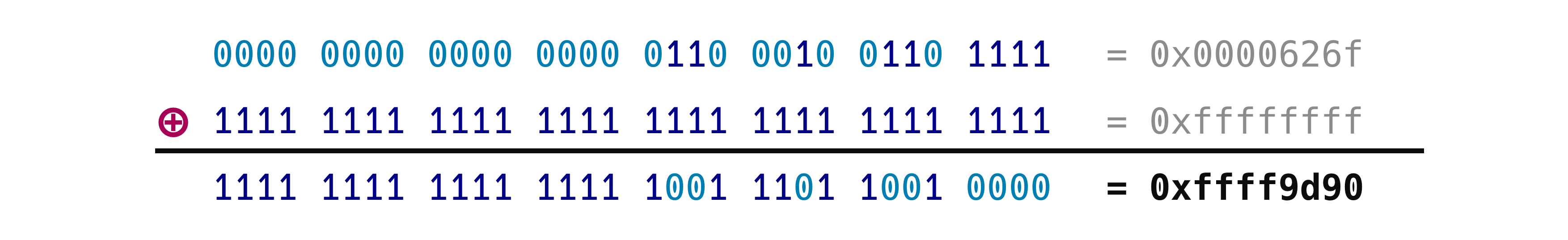}}
  \caption{The exclusive disjunction (XOR) between \texttt{\textbf{0x0000626f}}
  and \texttt{\textbf{0xffffffff}} results in \texttt{\textbf{0xffff9d90}}.}
\label{fig:zigzag-decoding-example-xor} \end{figure}

Finally, we calculate the exclusive disjunction (XOR) between
\texttt{\textbf{0x0000626f}} and \texttt{\textbf{0xffffffff}} as shown in
\autoref{fig:zigzag-decoding-example-xor}.

The result is \texttt{\textbf{0xffff9d90}} which is the Two's Complement
\cite{twos-complement} of the negative 32-bit signed integer $-25200$.

%% file: sections/formats/asn1.tex
\label{sec:asn1}

\begin{figure}[hb!]
\frame{\includegraphics[width=\linewidth]{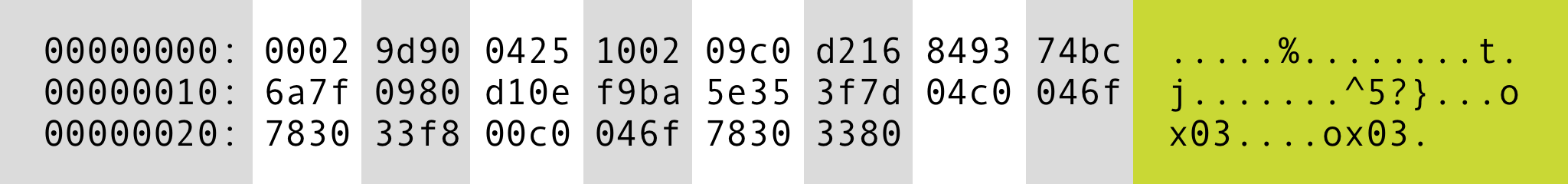}}
\caption{Hexadecimal output (\texttt{xxd}) of encoding
\autoref{lst:json-object-test} input data with ASN.1 PER Unaligned (44
bytes).} \label{lst:hex-asn1} \end{figure}

%%%%%%%%%%%%%%%%%%%%%%%%%%%%%%%%%%%%%%%%%%%%%%%%%
% HISTORY
%%%%%%%%%%%%%%%%%%%%%%%%%%%%%%%%%%%%%%%%%%%%%%%%%

\textbf{History.} ASN.1 \cite{asn1} is a standard schema language used to
serialize data structures using an extensible set of schema-driven encoding
rules. ASN.1 was originally developed in 1984 as a part of the \cite{asn1-old}
standard and became a International Telecommunication Union recommendation and
an ISO/IEC international standard \cite{asn1-iso} in 1988. The ASN.1
specification is publicly available
\footnote{\url{https://www.itu.int/rec/T-REC-X.680/en}} and there are
proprietary and open source implementations of its standard encoding rules. The
ASN.1 PER Unaligned encoding rules were designed to produce space-efficient
bit-strings while keeping the serialization and deserialization procedures
reasonably simple.

%%%%%%%%%%%%%%%%%%%%%%%%%%%%%%%%%%%%%%%%%%%%%%%%%
% ADVANTAGES
%%%%%%%%%%%%%%%%%%%%%%%%%%%%%%%%%%%%%%%%%%%%%%%%%

\textbf{Characteristics.} \begin{itemize}

  \item \textbf{Robustness.} ASN.1 is a mature technology that powers some of
    the highest-integrity communication systems in the world
    \footnote{\url{https://www.itu.int/en/ITU-T/asn1/Pages/Application-fields-of-ASN-1.aspx}}
    \footnote{\url{https://www.oss.com/asn1/resources/standards-use-asn1.html}}
    such as Space Link Extension Services (SLE) \cite{ccsds-sle} communication
    services for spaceflight and the LTE S1 signalling service application
    protocol \cite{s1ap}. Refer to \cite{10.1007/978-3-319-96142-2_25} for an
    example of formal verification of the encoding/decoding code produced by an
    ASN.1 PER Unaligned compiler (Galois \footnote{\url{https://galois.com}})
    in the automobile industry.

  \item \textbf{Standardization.} In comparison to informally documented
    serialization specification, ASN.1 is specified as a family of ITU
    Telecommunication Standardization Sector (ITU-T) recommendations and
    ISO/IEC international standards and has gone through extensive technical
    review.

  \item \textbf{Flexible Encoding Rules.} ASN.1 supports a wide range of
    standardised encodings for different use cases: BER (Basic Encoding Rules)
    based on tag-length-value (TLV) nested structures \cite{asn1-ber-cer-der},
    DER (Distinguished Encoding Rules) and CER (Canonical Encoding Rules)
    \cite{asn1-ber-cer-der} for restricted forms of BER
    \cite{asn1-ber-cer-der}, PER (Packed Encoding Rules) for space-efficiency
    \cite{asn1-per}, OER (Octet Encoding Rules) for runtime-efficiency
    \cite{asn1-oer}, JER (JSON Encoding Rules) for JSON encoding
    \cite{asn1-jer}, and XER (XML Encoding Rules) for XML encoding
    \cite{asn1-xer}.

\end{itemize}

%%%%%%%%%%%%%%%%%%%%%%%%%%%%%%%%%%%%%%%%%%%%%%%%%
% OVERALL STRUCTURE
%%%%%%%%%%%%%%%%%%%%%%%%%%%%%%%%%%%%%%%%%%%%%%%%%

\textbf{Layout.} An ASN.1 PER Unaligned bit-string is a sequence of untagged
values, sometimes nested where the ordering of the values is determined by the
schema. ASN.1 PER Unaligned encodes values in as few bits as reasonably
possible and does not align values to a multiple of 8 bits as the name of the
encoding implies. ASN.1 PER Unaligned only encodes runtime information that
cannot be inferred from the schema, such as the length of the lists or union
type.

ASN.1 PER Unaligned encodes unbounded data types and bounded data types whose
logical upper bound is greater than 65536 using a technique called
\emph{fragmentation} where the encoding of the value consists of one or more
consecutive fragments each consisting of a length prefix followed by a series
of items. The nature of each item depends on the type being encoded. For
example, an item might be a character, a bit, or a logical element of a list. A
value encoded using fragmentation consists of 0 or more fragments of either
16384, 32768, 49152, or 65536 items followed by a single fragment of 0 to 16383
items where each fragment is as large as possible and no fragment is larger
than the preceding fragment. Refer to \autoref{table:asn1-fragment-length} for
details on fragment length prefix encoding.

\begin{table}[hb!]
\caption{ASN.1 PER Unaligned fragment length prefixes depending on the number
  of items in the fragment as determined by the \emph{From} and \emph{To}
  ranges.}
\label{table:asn1-fragment-length}
  \begin{tabularx}{\linewidth}{l|l|X}
\toprule
\textbf{From} & \textbf{To} & \textbf{Fragment prefix}\\ \midrule
  0        & 127   & Length as an 8-bit unsigned integer \\ \hline
  128      & 16383 & Length as the 2-bits \texttt{10} followed by a Big Endian 14-bit unsigned integer \\ \hline
  16384    & 16384 & \texttt{1100 0001} \\ \hline
  32768    & 32768 & \texttt{1100 0010} \\ \hline
  49152    & 49152 & \texttt{1100 0011} \\ \hline
  65536    & 65536 & \texttt{1100 0100} \\
\bottomrule
\end{tabularx}
\end{table}

%%%%%%%%%%%%%%%%%%%%%%%%%%%%%%%%%%%%%%%%%%%%%%%%%
% NUMBERS
%%%%%%%%%%%%%%%%%%%%%%%%%%%%%%%%%%%%%%%%%%%%%%%%%

\textbf{Numbers.} ASN.1 PER Unaligned supports integer data types of arbitrary
widths. The schema-writer may constraint the integer data type with a lower and
upper bound:

\begin{itemize}

  \item \textbf{If the integer type has no bounds or if the integer type only
    has an upper bound.} ASN.1 PER Unaligned encodes the value as a Big Endian
    Two's Complement \cite{twos-complement} signed integer prefixed by its
    byte-length as a unsigned 8-bit integer.

\item \textbf{If the integer type has a lower bound but not an upper bound.}
  ASN.1 PER Unaligned substracts the lower bound from the value and encodes the
    result as a variable-length Big Endian unsigned integer prefixed by its
    byte-length as a unsigned 8-bit integer. For example, the value $-18$ of an
    integer type whose lower bound is $-20$ is encoded as the unsigned integer
    $2 = -18 - (-20)$ prefixed with the byte-length definition
    \texttt{\textbf{0x01}}.

\item \textbf{If the integer type has both a lower and an upper bound.} ASN.1
  encodes the difference between the value and the lower bound using the
    smallest possible fixed-length Big Endian unsigned integer that can encode
    the difference between the upper and the lower bound. For example, an
    integer type constrained between the values 5 and 6 encodes the values as
    an unsigned 2-bit integer where 0 corresponds to 5 and 1 corresponds to 6.

\end{itemize}

\begin{figure*}[hb!]
  \frame{\includegraphics[width=\linewidth]{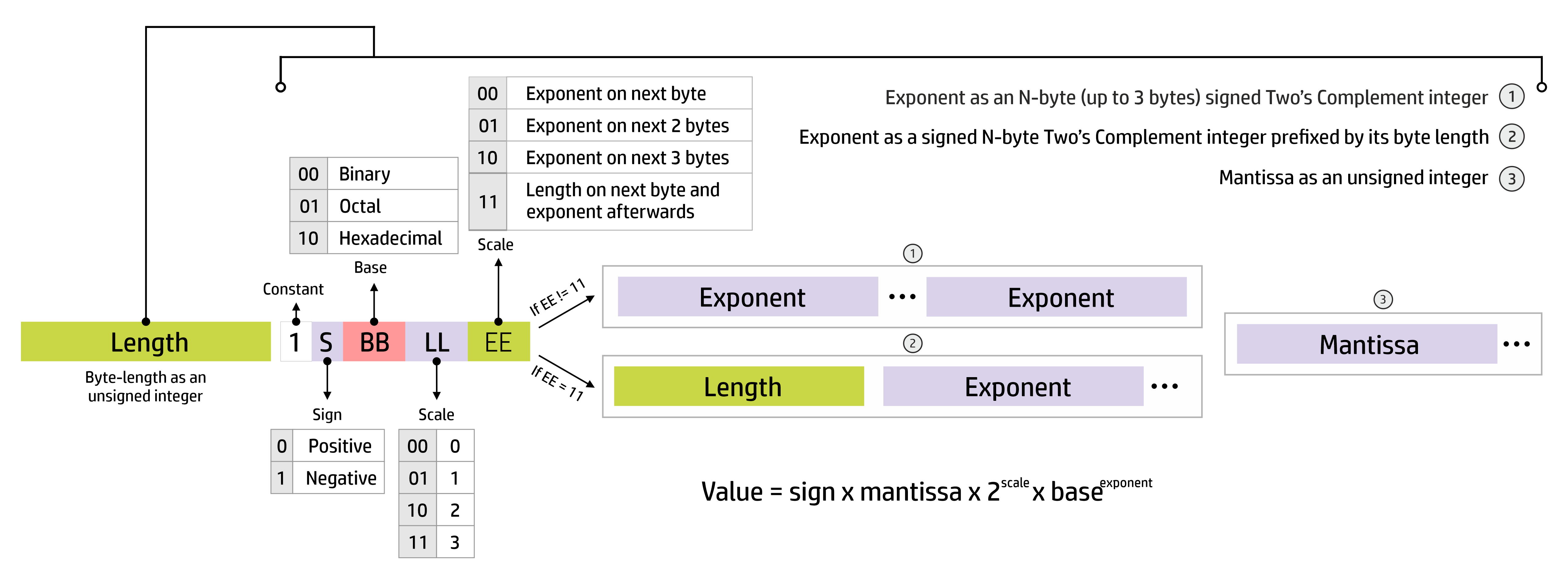}} \caption{A
  visual representation of the \texttt{\textbf{REAL}} data type encoding
  inspired by \cite{10.5555/358551}, page 401. ASN.1 PER Unaligned encodes a
  real number as an 8-bit unsigned integer representing the byte-length of the
  real number, the sign (positive or negative) as 1-bit, the base as 2-bits
  (binary, octal, or hexadecimal), the scale as 2-bits (0, 1, 2, or 3), the
exponent as a variable-length signed integer prefixed by its byte-length, and
the mantissa as a Big Endian variable-length unsigned integer whose width is
bounded by the data type length prefix.} \label{fig:asn1-reals}
\end{figure*}

In terms of real numbers, ASN.1 PER Unaligned does not support IEEE 764
floating-point numbers \cite{8766229}.  Instead, ASN.1 PER Unaligned encodes a
real numbers as concatenation of its sign, base, scale, exponent, and mantissa
where the real value equals {\boldmath $\textbf{sign} \times \textbf{mantissa}
\times 2^{\textbf{scale}} \times \textbf{base}^{\textbf{exponent}}$}.  Refer to
\autoref{fig:asn1-reals} for details on the encoding.

%%%%%%%%%%%%%%%%%%%%%%%%%%%%%%%%%%%%%%%%%%%%%%%%%
% STRINGS
%%%%%%%%%%%%%%%%%%%%%%%%%%%%%%%%%%%%%%%%%%%%%%%%%

\textbf{Strings.} ASN.1 supports a rich set of string types that are not
\emph{NUL}-delimited. The \texttt{\textbf{IA5String}} represents the full ASCII
\cite{STD80} range.  The \texttt{\textbf{VisibleString}} subtype represents the
subset of ASCII \cite{STD80} that does not include control characters.  The
\texttt{\textbf{NumericString}} subtype represents digits and spaces. Finally,
the \texttt{\textbf{UTF8String}} represents Unicode strings that are encoded in
UTF-8 \cite{UnicodeStandard}. Schema-authors may subtype the supported string
types to contrain the permitted alphabet. In the case of unconstrained string
types and constrained string subtypes whose permitted alphabet contains more
than 64 characters, each character is represented by its standard codepoint.
Otherwise, ASN.1 PER Unaligned creates an ordered list of permitted characters
(its alphabet) and encodes each character as an index of such list represented
using the smallest possible Big Endian unsigned integer that can represent the
set of permitted characters. ASN.1 PER Unaligned encodes strings using
\emph{fragmentation} when using string types in which the byte-length of the
string is not always a multiple of the logical length of the string like
\texttt{\textbf{UTF8String}}, when the string type has no size upper bound, or
when the string type has a size upper bound which is greater than 65536.
Otherwise, the string is prefixed with the string logical length as a bounded
integer encoded as described in the \emph{Numbers} section whose lower and
upper bounds correspond to the size bounds of the string type.

%%%%%%%%%%%%%%%%%%%%%%%%%%%%%%%%%%%%%%%%%%%%%%%%%
% BOOLEANS
%%%%%%%%%%%%%%%%%%%%%%%%%%%%%%%%%%%%%%%%%%%%%%%%%

\textbf{Booleans.} ASN.1 PER Unaligned encodes booleans as the bit constants
$0$ (False) and $1$ (True).

%%%%%%%%%%%%%%%%%%%%%%%%%%%%%%%%%%%%%%%%%%%%%%%%%
% ENUMS
%%%%%%%%%%%%%%%%%%%%%%%%%%%%%%%%%%%%%%%%%%%%%%%%%

\textbf{Enumerations.} ASN.1 PER Unaligned represents enumeration constants
using the smallest Big Endian unsigned integer width that can encode the range
of values in the enumeration.

%%%%%%%%%%%%%%%%%%%%%%%%%%%%%%%%%%%%%%%%%%%%%%%%%
% UNIONS
%%%%%%%%%%%%%%%%%%%%%%%%%%%%%%%%%%%%%%%%%%%%%%%%%

\textbf{Unions.} ASN.1 supports a union operator called
\texttt{\textbf{CHOICE}}. ASN.1 PER Unaligned prefixes the encoded value with
the index to the choice in the union data type as the smallest-width Big Endian
unsigned integer that can represent the available choices. ASN.1 also supports
the concept of an \emph{open type}. An open type is a container that holds an
arbitrary value of a type known by the serializer and the deserializer
applications. An open type is encoded as the encoding of the arbitrary value
using \emph{fragmentation}. ASN.1 PER Unaligned does not encode the type of the
arbitrary value. Therefore, the byte length information allows a deserializer
to skip the field if it does not know how to decode it.

%%%%%%%%%%%%%%%%%%%%%%%%%%%%%%%%%%%%%%%%%%%%%%%%%
% LISTS
%%%%%%%%%%%%%%%%%%%%%%%%%%%%%%%%%%%%%%%%%%%%%%%%%

\textbf{Lists.} ASN.1 PER Unaligned supports an heterogeneous list type called
\texttt{\textbf{SEQUENCE}} and an homogeneous list type called
\texttt{\textbf{SEQUENCE OF}}. Both sequence types can be bounded or unbounded.
Unbounded sequences are encoded using \emph{fragmentation} and empty unbounded
sequences are encoded as an empty fragment.  Bounded lists are encoded as the
sequence of its elements with no length metadata. If an heterogeneous sequence
(\texttt{\textbf{SEQUENCE}}) contains $N$ optional values, then the sequence is
prefixed by a sequence of $N$ bits that determine whether each optional value
is set.

\begin{figure}[hb!]
  \frame{\includegraphics[width=\linewidth]{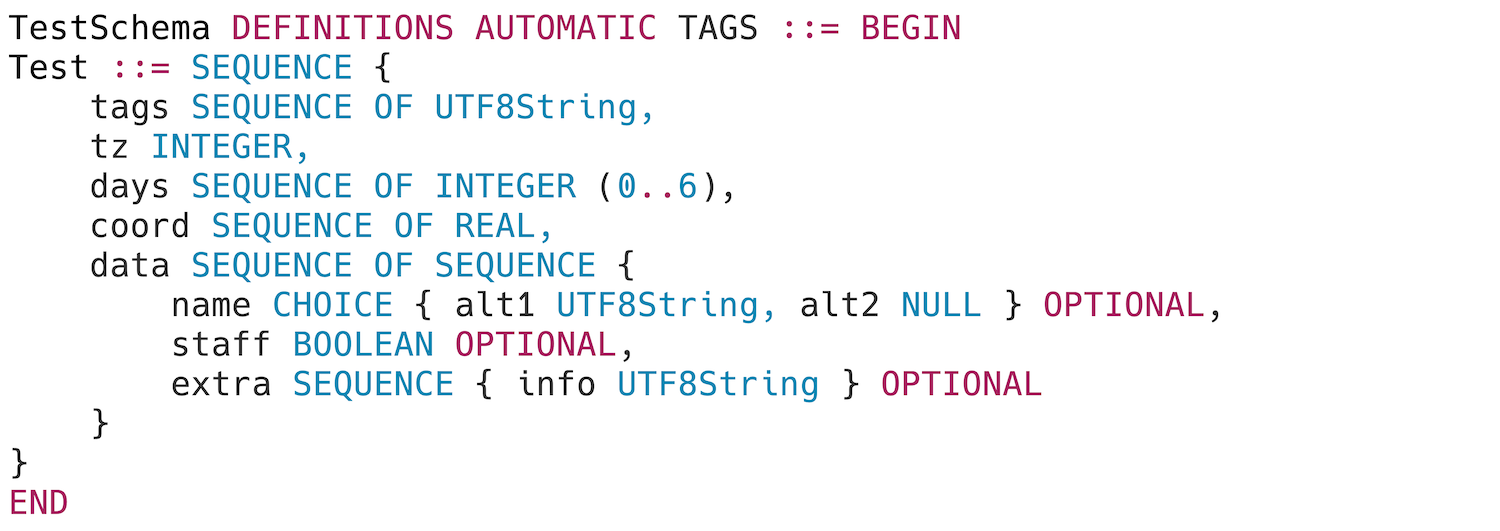}}
  \caption{ASN.1 schema to serialize the \autoref{lst:json-object-test} input data.}
  \label{lst:schema-asn1}
\end{figure}

\begin{figure*}[hb!]
  \frame{\includegraphics[width=\linewidth]{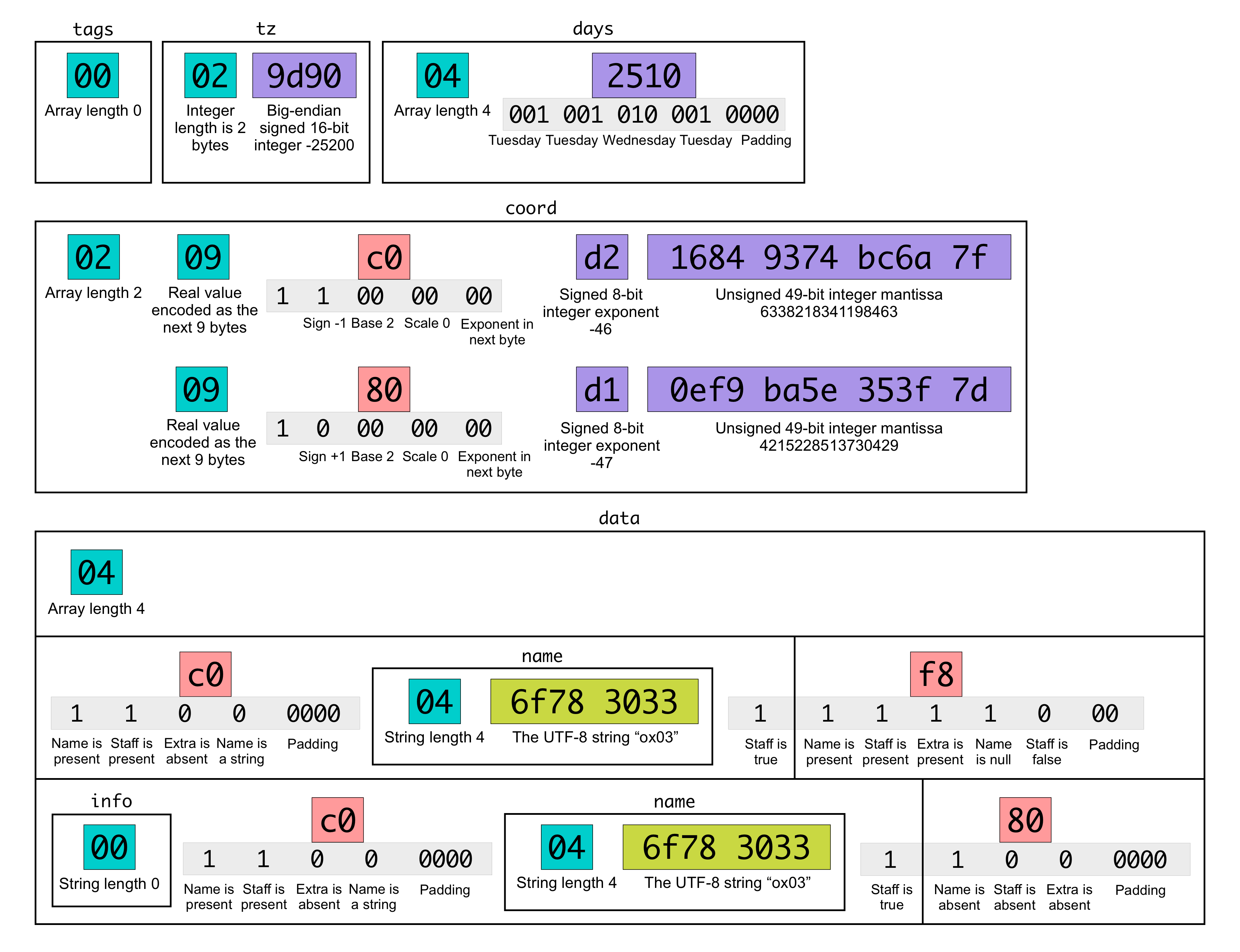}}
  \caption{Annotated hexadecimal output of serializing the \autoref{lst:json-object-test} input data with ASN.1 PER Unaligned.}
\end{figure*}

\begin{table}[hb!]
\caption{A high-level summary of the ASN.1 PER Unaligned schema-driven serialization specification.}
\label{table:asn1}
\begin{tabularx}{\linewidth}{l|X}
\toprule
  \textbf{Website} & \url{https://www.itu.int/rec/T-REC-X.680/en} \\ \hline
  \textbf{Company / Individual} & International Telecommunication Union \\ \hline
  \textbf{Year} & 1984 \\ \hline
  \textbf{Specification} & ITU-T X.680-X.693 \cite{asn1} \\ \hline
  \textbf{License} & Implementation-dependent \\ \hline
  \textbf{Schema Language} & ASN.1 \\ \hline
  \textbf{Layout} & Sequential and order-based \\ \hline
  \textbf{Languages} & C, C++, C\#, Java, Ada/Spark, Python, Erlang \\ \hline
  \multicolumn{2}{c}{\textbf{Types}} \\ \hline

  \textbf{Numeric}
  & Big Endian Two's Complement \cite{twos-complement} signed integers of user-defined length \par\smallskip
  Big Endian unsigned integers of user-defined length \par\smallskip
  Real numbers consisting of up to 255 bytes encoding the base, scale, exponent, and mantissa \par\smallskip
  Arbitrary-length ASCII-encoded decimal numbers \\ \hline

  \textbf{String} & ASCII \cite{STD80}, UTF-8 \cite{UnicodeStandard} \\ \hline
  \textbf{Composite} & Choice, Enum, Set, Sequence \\ \hline
  \textbf{Scalars} & Boolean, Null \\ \hline

  \textbf{Other} &
  Octet string (byte array) \par\smallskip
  Bit-string (arbitrary-length bit array) \par\smallskip
  Date, Time, Date-time \cite{ISO8601} \\
\bottomrule
\end{tabularx}
\end{table}

%% file: sections/formats/avro.tex
\label{sec:apache-avro}

\begin{figure}[hb!]
  \frame{\includegraphics[width=\linewidth]{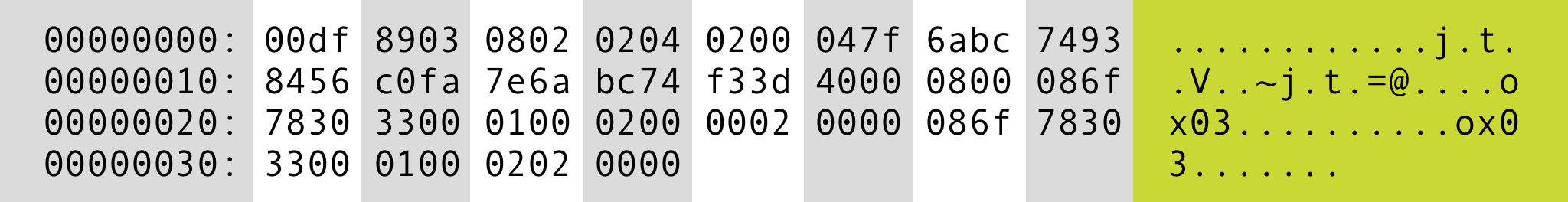}}
  \caption{Hexadecimal output (\texttt{xxd})
  of encoding \autoref{lst:json-object-test}
  input data with Apache Avro Binary Encoding
with no framing (56 bytes).}
\label{lst:hex-avro} \end{figure}

%%%%%%%%%%%%%%%%%%%%%%%%%%%%%%%%%%%%%%%%%%%%%%%%%
% HISTORY
%%%%%%%%%%%%%%%%%%%%%%%%%%%%%%%%%%%%%%%%%%%%%%%%%

\textbf{History.} Apache Avro \cite{avro} is a schema-driven binary
serialization specification introduced by Douglass Cutting
\footnote{\url{https://github.com/cutting}} during 2009 while working at Yahoo!
\footnote{\url{https://yahoo.com/}}. Apache Avro is part of the Apache Hadoop
\footnote{\url{https://hadoop.apache.org}} framework and is also deeply
integrated with other projects from the Big Data field such as Apache Spark
\footnote{\url{http://spark.apache.org/index.html}} and Apache Kafka
\footnote{\url{https://kafka.apache.org}}. Apache Avro is part of the Apache
Software Foundation \footnote{\url{https://www.apache.org}} and its released
under the Apache License 2.0
\footnote{\url{http://www.apache.org/licenses/LICENSE-2.0.html}}.
\cite{akhtar2020role} provides a detailed discussion of the role of the Apache
Software Foundation in the Big Data industry including Apache Avro.

%%%%%%%%%%%%%%%%%%%%%%%%%%%%%%%%%%%%%%%%%%%%%%%%%
% ADVANTAGES
%%%%%%%%%%%%%%%%%%%%%%%%%%%%%%%%%%%%%%%%%%%%%%%%%

\textbf{Characteristics.} \begin{itemize}

  \item \textbf{Optional Code Generation.} Apache Avro implementations can
    perform serialization and deserialization by taking the input schema at
    runtime. This enables programs to work with new types of data without a
    recompilation step. In comparison, schema-driven serialization
    specifications such as Apache Thrift \cite{slee2007thrift} require a
    code generation step at build time for each schema.

  \item \textbf{Compactness.} Apache Avro Binary Encoding with no framing
    produces considerably space-efficient bit-strings as it only encodes
    minimal metadata
    \footnote{\url{https://avro.apache.org/docs/current/spec.html}}.

\end{itemize}

%%%%%%%%%%%%%%%%%%%%%%%%%%%%%%%%%%%%%%%%%%%%%%%%%
% OVERALL STRUCTURE
%%%%%%%%%%%%%%%%%%%%%%%%%%%%%%%%%%%%%%%%%%%%%%%%%

\textbf{Layout.} An Apache Avro Binary Encoding with no framing bit-string is a
sequence of values, sometimes nested. The ordering of the values is determined
by the schema. Apache Avro Binary Encoding only encodes runtime information
that cannot be inferred from the schema, such as the length of the lists or
union type information. Fields are encoded even if they equal their default
values.

%%%%%%%%%%%%%%%%%%%%%%%%%%%%%%%%%%%%%%%%%%%%%%%%%
% NUMBERS
%%%%%%%%%%%%%%%%%%%%%%%%%%%%%%%%%%%%%%%%%%%%%%%%%

\textbf{Numbers.} Apache Avro Binary Encoding supports 32-bit and 64-bit
ZigZag-encoded (\ref{sec:zigzag-encoding}) Little Endian Base 128 (LEB128)
(\ref{sec:varints}) signed integers.  In terms of real numbers, Apache Avro
Binary Encoding supports Little Endian 32-bit and 64-bit IEEE 764
floating-point numbers \cite{8766229} and arbitrary-precision signed
decimal numbers.  Arbitrary-precision signed decimal numbers are encoded using
variable-length or fixed-length byte arrays. The byte array represents a Big
Endian signed integer using Two's Complement \cite{twos-complement} whose
width is defined by the byte array length prefix. The schema declares the scale
and precision as integers. The resulting decimal number equals the unscaled
integer multiplied by $10^{-\text{scale}}$.  The precision integer determines
the maximum number of decimals stored in the data type.

%%%%%%%%%%%%%%%%%%%%%%%%%%%%%%%%%%%%%%%%%%%%%%%%%
% STRINGS
%%%%%%%%%%%%%%%%%%%%%%%%%%%%%%%%%%%%%%%%%%%%%%%%%

\textbf{Strings.} Apache Avro Binary Encoding strings are encoded using UTF-8
\cite{UnicodeStandard} without \emph{NUL} delimiters. Apache Avro Binary
Encoding prefixes strings with 64-bit ZigZag-encoded
(\ref{sec:zigzag-encoding}) Little Endian Base 128 (LEB128) (\ref{sec:varints})
variable-length signed integers that represent the number of code-points in the
string. Empty strings are encoded with a length prefix of 0 with no following
characters. Apache Avro Binary Encoding does not attempt to deduplicate
multiple occurrences of the same string.

%%%%%%%%%%%%%%%%%%%%%%%%%%%%%%%%%%%%%%%%%%%%%%%%%
% BOOLEANS
%%%%%%%%%%%%%%%%%%%%%%%%%%%%%%%%%%%%%%%%%%%%%%%%%

\textbf{Booleans.} Apache Avro Binary Encoding encodes booleans using the byte
constants \texttt{\textbf{0x00}} (False) and \texttt{\textbf{0x01}} (True).

%%%%%%%%%%%%%%%%%%%%%%%%%%%%%%%%%%%%%%%%%%%%%%%%%
% ENUMS
%%%%%%%%%%%%%%%%%%%%%%%%%%%%%%%%%%%%%%%%%%%%%%%%%

\textbf{Enumerations.} Apache Avro Binary Encoding represents enumeration
constants using 32-bit ZigZag-encoded (\ref{sec:zigzag-encoding}) Little Endian
Base 128 (LEB128) (\ref{sec:varints}) variable-length signed integers.

%%%%%%%%%%%%%%%%%%%%%%%%%%%%%%%%%%%%%%%%%%%%%%%%%
% UNIONS
%%%%%%%%%%%%%%%%%%%%%%%%%%%%%%%%%%%%%%%%%%%%%%%%%

\begin{figure}[hb!]
  \frame{\includegraphics[width=\linewidth]{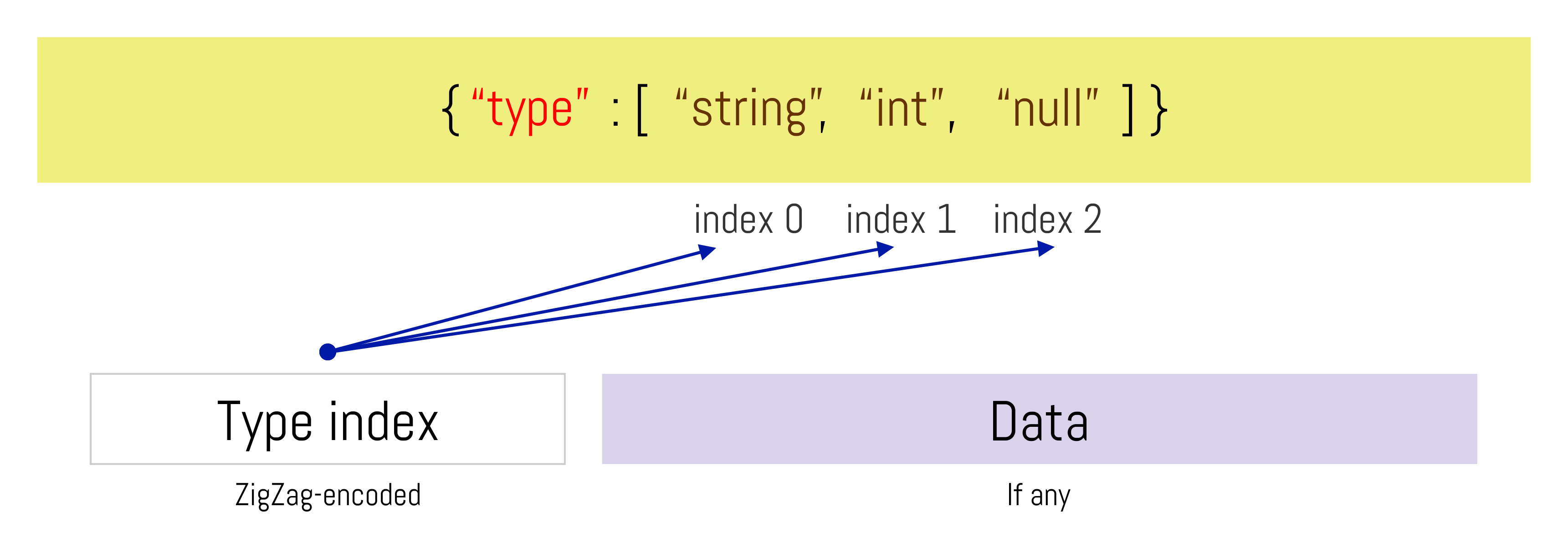}} \caption{A
  field consisting of more than one possible type is encoded as the type index
  followed by the data, if any. For example, if the schema defines a field
  whose type is \texttt{\textbf{[ "string", "int", "null" ]}}, then the value
  will be prefixed with \texttt{\textbf{0x00}} if the value is a string, with
  \texttt{\textbf{0x02}} (the ZigZag-encoded (\ref{sec:zigzag-encoding})
integer $1$) if the value is an integer, or with \texttt{\textbf{0x04}} (the
ZigZag-encoded (\ref{sec:zigzag-encoding}) integer $2$) if the value is null.}
\label{fig:avro-unions} \end{figure}

\textbf{Unions.} An Apache Avro schema may declare an ordered list of potential
types for a single field. In these cases, Apache Avro Binary Encoding prefixes
the value with a 32-bit ZigZag-encoded (\ref{sec:zigzag-encoding}) Little
Endian Base 128 (LEB128) (\ref{sec:varints}) variable-length signed integer
that corresponds to an index of the ordered list of types. Refer to
\autoref{fig:avro-unions} for a visual example.

%%%%%%%%%%%%%%%%%%%%%%%%%%%%%%%%%%%%%%%%%%%%%%%%%
% ARRAYS
%%%%%%%%%%%%%%%%%%%%%%%%%%%%%%%%%%%%%%%%%%%%%%%%%

\textbf{Lists.} Apache Avro Binary Encoding encodes a list (array) as a series
of blocks. Each block is prefixed with its logical size as a 64-bit
ZigZag-encoded (\ref{sec:zigzag-encoding}) Little Endian Base 128 (LEB128)
(\ref{sec:varints}) variable-length signed integer followed by its elements in
order. As a runtime optimization, if the logical size signed integer is
negative, then the block size is the absolute value of the integer and the
sequence of elements is further prefixed with the byte-length of the block as
another 64-bit ZigZag-encoded (\ref{sec:zigzag-encoding}) Little Endian Base
128 (LEB128) (\ref{sec:varints}) variable-length signed integer. Apache Avro
Binary Encoding does not attempt to deduplicate multiple occurrences of the
same element in an array.

\begin{figure}[hb!]
  \frame{\includegraphics[width=\linewidth]{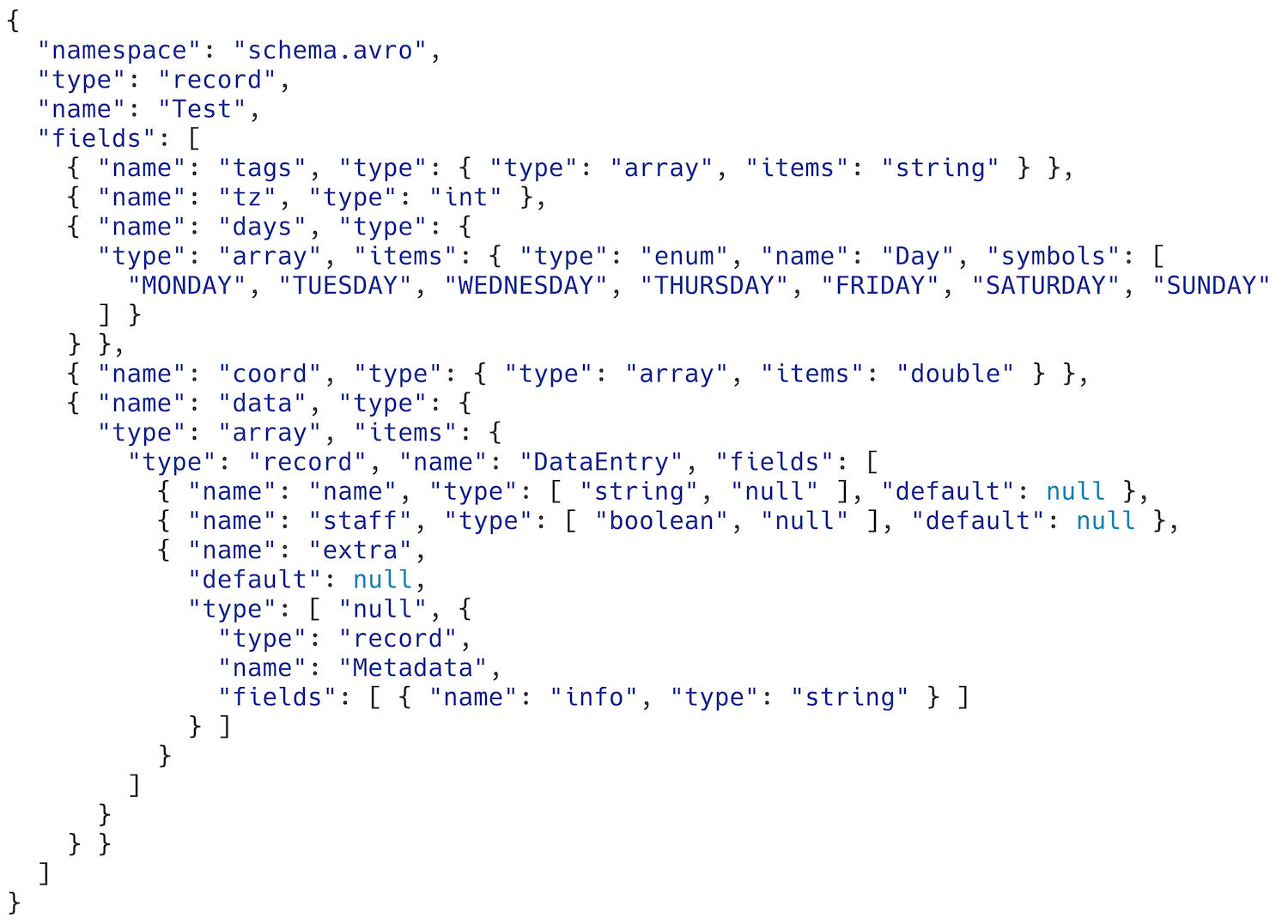}}
  \caption{Apache Avro schema to serialize the \autoref{lst:json-object-test} input data.}
  \label{lst:schema-avro}
\end{figure}

\begin{figure*}[hb!]
  \frame{\includegraphics[width=\linewidth]{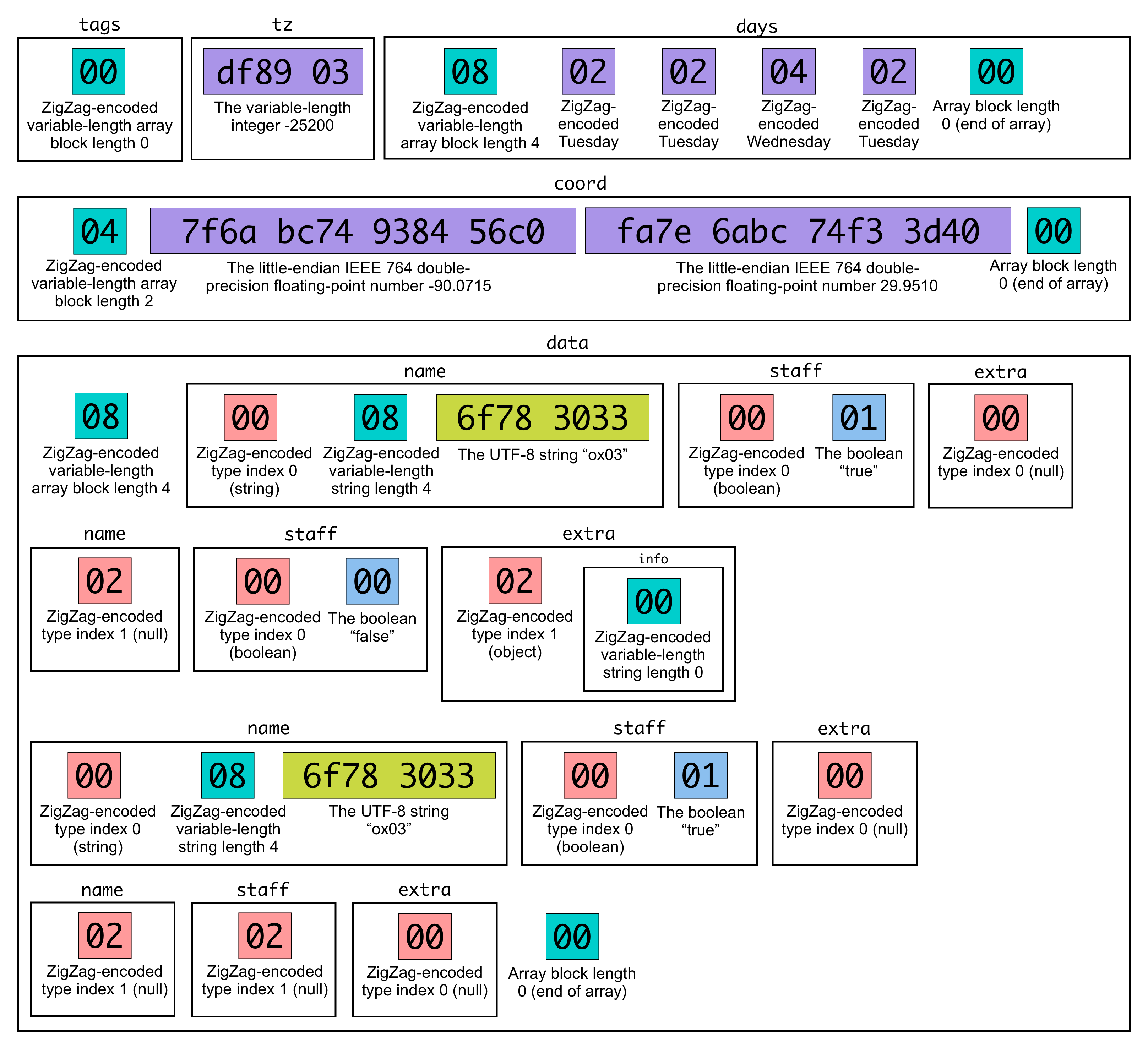}}
  \caption{Annotated hexadecimal output of serializing the
\autoref{lst:json-object-test} input data with Apache Avro Binary Encoding with no framing.}
\end{figure*}

\begin{table}[hb!]
\caption{A high-level summary of the Apache Avro Binary Encoding with no framing schema-driven serialization specification.}
\label{table:avro}
\begin{tabularx}{\linewidth}{l|X}
  \toprule
  \textbf{Website} & \url{https://avro.apache.org} \\ \hline
  \textbf{Company / Individual} & Apache Software Foundation \\ \hline
  \textbf{Year} & 2009 \\ \hline
  \textbf{Specification} & \url{https://avro.apache.org/docs/current/spec.html} \\ \hline
  \textbf{License} & Apache License 2.0 \\ \hline
  \textbf{Schema Language} & Avro IDL \\ \hline
  \textbf{Layout} & Sequential and order-based \\ \hline
  \textbf{Languages} & C, C++, C\#, Java \\ \hline
  \multicolumn{2}{c}{\textbf{Types}} \\ \hline

  \textbf{Numeric} &
  32-bit and 64-bit ZigZag-encoded (\ref{sec:zigzag-encoding}) Little Endian Base 128 (LEB128) (\ref{sec:varints}) variable-length integers \par\smallskip
  Arbitrary-precision Two's Complement \cite{twos-complement} signed decimal numbers \par\smallskip
  Little Endian 32-bit and 64-bit IEEE 764 floating-point numbers \cite{8766229} \\ \hline

  \textbf{String} & UTF-8 \cite{UnicodeStandard} \\ \hline
  \textbf{Composite} & Array, Enum, Map, Record, Union \\ \hline
  \textbf{Scalars} & Boolean, Null \\ \hline

  \textbf{Other} &
  Bytes (variable-length byte array) \par\smallskip
  Fixed (fixed-length byte array) \par\smallskip
  UUID \cite{RFC4122} \par\smallskip
  Date (days from the UNIX Epoch) \cite{RFC8877} \par\smallskip
  Time (milliseconds and microseconds) \cite{ISO8601} \par\smallskip
  Timestamp (milliseconds and microseconds) \cite{RFC8877} \par\smallskip
  Duration \footnotemark \\
  \bottomrule
\end{tabularx}
\end{table}

\footnotetext{\url{https://avro.apache.org/docs/current/spec.html\#Duration}}

%% file: sections/formats/bond.tex
\label{sec:microsoft-bond}

\begin{figure}[hb!]
  \frame{\includegraphics[width=\linewidth]{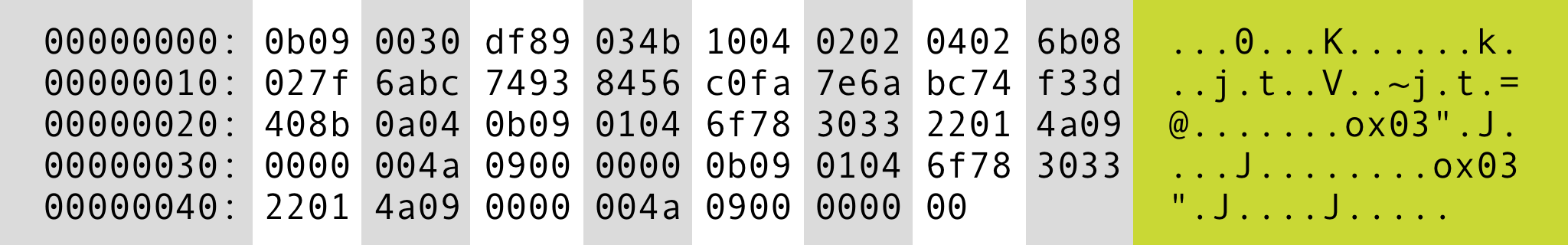}}
  \caption{Hexadecimal output (\texttt{xxd})
  of encoding \autoref{lst:json-object-test}
input data with Microsoft Bond Compact Binary
version 1 (77 bytes).} \label{lst:hex-bond}
\end{figure}

%%%%%%%%%%%%%%%%%%%%%%%%%%%%%%%%%%%%%%%%%%%%%%%%%
% HISTORY
%%%%%%%%%%%%%%%%%%%%%%%%%%%%%%%%%%%%%%%%%%%%%%%%%

\textbf{History.} Microsoft Bond \cite{microsoft-bond} is an RPC protocol and
schema-driven serialization specification developed by Microsoft in 2011 and was made
open-source in 2015. The Microsoft Bond project was started by Adam Sapek
\footnote{\url{https://github.com/sapek}}, a Principal Software Engineer at
Microsoft Research, while working on the Microsoft Bing
\footnote{\url{https://www.bing.com}} search engine. Microsoft Bond is used at
the core of many Microsoft services and it is released under the MIT license
\footnote{\url{https://opensource.org/licenses/MIT}}.

%%%%%%%%%%%%%%%%%%%%%%%%%%%%%%%%%%%%%%%%%%%%%%%%%
% ADVANTAGES
%%%%%%%%%%%%%%%%%%%%%%%%%%%%%%%%%%%%%%%%%%%%%%%%%

\textbf{Characteristics.} \begin{itemize}

  \item \textbf{Rich Type System.} The Microsoft Bond schema language supports
    generic types, inheritance and a wide range of scalar and composite data
    types such as sets and nullable types.

  \item \textbf{Custom Type Mappings.} To ease integration, Microsoft Bond
    supports statically mapping the types supported by the schema language to
    any compatible programming language type.

  \item \textbf{Opt-in Lazy Deserialization.} For runtime-performance reasons,
    the Microsoft Bond schema language supports marking certain fields to not
    be de-serialized automatically. These fields can be de-serialized when
    needed or omitted altogether.

\end{itemize}

%%%%%%%%%%%%%%%%%%%%%%%%%%%%%%%%%%%%%%%%%%%%%%%%%
% OVERALL STRUCTURE
%%%%%%%%%%%%%%%%%%%%%%%%%%%%%%%%%%%%%%%%%%%%%%%%%

\textbf{Layout.} A Microsoft Bond Compact Binary version 1 (v1) bit-string is a
sequence of values, sometimes nested. Each value is prefixed with a type
definition and the length of the value, where applicable. Each value has a
numeric identifier that must be unique on the current nesting level. Microsoft
Bond Compact Binary v1 only encodes a field if the field is required or if its
value is not equal to the default hence resulting in efficient use of space.

%%%%%%%%%%%%%%%%%%%%%%%%%%%%%%%%%%%%%%%%%%%%%%%%%
% KEYS
%%%%%%%%%%%%%%%%%%%%%%%%%%%%%%%%%%%%%%%%%%%%%%%%%

\begin{figure}[hb!]
\frame{\includegraphics[width=\linewidth]{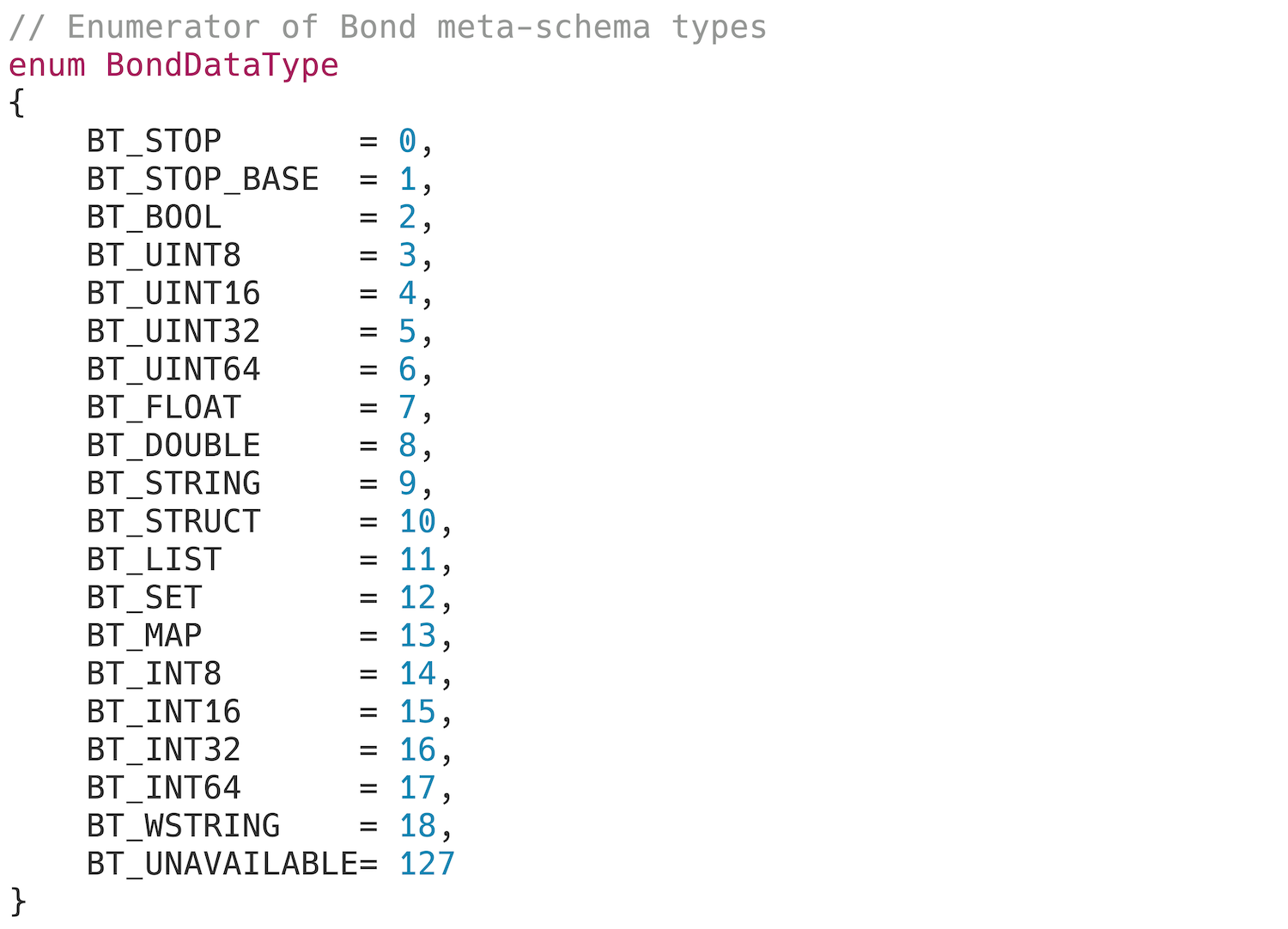}}
\caption{Microsoft Bond type identifiers definition \protect\footnotemark.}
\label{lst:bond-types}
\end{figure}

\textbf{Types.} A Microsoft Bond Compact Binary v1 type definition consist of
an positive absolute unique identifier integer and a type identifier constant
as shown in \autoref{lst:bond-types}. Microsoft Bond Compact Binary v1 defines
three type definition encodings depending on the length of the unique
identifier. Refer to \autoref{table:bond-type-definitions}.

\begin{table}[hb!]
\caption{The three type definition encodings that Microsoft Bond Compact Binary v1 supports
  depending on the value of the unique field identifier as determined by the
  \emph{From} and \emph{To} ranges. Note that Microsoft Bond Compact Binary v1 cannot encode
  unique field identifiers larger than 65535.}

\label{table:bond-type-definitions}
  \begin{tabularx}{\linewidth}{l|l|l|X|X|X}
  \toprule
  \textbf{From} & \textbf{To} & \textbf{Size} & \textbf{First 3-bits} & \textbf{Next 5-bits} & \textbf{Remaining bits} \\ \midrule
  0   & 5     & 1 byte  & The unique identifier as a 3-bit unsigned integer & The field type identifier constant as shown in \autoref{lst:bond-types} & None \\ \hline
  6   & 255   & 2 bytes & \texttt{110} & The field type identifier constant as shown in \autoref{lst:bond-types} & The unique identifier as an 8-bit unsigned integer \\ \hline
  256 & 65535 & 3 bytes & \texttt{111} & The field type identifier constant as shown in \autoref{lst:bond-types} & The unique identifier as a Big Endian 16-bit unsigned integer \\
  \bottomrule
\end{tabularx}
\end{table}

\footnotetext{\url{https://github.com/microsoft/bond/blob/8d0fe6c00cbcd7ea9c54b1f1e947174caff596e4/idl/bond/core/bond\_const.bond}}

Microsoft Bond schemas can instruct implementations to not de-serialize a field
automatically by marking the field as \emph{bonded} at the schema level.
Instead, the consumer de-serializes \emph{bonded} fields when and if needed by
the application. This results in runtime efficiency.

%%%%%%%%%%%%%%%%%%%%%%%%%%%%%%%%%%%%%%%%%%%%%%%%%
% NUMBERS
%%%%%%%%%%%%%%%%%%%%%%%%%%%%%%%%%%%%%%%%%%%%%%%%%

\textbf{Numbers.} Microsoft Bond Compact Binary v1 supports Little Endian IEEE
764 32-bit and 64-bit floating point-numbers \cite{8766229}. In terms of
integers, Microsoft Bond Compact Binary v1 supports Little Endian 8-bit
fixed-length unsigned integers, 16-bit, 32-bit, and 64-bit Little Endian Base
128 (LEB128) (\ref{sec:varints}) variable-length unsigned integers, 8-bit
fixed-length signed integers with Two's Complement \cite{twos-complement},
and 16-bit, 32-bit, and 64-bit Little Endian Base 128 (LEB128)
(\ref{sec:varints}) variable-length ZigZag-encoded (\ref{sec:zigzag-encoding})
signed integers.

%%%%%%%%%%%%%%%%%%%%%%%%%%%%%%%%%%%%%%%%%%%%%%%%%
% STRINGS
%%%%%%%%%%%%%%%%%%%%%%%%%%%%%%%%%%%%%%%%%%%%%%%%%

\textbf{Strings.} Microsoft Bond can produce strings with UTF-8 and Little
Endian UTF-16 encodings \cite{UnicodeStandard} without \emph{NUL}
delimiters. Refer to the \texttt{\textbf{BT\_STRING}} and
\texttt{\textbf{BT\_WSTRING}} data types from \autoref{lst:bond-types}. Each
string is prefixed with a 32-bit Little Endian Base 128 (LEB128)
(\ref{sec:varints}) variable-length unsigned integer determining the number of
code-points in the string. Microsoft Bond Compact Binary v1 does not attempt to
deduplicate multiple occurrences of the same string.

%%%%%%%%%%%%%%%%%%%%%%%%%%%%%%%%%%%%%%%%%%%%%%%%%
% BOOLEANS
%%%%%%%%%%%%%%%%%%%%%%%%%%%%%%%%%%%%%%%%%%%%%%%%%

\textbf{Booleans.} Microsoft Bond Compact Binary v1 encodes booleans using the
1-byte constants \texttt{\textbf{0x00}} (False) and \texttt{\textbf{0x01}}
(True).

%%%%%%%%%%%%%%%%%%%%%%%%%%%%%%%%%%%%%%%%%%%%%%%%%
% ENUMS
%%%%%%%%%%%%%%%%%%%%%%%%%%%%%%%%%%%%%%%%%%%%%%%%%

\textbf{Enumerations.} Microsoft Bond Compact Binary v1 represents enumeration
constants using 32-bit ZigZag-encoded (\ref{sec:zigzag-encoding}) Little Endian
Base 128 (LEB128) (\ref{sec:varints}) variable-length signed integers.

%%%%%%%%%%%%%%%%%%%%%%%%%%%%%%%%%%%%%%%%%%%%%%%%%
% UNIONS
%%%%%%%%%%%%%%%%%%%%%%%%%%%%%%%%%%%%%%%%%%%%%%%%%

\begin{figure}[hb!]
  \frame{\includegraphics[width=\linewidth]{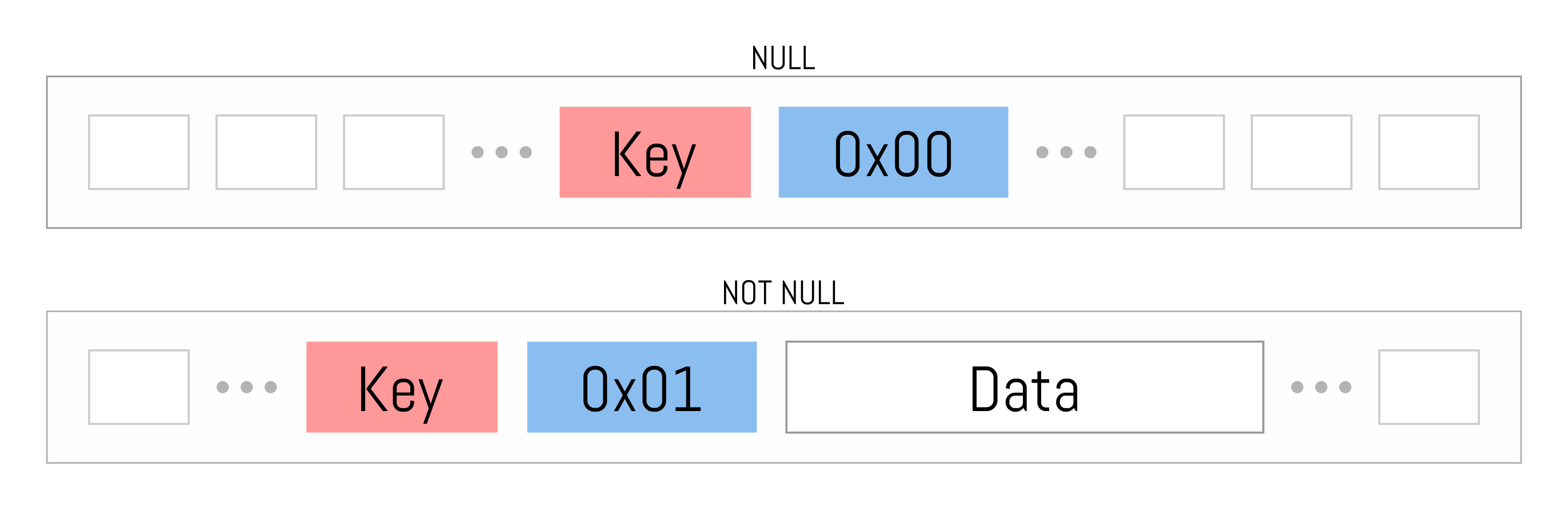}}
  \caption{A visual representation of Microsoft Bond Compact Binary v1
  \emph{nullable} types.  If the value is \emph{NULL} (top), then it is encoded
  as the type definition followed by the length \texttt{\textbf{0x00}}. If the
  value is not null (bottom), then it is encoded as the type definition
followed by the length \texttt{\textbf{0x01}}, followed by the value.}
\label{fig:bond-nullable} \end{figure}

\begin{figure}[hb!]
\frame{\includegraphics[width=\linewidth]{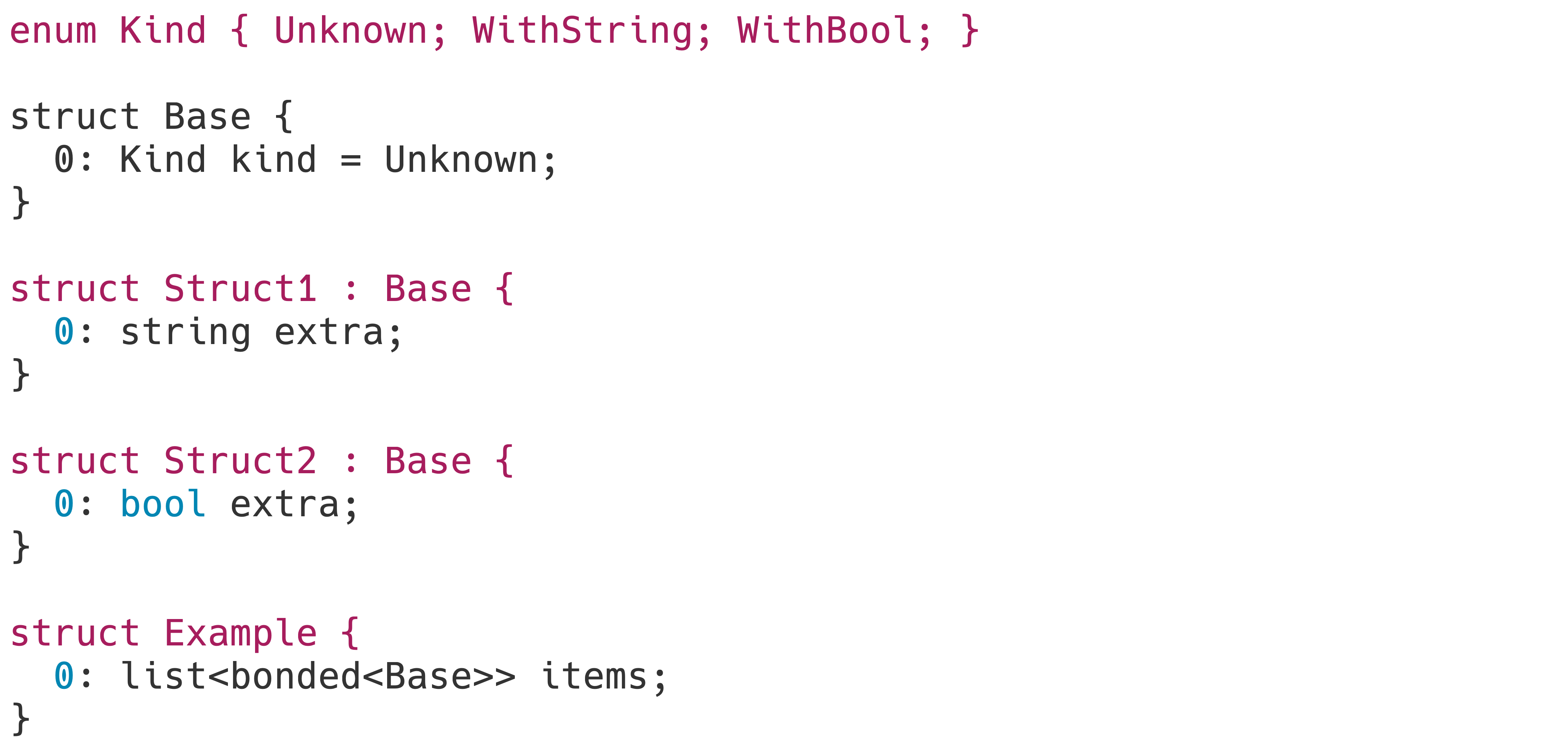}}
\caption{An adapted example of a polymorphic list definition
\protect\footnotemark.} \label{lst:bond-polymorphic-unions} \end{figure}

\textbf{Unions.} Microsoft Bond supports the \emph{nullable} type union.
Nullable fields are encoded as lists of zero or one value. Refer to
\autoref{fig:bond-nullable} for a visual example. Microsoft Bond Compact Binary
v1 does not encode unset optional fields. Therefore schema-writers can
approximate unions by defining a structure containing multiple fields and
ensuring that only one of them is set at a time. Another common pattern to
approximate unions is to rely on polymorphism and bonded types. In this case, a
schema-writer could define a base structure including a single field: an
enumeration denoting the union choice and extend the base structure with
subclasses defining each of the choices. Other structures refer to the union
structure as a bonded field, so that the client can first deserialize the
enumeration and then deserialize the rest of the fields depending on the
enumeration constant value. The serializer program is responsible for correctly
setting the enumeration constant. In \autoref{lst:bond-polymorphic-unions}, a
heterogeneous list using this technique is shown.

\footnotetext{\url{https://github.com/microsoft/bond/blob/8d0fe6c00cbcd7ea9c54b1f1e947174caff596e4/examples/cpp/core/polymorphic\_container/polymorphic\_container.bond}}

%%%%%%%%%%%%%%%%%%%%%%%%%%%%%%%%%%%%%%%%%%%%%%%%%
% ARRAYS
%%%%%%%%%%%%%%%%%%%%%%%%%%%%%%%%%%%%%%%%%%%%%%%%%

\textbf{Lists.} Microsoft Bond Compact Binary v1 encodes a list as the type
definition, followed by the list definition, followed by the elements encoded
in order. The list definition consists in a byte where the five
least-significant bits encode the element type identifier constant as shown in
\autoref{lst:bond-types} followed by the length of the list as a 32-bit Little
Endian Base 128 (LEB128) (\ref{sec:varints}) variable-length unsigned integer.
Microsoft Bond Compact Binary v1 does not attempt to deduplicate multiple
occurrences of the same element in a list.

\begin{figure}[hb!]
  \frame{\includegraphics[width=\linewidth]{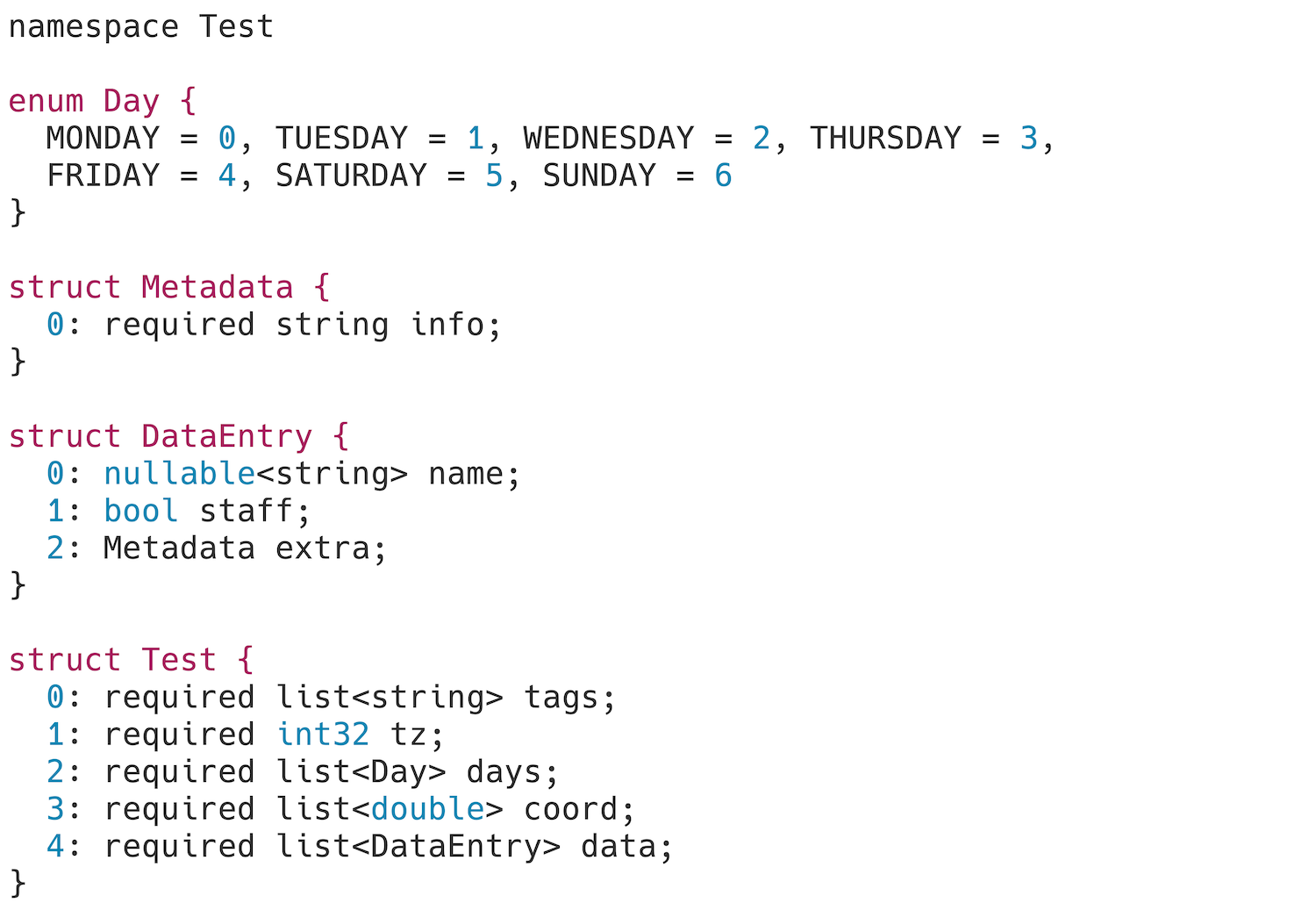}}
  \caption{Microsoft Bond schema to serialize the \autoref{lst:json-object-test} input data.}
  \label{lst:schema-bond}
\end{figure}

\begin{figure*}[hb!]
\frame{\includegraphics[width=\linewidth]{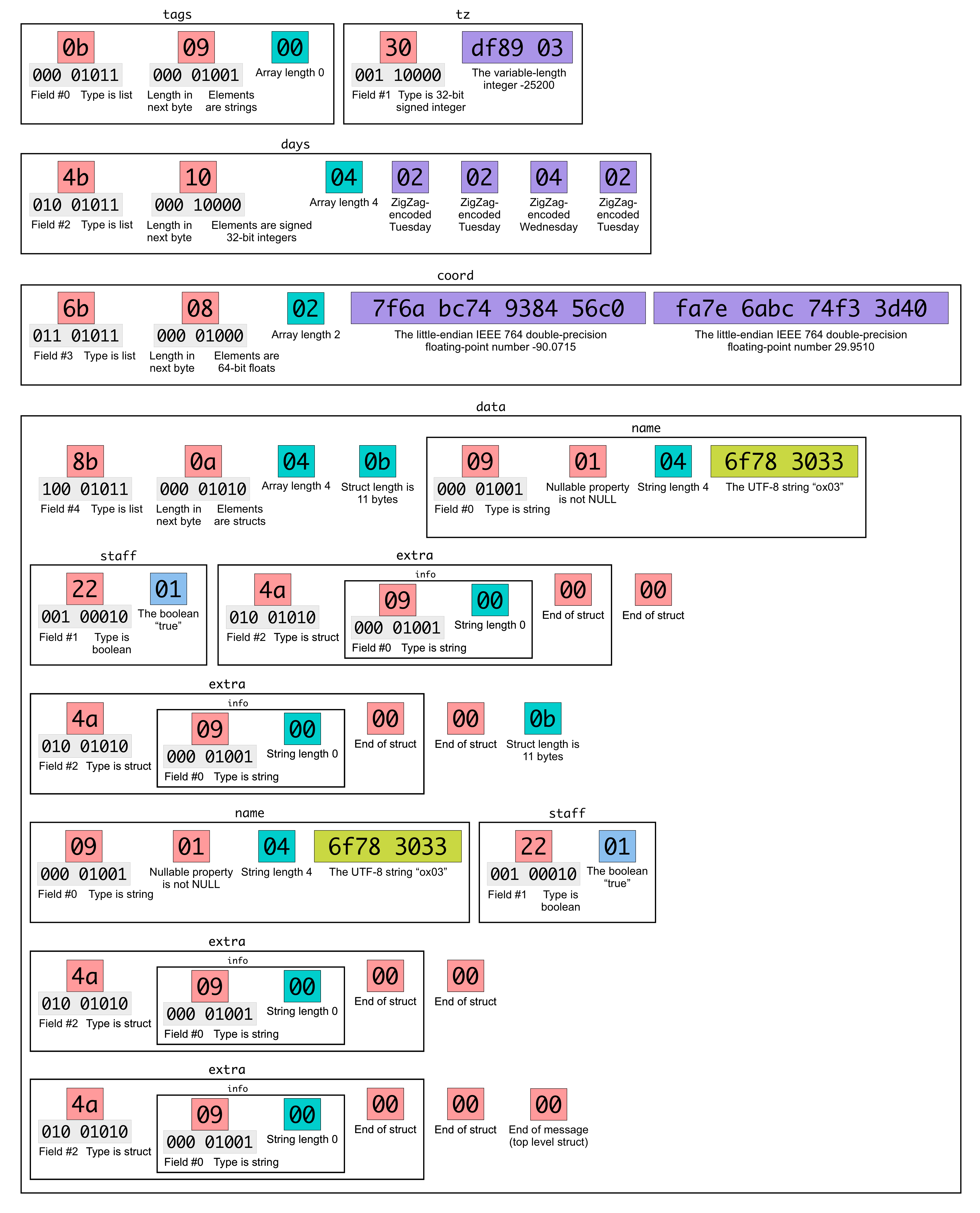}}
\caption{Annotated hexadecimal output of serializing the
\autoref{lst:json-object-test} input data with Microsoft Bond Compact Binary
v1.} \end{figure*}

\begin{table}[hb!]
\caption{A high-level summary of the Microsoft Bond Compact Binary v1 schema-driven serialization specification.}
\label{table:bond}
  \begin{tabularx}{\linewidth}{l|X}
    \toprule
  \textbf{Website} & \url{https://microsoft.github.io/bond/} \\ \hline
  \textbf{Company / Individual} & Microsoft \\ \hline
  \textbf{Year} & 2011 \\ \hline
  \textbf{Specification} & \url{https://microsoft.github.io/bond/reference/cpp/compact\_\_binary\_8h\_source.html} \\ \hline
  \textbf{License} & MIT \\ \hline
  \textbf{Schema Language} & Bond IDL \\ \hline
  \textbf{Layout} & Sequential with field identifiers \\ \hline
  \textbf{Languages} & C++, C\#, Java, Python \\ \hline
  \multicolumn{2}{c}{\textbf{Types}} \\ \hline

  \textbf{Numeric} &
  16-bit, 32-bit, and 64-bit Little Endian Base 128 (LEB128) (\ref{sec:varints}) variable-length unsigned integers \par\smallskip
  16-bit, 32-bit, and 64-bit ZigZag-encoded (\ref{sec:zigzag-encoding}) Little Endian Base 128 (LEB128) (\ref{sec:varints}) variable-length signed integers \par\smallskip
  Fixed-length 8-bit unsigned integers \par\smallskip
  Fixed-length 8-bit Two's Complement \cite{twos-complement} signed integers \par\smallskip
  Little Endian 32-bit and 64-bit IEEE 764 floating-point numbers \cite{8766229} \\ \hline

  \textbf{String} & UTF-8, Little Endian UTF-16 \cite{UnicodeStandard} \\ \hline
  \textbf{Composite} & List, Maybe, Nullable, Set, Struct, Vector \\ \hline
  \textbf{Scalars} & Boolean \\ \hline

  \textbf{Other} & Blob (byte array) \\ \bottomrule
\end{tabularx}
\end{table}

%% file: sections/formats/capnproto.tex
\label{sec:capnproto}

\begin{figure}[hb!]
  \frame{\includegraphics[width=\linewidth]{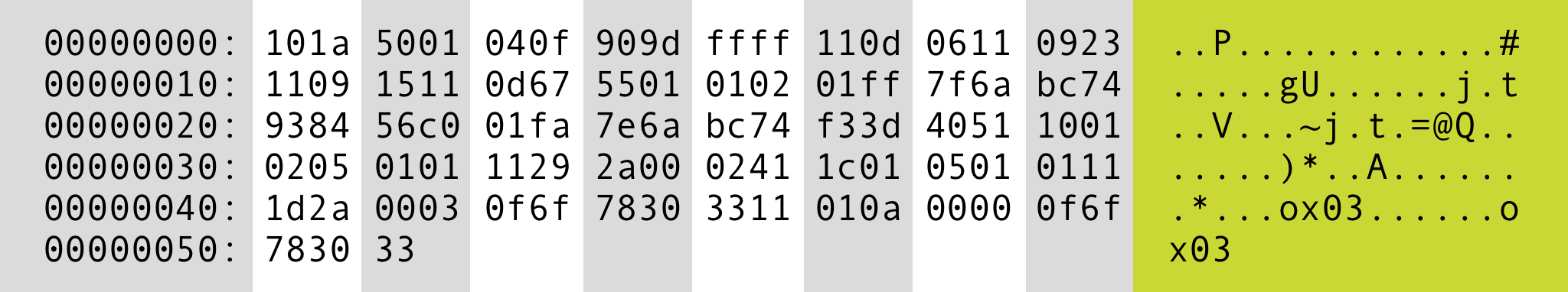}}
  \caption{Hexadecimal output (\texttt{xxd})
  of encoding \autoref{lst:json-object-test}
input data with Cap'n Proto Packed Encoding
(83 bytes).} \label{lst:hex-capnproto}
\end{figure}

%%%%%%%%%%%%%%%%%%%%%%%%%%%%%%%%%%%%%%%%%%%%%%%%%
% HISTORY
%%%%%%%%%%%%%%%%%%%%%%%%%%%%%%%%%%%%%%%%%%%%%%%%%

\textbf{History.} Cap'n Proto \cite{capnproto} is an RPC protocol and
schema-driven binary serialization specification created in 2013 by Kenton
Varda \footnote{\url{https://github.com/kentonv}}, the primary author of
Protocol Buffers \cite{protocolbuffers} version 2, while working as a
Technology Lead at Sandstorm \footnote{\url{https://sandstorm.io}}. Cap'n Proto
is designed to support memory-efficient serialization and deserialization.
Cap'n Proto is extensively used at Sandstorm and at high-profile companies such
as Cloudflare \footnote{\url{https://www.cloudflare.com}}, where Kenton Varda
is currently employed as a Principal Engineer. Cap'n Proto is released under
the MIT license \footnote{\url{https://opensource.org/licenses/MIT}}.

%%%%%%%%%%%%%%%%%%%%%%%%%%%%%%%%%%%%%%%%%%%%%%%%%
% ADVANTAGES
%%%%%%%%%%%%%%%%%%%%%%%%%%%%%%%%%%%%%%%%%%%%%%%%%

\textbf{Characteristics.} \begin{itemize}

  \item \textbf{Efficient Reads.} Cap'n Proto produces implementations that
    perform runtime-efficiency and memory-efficient incremental and
    random-access reads as noted by \cite{zaluzhnyi2016serialization}.

  \item \textbf{Small Code Footprint.} Cap'n Proto includes a small runtime
    library with minimal dependencies and generates small amounts of
    serialization and deserialization code.

\end{itemize}

%%%%%%%%%%%%%%%%%%%%%%%%%%%%%%%%%%%%%%%%%%%%%%%%%
% OVERALL STRUCTURE
%%%%%%%%%%%%%%%%%%%%%%%%%%%%%%%%%%%%%%%%%%%%%%%%%

\begin{figure}[hb!]
  \frame{\includegraphics[width=\linewidth]{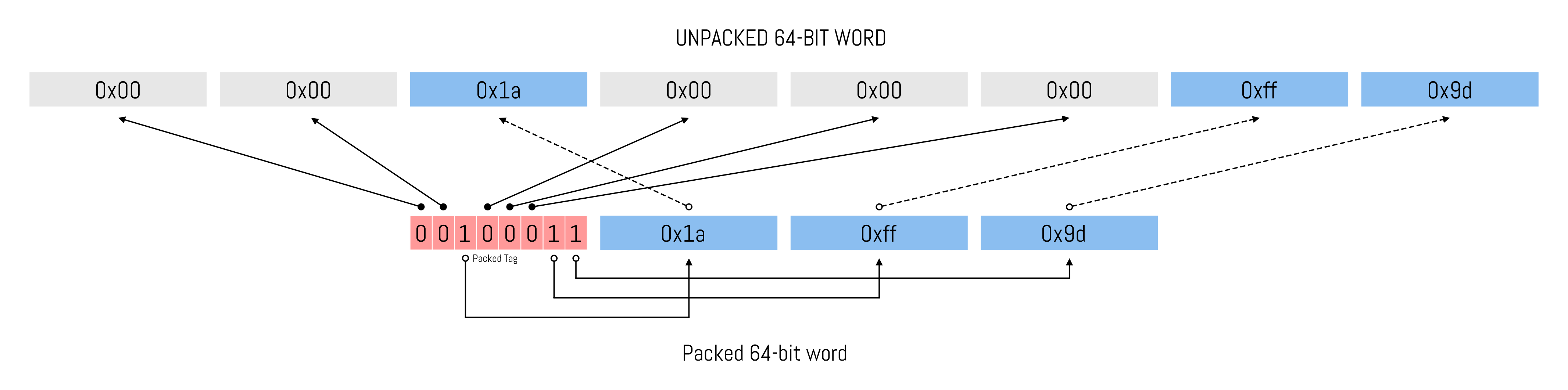}}
  \caption{Cap'n Proto Packed Encoding compresses a 64-bit word by encoding a
  tag byte where its bits represent the unpacked bytes followed by up to 8
  content bytes. The number of content bytes equal the number if bits set in
  the tag byte. If the bit from the tag byte is zero, the corresponding
  unpacked byte is zero. If the bit from the tag byte is set, then the byte is
  the corresponding unpacked byte following the tag byte.}
\label{fig:capnproto-packed-encoding-1} \end{figure}

\textbf{Layout.} A Cap'n Proto bit-string consist of a tree hierarchy of
pointers that eventually points at scalar types. These pointers are scattered
across the bit-string close to the data that they point to for cache locality
purposes. For runtime-performance reasons, Cap'n Proto values are aligned to
64-bit words. As a consequence, Cap'n Proto bit-strings tend to contain
significant zero-byte padding. As a solution, Cap'n Proto defines a simple
compression scheme called \emph{Packed Encoding} where each 64-bit word in the
bit-string is replaced by a tag byte followed by up to 8 content bytes. The
position of the bits set in the tag byte determines the location of each
content byte in the uncompressed 64-bit word. Refer to
\autoref{fig:capnproto-packed-encoding-1} for a visual example. Additionally,
Cap'n Proto Packed Encoding compresses sequences of zero-byte 64-bit using the
\texttt{\textbf{0x00}} byte followed by the amount of zero-byte 64-bit words
minus 1 as an 8-bit unsigned integer. The compression scheme can encode
unpacked data prefixing the unpacked 64-bit word with the
\texttt{\textbf{0xff}} byte and suffixing it with the amount of unpacked words
to follow as an 8-bit unsigned integer.

Cap'n Proto structures consist of a 64-bit type definition, followed by
$\textbf{N}$ 64-bit words of scalar values, followed by $\textbf{M}$ 64-bit
pointers to composite values. The structure type definition and the remaining
64-bit words do not need to be contiguous in memory as the most-significant
30-bits of the structure type definition consists of a pointer to the data
section. The next 2-bits equal $00$ to declare that the type definition
corresponds to a structure. The remaining 32-bits encode two Little Endian
16-bit unsigned integers corresponding to the word-lengths of the data and
pointer sections, respectively.

%%%%%%%%%%%%%%%%%%%%%%%%%%%%%%%%%%%%%%%%%%%%%%%%%
% NUMBERS
%%%%%%%%%%%%%%%%%%%%%%%%%%%%%%%%%%%%%%%%%%%%%%%%%

\textbf{Numbers.} Cap'n Proto supports Little Endian IEEE 754 32-bit and 64-bit
floating-point numbers \cite{8766229}. In terms of integers, Cap'n Proto
supports Little Endian 8-bit, 16-bit, 32-bit, and 64-bit unsigned integers and
Little Endian 8-bit, 16-bit, 32-bit, and 64-bit Two's Complement
\cite{twos-complement} signed integers.

%%%%%%%%%%%%%%%%%%%%%%%%%%%%%%%%%%%%%%%%%%%%%%%%%
% STRINGS
%%%%%%%%%%%%%%%%%%%%%%%%%%%%%%%%%%%%%%%%%%%%%%%%%

\textbf{Strings.} Cap'n Proto encodes strings as lists of UTF-8
\cite{UnicodeStandard} characters. Cap'n Proto strings are delimited with
the \emph{NUL} ASCII \cite{STD80} character. However, the \emph{NUL}
character is typically packed by the Cap'n Proto word compression scheme. The
list definition corresponding to the string encodes the byte-length of the
string as a Little Endian 30-bit unsigned integer. Cap'n Proto does not attempt
to deduplicate multiple occurrences of the same string.

%%%%%%%%%%%%%%%%%%%%%%%%%%%%%%%%%%%%%%%%%%%%%%%%%
% BOOLEANS
%%%%%%%%%%%%%%%%%%%%%%%%%%%%%%%%%%%%%%%%%%%%%%%%%

\textbf{Booleans.} Cap'n Proto encodes booleans as the bits $0$ (False) or $1$
(True) aligned to a multiple of their size on the structure they are defined
in.

%%%%%%%%%%%%%%%%%%%%%%%%%%%%%%%%%%%%%%%%%%%%%%%%%
% ENUMS
%%%%%%%%%%%%%%%%%%%%%%%%%%%%%%%%%%%%%%%%%%%%%%%%%

\textbf{Enumerations.} Cap'n Proto represents enumeration constants using
aligned Little Endian 16-bit unsigned integers. As a consequence, Cap'n Proto
does not support negative enumeration constants.

%%%%%%%%%%%%%%%%%%%%%%%%%%%%%%%%%%%%%%%%%%%%%%%%%
% UNIONS
%%%%%%%%%%%%%%%%%%%%%%%%%%%%%%%%%%%%%%%%%%%%%%%%%

\textbf{Unions.} Cap'n Proto relies on unique field identifiers to implement
union types. Each alternative in the union type must have a different field
identifier and Cap'n Proto enforces that only one of such unique field
identifiers is present at a given time. Cap'n Proto prefixes the encoded value
with an 8-bit unsigned integer that determines the corresponding union field
identifier.

%%%%%%%%%%%%%%%%%%%%%%%%%%%%%%%%%%%%%%%%%%%%%%%%%
% LISTS
%%%%%%%%%%%%%%%%%%%%%%%%%%%%%%%%%%%%%%%%%%%%%%%%%

\textbf{Lists.} Cap'n Proto encodes lists as a 64-bit list type definition with
the elements encoded in order. The list type definition and the list elements
do not need to be contiguous in memory. Cap'n Proto does not attempt to
deduplicate multiple occurrences of the same value in a list. Cap'n Proto
defines two list type definition encodings depending on whether the elements
are scalar or composite as shown in \autoref{table:capnproto-list-definitions}.

\begin{table}[hb!]
\caption{The two 64-bit list type definition encodings supported by Cap'n Proto
  depending on whether the elements of the list are scalar or composite
  values.} \label{table:capnproto-list-definitions}
  \begin{tabularx}{\linewidth}{X|X|p{0.8cm}|X|X}
  \toprule
  \textbf{Element type} & \textbf{First 30-bits} & \textbf{Next 2-bits} & \textbf{Next 29-bits} & \textbf{Remaining 3-bits} \\ \midrule
  Scalar    & Pointer to the start of the list & \texttt{01} & Number of elements in the list as a Little Endian 29-bit unsigned integer & The element length definition as a 3-bit unsigned integer as shown in \autoref{lst:capnproto-element-size} \\ \hline
  Composite & Pointer to the composite tag word definition & \texttt{01} & Number of 64-bit words in the list as a Little Endian 29-bit unsigned integer & \texttt{111} \\
  \bottomrule
\end{tabularx}
\end{table}

If the list consists of composite elements, then the list definitions points at
a 64-bit word that describes each element. This 64-bit word starts with the
number of elements in the list as a Little Endian 30-bit unsigned integer,
followed by the 2-bit constant \texttt{\textbf{00}}, followed by the number of
64-bit scalar words in the element as a Little Endian 16-bit unsigned integer,
followed by the number of 64-bit pointers in the element as another Little
Endian 16-bit unsigned integer.

\begin{figure}[hb!]
\frame{\includegraphics[width=\linewidth]{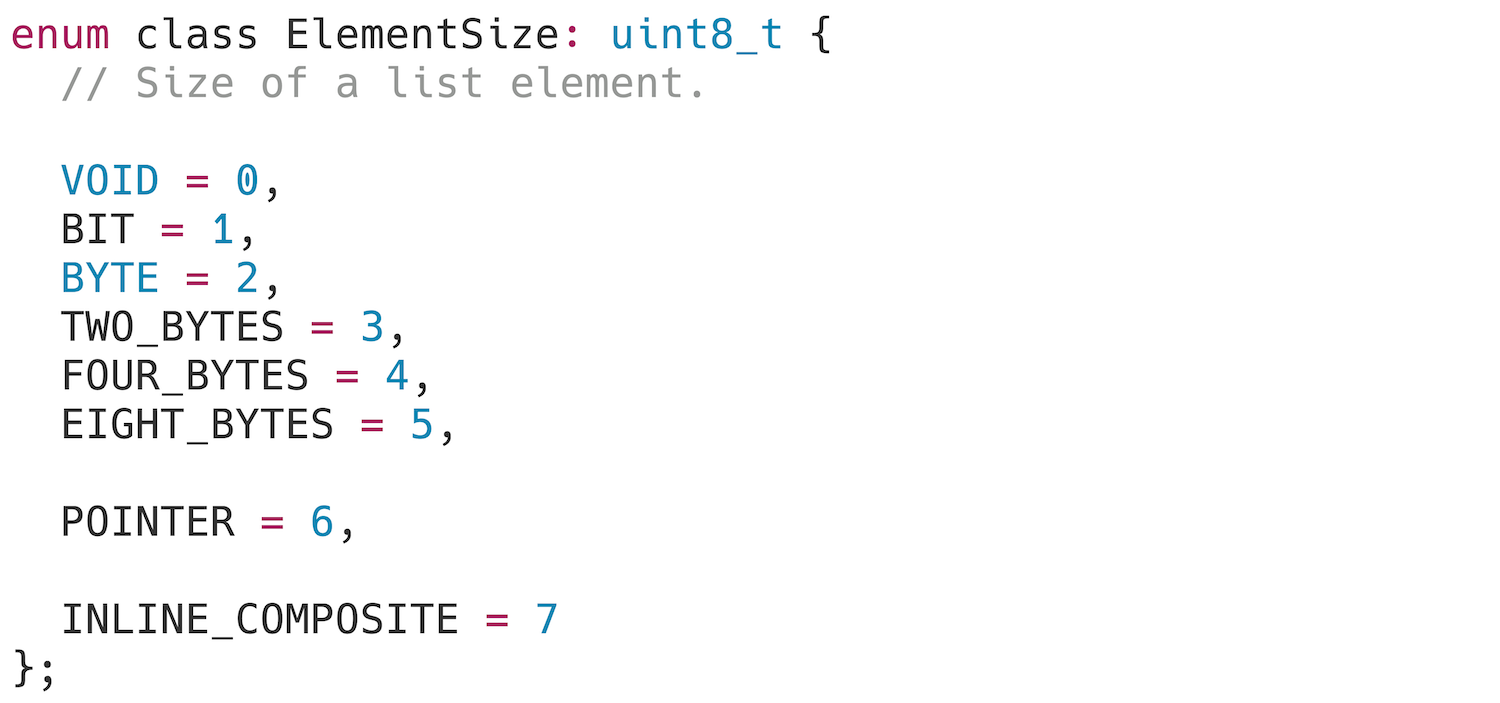}}
\caption{Cap'n Proto list element size definitions \protect\footnotemark.}
\label{lst:capnproto-element-size} \end{figure}

\footnotetext{\url{https://github.com/capnproto/capnproto/blob/bbea19f0e0b2ee1ed28d0836b778d8cf3995597d/c\%2B\%2B/src/capnp/common.h}}

\begin{figure}[hb!]
  \frame{\includegraphics[width=\linewidth]{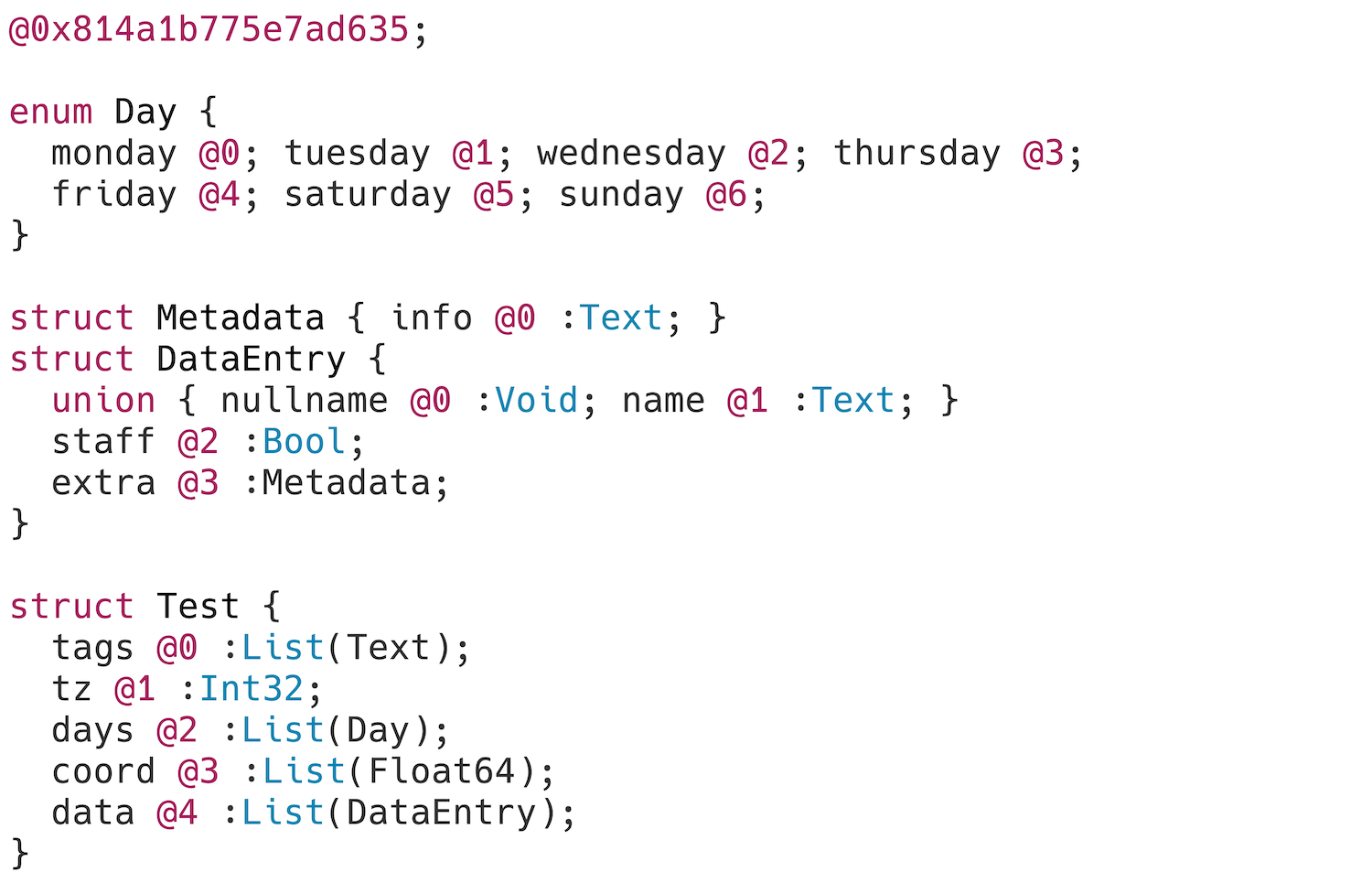}}
  \caption{Cap'n Proto schema to serialize the \autoref{lst:json-object-test} input data.}
  \label{lst:schema-capnproto}
\end{figure}

\begin{figure*}[hb!]
  \frame{\includegraphics[width=\linewidth]{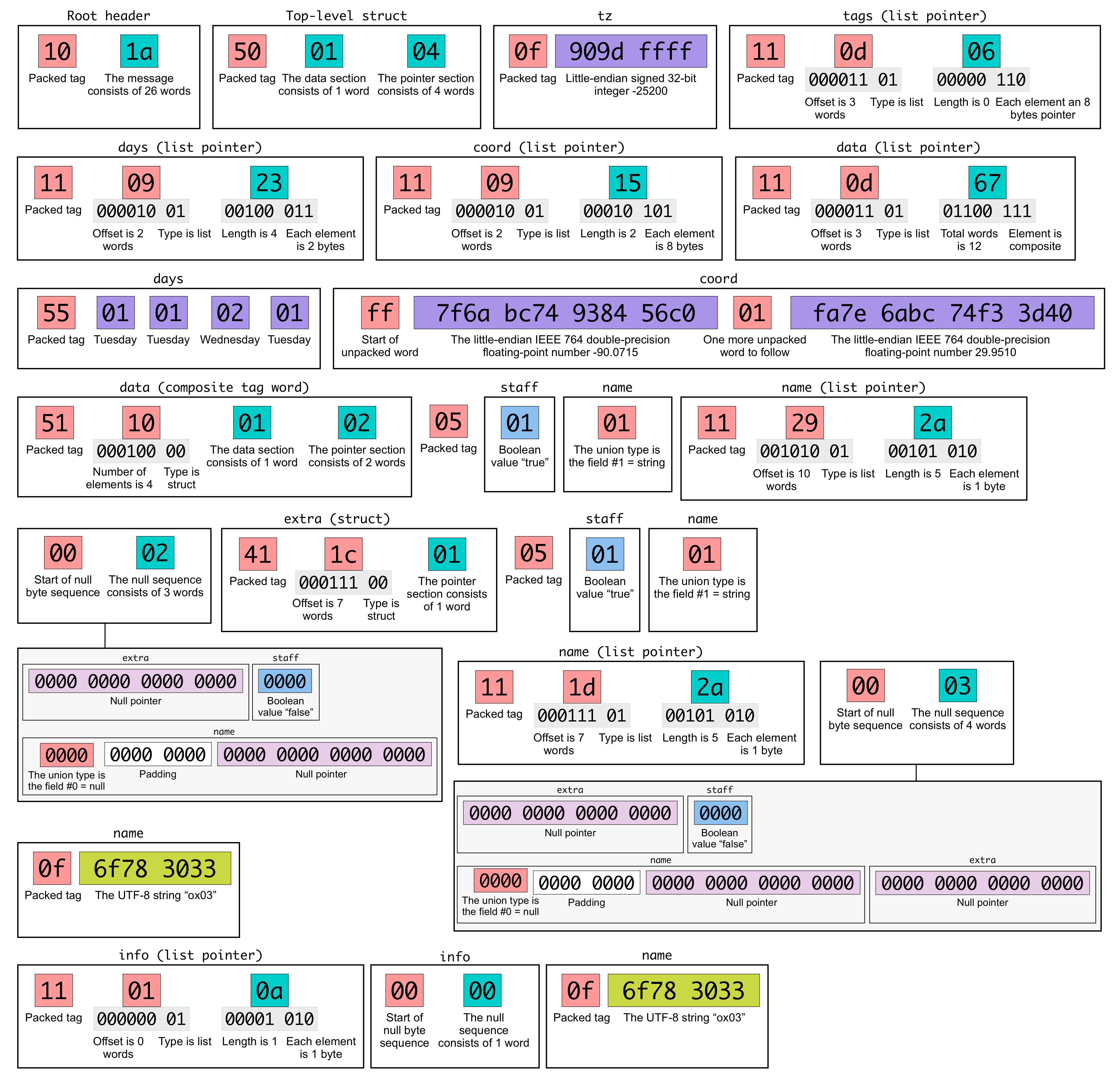}}
  \caption{Annotated hexadecimal output of serializing the
\autoref{lst:json-object-test} input data with Cap'n Proto Packed Encoding.}
\label{lst:capnproto-hex}
\end{figure*}

\begin{table}[hb!]
\caption{A high-level summary of the Cap'n Proto schema-driven serialization specification.}
\label{table:capnproto}
  \begin{tabularx}{\linewidth}{l|X}
  \toprule
  \textbf{Website} & \url{https://capnproto.org} \\ \hline
  \textbf{Company / Individual} & Sandstorm \\ \hline
  \textbf{Year} & 2013 \\ \hline
  \textbf{Specification} & \url{https://capnproto.org/encoding.html} \\ \hline
  \textbf{License} & MIT \\ \hline
  \textbf{Schema Language} & Cap'n Proto IDL \\ \hline
  \textbf{Layout} & Pointed-based and order-based \\ \hline
  \textbf{Languages} & C, C++, C\#, D, Erlang, Go, Haskell, Java, JavaScript, Lua, Nim, OCaml, Python, Ruby, Rust, Scala \\ \hline
  \multicolumn{2}{c}{\textbf{Types}} \\ \hline

  \textbf{Numeric}
  & Little Endian 8-bit, 16-bit, 32-bit, and 64-bit Two's Complement \cite{twos-complement} signed integers \par\smallskip
  Little Endian 8-bit, 16-bit, 32-bit, and 64-bit unsigned integers \par\smallskip
  Little Endian 32-bit and 64-bit IEEE 764 floating-point numbers \cite{8766229} \\ \hline

  \textbf{String} & UTF-8 \cite{UnicodeStandard} \\ \hline
  \textbf{Composite} & Enum, List, Struct, Union \\ \hline
  \textbf{Scalars} & Bool, Void \\ \hline
  \textbf{Other} & Data (byte array) \\
  \bottomrule
\end{tabularx}
\end{table}

%% file: sections/formats/flatbuffers.tex
\label{sec:flatbuffers}

\begin{figure}[hb!]
  \frame{\includegraphics[width=\linewidth]{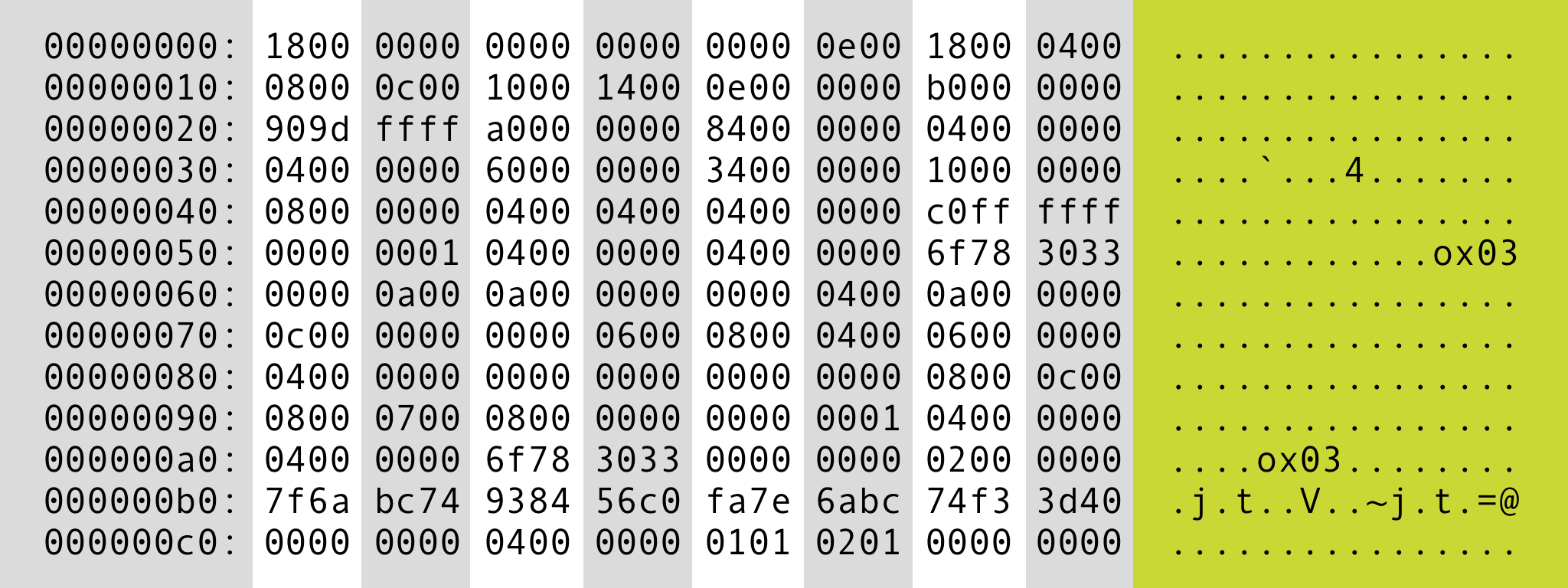}}
  \caption{Hexadecimal output (\texttt{xxd})
  of encoding \autoref{lst:json-object-test}
input data with FlatBuffers Binary Wire Format
(208 bytes).} \label{lst:hex-flatbuffers}
\end{figure}

%%%%%%%%%%%%%%%%%%%%%%%%%%%%%%%%%%%%%%%%%%%%%%%%%
% HISTORY
%%%%%%%%%%%%%%%%%%%%%%%%%%%%%%%%%%%%%%%%%%%%%%%%%

\textbf{History.} FlatBuffers \cite{flatbuffers} is a schema-driven
serialization specification created at Google in 2014 by the \emph{Fun Propulsion
Labs} (FPL) group whose mission was to improve game-related technologies for
Android. FlatBuffers has been designed to support memory-efficient
serialization and deserialization in the context of games and mobile.  The
project was started by Wouter van Oortmerssen
\footnote{\url{https://github.com/aardappel}}, a Software Engineer at Google,
and was released under the Apache License 2.0
\footnote{\url{http://www.apache.org/licenses/LICENSE-2.0.html}}. FlatBuffers
is also used in the context Machine Learning as part of the TensorFlow Lite
\footnote{\url{https://www.tensorflow.org/lite/api\_docs/cc/class/tflite/flat-buffer-model}}
framework for mobile and IoT devices developed by Google \cite{8612133} and in
the DOS spatial system \cite{10.1145/3274895.3274898}.

%%%%%%%%%%%%%%%%%%%%%%%%%%%%%%%%%%%%%%%%%%%%%%%%%
% ADVANTAGES
%%%%%%%%%%%%%%%%%%%%%%%%%%%%%%%%%%%%%%%%%%%%%%%%%

\textbf{Characteristics.} \begin{itemize}

  \item \textbf{Efficient Reads.} FlatBuffers produces implementations that
    perform runtime-efficiency and memory-efficient incremental and
    random-access reads as noted by \cite{9142787}, \cite{10589/150617},
    \cite{8977050}, \cite{10.1007/978-981-15-8697-2_23}, and \cite{8876986}.
    Given its efficient deserialization process,
    \cite{10.1007/978-3-030-02931-9_20} proposes FlatBuffers as the specification in a
    system architecture for data and video streaming with unmanned aerial
    vehicles for natural disaster management.

  \item \textbf{Small Code Footprint.} FlatBuffers includes a small runtime
    library with minimal dependencies and generates small amounts of
    serialization and deserialization code.

\end{itemize}

%%%%%%%%%%%%%%%%%%%%%%%%%%%%%%%%%%%%%%%%%%%%%%%%%
% OVERALL STRUCTURE
%%%%%%%%%%%%%%%%%%%%%%%%%%%%%%%%%%%%%%%%%%%%%%%%%

\begin{figure}[hb!]
  \frame{\includegraphics[width=\linewidth]{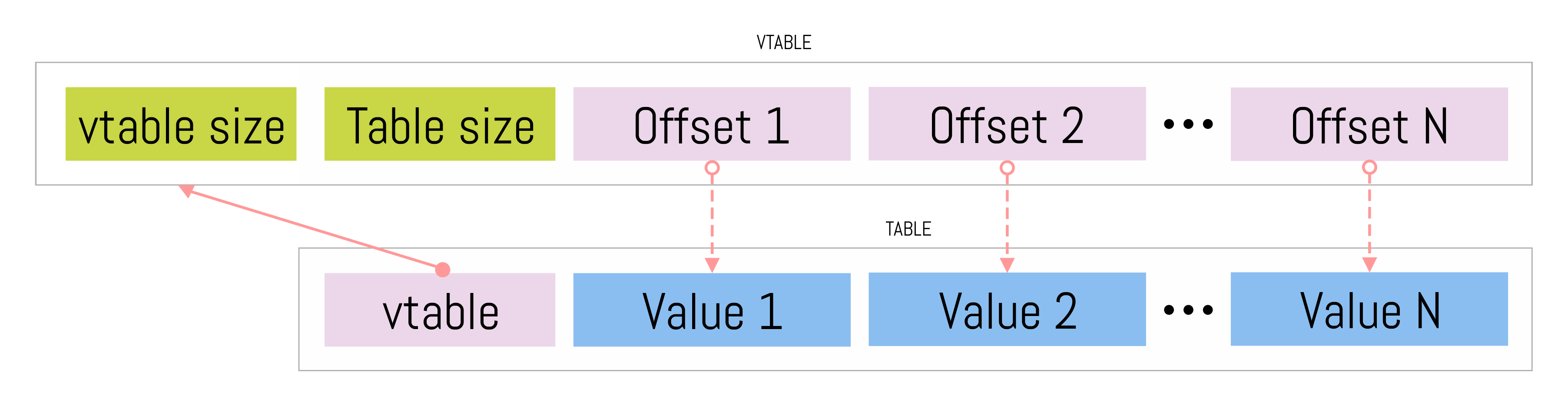}}
  \caption{A Table is an aligned sequence of values. Tables are prefixed with a
  pointer to a \emph{vTable} structure that defines the layout of the Table by
specifying the offsets to each element.} \label{fig:flatbuffers-tables}
\end{figure}

\textbf{Layout.} A FlatBuffers Binary Wire Format bit-string consist of a tree
hierarchy of 32-bit relative pointers that eventually point at scalar types. A
FlatBuffers bit-string starts with a pointer to the root element. The
FlatBuffers core data structure is a \emph{Table}. A FlatBuffers Table is an
ordered sequence of aligned values prefixed with a pointer to a \emph{vTable}
structure that defines the layout of the Table. A vTable consist of two Little
Endian 16-bit unsigned integers describing the byte-lengths of the vTable and
the Table.  The size declarations are followed by a sequence of Little Endian
16-bit unsigned integer offsets to each element in the Table relative to the
vTable pointer.  Refer to \autoref{fig:flatbuffers-tables} for a visual
example. As a space optimisation, multiple Tables sharing the same layout may
point to the same vTable. FlatBuffers does not encode values that equal their
default.

As an alternative to Tables, FlatBuffers supports the concept of
\emph{structs}. A struct is a more space-efficient alternative to a Table,
however a struct can only include scalar values and other structs, and lacks
the versioning and extensibility features of a Table. FlatBuffers encodes
structs as the sequence of its members aligned to the largest scalar element it
contains.

%%%%%%%%%%%%%%%%%%%%%%%%%%%%%%%%%%%%%%%%%%%%%%%%%
% NUMBERS
%%%%%%%%%%%%%%%%%%%%%%%%%%%%%%%%%%%%%%%%%%%%%%%%%

\textbf{Numbers.} FlatBuffers supports Little Endian IEEE 754 32-bit and 64-bit
floating-point numbers \cite{8766229}. In terms of integers, FlatBuffers
supports Little Endian 8-bit, 16-bit, 32-bit and 64-bit unsigned integers and
Little Endian 8-bit, 16-bit, 32-bit and 64-bit Two's Complement
\cite{twos-complement} signed integers.

%%%%%%%%%%%%%%%%%%%%%%%%%%%%%%%%%%%%%%%%%%%%%%%%%
% STRINGS
%%%%%%%%%%%%%%%%%%%%%%%%%%%%%%%%%%%%%%%%%%%%%%%%%

\textbf{Strings.} FlatBuffers produces \emph{NUL}-delimited UTF-8
\cite{UnicodeStandard} strings. FlatBuffers strings are prefixed with a
Little Endian 32-bit unsigned integer that represents the byte-length of the
string without taking the \emph{NUL} delimiter into consideration. Empty
strings are encoded with a length 0 followed by the \emph{NUL} character. By
default, FlatBuffers does not attempt to deduplicate multiple occurrences of
the same string. However, its serialization interface allows the application to
track and share duplicated string values.

%%%%%%%%%%%%%%%%%%%%%%%%%%%%%%%%%%%%%%%%%%%%%%%%%
% BOOLEANS
%%%%%%%%%%%%%%%%%%%%%%%%%%%%%%%%%%%%%%%%%%%%%%%%%

\textbf{Booleans.} FlatBuffers encodes booleans as the Little Endian unsigned
integers $0$ (False) and $1$ (True) aligned to their own size.

%%%%%%%%%%%%%%%%%%%%%%%%%%%%%%%%%%%%%%%%%%%%%%%%%
% ENUMS
%%%%%%%%%%%%%%%%%%%%%%%%%%%%%%%%%%%%%%%%%%%%%%%%%

\textbf{Enumerations.} FlatBuffers lets the schema-writer decide the data type
to represent enumeration constants. A common choice is the
\texttt{\textbf{byte}} type that represents an 8-bit signed Two's Complement
\cite{twos-complement} integer.

%%%%%%%%%%%%%%%%%%%%%%%%%%%%%%%%%%%%%%%%%%%%%%%%%
% UNIONS
%%%%%%%%%%%%%%%%%%%%%%%%%%%%%%%%%%%%%%%%%%%%%%%%%

\textbf{Unions.} FlatBuffers encodes union data types as the combination of two
fields: an enumeration that represents the union alternative choices and the
offset to the union value. FlatBuffers reserves the union identifier $0$ to
mean that the value is not set. FlatBuffers unions do not support scalar data
types. However, a FlatBuffers union may include a \emph{struct} consisting of a
single scalar value encoded with no space overhead.

%%%%%%%%%%%%%%%%%%%%%%%%%%%%%%%%%%%%%%%%%%%%%%%%%
% LISTS
%%%%%%%%%%%%%%%%%%%%%%%%%%%%%%%%%%%%%%%%%%%%%%%%%

\textbf{Lists.} A FlatBuffers list (vector) consists of the concatenation of
its elements prefixed by the logical size of the vector as a Little Endian
32-bit unsigned integer. A vector of composite elements is encoded as a list of
32-bit pointers. By default, FlatBuffers does not attempt to deduplicate
multiple occurrences of the same element in a vector. However, its
serialization interface allows the application to track and share duplicated
vector composite elements.

\begin{figure}[hb!]
  \frame{\includegraphics[width=\linewidth]{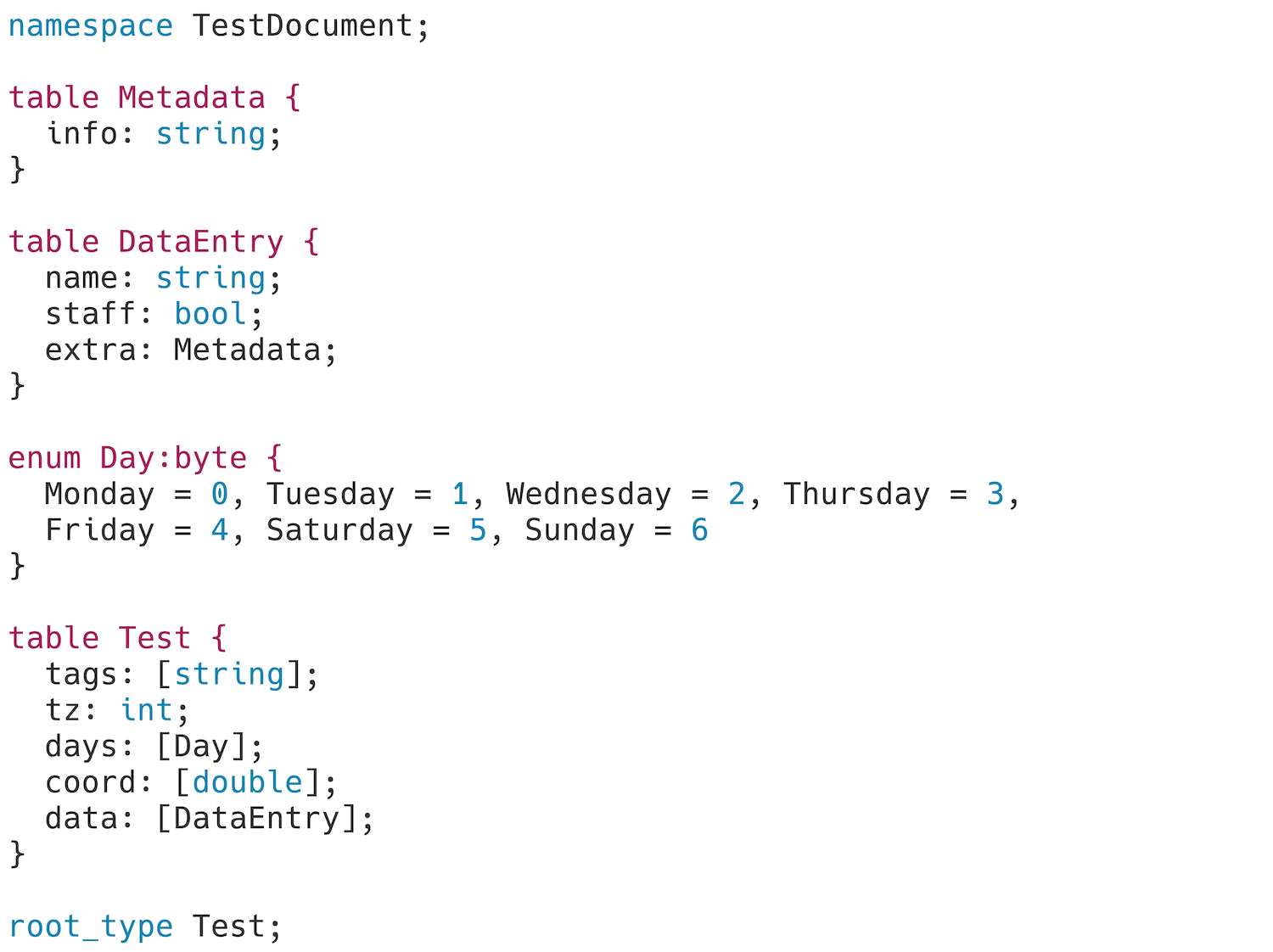}}
  \caption{FlatBuffers schema to serialize the \autoref{lst:json-object-test} input data.}
  \label{lst:schema-flatbuffers}
\end{figure}

\begin{figure*}[hb!]
  \frame{\includegraphics[width=\linewidth]{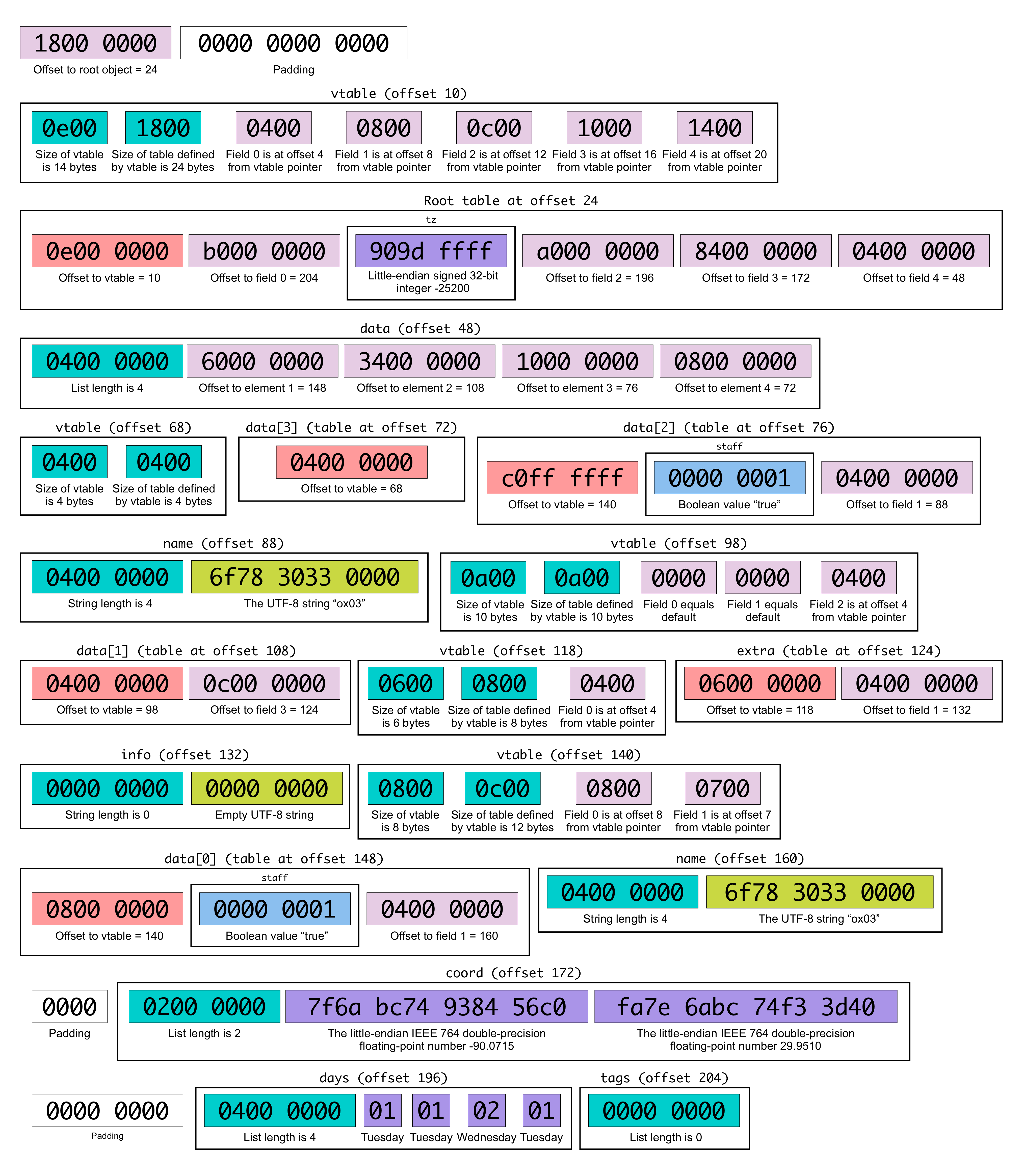}}
  \caption{Annotated hexadecimal output of serializing the
\autoref{lst:json-object-test} input data with FlatBuffers Binary Wire Format.}
\label{lst:flatbuffers-hex}
\end{figure*}

\begin{table}[hb!]
\caption{A high-level summary of the FlatBuffers Binary Wire Format schema-driven serialization specification.}
\label{table:flatbuffers}
\begin{tabularx}{\linewidth}{l|X}
  \toprule
  \textbf{Website} & \url{https://google.github.io/flatbuffers/} \\ \hline
  \textbf{Company / Individual} & Google \\ \hline
  \textbf{Year} & 2014 \\ \hline
  \textbf{Specification} & \url{https://google.github.io/flatbuffers/flatbuffers\_internals.html} \\ \hline
  \textbf{License} & Apache License 2.0 \\ \hline
  \textbf{Schema Language} & FlatBuffers IDL \\ \hline
  \textbf{Layout} & Pointer-based and order-based \\ \hline
  \textbf{Languages} & C, C++, C\#, Go, Java, Kotlin, JavaScript, Lobster, Lua, TypeScript, PHP, Python, Rust, Swift \\ \hline
  \multicolumn{2}{c}{\textbf{Types}} \\ \hline

  \textbf{Numeric} &
  Little Endian 8-bit, 16-bit, 32-bit, and 64-bit unsigned integers \par\smallskip
  Little Endian 8-bit, 16-bit, 32-bit, and 64-bit Two's Complement \cite{twos-complement} signed integers \par\smallskip
  Little Endian 32-bit and 64-bit IEEE 764 floating-point numbers \cite{8766229} \\ \hline

  \textbf{String} & UTF-8 \cite{UnicodeStandard} \\ \hline
  \textbf{Composite} & Array, Enum, Struct, Table, Union, Vector \\ \hline
  \textbf{Scalars} & Boolean \\
  \bottomrule
\end{tabularx}
\end{table}

%% file: sections/formats/protocolbuffers.tex
\label{sec:protocolbuffers}

\begin{figure}[hb!]
  \frame{\includegraphics[width=\linewidth]{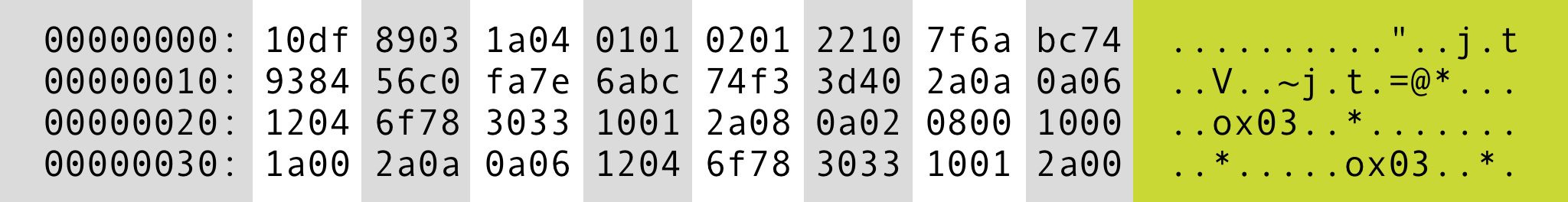}}
  \caption{Hexadecimal output (\texttt{xxd})
  of encoding \autoref{lst:json-object-test}
  input data with Protocol Buffers Binary Wire
  Format (64 bytes).}
  \label{lst:hex-protocolbuffers} \end{figure}

%%%%%%%%%%%%%%%%%%%%%%%%%%%%%%%%%%%%%%%%%%%%%%%%%
% HISTORY
%%%%%%%%%%%%%%%%%%%%%%%%%%%%%%%%%%%%%%%%%%%%%%%%%

\textbf{History.} Protocol Buffers \cite{protocolbuffers} is an RPC protocol
and schema-driven binary serialization specification developed by Google in 2001
\footnote{\url{https://developers.google.com/protocol-buffers/docs/faq}} and
open-sourced under the 3-clause BSD license
\footnote{\url{https://opensource.org/licenses/BSD-3-Clause}} in 2008
\footnote{\url{https://opensource.googleblog.com/2008/07/protocol-buffers-googles-data.html}}.
Protocol Buffers was initially maintained by Jon Skeet
\footnote{\url{https://github.com/jskeet}}, a Staff Software Engineer at
Google.  Google uses Protocol Buffers for nearly all of its storage and
transmission needs.

\textbf{Related Literature.} Protocol Buffers has been extensively optimized in
the context of data centers.  \cite{parimidatacenter} further improve runtime
efficiency of the official Protocol Buffers C++ implementation using hardware
acceleration through a programmed co-processor. \cite{6750189} describes
\emph{ProtoML}, a tool that takes complex constrains definitions and generates
code to validate and potentially correct Protocol Buffers messages at runtime.
\cite{10.1145/3293880.3294105} successfully developed a formally verified
subset (most notably missing unions and recursive messages) of Protocol Buffers
version 3 using the Coq \footnote{\url{https://coq.inria.fr}} formal proof
assistant. Protocol Buffers is used to encode certain parts of the \emph{ProMC}
specification to store representations of high-energy physics data for
space-efficiency reasons \cite{chekanov2013generation}. Protocol Buffers is
also a core component of the Caffe \cite{jia2014caffe} deep learning framework
\footnote{\url{https://caffe.berkeleyvision.org/tutorial/layers.html}} used by
popular projects such as OpenPose, an open-source real-time multi-person
keypoint detection library \cite{8765346}.

%%%%%%%%%%%%%%%%%%%%%%%%%%%%%%%%%%%%%%%%%%%%%%%%%
% ADVANTAGES
%%%%%%%%%%%%%%%%%%%%%%%%%%%%%%%%%%%%%%%%%%%%%%%%%

\textbf{Characteristics.} \begin{itemize}

  \item \textbf{Robustness.} The Protocol Buffers schema-driven serialization
    specification and official implementations have been battle-tested by Google in
    high-scale production environments.

  \item \textbf{Security.} Protocol Buffers has been designed with security in
    mind and has undergone reviews by the Google security team.

  \item \textbf{Popularity.} Protocol Buffers is one of the most popular
    schema-driven binary serialization specifications. As a result, it features
    excellent documentation, relevant tools and an active open source
    community.

\end{itemize}

%%%%%%%%%%%%%%%%%%%%%%%%%%%%%%%%%%%%%%%%%%%%%%%%%
% OVERALL STRUCTURE
%%%%%%%%%%%%%%%%%%%%%%%%%%%%%%%%%%%%%%%%%%%%%%%%%

\textbf{Layout.} A Protocol Buffers Binary Wire Format bit-string is a sequence
of values, sometimes nested. Each value is prefixed with a type definition and
the length of the value, if applicable. Each value has a numeric identifier
that must be unique on the current nesting level. As a space-optimization,
Protocol Buffers only encodes a field if its value is not equal to the default
or if the field is explicitly marked as optional. The order of fields in a
Protocol Buffers Binary Wire Format bit-string is non-deterministic.

%%%%%%%%%%%%%%%%%%%%%%%%%%%%%%%%%%%%%%%%%%%%%%%%%
% KEYS
%%%%%%%%%%%%%%%%%%%%%%%%%%%%%%%%%%%%%%%%%%%%%%%%%

\textbf{Types.} A Protocol Buffers Binary Wire Format type definition is a
32-bit Little Endian Base 128 (LEB128) (\ref{sec:varints}) variable-length
integer encoding of the concatenation of a Big Endian arbitrary-length unsigned
integer identifier and its 3-bit type category identifier. Protocol Buffers
groups the data types it supports into a set of type categories depending on
their length characteristics as shown in \autoref{table:proto-wire-types}. The
de-serializer refers to the schema for the specific data type.

\begin{table}[hb!] \caption{The type categories (wire types) that Protocol
  Buffers supports. The two other wire types that are deprecated and
  unused at the time of this writing are not considered.}

\label{table:proto-wire-types}
  \begin{tabularx}{\linewidth}{p{2cm}|l|X}
    \toprule
\textbf{Name} & \textbf{Identifier} & \textbf{Data types} \\ \midrule
  Variable-length integer values & \texttt{000} & 32-bit and 64-bit Little Endian Base 128 (LEB128) (\ref{sec:varints}) variable-length signed and unsigned integers, booleans, enumerations \\ \hline
  64-bit values                  & \texttt{001} & 64-bit fixed-length signed and unsigned integers, IEEE 764 64-bit floating-point numbers \cite{8766229} \\ \hline
  Length-delimited values        & \texttt{010} & Strings, bytes, messages, lists \\ \hline
  32-bit values                  & \texttt{101} & 32-bit fixed-length signed and unsigned integers, IEEE 764 32-bit floating-point numbers \cite{8766229} \\
  \bottomrule
\end{tabularx}
\end{table}

For example, a type definition consisting of 64-bit value with a unique field
identifier $5$ is encoded as \texttt{\textbf{0x2a = 00101 (5) 001 (wire type)}}
and a type definition consisting of a length-delimited value with a unique
field identifier $17$ is encoded as \texttt{\textbf{0x9a 0x02}}, the Little
Endian Base 128 (LEB128) (\ref{sec:varints}) variable-length integer encoding
of \texttt{\textbf{100011 (35) 010 (wire type)}}.

%%%%%%%%%%%%%%%%%%%%%%%%%%%%%%%%%%%%%%%%%%%%%%%%%
% NUMBERS
%%%%%%%%%%%%%%%%%%%%%%%%%%%%%%%%%%%%%%%%%%%%%%%%%

\textbf{Numbers.} Protocol Buffers supports Little Endian IEEE 754 32-bit and
64-bit floating-point numbers \cite{8766229}. In terms of integers,
Protocol Buffers supports 32-bit and 64-bit fixed-length and Little Endian Base
128 (LEB128) (\ref{sec:varints}) variable-length unsigned integers, 32-bit and
64-bit fixed-length and Little Endian Base 128 (LEB128) (\ref{sec:varints})
variable-length Two's Complement \cite{twos-complement} signed integers,
and 32-bit and 64-bit Little Endian Base 128 (LEB128) (\ref{sec:varints})
variable-length ZigZag-encoded (\ref{sec:zigzag-encoding}) signed integers.
Fixed-length numbers make use of the \emph{32-bit values} or \emph{64-bit
values} wire types while variable-length integers make use of the
\emph{variable-length integer values} wire type as shown in
\autoref{table:proto-wire-types}.

%%%%%%%%%%%%%%%%%%%%%%%%%%%%%%%%%%%%%%%%%%%%%%%%%
% STRINGS
%%%%%%%%%%%%%%%%%%%%%%%%%%%%%%%%%%%%%%%%%%%%%%%%%

\textbf{Strings.} Protocol Buffers Binary Wire Format produces UTF-8
\cite{UnicodeStandard} strings that are not \emph{NUL} delimited encoded
using the \emph{length-delimited} wire type as shown in
\autoref{table:proto-wire-types}.  A Protocol Buffers Binary Wire Format string
is prefixed with a 32-bit Little Endian Base 128 (LEB128) (\ref{sec:varints})
variable-length unsigned integer that declares the byte-length of the string.
Protocol Buffers Binary Wire Format encodes empty strings with a zero-length
prefix and no additional data.  Protocol Buffers Binary Wire Format does not
deduplicate multiple occurrences of the same string.

%%%%%%%%%%%%%%%%%%%%%%%%%%%%%%%%%%%%%%%%%%%%%%%%%
% BOOLEANS
%%%%%%%%%%%%%%%%%%%%%%%%%%%%%%%%%%%%%%%%%%%%%%%%%

\textbf{Booleans} Protocol Buffers Binary Wire Format encodes booleans using
the \emph{variable-length integer} wire type as shown in
\autoref{table:proto-wire-types} and the 32-bit Little Endian Base 128 (LEB128)
(\ref{sec:varints}) variable-length integers $0$ (False) and $1$ (True). In
practice, these variable-length integers are encoded as the 8-bit constants
\texttt{\textbf{0x00}} and \texttt{\textbf{0x01}}, respectively.

%%%%%%%%%%%%%%%%%%%%%%%%%%%%%%%%%%%%%%%%%%%%%%%%%
% ENUMS
%%%%%%%%%%%%%%%%%%%%%%%%%%%%%%%%%%%%%%%%%%%%%%%%%

\textbf{Enumerations.} Protocol Buffers Binary Wire Format represents
enumeration constants using 32-bit Little Endian Base 128 (LEB128)
(\ref{sec:varints}) variable-length Two's Complement \cite{twos-complement}
signed integers.

%%%%%%%%%%%%%%%%%%%%%%%%%%%%%%%%%%%%%%%%%%%%%%%%%
% UNIONS
%%%%%%%%%%%%%%%%%%%%%%%%%%%%%%%%%%%%%%%%%%%%%%%%%

\textbf{Unions.} Protocol Buffers relies on unique field identifiers to
implement union types called \emph{oneof}. Each alternative in the union type
must have a different field identifier and Protocol Buffers enforces that only
one of such unique field identifiers is present at a given time. The
de-serializer knows the type of the encoded value by comparing its field
identifier against the union definition.

%%%%%%%%%%%%%%%%%%%%%%%%%%%%%%%%%%%%%%%%%%%%%%%%%
% ARRAYS
%%%%%%%%%%%%%%%%%%%%%%%%%%%%%%%%%%%%%%%%%%%%%%%%%

\textbf{Lists.} Protocol Buffers Binary Wire Format encodes a list (a
\emph{repeated} field) by encoding more than one value with the same unique
field identifier. The type definitions corresponding to each element in the
list use the \emph{length-delimited} wire type as shown in
\autoref{table:proto-wire-types}, and include the byte-length of the value as a
32-bit Little Endian Base 128 (LEB128) (\ref{sec:varints}) variable-length
unsigned integer followed by the value. Lists that consist of scalar values of
the same type can be encoded as a single \emph{length-delimited} field,
followed by the cummulative byte-length of the elements as a 32-bit Little
Endian Base 128 (LEB128) (\ref{sec:varints}) variable-length unsigned integer,
followed by the elements encoded in order.  Protocol Buffers does not natively
support heterogeneous or multi-dimensional lists. As a workaround,
schema-writers may define lists of structures including unions or lists.
Protocol Buffers Binary Wire Format does not deduplicate multiple occurrences
of the same element in a list.

\begin{figure}[hb!]
  \frame{\includegraphics[width=\linewidth]{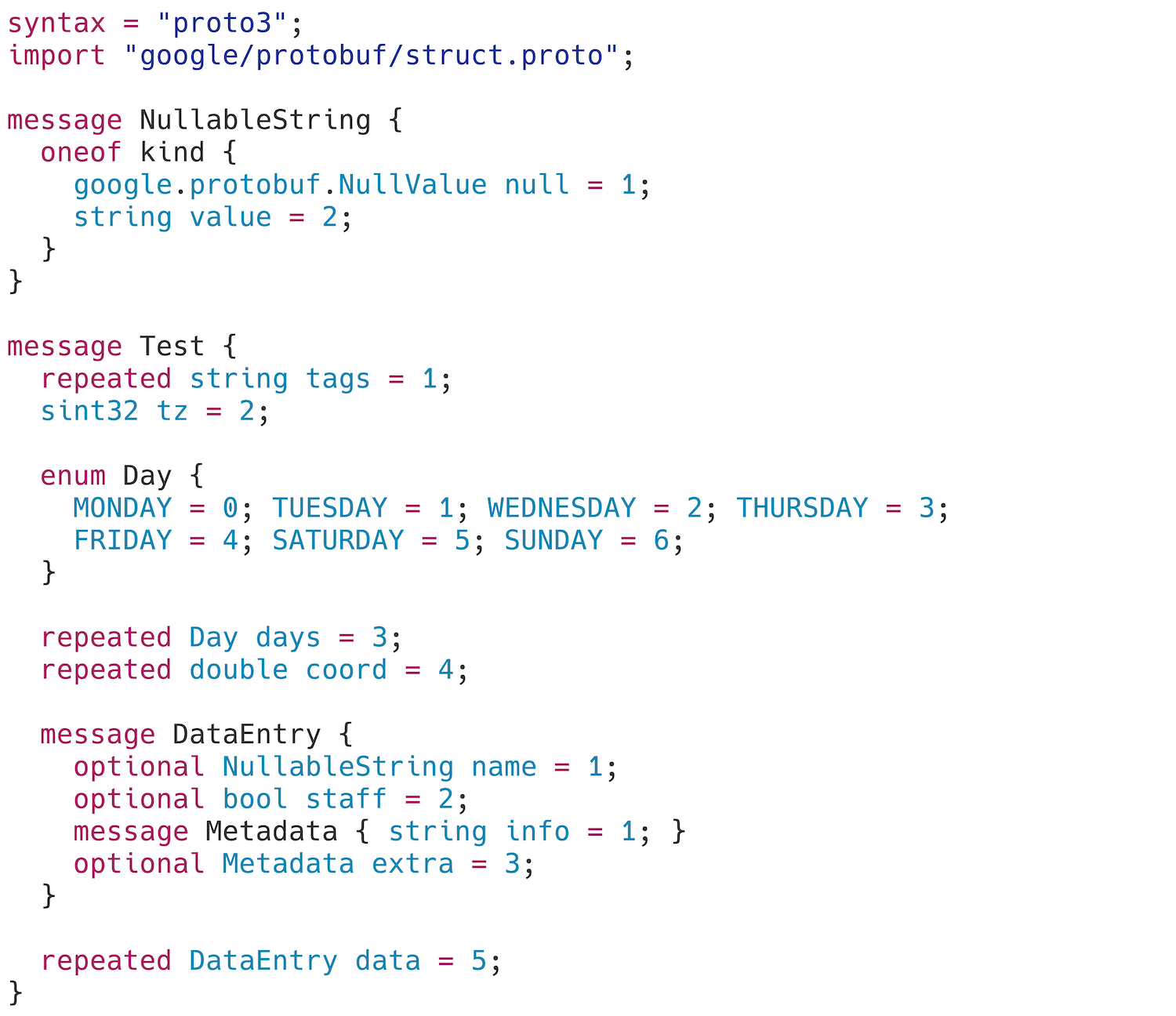}}
  \caption{Protocol Buffers schema to serialize the \autoref{lst:json-object-test} input data.}
  \label{lst:schema-protocolbuffers}
\end{figure}

\begin{figure*}[hb!]
  \frame{\includegraphics[width=\linewidth]{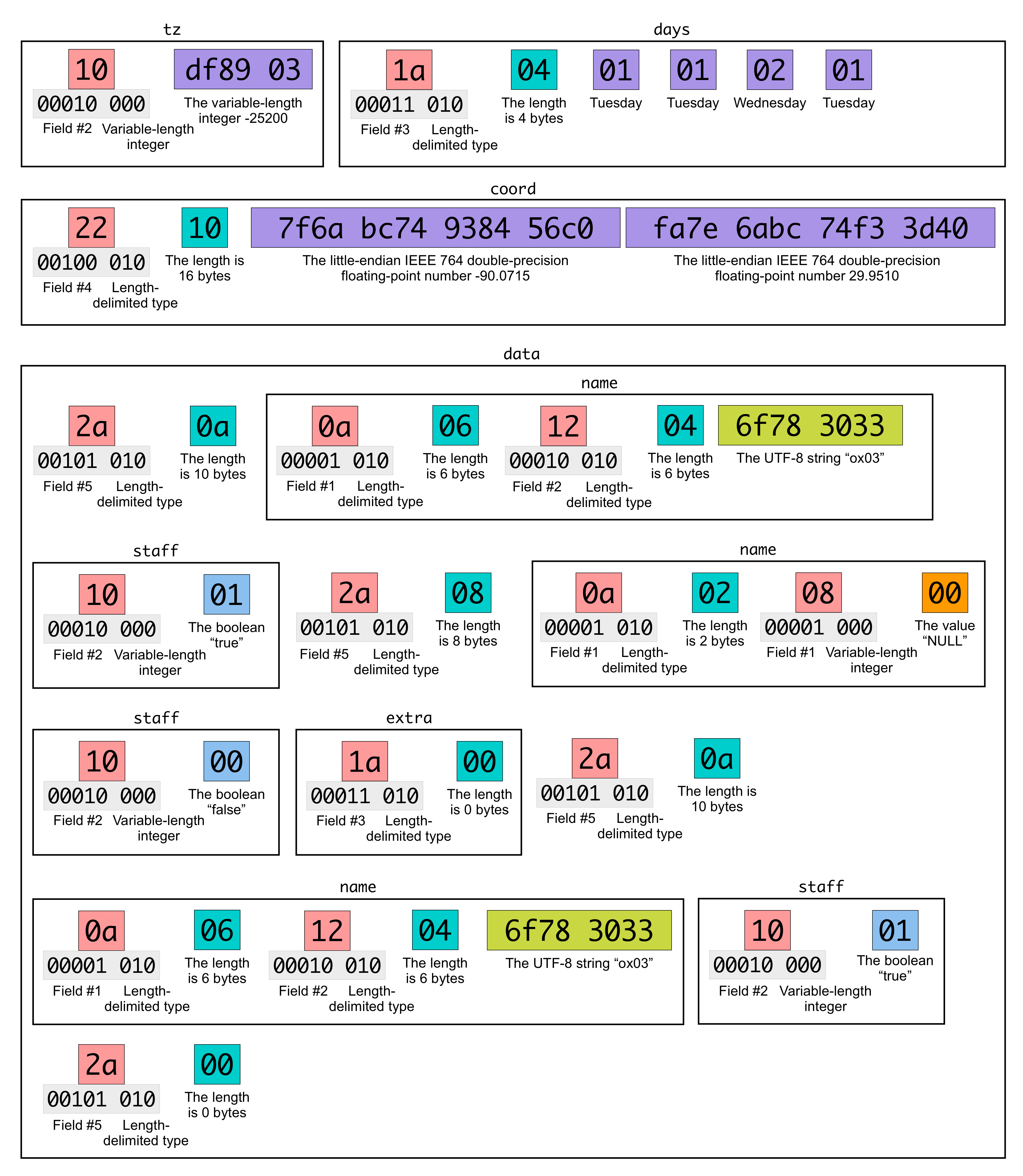}}
  \caption{Annotated hexadecimal output of serializing the
\autoref{lst:json-object-test} input data with Protocol Buffers Binary Wire Format.}
\end{figure*}

\begin{table}[hb!]
\caption{A high-level summary of the Protocol Buffers Binary Wire Format schema-driven serialization specification.}
\label{table:protocolbuffers}
  \begin{tabularx}{\linewidth}{l|X}
  \toprule
  \textbf{Website} & \url{https://developers.google.com/protocol-buffers/} \\ \hline
  \textbf{Company / Individual} & Google \\ \hline
  \textbf{Year} & 2001 \\ \hline
  \textbf{Specification} & \url{https://developers.google.com/protocol-buffers/docs/encoding} \\ \hline
  \textbf{License} & 3-clause BSD \\ \hline
  \textbf{Schema Language} & Proto3 \\ \hline
  \textbf{Layout} & Sequential with field identifiers \\ \hline
  \textbf{Languages} & C++, C\#, Dart, Go, Java, Objective-C, Python, Ruby \\ \hline
  \multicolumn{2}{c}{\textbf{Types}} \\ \hline

  \textbf{Numeric} &
  32-bit and 64-bit Two’s Complement \cite{twos-complement} Little Endian Base 128 (LEB128) (\ref{sec:varints}) variable-length signed integers \par\smallskip
  32-bit and 64-bit Little Endian Base 128 (LEB128) (\ref{sec:varints}) variable-length unsigned integers \par\smallskip
  32-bit and 64-bit ZigZag-encoded (\ref{sec:zigzag-encoding}) Little Endian Base 128 (\ref{sec:varints}) variable-length signed integers \par\smallskip
  Little Endian 32-bit and 64-bit fixed-length unsigned integers \par\smallskip
  Little Endian 32-bit and 64-bit fixed-length Two's Complement \cite{twos-complement} signed integers \par\smallskip
  Little Endian 32-bit and 64-bit IEEE 764 floating-point numbers \cite{8766229} \\ \hline

  \textbf{String} & UTF-8 \cite{UnicodeStandard} \\ \hline
  \textbf{Composite} & Any, Enum, List, Map, Message, Oneof, Struct \\ \hline
  \textbf{Scalars} & Boolean \\ \hline

  \textbf{Other} &
  Bytes (byte array) \par\smallskip
  Timestamp \cite{RFC3339} \par\smallskip
  Duration \footnotemark \\
  \bottomrule
\end{tabularx}
\end{table}

\footnotetext{\url{https://developers.google.com/protocol-buffers/docs/proto3\#json}}

%% file: sections/formats/thrift.tex
\label{sec:apache-thrift}

\begin{figure}[hb!]
  \frame{\includegraphics[width=\linewidth]{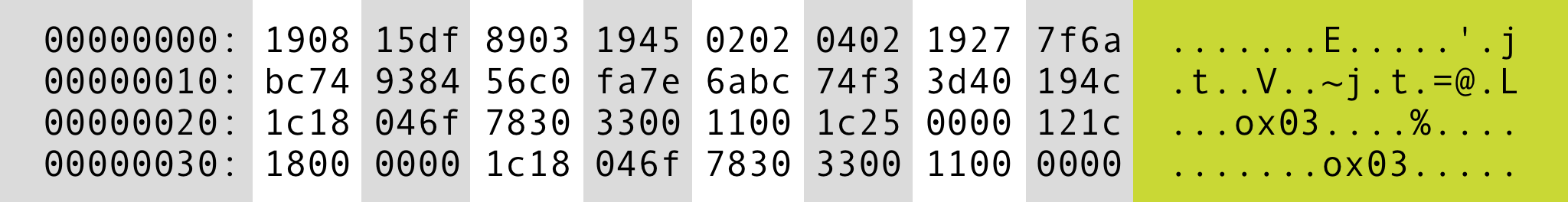}}
  \caption{Hexadecimal output (\texttt{xxd})
  of encoding \autoref{lst:json-object-test}
input data with Apache Thrift Compact Protocol
(64 bytes).} \label{lst:hex-thrift}
\end{figure}

%%%%%%%%%%%%%%%%%%%%%%%%%%%%%%%%%%%%%%%%%%%%%%%%%
% HISTORY
%%%%%%%%%%%%%%%%%%%%%%%%%%%%%%%%%%%%%%%%%%%%%%%%%

\textbf{History.} Apache Thrift \cite{slee2007thrift} is an RPC protocol and
schema-driven binary serialization specification developed at Facebook in 2006 and
donated to the Apache Software Foundation
\footnote{\url{https://www.apache.org}}.  Apache Thrift graduated from the
Apache Incubator in 2010 and is released under the Apache License 2.0
\footnote{\url{http://www.apache.org/licenses/LICENSE-2.0.html}}. Apache Thrift
is used in a large number of scalable backend services at Facebook. Apache
Thrift is also used as the transmission format of the Carat
\cite{10.1145/2517351.2517354} large-scale research project to collect
energy-related analytics from iOS and Android devices which collected 1.5 TB of
data as of 2016 \cite{7840871}.

\textbf{Related Literature.} \cite{madani2016re} attempts to re-implement
Apache Thrift using model-driven engineering technologies such as Xtext
\footnote{\url{https://www.eclipse.org/Xtext/}}, Eclipse Modeling Framework
(EMF) \footnote{\url{https://www.eclipse.org/modeling/emf/}}, and Eclipse
Epsilon \footnote{\url{https://www.eclipse.org/epsilon/}} resulting in a
significantly more concise implementation in terms of lines of code. The
results are published on GitHub
\footnote{\url{https://github.com/SMadani/ThriftMDE/}}. \cite{7863681} explores
automatically generating Apache Thrift service definitions as space and runtime
efficient proxies to XML-based \cite{Paoli:06:EML} SOAP \cite{gudgin2003soap}
web services. \cite{10.1145/3269961.3269973} proposes an architecture to
compose Apache Thrift services with with other Apache Thrift, SOAP
\cite{gudgin2003soap}, and REST \cite{fielding2000architectural} services using
the Web Services Business Process Execution Language (BPEL) \cite{wsbpel}.
\cite{10.1145/2818869.2818892} performs high-performance large-scale datacenter
backups using Apache Thrift on the Apache HBase
\footnote{\url{https://hbase.apache.org}} and Apache Cassandra
\footnote{\url{http://cassandra.apache.org}} NoSQL databases. \cite{8362750}
explores the implications of a microservices architecture based on Apache
Thrift for a movie renting, streaming, and reviewing system comprised of 33
microservice. \cite{8090961} proposes an offline and online database system
based on Conflict-free Replicated Data Types (CRDT)
\cite{10.1007/978-3-642-24550-3_29} which uses Apache Thrift as the middleware
serialization specification.

%%%%%%%%%%%%%%%%%%%%%%%%%%%%%%%%%%%%%%%%%%%%%%%%%
% ADVANTAGES
%%%%%%%%%%%%%%%%%%%%%%%%%%%%%%%%%%%%%%%%%%%%%%%%%

\textbf{Characteristics.} \begin{itemize}

  \item \textbf{Native Type Mappings.} To ease integration, Apache Thrift
    implementations do not introduce Apache Thrift-specific types or wrapper
    objects. Instead, the implementations make use of programming language
    native types.

  \item \textbf{Portability.} Apache Thrift has well-maintained official
    implementations for a large number of programming languages.

\end{itemize}

%%%%%%%%%%%%%%%%%%%%%%%%%%%%%%%%%%%%%%%%%%%%%%%%%
% OVERALL STRUCTURE
%%%%%%%%%%%%%%%%%%%%%%%%%%%%%%%%%%%%%%%%%%%%%%%%%

\textbf{Layout.} An Apache Thrift Compact Protocol bit-string is a sequence of
values, sometimes nested. Each value is prefixed with a type definition and the
length of the value, if applicable. Each value has a numeric identifier that
must be unique on the current nesting level. Based on \our observations, Apache
Thrift Compact Protocol encodes fields even if their values equal their
explicitly-set defaults.  Apache Thrift Compact Protocol structures are
suffixed with the constant byte \texttt{\textbf{0x00}}.

%%%%%%%%%%%%%%%%%%%%%%%%%%%%%%%%%%%%%%%%%%%%%%%%%
% KEYS
%%%%%%%%%%%%%%%%%%%%%%%%%%%%%%%%%%%%%%%%%%%%%%%%%

\begin{figure}[hb!]
\frame{\includegraphics[width=\linewidth]{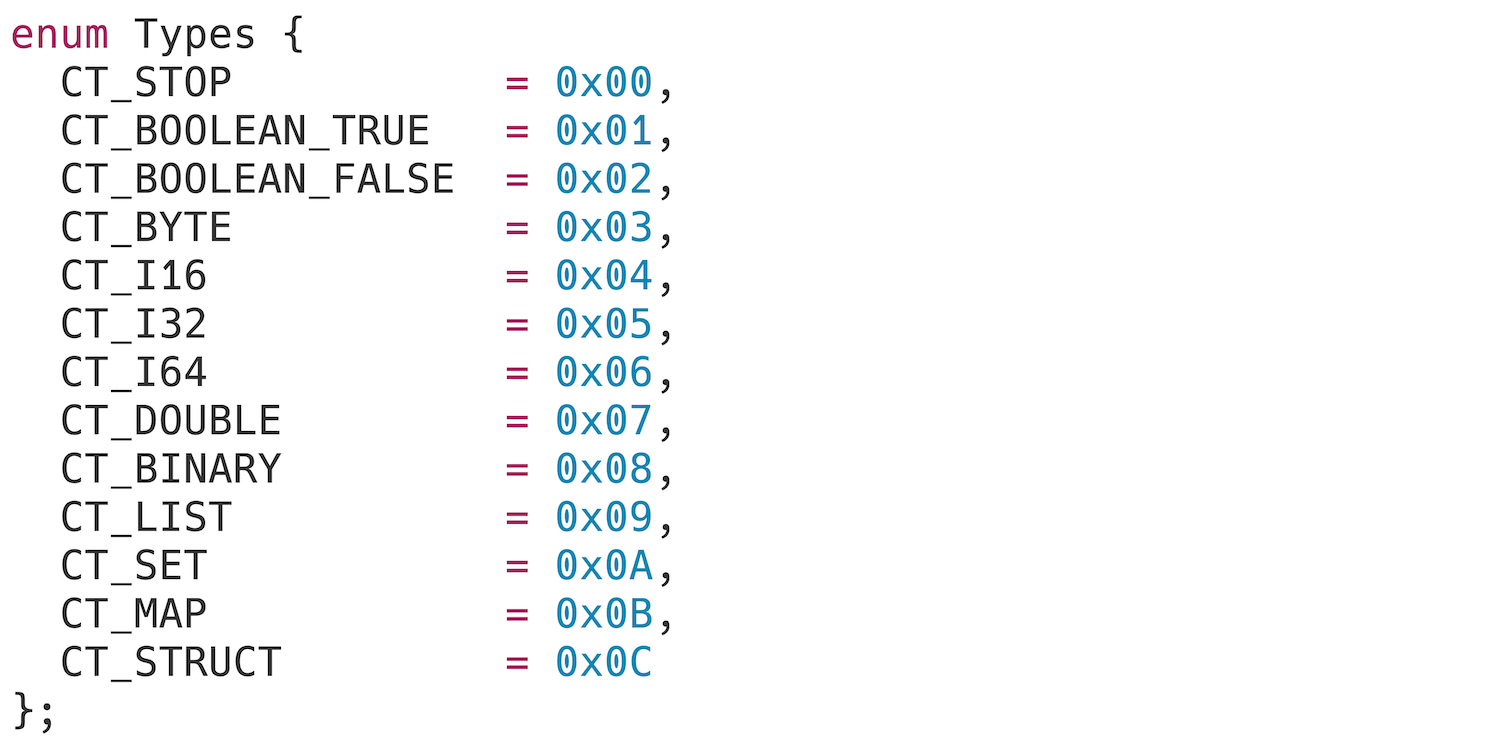}}
\caption{Apache Thrift Compact Protocol struct type identifiers definition
\protect\footnotemark.} \label{lst:thrift-struct-types} \end{figure}

\textbf{Keys.} Apache Thrift Compact Protocol type definitions consist in a
unique field identifier and a type identifier as shown in
\autoref{lst:thrift-struct-types}.  Apache Thrift Compact Protocol supports two
type definition encodings: the \emph{Short form} and the \emph{Long form}
depending on whether the field identifiers are encoded in a relative or
absolute manner. Refer to \autoref{table:thrift-type-definitions} for details.

\begin{table}[hb!]
\caption{Apache Thrift Compact Protocol supports two structure type definition
  encodings. The \emph{Long form} encoding represents the field identifier as
  an absolute integer while the \emph{Short form} encoding represents the field
  identifier as the difference from the previous field identifier as a space
  optimization.}

\label{table:thrift-type-definitions}
  \begin{tabularx}{\linewidth}{p{1cm}|p{1cm}|X|X|X}
  \toprule
  \textbf{Name} & \textbf{Size} & \textbf{First 4-bits} & \textbf{Next 4-bits} & \textbf{Remaining bits} \\ \midrule
  Short form & 1 byte & The unique identifier delta as a 4-bit unsigned integer greater than 0 & The field type identifier constant as shown in \autoref{lst:thrift-struct-types} & None \\ \hline
  Long form  & 2 to 4 bytes & \texttt{0000} & The field type identifier constant as shown in \autoref{lst:thrift-struct-types} & The unique identifier as a 16-bit Little Endian Base 128 (LEB128) (\ref{sec:varints}) variable-length ZigZag-encoded (\ref{sec:zigzag-encoding}) signed integer \\
  \bottomrule
\end{tabularx}
\end{table}

\begin{figure}[hb!]
  \frame{\includegraphics[width=\linewidth]{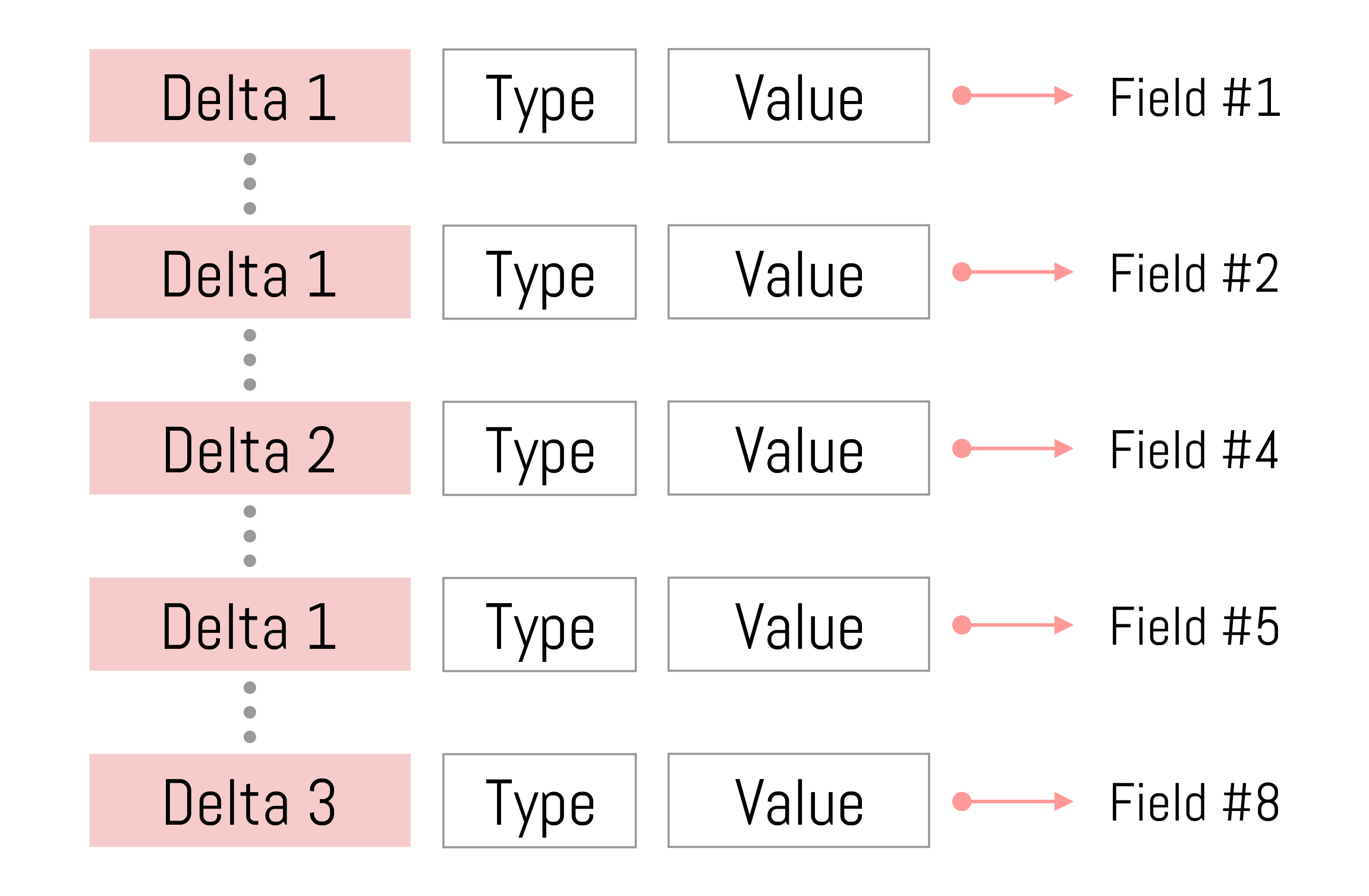}}
  \caption{We can think of Apache Thrift fields as triplets consisting of a
  field delta, a field type, and a field value. Field deltas determine the
  unique identifier of the current field based on the previous delta values. A
  field identifier is the sum of all the deltas up to that field.}
\label{fig:thrift-deltas} \end{figure}

\autoref{fig:thrift-deltas} illustrates a visual representation of the field
identifier delta approach from the \emph{Short form} encoding. Apache Thrift
Compact Protocol implementations will choose the \emph{Short form} delta-based
encoding unless the unique identifier delta exceeds the value $15$ or if the
unique identifier is negative. Apache Thrift encourages schema-writer to set
explicit unique field identifiers, which must be positive. However, Apache
Thrift will automatically assign negative unique field identifiers by default.

\footnotetext{\url{https://github.com/apache/thrift/blob/05bb55148608b4324a8c91c21cf9a6a0dff282fa/lib/cpp/src/thrift/protocol/TCompactProtocol.tcc}}

%%%%%%%%%%%%%%%%%%%%%%%%%%%%%%%%%%%%%%%%%%%%%%%%%
% NUMBERS
%%%%%%%%%%%%%%%%%%%%%%%%%%%%%%%%%%%%%%%%%%%%%%%%%

\textbf{Numbers.} Apache Thrift supports 16-bit, 32-bit, and 64-bit Little
Endian Base 128 (LEB128) (\ref{sec:varints}) variable-length ZigZag-encoded
(\ref{sec:zigzag-encoding}) signed integers. Apache Thrift does not support
fixed-length integer types. However, Apache Thrift supports a
\texttt{\textbf{byte}} type that can represent a fixed-length 8-bit
ZigZag-encoded (\ref{sec:zigzag-encoding}) signed integer. In terms of real
numbers, Apache Thrift supports Little Endian IEEE 764 64-bit floating-point
numbers \cite{8766229}.

%%%%%%%%%%%%%%%%%%%%%%%%%%%%%%%%%%%%%%%%%%%%%%%%%
% STRINGS
%%%%%%%%%%%%%%%%%%%%%%%%%%%%%%%%%%%%%%%%%%%%%%%%%

\textbf{Strings.} Apache Thrift Compact Protocol produces UTF-8
\cite{UnicodeStandard} strings that are not \emph{NUL} delimited. Apache
Thrift Compact Protocol strings are prefixed with positive 32-bit Little Endian
Base 128 (\ref{sec:varints}) variable-length ZigZag-encoded
(\ref{sec:zigzag-encoding}) signed integers declaring the byte-length of the
string. Apache Thrift Compact Protocol encodes empty strings with a zero-length
prefix and no additional data. Apache Thrift Compact Protocol does not attempt
to deduplicate multiple occurrences of the same string.

%%%%%%%%%%%%%%%%%%%%%%%%%%%%%%%%%%%%%%%%%%%%%%%%%
% BOOLEANS
%%%%%%%%%%%%%%%%%%%%%%%%%%%%%%%%%%%%%%%%%%%%%%%%%

\textbf{Booleans.} Apache Thrift Compact Protocol encodes a boolean field and a
list of booleans in different manners. If the value is a standalone field, then
Apache Thrift Compact Protocol encodes the boolean value at the type definition
level using the \texttt{\textbf{CT\_BOOLEAN\_TRUE}} and
\texttt{\textbf{CT\_BOOLEAN\_FALSE}} data types shown in
\autoref{lst:thrift-struct-types}. If the boolean value is a member of a list,
then it is encoded using the 8-bit ZigZag-encoded (\ref{sec:zigzag-encoding})
signed integers \texttt{\textbf{0x02}} (True) and \texttt{\textbf{0x00}}
(False).

%%%%%%%%%%%%%%%%%%%%%%%%%%%%%%%%%%%%%%%%%%%%%%%%%
% ENUMS
%%%%%%%%%%%%%%%%%%%%%%%%%%%%%%%%%%%%%%%%%%%%%%%%%

\textbf{Enumerations.} Apache Thrift Compact Protocol represents enumeration
constants using positive 32-bit ZigZag-encoded (\ref{sec:zigzag-encoding})
Little Endian Base 128 (\ref{sec:varints}) variable-length signed integers.

%%%%%%%%%%%%%%%%%%%%%%%%%%%%%%%%%%%%%%%%%%%%%%%%%
% UNIONS
%%%%%%%%%%%%%%%%%%%%%%%%%%%%%%%%%%%%%%%%%%%%%%%%%

\textbf{Unions.} Apache Thrift relies on unique field identifiers to implement
union types. Each alternative in the union type must have a different field
identifier and Apache Thrift enforces that only one of such unique field
identifiers is present at a given time. The de-serializer knows the type of the
encoded value by comparing its field identifier against the union definition.

%%%%%%%%%%%%%%%%%%%%%%%%%%%%%%%%%%%%%%%%%%%%%%%%%
% ARRAYS
%%%%%%%%%%%%%%%%%%%%%%%%%%%%%%%%%%%%%%%%%%%%%%%%%

\textbf{Lists.} Apache Thrift Compact Protocol encodes a list as the type
definition, followed by the list definition, followed by the elements encoded
in order. Apache Thrift Compact Protocol specifies two encodings for the list
definition depending on the length of the list as shown in
\autoref{table:thrift-list-types}. Apache Thrift Compact Protocol does not
attempt to deduplicate multiple occurrences of the same element in a list.
Apache Thrift natively supports lists of lists and list of unions as a
mechanism to support heterogeneous lists.

\begin{table}[hb!]
\caption{The two list definition encodings supported by Apache Thrift Compact
  Protocol depending on the list length as determined by the \emph{From} and
  \emph{To} ranges.  Note that Apache Thrift Compact Protocol cannot encode
  lists containing more than $2^{32} - 1$ elements.}
  \label{table:thrift-list-types}
  \begin{tabularx}{\linewidth}{l|l|l|X|X|X}
  \toprule
  \textbf{From} & \textbf{To} & \textbf{Size} & \textbf{First 4-bits} & \textbf{Next 4-bits} & \textbf{Remaining bits} \\ \midrule
  0 & 14         & 1 byte       & Length of list as a 4-bit unsigned integer & Element type identifier constant as shown in \autoref{lst:thrift-list-types} & None \\ \hline
  15 & $2^{32} - 1$ & 2 to 6 bytes & \texttt{1111} & Element type identifier constant as shown in \autoref{lst:thrift-list-types} & Length of list as a positive 32-bit Little Endian Base 128 (\ref{sec:varints}) variable-length ZigZag-encoded (\ref{sec:zigzag-encoding}) signed integer \\
  \bottomrule
\end{tabularx}
\end{table}

\begin{figure}[hb!]
  \frame{\includegraphics[width=\linewidth]{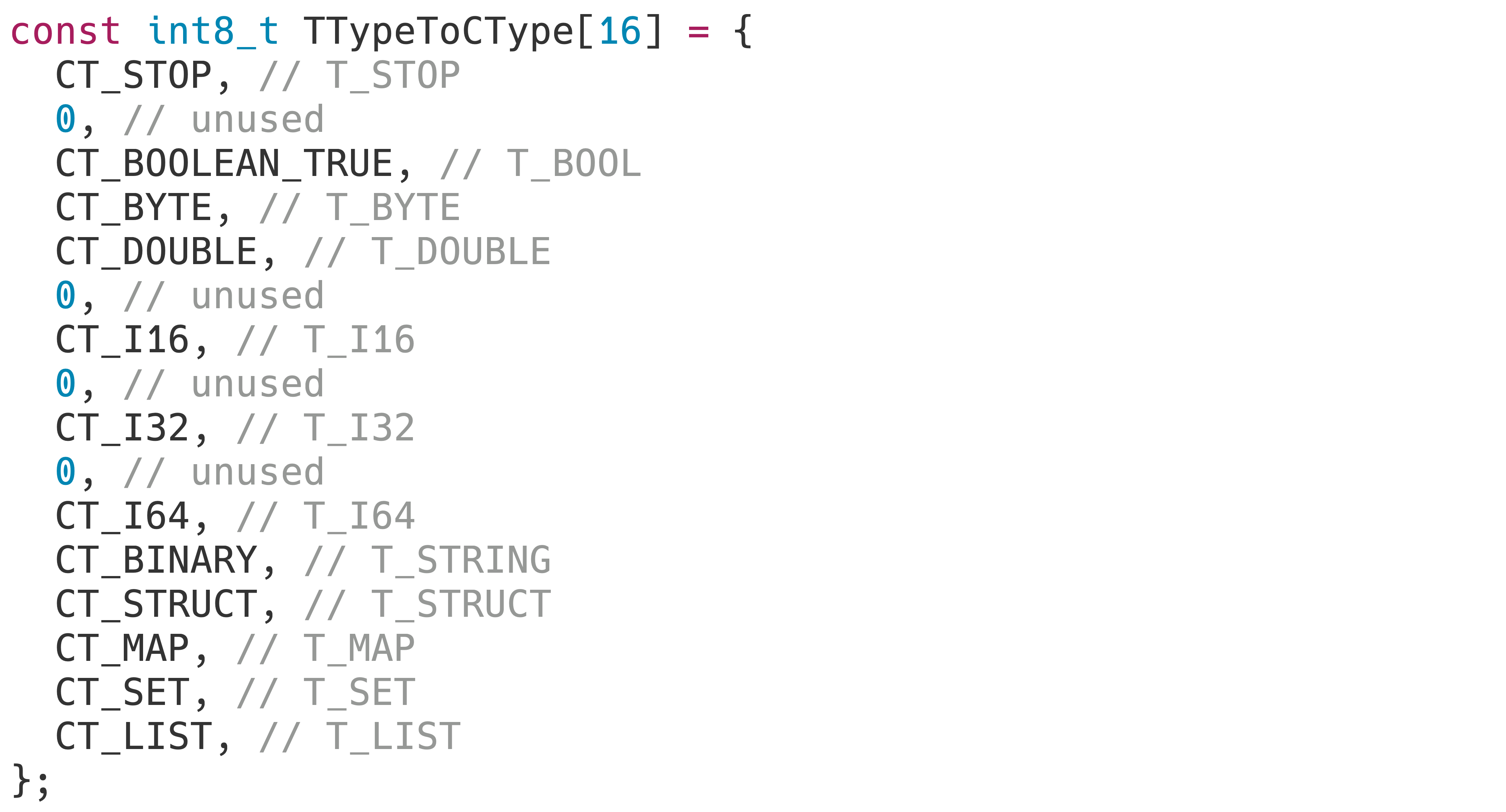}}
  \caption{Apache Thrift Compact Protocol list type identifiers definition
  \protect\footnotemark.  The indexes represent the type identifier constants.
  For example, the \texttt{\textbf{CT\_DOUBLE}} type is encoded as
  \texttt{\textbf{0x04}} because it is the fourth element of the
  \texttt{\textbf{TTypeToCType}} array.} \label{lst:thrift-list-types}
\end{figure}

\footnotetext{\url{https://github.com/apache/thrift/blob/05bb55148608b4324a8c91c21cf9a6a0dff282fa/lib/cpp/src/thrift/protocol/TCompactProtocol.tcc}}

\begin{figure}[hb!]
  \frame{\includegraphics[width=\linewidth]{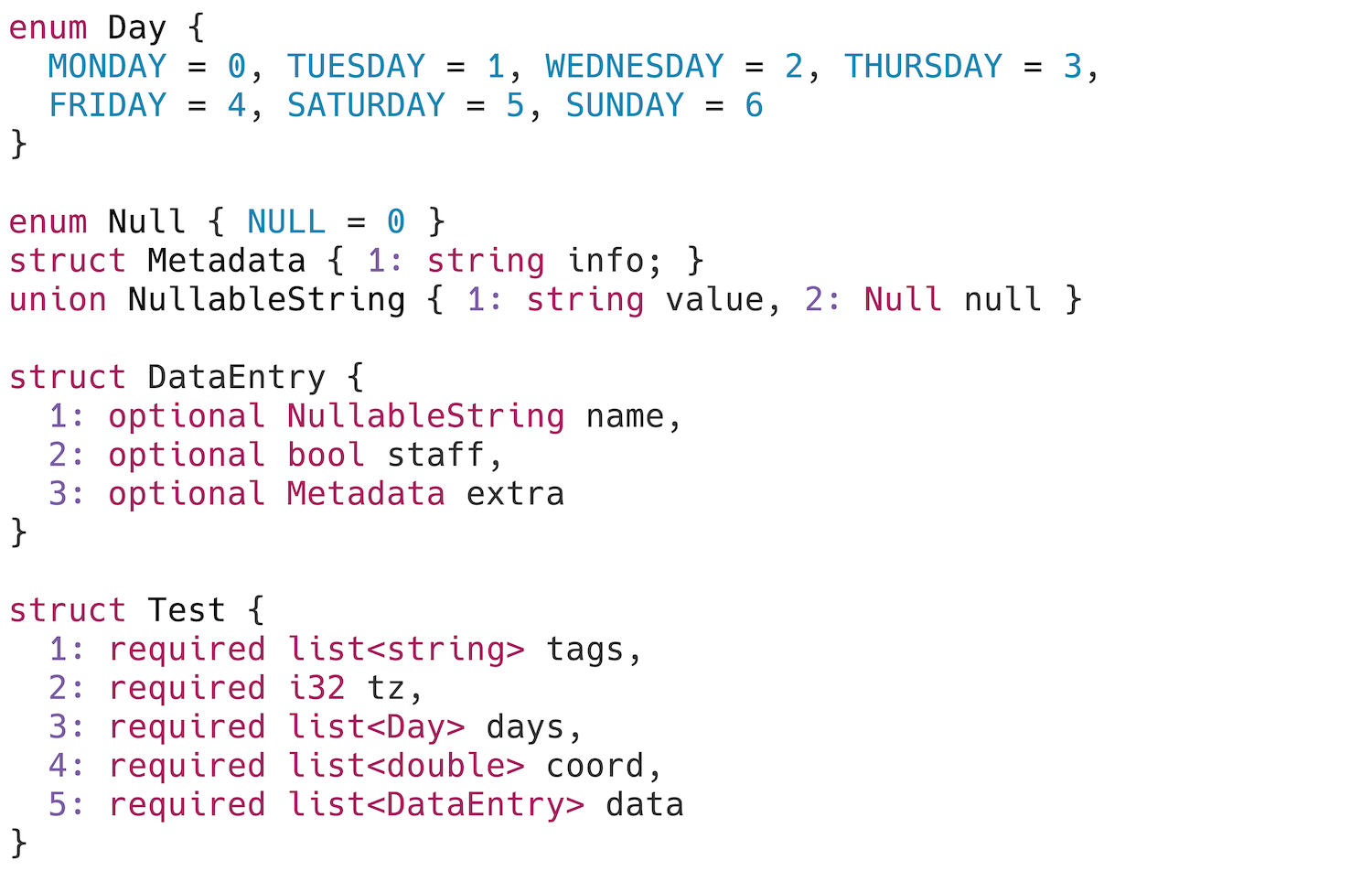}}
  \caption{Apache Thrift schema to serialize the \autoref{lst:json-object-test} input data.}
  \label{lst:schema-thrift}
\end{figure}

\begin{figure*}[hb!]
  \frame{\includegraphics[width=\linewidth]{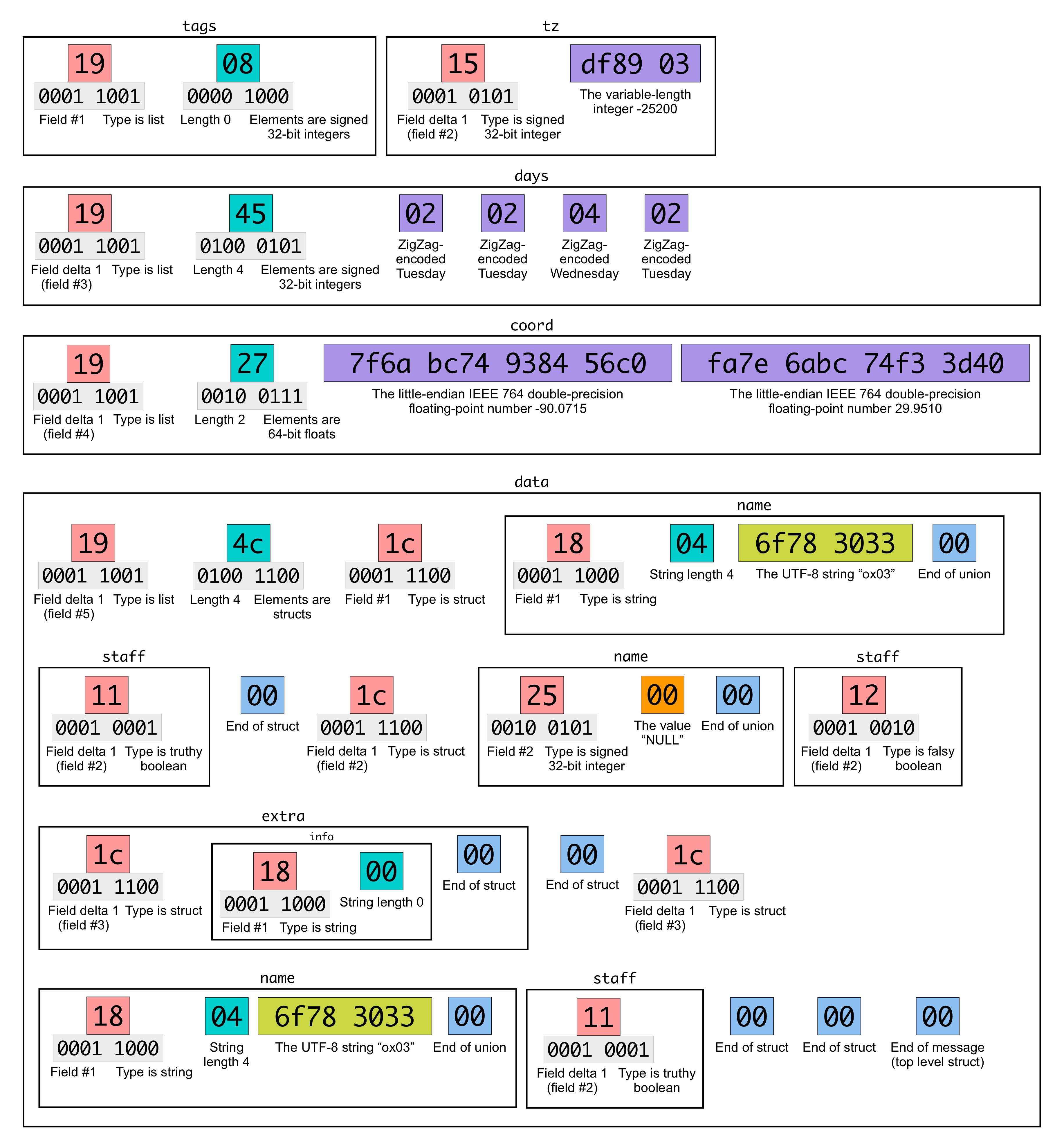}}
  \caption{Annotated hexadecimal output of serializing the
\autoref{lst:json-object-test} input data with Apache Thrift Compact Protocol.}
\end{figure*}

\begin{table}[hb!]
\caption{A high-level summary of the Apache Thrift Compact Protocol schema-driven serialization specification.}
\label{table:thrift}
  \begin{tabularx}{\linewidth}{l|X}
    \toprule
  \textbf{Website} & \url{https://thrift.apache.org} \\ \hline
  \textbf{Company / Individual} & Apache Software Foundation \\ \hline
  \textbf{Year} & 2006 \\ \hline
  \textbf{Specification} & \url{https://github.com/apache/thrift/tree/master/doc/specs} \\ \hline
  \textbf{License} & Apache License 2.0 \\ \hline
  \textbf{Schema Language} & Thrift IDL \\ \hline
  \textbf{Layout} & Sequential with field identifiers \\ \hline
  \textbf{Languages} & ActionScript, C, C++, C\#, Common LISP, D, Dart, Erlang, Haskell, Haxe, Go, Java, JavaScript, Lua, OCaml, Perl, PHP, Python, Ruby, Rust, Smalltalk, Swift \\ \hline
  \multicolumn{2}{c}{\textbf{Types}} \\ \hline

  \textbf{Numeric} &
  16-bit, 32-bit, and 64-bit ZigZag-encoded (\ref{sec:zigzag-encoding}) Little Endian Base 128 (\ref{sec:varints}) variable-length signed integers \par\smallskip
  Little Endian 64-bit IEEE 764 floating-point numbers \cite{8766229} \\ \hline

  \textbf{String} & UTF-8 \cite{UnicodeStandard} \\ \hline
  \textbf{Composite} & List, Map, Set, Struct, Union \\ \hline
  \textbf{Scalars} & Boolean \\ \hline

  \textbf{Other} &
  Binary (byte array) \par\smallskip
  Byte \\
  \bottomrule
\end{tabularx}
\end{table}

%% file: sections/schema-less-intro.tex
In this section, \we discuss the history and characteristics of JSON-compatible
schema-less binary serialization specifications: BSON \cite{bson}, CBOR
\cite{RFC7049}, FlexBuffers \cite{flexbuffers}, MessagePack \cite{messagepack},
Smile \cite{smile} and UBJSON \cite{ubjson}.

%% file: sections/formats/bson.tex
\label{sec:bson}

\begin{figure}[hb!]
  \frame{\includegraphics[width=\linewidth]{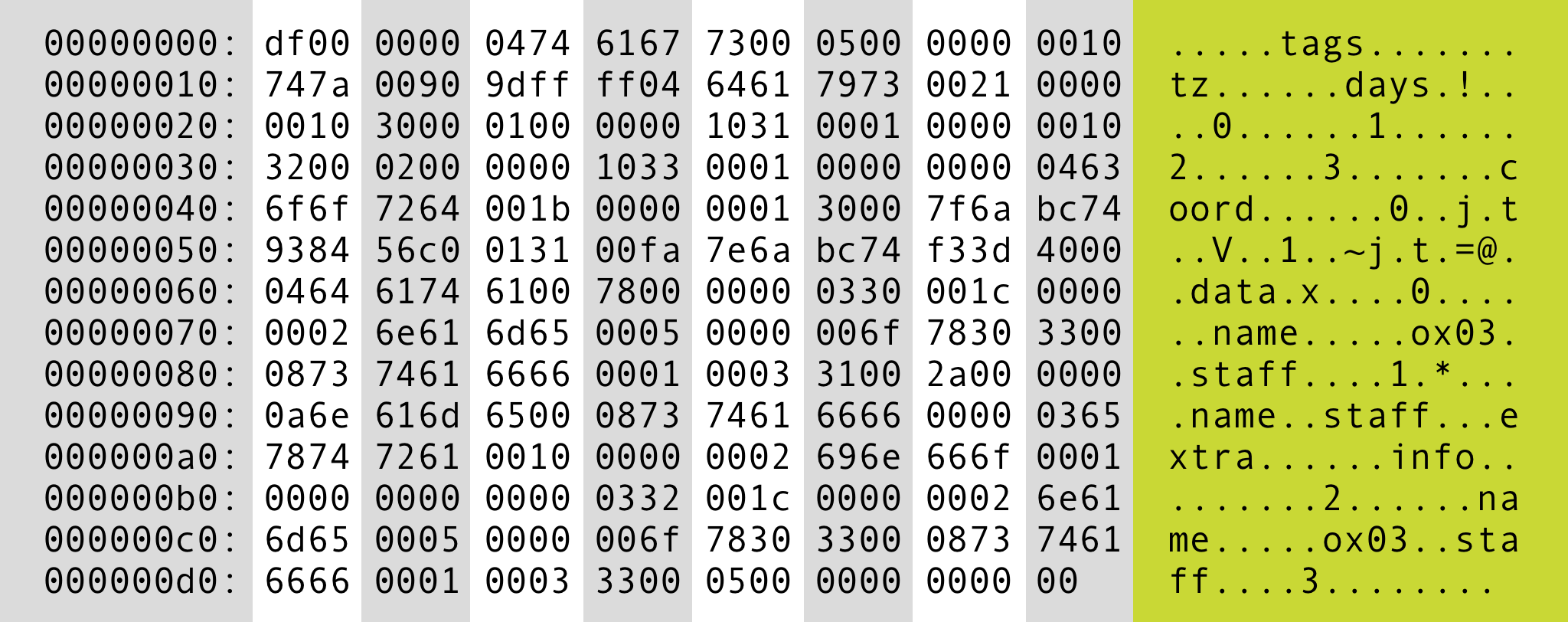}}
  \caption{Hexadecimal output (\texttt{xxd})
  of encoding \autoref{lst:json-object-test}
input data with BSON (223 bytes).}
\label{lst:hex-bson} \end{figure}

%%%%%%%%%%%%%%%%%%%%%%%%%%%%%%%%%%%%%%%%%%%%%%%%%
% HISTORY
%%%%%%%%%%%%%%%%%%%%%%%%%%%%%%%%%%%%%%%%%%%%%%%%%

\textbf{History.} BSON \cite{bson} is a schema-less binary serialization specification
created by MongoDB, Inc \footnote{\url{https://www.mongodb.com}} in 2009, which
was called \emph{10gen} at that time
\footnote{\url{https://www.mongodb.com/press/10gen-announces-company-name-change-mongodb-inc}}.
BSON is a superset of JSON \cite{RFC8259} as it includes additional data types
relevant to the MongoDB NoSQL database. BSON is a core component of the MongoDB
NoSQL database storage layer and drivers.  The BSON project was started by Mike
Dirolf \footnote{\url{https://github.com/mdirolf}}, a Software Engineer at
10gen. The BSON specification is released to the public domain under the
Creative Commons 1.0 license
\footnote{\url{https://creativecommons.org/publicdomain/zero/1.0/}}. There is
also a minimal BSON C third-party implementation targetted at embedded devices
called \emph{BINSON} \footnote{\url{https://github.com/alialavia/binson}}.

%%%%%%%%%%%%%%%%%%%%%%%%%%%%%%%%%%%%%%%%%%%%%%%%%
% ADVANTAGES
%%%%%%%%%%%%%%%%%%%%%%%%%%%%%%%%%%%%%%%%%%%%%%%%%

\textbf{Characteristics.} \begin{itemize}

  \item \textbf{Traversal Runtime-performance.} The MongoDB NoSQL database
    stores documents using BSON. Database data access patterns typically
    involve searching for fields and values within documents and
    \cite{whittaker2013improving} shows that BSON documents can be traversed at
    least four times faster than JSON \cite{ECMA-404}. However, in the context
    of Big Data, \cite{10.1007/978-981-10-5523-2_29} note that BSON collections
    are often larger than their JSON \cite{ECMA-404} and CSV \cite{RFC4180}
    counterparts due to the additional metadata included to support fast
    traversals.

  \item \textbf{In-place Updates.} BSON has been designed to support updates
    that do not involve re-serializing the entire document.

\end{itemize}

%%%%%%%%%%%%%%%%%%%%%%%%%%%%%%%%%%%%%%%%%%%%%%%%%
% OVERALL STRUCTURE
%%%%%%%%%%%%%%%%%%%%%%%%%%%%%%%%%%%%%%%%%%%%%%%%%

\textbf{Layout.} A BSON bit-string (a document) is a sequence of key-value
pairs prefixed with the byte-size of the rest of the document as a positive
Little Endian signed Two's Complement \cite{twos-complement} 32-bit
integer. Each key-value pair consists of a 1-byte type definition, followed by
the property name as a \emph{NUL}-delimited UTF-8 string
\cite{UnicodeStandard}, followed by the value. BSON documents are suffixed
with the \texttt{\textbf{0x00}} byte.

%%%%%%%%%%%%%%%%%%%%%%%%%%%%%%%%%%%%%%%%%%%%%%%%%
% TYPES
%%%%%%%%%%%%%%%%%%%%%%%%%%%%%%%%%%%%%%%%%%%%%%%%%

\begin{figure}[hb!]
\frame{\includegraphics[width=\linewidth]{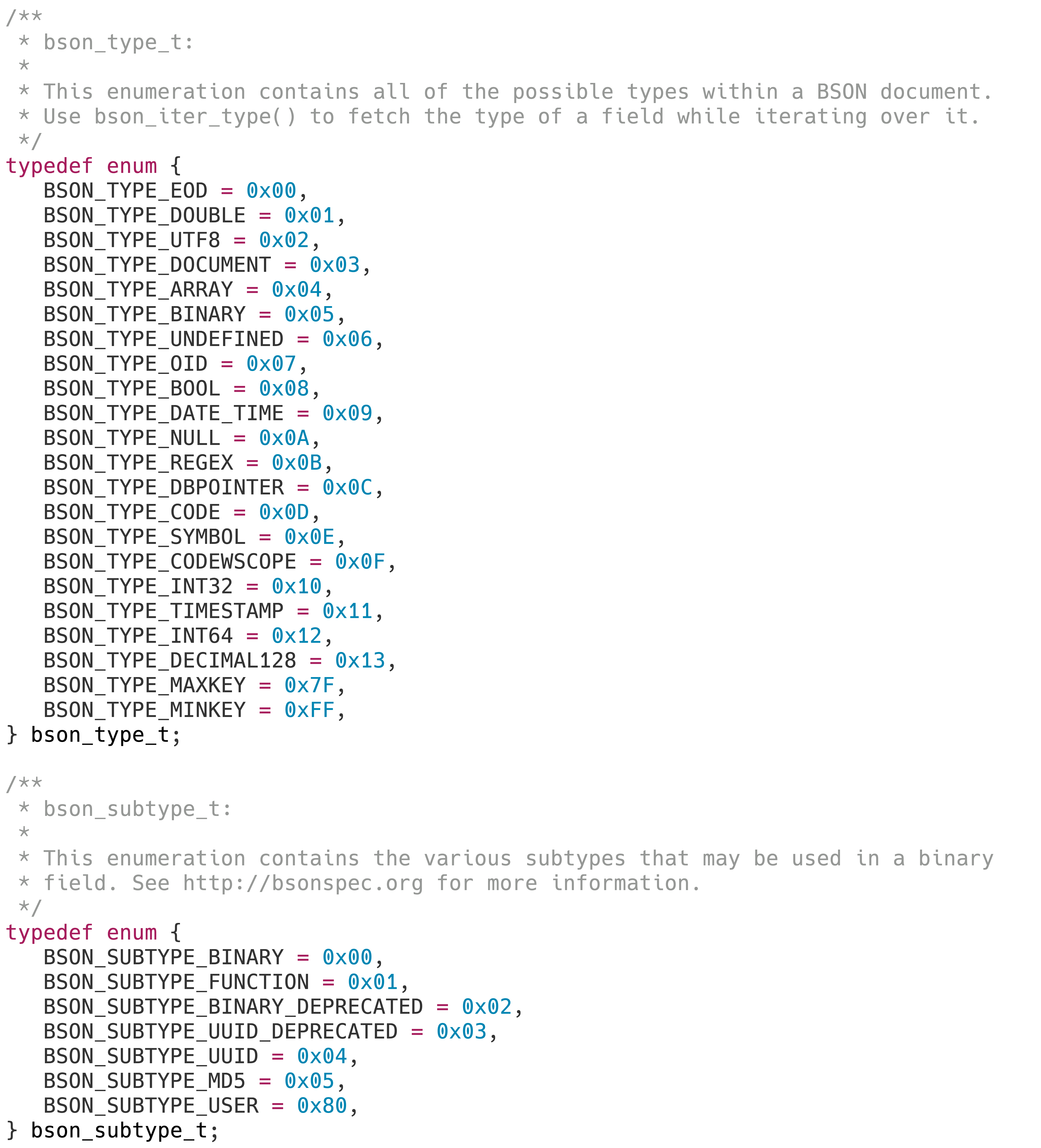}}
\caption{BSON type definitions \protect\footnotemark.} \label{lst:bson-types}
\end{figure}

\textbf{Types.} BSON declares 29 1-byte type definitions as shown in
\autoref{lst:bson-types}. Many of these type definitions correspond to
MongoDB-specific extensions to JSON \cite{RFC8259} such as the
\texttt{\textbf{DBPointer}} and \texttt{\textbf{ObjectId}} (OID)
\footnote{\url{https://docs.mongodb.com/manual/reference/method/ObjectId/}}
data types. BSON also supports binary values annotated with a sub-type such as
UUID \cite{RFC4122} and MD5 \cite{RFC1321}.

\footnotetext{\url{https://github.com/mongodb/libbson/blob/ffc8d983ecf6b46d5404f5cc20e756a85dfcbfd2/src/bson/bson-types.h}}

%%%%%%%%%%%%%%%%%%%%%%%%%%%%%%%%%%%%%%%%%%%%%%%%%
% NUMBERS
%%%%%%%%%%%%%%%%%%%%%%%%%%%%%%%%%%%%%%%%%%%%%%%%%

\textbf{Numbers.} BSON supports Little Endian 64-bit and 128-bit IEEE 764
floating-point numbers \cite{8766229}. In terms of integers, BSON supports
Little Endian 32-bit and 64-bit signed Two's Complement
\cite{twos-complement} integers and Little Endian 64-bit unsigned integers.
BSON implementations pick the smallest data type that can encode the given
number.

%%%%%%%%%%%%%%%%%%%%%%%%%%%%%%%%%%%%%%%%%%%%%%%%%
% STRINGS
%%%%%%%%%%%%%%%%%%%%%%%%%%%%%%%%%%%%%%%%%%%%%%%%%

\textbf{Strings.} BSON produces \emph{NUL}-delimited UTF-8
\cite{UnicodeStandard} strings for property names and for string values. In
comparison to property names, string values are prefixed with the byte-length
of the string as a positive Little Endian signed Two's Complement
\cite{twos-complement} 32-bit integer. BSON does not attempt to
deduplicate multiple occurrences of the same property name or string value.

%%%%%%%%%%%%%%%%%%%%%%%%%%%%%%%%%%%%%%%%%%%%%%%%%
% BOOLEANS
%%%%%%%%%%%%%%%%%%%%%%%%%%%%%%%%%%%%%%%%%%%%%%%%%

\textbf{Booleans.} Booleans values are encoded with the 1-byte type tag
\texttt{\textbf{0x08}} as shown in \autoref{lst:bson-types} followed by the
byte constants \texttt{\textbf{0x00}} (False) or \texttt{\textbf{0x01}} (True).

%%%%%%%%%%%%%%%%%%%%%%%%%%%%%%%%%%%%%%%%%%%%%%%%%
% ARRAYS
%%%%%%%%%%%%%%%%%%%%%%%%%%%%%%%%%%%%%%%%%%%%%%%%%

\textbf{Lists.} BSON treats arrays as JSON objects \cite{RFC8259} with
stringified integral keys. For example, the JSON document \texttt{\textbf{\{
  "foo": [ true, false ] \} }} is encoded as \texttt{\textbf{\{ "foo": \{ "1":
true, "2": false \} \}}}. BSON can distinguish an array from a user-supplied
JSON object with stringified integral keys by their different 1-byte type
definition prefixes.  BSON does not attempt to deduplicated multiple
occurrences of the same value in an array.

\begin{figure*}[hb!]
  \frame{\includegraphics[width=\linewidth]{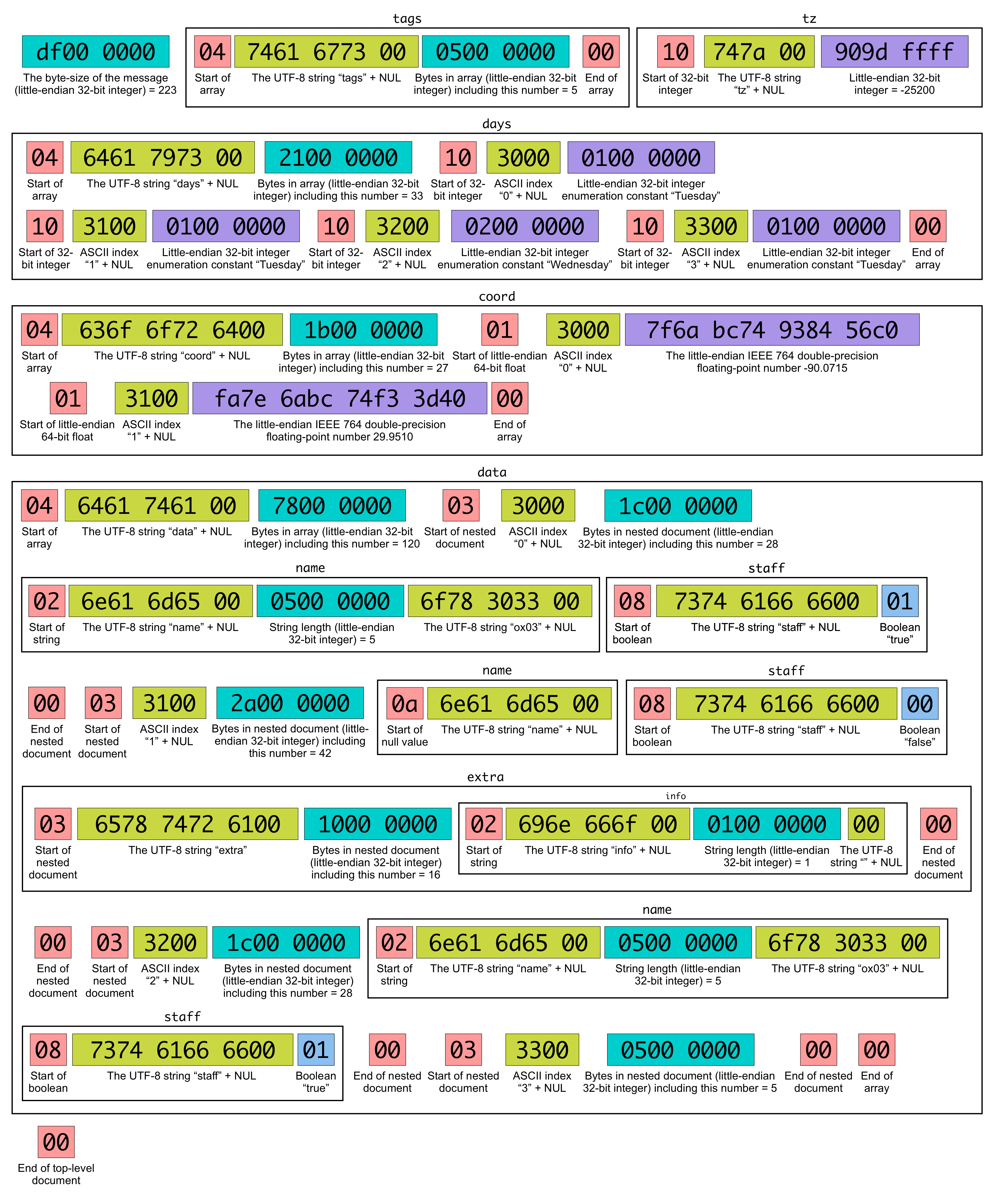}}
  \caption{Annotated hexadecimal output of serializing the
\autoref{lst:json-object-test} input data with BSON.}
\end{figure*}

\begin{table}[hb!]
\caption{A high-level summary of the BSON schema-less serialization specification.}
\label{table:bson}
  \begin{tabularx}{\linewidth}{l|X}
  \toprule
  \textbf{Website} & \url{http://bsonspec.org} \\ \hline
  \textbf{Company / Individual} & MongoDB \\ \hline
  \textbf{Year} & 2010 \\ \hline
  \textbf{Specification} & \url{http://bsonspec.org/spec.html} \\ \hline
  \textbf{License} & Creative Commons 1.0 \\ \hline
  \textbf{Layout} & Sequential \\ \hline
  \textbf{Languages} & C, C++, C\#, Go, Java, Node.js, Perl, PHP, Python, Ruby, Scala \\ \hline
  \multicolumn{2}{c}{\textbf{Types}} \\ \hline

  \textbf{Numeric} &
  Little Endian Two's Complement \cite{twos-complement} signed 32-bit and 64-bit integers \par\smallskip
  Little Endian unsigned 64-bit integers \par\smallskip
  Little Endian 64-bit and 128-bit IEEE 764 floating-point numbers \cite{8766229} \\ \hline

  \textbf{String} & UTF-8 \cite{UnicodeStandard} \\ \hline
  \textbf{Composite} & Array, Document (object) \\ \hline
  \textbf{Scalars} & Boolean, Null, Undefined \\ \hline

  \textbf{Other} &
  Binary data (byte array) \par\smallskip
  ObjectId \footnotemark \par\smallskip
  Epoch \cite{RFC8877} \par\smallskip
  DBPointer \par\smallskip
  JavaScript code \par\smallskip
  Function \par\smallskip
  UUID \cite{RFC4122} \par\smallskip
  MD5 \cite{RFC1321} \par\smallskip
  Symbol \par\smallskip
  MongoDB Timestamp \par\smallskip
  Regular expression \\
  \bottomrule
\end{tabularx}
\end{table}

\footnotetext{\url{http://dochub.mongodb.org/core/objectids}}

%% file: sections/formats/cbor.tex
\label{sec:cbor}

\begin{figure}[hb!]
\frame{\includegraphics[width=\linewidth]{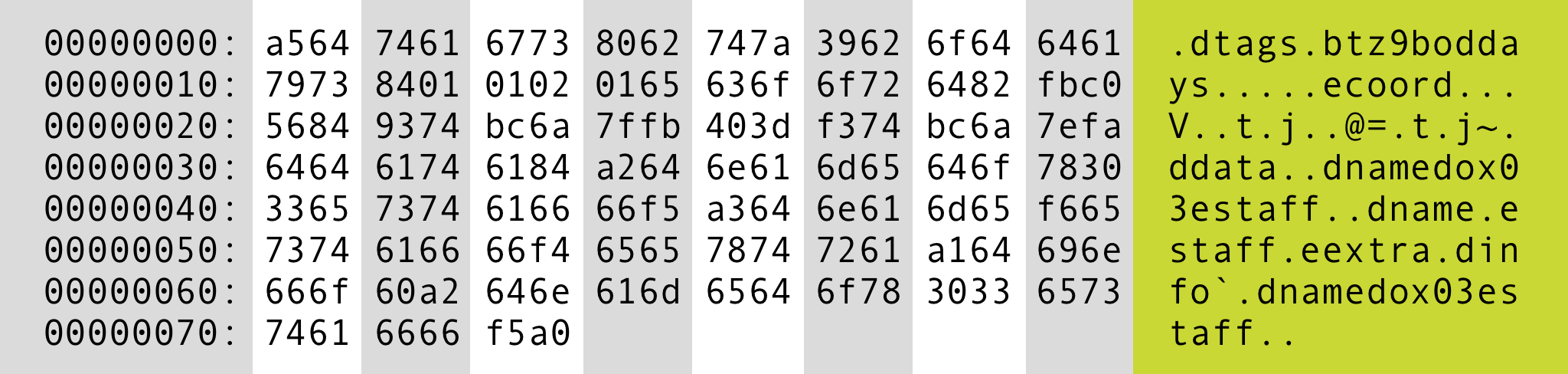}}
\caption{Hexadecimal output (\texttt{xxd}) of encoding
\autoref{lst:json-object-test} input data with CBOR (118 bytes).}
\label{lst:hex-cbor} \end{figure}

%%%%%%%%%%%%%%%%%%%%%%%%%%%%%%%%%%%%%%%%%%%%%%%%%
% HISTORY
%%%%%%%%%%%%%%%%%%%%%%%%%%%%%%%%%%%%%%%%%%%%%%%%%

\textbf{History.} Concise Binary Object Representation (CBOR) \cite{RFC7049} is
a schema-less binary serialization specification first published by Carsten
Bormann \footnote{\url{https://www.linkedin.com/in/cabo/}} and Paul Hoffman
\footnote{\url{https://datatracker.ietf.org/person/paul.hoffman@vpnc.org}} as
an Internet Engineering Task Force (IETF) \footnote{\url{https://www.ietf.org}}
document in 2013. CBOR has been primarily designed for the Internet of Things.
In fact, CBOR is the recommended serialization layer for the CoAP
\cite{RFC7252} Internet of Things transfer protocol. CBOR is natively supported
as part of the RIOT operating system for the Internet of Things
\cite{borgohain2015survey}. Outside of IoT, CBOR is also used as the
serialization specification behind the \emph{piChain} fault-tolerant
distributed state machine \cite{morath2018implementing}.  \cite{7805797}
provides an alternative introduction to the CBOR specification and describes
how to translate XML documents \cite{Paoli:06:EML} into CBOR.

%%%%%%%%%%%%%%%%%%%%%%%%%%%%%%%%%%%%%%%%%%%%%%%%%
% ADVANTAGES
%%%%%%%%%%%%%%%%%%%%%%%%%%%%%%%%%%%%%%%%%%%%%%%%%

\textbf{Characteristics.} \begin{itemize}

  \item \textbf{Resource Efficient.} CBOR has been designed to produce
    implementations to run on memory and processor constrained devices
    \footnote{\url{https://tools.ietf.org/html/rfc7049\#section-1.1}}.

  \item \textbf{Standardization.} In comparison to informally documented
    serialization specifications, CBOR is specified as an IETF RFC document
    \cite{RFC7049} and has gone through extensive technical review as a
    result.

\end{itemize}

%%%%%%%%%%%%%%%%%%%%%%%%%%%%%%%%%%%%%%%%%%%%%%%%%
% OVERALL STRUCTURE
%%%%%%%%%%%%%%%%%%%%%%%%%%%%%%%%%%%%%%%%%%%%%%%%%

\textbf{Layout.} A CBOR bit-string is a sequence of key-value pairs, sometimes
nested. Each CBOR key or value starts with a type definition. A CBOR key-value
pair is a concatenation of the key followed by the value and a CBOR map is a
concatenation of key-value pairs.

%%%%%%%%%%%%%%%%%%%%%%%%%%%%%%%%%%%%%%%%%%%%%%%%%
% TYPES
%%%%%%%%%%%%%%%%%%%%%%%%%%%%%%%%%%%%%%%%%%%%%%%%%

\begin{figure}[hb!]
\frame{\includegraphics[width=\linewidth]{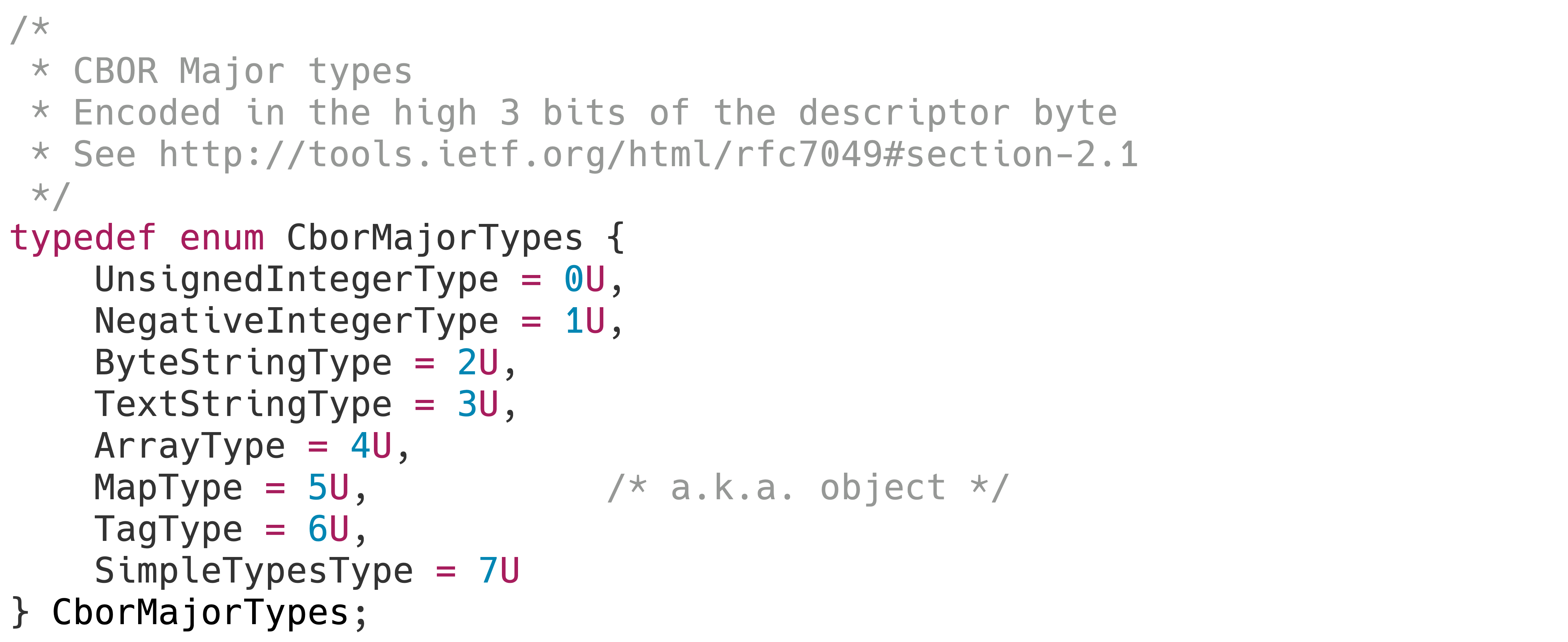}}
\caption{CBOR major types definition \protect\footnotemark.}
\label{lst:cbor-major-types} \end{figure}

\textbf{Types.} CBOR groups the types it supports into 8 \emph{major types} as
shown in \autoref{lst:cbor-major-types}. A CBOR type definition consists of
1-byte whose most-significant 3-bits encode the major type. The remaining
5-bits encode type-specific information as shown in
\autoref{table:cbor-major-types}.

\begin{table}[hb!]
\caption{A definition of CBOR major types as specified in \cite{RFC7049}.}
\label{table:cbor-major-types}
  \begin{tabularx}{\linewidth}{p{0.8cm}|p{2cm}|X}
  \toprule
\textbf{Major type} & \textbf{Description} & \textbf{Remaining 5-bits}\\ \midrule
  0 & Unsigned integers & The integer itself or the byte-length of the integer to follow \\ \hline
  1 & Negative integers & The integer itself or the byte-length of the integer to follow \\ \hline
  2 & Byte strings & The length of the byte-string to follow \\ \hline
  3 & Text strings & The length of the string to follow \\ \hline
  4 & Arrays & The logical length of the array or the byte-length of the integer representing the logical length of the array following the type definition \\ \hline
  5 & Maps & The number of keys in the map or the byte-length of the integer representing the number of keys in the map following the type definition \\ \hline
  6 & Optional metadata & Additional semantic information \\ \hline
  7 & Floats and types without content & The sub-type followed by the content if the value is a floating-point number \\
  \bottomrule
\end{tabularx}
\end{table}

\footnotetext{\url{https://github.com/intel/tinycbor/blob/fc42a049853b802e45f49588f8148fc29d7b4d9c/src/cborinternal\_p.h}}

%%%%%%%%%%%%%%%%%%%%%%%%%%%%%%%%%%%%%%%%%%%%%%%%%
% NUMBERS
%%%%%%%%%%%%%%%%%%%%%%%%%%%%%%%%%%%%%%%%%%%%%%%%%

\textbf{Numbers.} CBOR supports Big Endian IEEE 764 16-bit, 32-bit, and 64-bit
floating-point numbers \cite{8766229}. Floating-point numbers are defined
with the \textbf{major type 7} as shown in \autoref{table:cbor-major-types} and
with the floating-point precision as a sub-type. CBOR makes a distinction
between positive integers (\textbf{major type 0}) and negative integers
(\textbf{major type 1}) at the type level. In terms of positive integers, CBOR
supports Big Endian 5-bit, 8-bit, 16-bit, 32-bit and 64-bit unsigned integers.
Negative integers do not use Two's Complement \cite{twos-complement} nor
ZigZag encoding (\ref{sec:zigzag-encoding}). Instead, CBOR encodes the negative
number as a Big Endian 5-bit, 8-bit, 16-bit, 32-bit, or 64-bit unsigned integer
where the final value is equal to minus one minus the unsigned integer. For
example, the negative integer $\mathbf{-25200}$ is encoded as $\mathbf{25199}$
given that $\mathbf{-1 - 25199 = -25200}$.

Additionally, CBOR supports arbitrary-length positive and negative integers and
arbitrary-precision decimal numbers.  Arbitrary-length integers are encoded as
unsigned integers represented as potentially-indefinite byte strings prefixed
with a \textbf{major type 6} as shown in \autoref{table:cbor-major-types} and a
positive or negative sub-type annotation.  Negative arbitrary-length integers
are encoded similarly to fixed-length negative numbers: as unsigned integers
where the value is equal to minus one minus the unsigned integer.
Arbitrary-precision decimal numbers are encoded as arrays of exactly two
integer elements: the scale as a fixed-length positive or negative integer, and
the mantissa as a fixed-length or arbitrary-length positive or negative
integer. The resulting decimal number is equal to $\mathbf{\textbf{mantissa}
\times 10^{\textbf{scale}}}$.

%%%%%%%%%%%%%%%%%%%%%%%%%%%%%%%%%%%%%%%%%%%%%%%%%
% STRINGS
%%%%%%%%%%%%%%%%%%%%%%%%%%%%%%%%%%%%%%%%%%%%%%%%%

\textbf{Strings.} CBOR property names and string values are encoded using UTF-8
\cite{UnicodeStandard} without \emph{NUL} delimiters. CBOR can encode
fixed-length strings where the type definition prefix includes the byte-length
of the string as a Big Endian 5-bit, 16-bit, 32-bit, or 64-bit unsigned
integer. Additionally, CBOR can encode arbitrary-length strings by splitting
the input string into a sequence of fixed-length string values suffixed with
the \texttt{\textbf{0xff}} stop code. CBOR does not attempt to deduplicate
multiple occurrences of the same property name or string value.

%%%%%%%%%%%%%%%%%%%%%%%%%%%%%%%%%%%%%%%%%%%%%%%%%
% BOOLEANS
%%%%%%%%%%%%%%%%%%%%%%%%%%%%%%%%%%%%%%%%%%%%%%%%%

\textbf{Booleans} CBOR encodes boolean values at the type level through a type
definition byte that consists of the \textbf{major type 7} as shown in
\autoref{table:cbor-major-types} and the Big Endian 5-bit unsigned integer
sub-types 20 (False) or 21 (True). For example, the boolean value \emph{True}
is encoded as \texttt{\textbf{0xf5 = 111 (major type 7) 10101 = 21}}.

%%%%%%%%%%%%%%%%%%%%%%%%%%%%%%%%%%%%%%%%%%%%%%%%%
% ARRAYS
%%%%%%%%%%%%%%%%%%%%%%%%%%%%%%%%%%%%%%%%%%%%%%%%%

\textbf{Lists.} CBOR supports fixed-length and arbitrary-length arrays both
constisting of a \textbf{major type 4} as shown in
\autoref{table:cbor-major-types} type definition and the array elements encoded
in order. Fixed-length arrays use the least-significant 5-bits of the type
definition to store the logical length of the array as a Big Endian 5-bit
unsigned integer or to store the byte-length of the logical length of the array
following the type definition represented as a Big Endian unsigned 16-bit,
32-bit, or 64-bit integer.

In comparison to fixed-length arrays, arbitrary-length arrays do not encode
the logical length of the array. Instead, all the least-significant 5-bits of
the type definition are set and the array is suffixed with the
\texttt{\textbf{0xff}} stop code.  CBOR does not attempt to deduplicate
multiple occurrences of the same scalar or composite value in an array.

\begin{figure*}[hb!]
  \frame{\includegraphics[width=\linewidth]{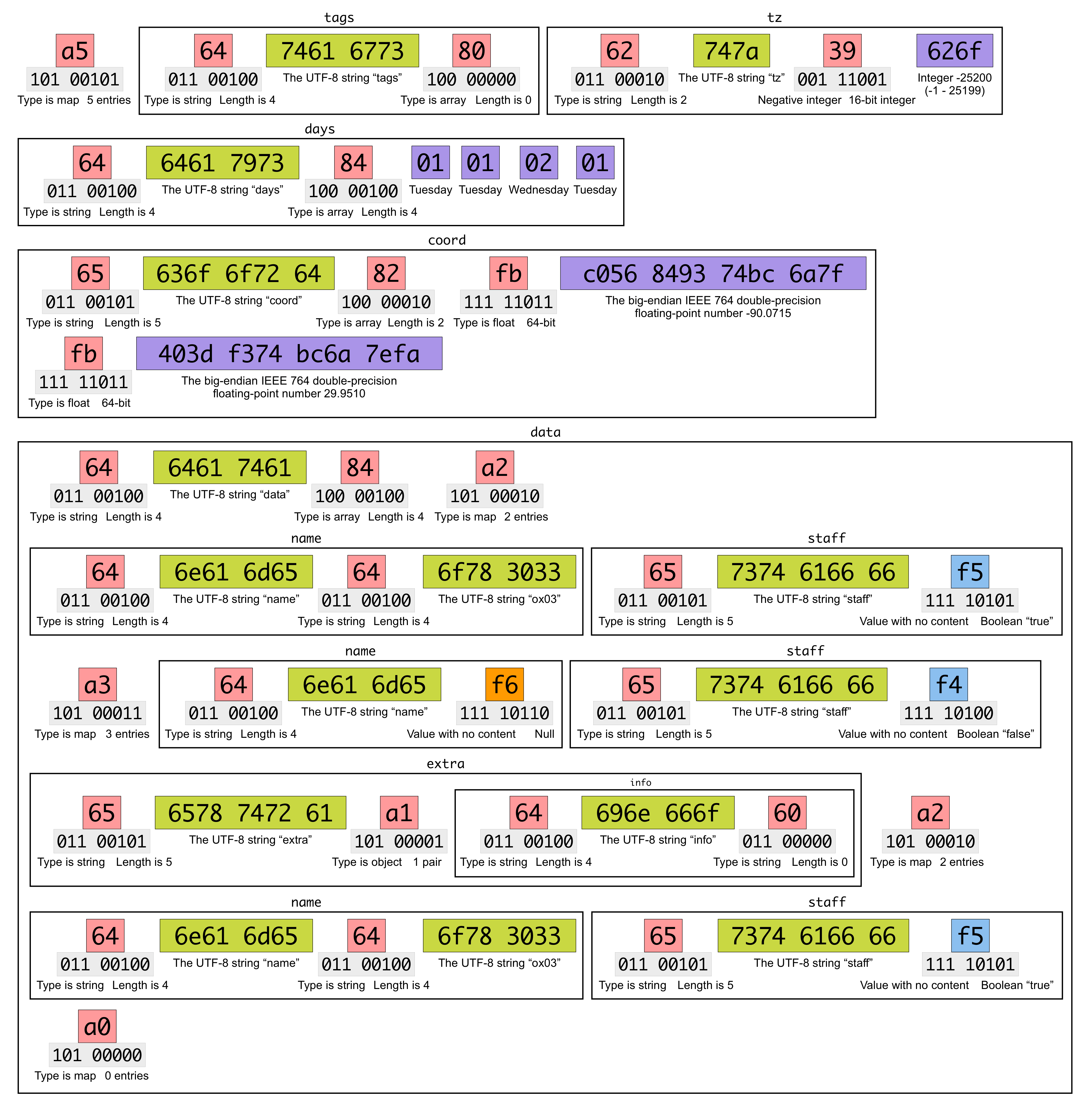}}
  \caption{Annotated hexadecimal output of serializing the
\autoref{lst:json-object-test} input data with CBOR.}
\end{figure*}

\begin{table}[hb!]
\caption{A high-level summary of the CBOR schema-less serialization specification.}
\label{table:cbor}
  \begin{tabularx}{\linewidth}{l|X}
  \toprule
  \textbf{Website} & \url{http://cbor.io} \\ \hline
  \textbf{Company / Individual} & Carsten Bormann and Paul Hoffman \\ \hline
  \textbf{Year} & 2013 \\ \hline
  \textbf{Specification} & RFC-7049 \cite{RFC7049} \\ \hline
  \textbf{License} & Implementation-dependent \\ \hline
  \textbf{Layout} & Sequential \\ \hline
  \textbf{Languages} & C, C++, C\#, Clojure, Crystal, D, Dart, Elixir, Erlang, Go, Haskell, Java, JavaScript, Julia, Lua, OCaml, Perl, PHP, Python, Ruby, Rust, Scala, Scala, Swift \\ \hline
  \multicolumn{2}{c}{\textbf{Types}} \\ \hline

  \textbf{Numeric} &
  Big Endian 5-bit, 8-bit, 16-bit, 32-bit, and 64-big unsigned integers \par\smallskip
  Big Endian 5-bit, 8-bit, 16-bit, 32-bit, and 64-big negative integers (encoded as $-1$ minus the value) \par\smallskip
  Big Endian 32-bit and 64-bit IEEE 764 floating-point numbers \cite{8766229} \par\smallskip
  Big Endian arbitrary-length positive and negative integers \par\smallskip
  Arbitrary-precision signed decimals (mantissa and scale-based) \\ \hline

  \textbf{String} & UTF-8 \cite{UnicodeStandard} \\ \hline
  \textbf{Composite} & Array, Map \\ \hline
  \textbf{Scalars} & Boolean, Null, Undefined \\ \hline

  \textbf{Other} &
  Byte string \par\smallskip
  Datetime \cite{RFC3339} \par\smallskip
  Epoch \cite{RFC8877} \par\smallskip
  Base64 \cite{RFC4648} \par\smallskip
  Regular expression \par\smallskip
  MIME message \cite{RFC2045} \\
  \bottomrule
\end{tabularx}
\end{table}

%% file: sections/formats/flexbuffers.tex
\label{sec:flexbuffers}

\begin{figure}[hb!]
\frame{\includegraphics[width=\linewidth]{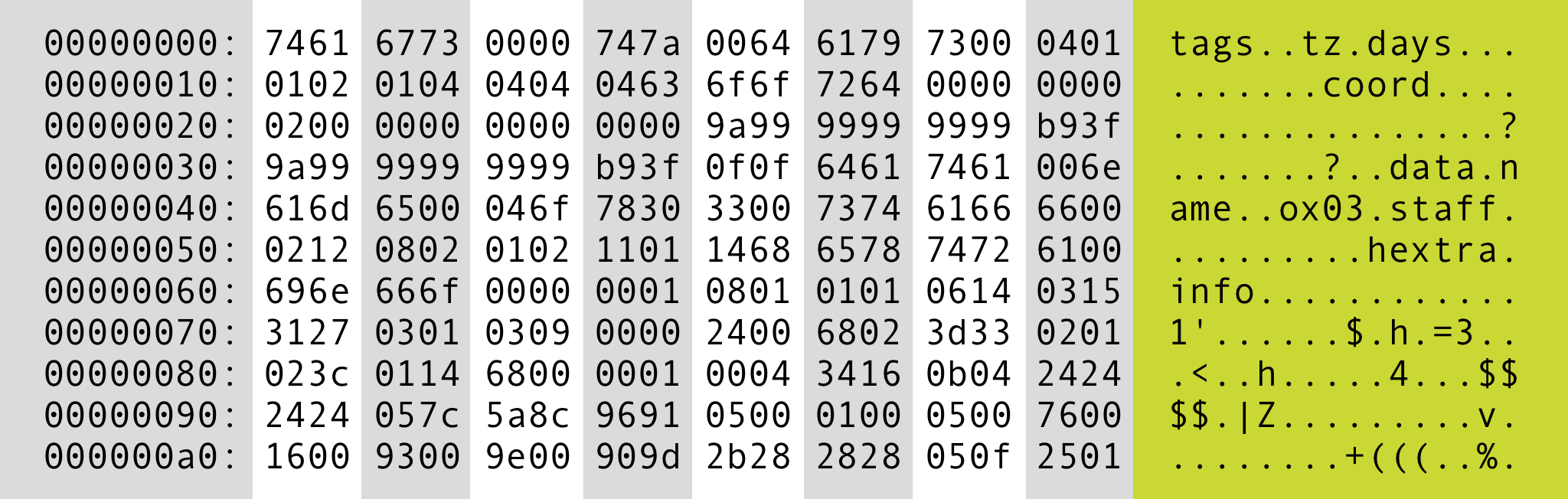}}
\caption{Hexadecimal output (\texttt{xxd}) of encoding
\autoref{lst:json-object-test} input data with FlexBuffers (176 bytes).}
\label{lst:hex-flexbuffers} \end{figure}

%%%%%%%%%%%%%%%%%%%%%%%%%%%%%%%%%%%%%%%%%%%%%%%%%
% HISTORY
%%%%%%%%%%%%%%%%%%%%%%%%%%%%%%%%%%%%%%%%%%%%%%%%%

\textbf{History.} FlexBuffers \cite{flexbuffers} is the official schema-less
variant of the FlatBuffers \cite{flatbuffers} schema-driven binary
serialization specification developed by Google.  The FlexBuffers serialization
specification was developed in 2016 by Wouter van Oortmerssen
\footnote{\url{https://github.com/aardappel}}, the main author of FlatBuffers.
FlexBuffers has the same use cases as FlatBuffers: memory-efficient
serialization and deserialization mainly in the context of games and mobile.
FlexBuffers is also released under the Apache License 2.0
\footnote{\url{http://www.apache.org/licenses/LICENSE-2.0.html}}.

%%%%%%%%%%%%%%%%%%%%%%%%%%%%%%%%%%%%%%%%%%%%%%%%%
% ADVANTAGES
%%%%%%%%%%%%%%%%%%%%%%%%%%%%%%%%%%%%%%%%%%%%%%%%%

\textbf{Characteristics.} \begin{itemize}

  \item \textbf{Efficient Reads.} FlexBuffers also produces implementations
    that perform runtime-efficiency and memory-efficient incremental and
    random-access reads.

  \item \textbf{Embeddable.} FlexBuffers can be used in conjunction with
    FlatBuffers \cite{flatbuffers} by storing a part of a FlatBuffers
    bit-string in the FlexBuffers specification.

\end{itemize}

%%%%%%%%%%%%%%%%%%%%%%%%%%%%%%%%%%%%%%%%%%%%%%%%%
% OVERALL STRUCTURE
%%%%%%%%%%%%%%%%%%%%%%%%%%%%%%%%%%%%%%%%%%%%%%%%%

\begin{figure}[hb!]
  \frame{\includegraphics[width=\linewidth]{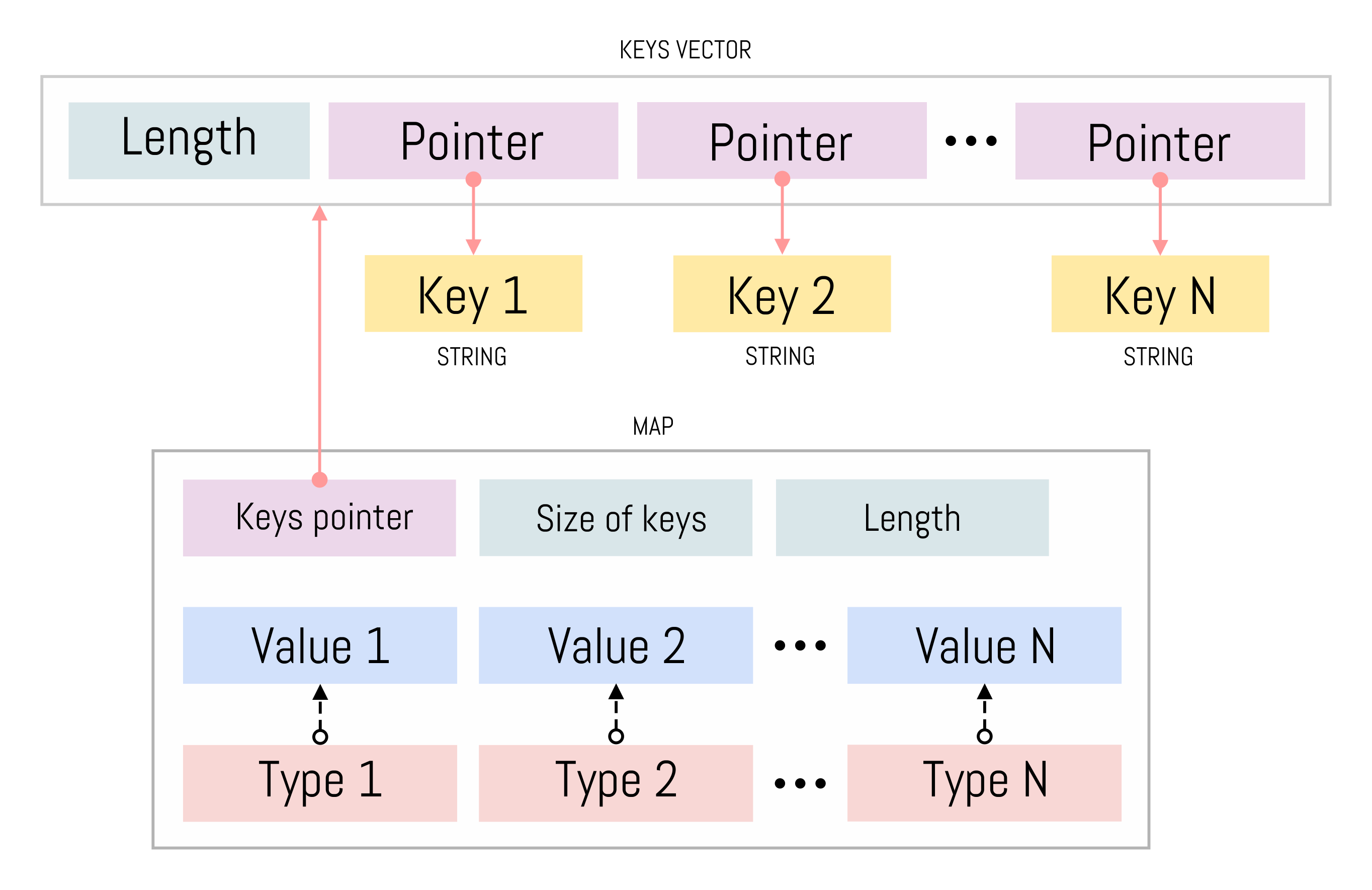}}
  \caption{FlexBuffers maps consist of values vectors that include $N$ number
  of elements followed by $N$ type definitions. These vectors point to keys
  vectors consisting of $N$ elements that typically consist of pointers to
strings.} \label{fig:flexbuffers-maps} \end{figure}

\textbf{Layout.} A FlexBuffers bit-string consist of a tree hierarchy of
pointers that eventually point at scalar types. The FlexBuffers core data
structure is the \emph{map}. A FlexBuffers map consist of a values vector and a
keys vector as shown in \autoref{fig:flexbuffers-maps} and as follows:

\begin{itemize}

  \item \textbf{Values Vector.} The values vector starts with a pointer to its
    corresponding keys vector, the byte-length of each key in the keys vector
    as a Little Endian unsigned integer, and the logical size of the values
    vector as a Little Endian unsigned integer. The values vector then
    sequentially encodes its elements followed by a sequence of 1-byte type
    definitions that correspond to each value in the map.

  \item \textbf{Keys Vector.} The keys vector consists of the logical size of
    the vector as a Little Endian unsigned integer followed by the sequence of
    keys, which are typically pointers to string values. Individual keys may be
    shared and multiple values vectors can point to the same keys vector if
    they share the same structure resulting in efficient use of space.

\end{itemize}

Every vector element is aligned to the largest element that the vector
contains. The parent structure that points to the vector encodes the size of
the elements. Scalar values are encoded at the beginning of the bit-string
while map definitions are encoded at the end of the bit-string.  FlexBuffers
bit-strings end with a footer vector that consists of a pointer to the root
element, the type definition of the root element, and the byte-length of each
element in the footer vector as a Little Endian 8-bit unsigned integer.

%%%%%%%%%%%%%%%%%%%%%%%%%%%%%%%%%%%%%%%%%%%%%%%%%
% TYPES
%%%%%%%%%%%%%%%%%%%%%%%%%%%%%%%%%%%%%%%%%%%%%%%%%

\begin{figure}[hb!]
  \frame{\includegraphics[width=\linewidth]{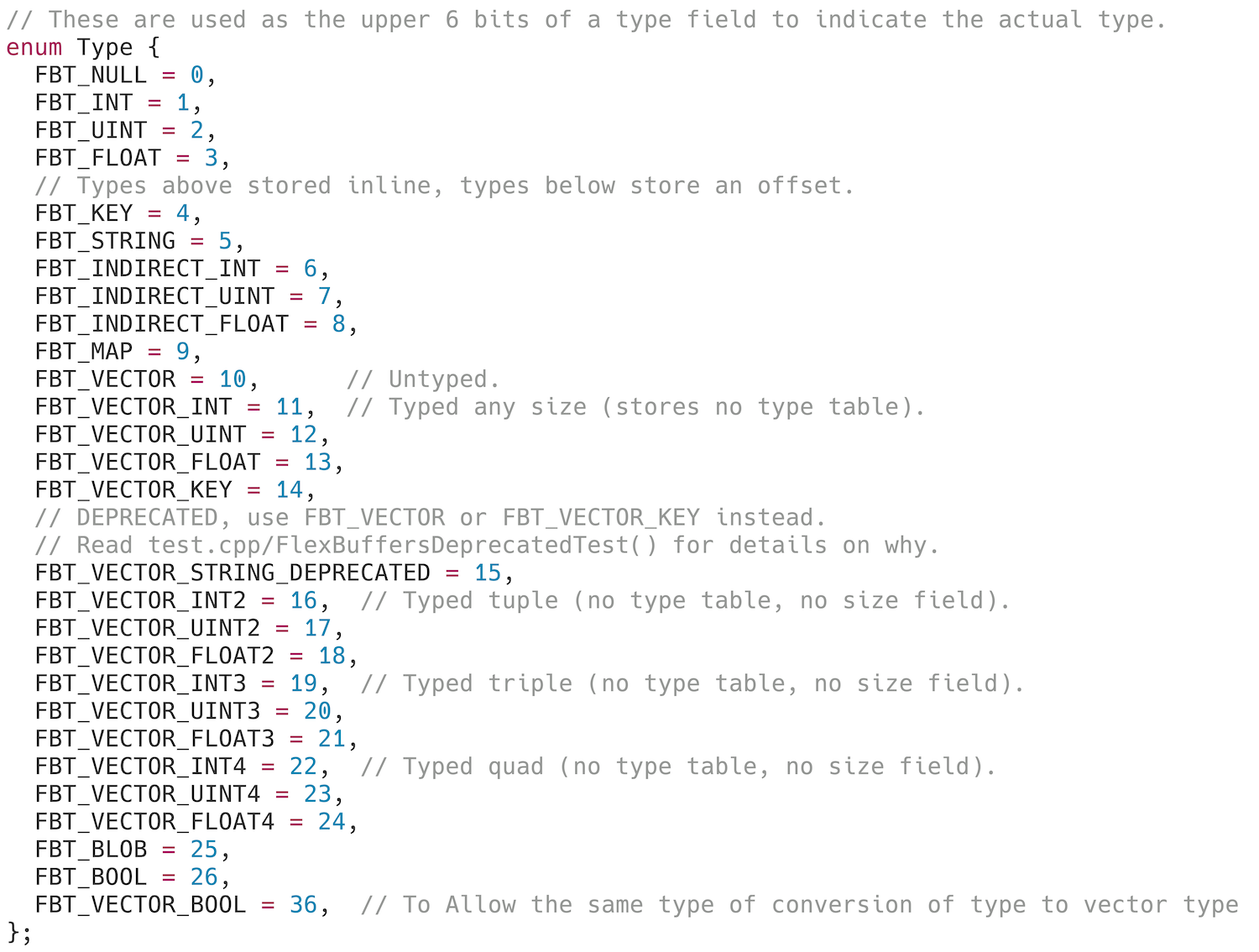}}
  \caption{FlexBuffers type identifiers definition
  \protect\footnotemark.}
\label{lst:flexbuffers-types} \end{figure}

\begin{figure}[hb!]
  \frame{\includegraphics[width=\linewidth]{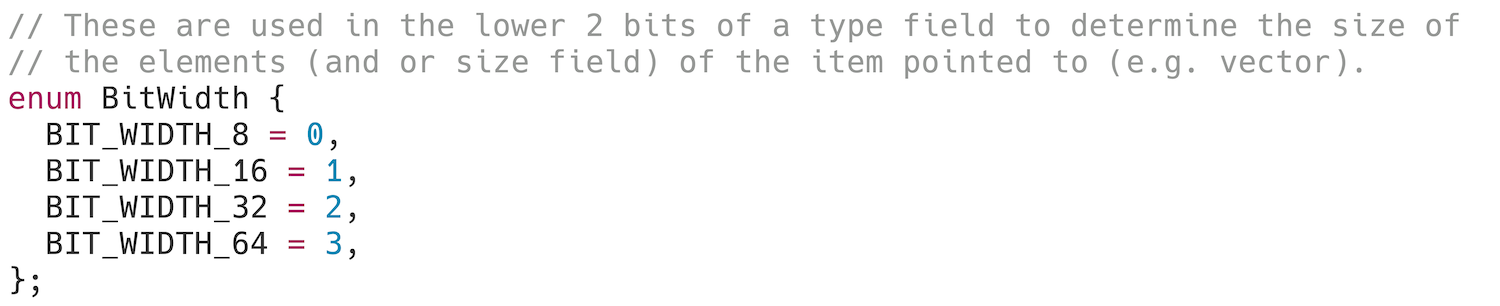}}
  \caption{FlexBuffers bit-width definition
  \protect\footnotemark[\value{footnote}].}
\label{lst:flexbuffers-bit-width} \end{figure}

\textbf{Types.} FlexBuffers uses 1-byte type definitions. The most-significant
6-bits encode the data type as an unsigned integer as shown in
\autoref{lst:flexbuffers-types}. The least-significant 2-bits encode the
bit-width of the data type as shown in \autoref{lst:flexbuffers-bit-width}.

\footnotetext{\url{https://github.com/google/flatbuffers/blob/8778dc7c2bc20b3165a86d62e2e499d2b86912f0/include/flatbuffers/flexbuffers.h}}

\footnotetext{\url{https://github.com/google/flatbuffers/blob/8778dc7c2bc20b3165a86d62e2e499d2b86912f0/include/flatbuffers/flexbuffers.h}}

%%%%%%%%%%%%%%%%%%%%%%%%%%%%%%%%%%%%%%%%%%%%%%%%%
% NUMBERS
%%%%%%%%%%%%%%%%%%%%%%%%%%%%%%%%%%%%%%%%%%%%%%%%%

\textbf{Numbers.} FlexBuffers supports Little Endian 8-bit, 16-bit, 32-bit, and
64-bit unsigned and signed integers. Signed integers use Two's Complement
\cite{twos-complement}. FlexBuffers defines four bit-widths for all its
types as shown in \autoref{lst:flexbuffers-bit-width}. Therefore, FlexBuffers
theoretically supports Little Endian 8-bit, 16-bit, 32-bit, and 64-bit IEEE 764
\cite{8766229} floating-point numbers. While \cite{8766229} does not define
8-bit floating-point numbers, these type of reduced-precision floats are useful
in the context of artificial intelligence \cite{NEURIPS2018_335d3d1c}.

%%%%%%%%%%%%%%%%%%%%%%%%%%%%%%%%%%%%%%%%%%%%%%%%%
% STRINGS
%%%%%%%%%%%%%%%%%%%%%%%%%%%%%%%%%%%%%%%%%%%%%%%%%

\textbf{Strings.} FlexBuffers produces \emph{NUL}-delimited UTF-8
\cite{UnicodeStandard} strings for property names and for string values. In
comparison to property names, string values are prefixed with the byte-length
of the string as an 8-bit unsigned integer. By default, FlexBuffers does not
attempt to deduplicate multiple occurrences of the same string. However, its
serialization interface allows the application to track and share duplicated
string values.

%%%%%%%%%%%%%%%%%%%%%%%%%%%%%%%%%%%%%%%%%%%%%%%%%
% BOOLEANS
%%%%%%%%%%%%%%%%%%%%%%%%%%%%%%%%%%%%%%%%%%%%%%%%%

\textbf{Booleans.} FlexBuffers encodes booleans as the Little Endian unsigned
integers $0$ (False) and $1$ (True). A type definition byte that describes a
boolean value consists of the type $26$ (\texttt{\textbf{FBT\_BOOL}}) as shown
in \autoref{lst:flexbuffers-types}. The bit-width declared in the
least-significant 2-bits of the type definition byte defines the width of the
unsigned integer that represents the boolean value depending on the alignment
of the vector that includes the boolean value. For example, a truthy boolean
that is part of a 16-bit elements vector is encoded as \texttt{\textbf{0x00
0x01}} with the type definition byte \texttt{\textbf{0x69 = 011010 (FBT\_BOOL)
01 (16-bits)}}.

%%%%%%%%%%%%%%%%%%%%%%%%%%%%%%%%%%%%%%%%%%%%%%%%%
% LISTS
%%%%%%%%%%%%%%%%%%%%%%%%%%%%%%%%%%%%%%%%%%%%%%%%%

\begin{figure}[hb!]
  \frame{\includegraphics[width=\linewidth]{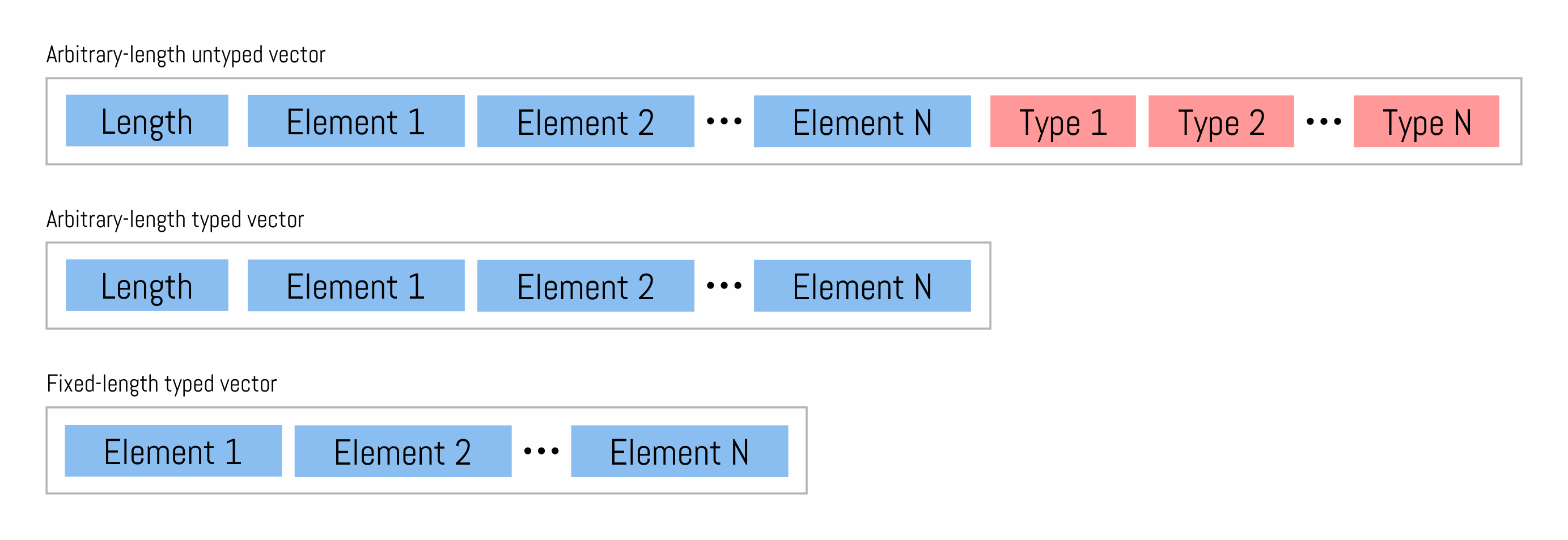}}
  \caption{A visual representation of arbitrary-length untyped vectors (top),
  arbitrary-length typed vectors (middle), and fixed-length typed vectors
  (bottom).} \label{fig:flexbuffers-vectors} \end{figure}

\textbf{Lists.} A FlexBuffers list (vector) consists of the concatenation of
its elements and a variable amount of surrounding metadata.  FlexBuffers
supports arbitrary-length untyped vectors, arbitrary-length typed vectors of
certain scalar types, and fixed-length (of 2, 3, or 4 elements) typed vectors
of certain scalar types as shown in \autoref{fig:flexbuffers-vectors}.

\begin{itemize}

  \item \textbf{Arbitrary-length Untyped Vectors.} These type of vectors are
    prefixed with their logical size as a Little Endian unsigned integer
    aligned to the size of the largest element of the vector. The concatenation
    of elements is suffixed with a sequence of 1-byte type definitions
    corresponding to each of the vector elements. The parent element pointing
    at the vector declares the generic type \texttt{\textbf{FBT\_VECTOR}} as
    shown in \autoref{lst:flexbuffers-types}.

  \item \textbf{Arbitrary-length Typed Vectors.} As the \emph{Arbitrary-length
    untyped vectors}, these type of vectors are prefixed with their logical
    size as a Little Endian unsigned integer aligned to the size of the largest
    element of the vector. However, the concatenation of elements is not
    suffixed with a list of type definitions. Instead, the parent element
    pointing at the vector declares one of the available typed vector types
    such as \texttt{\textbf{FBT\_VECTOR\_INT}} or
    \texttt{\textbf{FBT\_VECTOR\_FLOAT}} as shown in
    \autoref{lst:flexbuffers-types}.

  \item \textbf{Fixed-length Typed Vectors.} These type of vectors do not
    include the logical size unsigned integer nor a list of type definitions.
    Instead, the parent element pointing at the vector declares one of the
    available fixed-length typed vector types such as
    \texttt{\textbf{FBT\_VECTOR\_UINT4}} or
    \texttt{\textbf{FBT\_VECTOR\_FLOAT3}} as shown in
    \autoref{lst:flexbuffers-types}.

\end{itemize}

The elements of a vector are aligned to the largest element and the parent
element pointing at the vector declares the byte-length of the elements.
FlexBuffers encodes empty vectors with the 8-bit unsigned integer vector length
$0$ and no additional information. FlexBuffers does not attempt to deduplicate
multiple occurences of the same element in a vector but a vector may include
more than one pointer to the same value.

\begin{figure*}[hb!]
  \frame{\includegraphics[width=\linewidth]{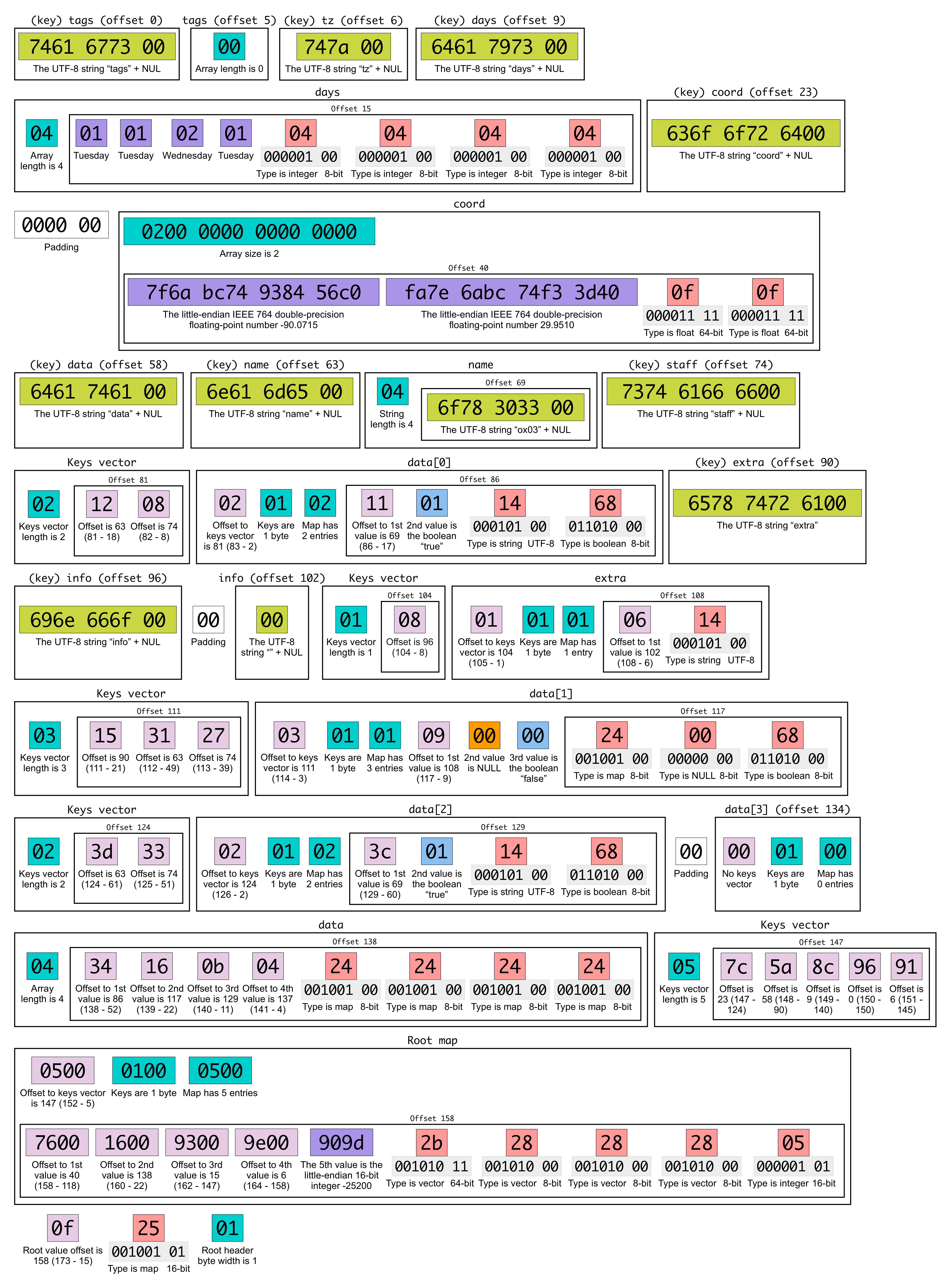}}
  \caption{Annotated hexadecimal output of serializing the
\autoref{lst:json-object-test} input data with FlexBuffers.}
\label{lst:flexbuffers-hex}
\end{figure*}

\begin{table}[hb!]
\caption{A high-level summary of the FlexBuffers schema-less serialization specification.}
\label{table:flexbuffers}
  \begin{tabularx}{\linewidth}{l|X}
  \toprule
  \textbf{Website} & \url{https://google.github.io/flatbuffers/flexbuffers.html} \\ \hline
  \textbf{Company / Individual} & Google \\ \hline
  \textbf{Year} & 2016 \\ \hline
  \textbf{Specification} & \url{https://google.github.io/flatbuffers/flatbuffers\_internals.html} \\ \hline
  \textbf{License} & Apache License 2.0 \\ \hline
  \textbf{Layout} & Pointer-based \\ \hline
  \textbf{Languages} & C++, Java \\ \hline
  \multicolumn{2}{c}{\textbf{Types}} \\ \hline

  \textbf{Numeric} &
  Little Endian 8-bit, 16-bit, 32-bit, and 64-bit Two's Complement \cite{twos-complement} signed integers \par\smallskip
  Little Endian 8-bit, 16-bit, 32-bit, and 64-bit unsigned integers \par\smallskip
  Little Endian 16-bit, 32-bit, and 64-bit IEEE 764 floating-point numbers \cite{8766229} \\ \hline

  \textbf{String} & UTF-8 \cite{UnicodeStandard} \\ \hline
  \textbf{Composite} & Vector, Map \\ \hline
  \textbf{Scalars} & Boolean, Null \\ \hline
  \textbf{Other} & Blob (byte array) \\
  \bottomrule
\end{tabularx}
\end{table}

%% file: sections/formats/messagepack.tex
\label{sec:messagepack}

\begin{figure}[hb!]
  \frame{\includegraphics[width=\linewidth]{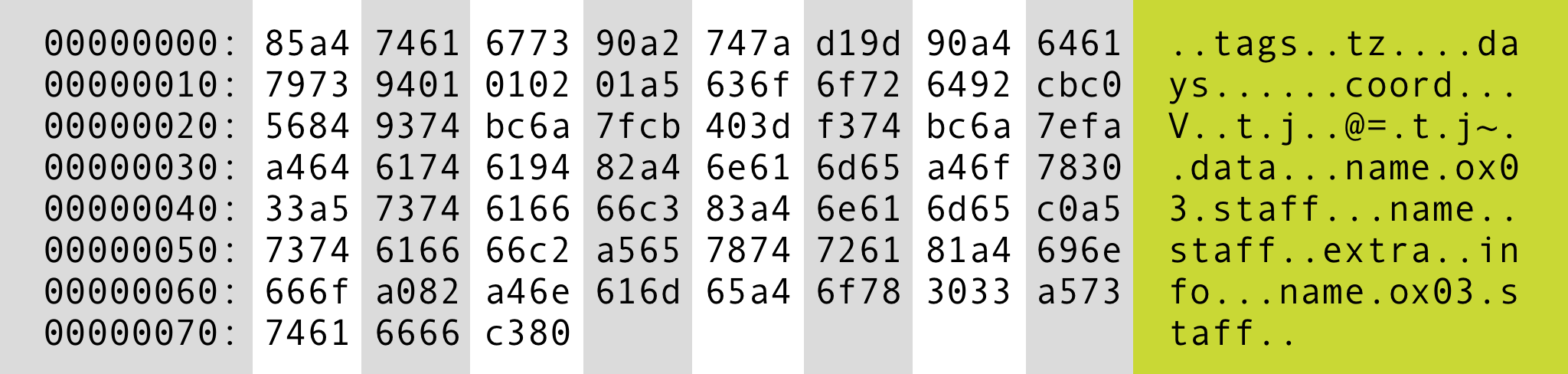}}
  \caption{Hexadecimal output (\texttt{xxd})
  of encoding \autoref{lst:json-object-test}
input data with MessagePack (118 bytes).}
\label{lst:hex-messagepack} \end{figure}

%%%%%%%%%%%%%%%%%%%%%%%%%%%%%%%%%%%%%%%%%%%%%%%%%
% HISTORY
%%%%%%%%%%%%%%%%%%%%%%%%%%%%%%%%%%%%%%%%%%%%%%%%%

\textbf{History.} MessagePack \cite{messagepack} is a schema-less binary
serialization specification designed by Sadayuki Furuhashi
\footnote{\url{https://github.com/frsyuki}} in 2009 while working on the Kumofs
\footnote{\url{https://github.com/etolabo/kumofs}} distributed key-value store.
MessagePack is used at the core of services such as Fluentd
\footnote{\url{https://www.fluentd.org}}, another popular project by the same
author, and Pinterest \footnote{\url{https://www.pinterest.com}}.  MessagePack
has been released under the Apache License 2.0
\footnote{\url{http://www.apache.org/licenses/LICENSE-2.0.html}}. There also
exists third-party MessagePack C implementations suited for embedded
development such as \footnote{\url{https://github.com/clwi/CWPack}},
\footnote{\url{https://github.com/ludocode/mpack}}, and
\footnote{\url{https://github.com/rtsisyk/msgpuck}}.

%%%%%%%%%%%%%%%%%%%%%%%%%%%%%%%%%%%%%%%%%%%%%%%%%
% ADVANTAGES
%%%%%%%%%%%%%%%%%%%%%%%%%%%%%%%%%%%%%%%%%%%%%%%%%

\textbf{Characteristics.} \begin{itemize}

  \item \textbf{Simplicity.} The MessagePack specification is easy to
    understand and implement from a developer point of view.
    \cite{10.1007/978-3-030-24305-0_58} cites MessagePack's ease of use as the
    main reason why they chose MessagePack over Protocol Buffers
    \cite{protocolbuffers} to transmit information obtained from the analysis
    of the video signal of their real-time position tracking system design.

  \item \textbf{Portability.} MessagePack popularity and simplicity resulted in
    a large amount of official and third-party implementations covering over
    forty programming languages. In comparison, FlexBuffers \cite{flexbuffers}
    and Microsoft Bond \cite{microsoft-bond} support two and four programming
    languages, respectively.

\end{itemize}

%%%%%%%%%%%%%%%%%%%%%%%%%%%%%%%%%%%%%%%%%%%%%%%%%
% OVERALL STRUCTURE
%%%%%%%%%%%%%%%%%%%%%%%%%%%%%%%%%%%%%%%%%%%%%%%%%

\textbf{Layout.} A MessagePack bit-string is a sequence of key-value pairs,
sometimes nested. Each key or value (an element) is prefixed with a type
definition. A key-value pair is a concatenation of the key element followed
value element. A map (an object) is the concatenation of its key-value pairs.

%%%%%%%%%%%%%%%%%%%%%%%%%%%%%%%%%%%%%%%%%%%%%%%%%
% TYPES
%%%%%%%%%%%%%%%%%%%%%%%%%%%%%%%%%%%%%%%%%%%%%%%%%

\textbf{Types.} MessagePack elements are prefixed with a type definition that
occupies from 1 to 9 bytes depending on the type and the length of the element
as shown in \autoref{table:messagepack-types}.  If applicable, the type
definition includes a Big Endian unsigned integer representing the logical size
or byte-length of the element. The width of the unsigned integer is determined
by the first part of the type definition.

\begin{table}[hb!]
\caption{A list of MessagePack type definitions as specified in \cite{messagepack}. This table omits extension types which are not discussed in this paper.}
\label{table:messagepack-types}
  \begin{tabularx}{\linewidth}{l|l|X|X}
\toprule
\textbf{Type} & \textbf{Type byte} & \textbf{Embedded in type byte} & \textbf{Length}\\ \midrule
Nil & \texttt{0xc0} & Nil & None \\ \hline
Boolean & \texttt{0xc2} to \texttt{0xc3} & False, True & None \\ \hline
Unsigned integer & \texttt{0x00} to \texttt{0x7f} & 7-bit unsigned integer & In type byte  \\ \hline
Unsigned integer  & \texttt{0xcc} to \texttt{0xcf} & Integer width: 8-bit, 16-bit, 32-bit, 64-bit & In type byte  \\ \hline
Signed integer & \texttt{0xe0} to \texttt{0xff} & 5-bit signed integer & In type byte  \\ \hline
Signed integer & \texttt{0xd0} to \texttt{0xd3} & Integer width: 8-bit, 16-bit, 32-bit, 64-bit & In type byte  \\ \hline
Float & \texttt{0xca} to \texttt{0xcb} & Float precision: 32-bit, 64-bit & In type byte  \\ \hline
String & \texttt{0xa0} to \texttt{0xbf} & 5-bit unsigned integer & In type byte \\ \hline
String & \texttt{0xd9} to \texttt{0xdb} & String byte-length integer width: 8-bit, 16-bit, 32-bit, 64-bit & Big Endian 8-bit, 16-bit, or 64-bit unsigned integer \\ \hline
Byte array & \texttt{0xc4} to \texttt{0xc6} & Byte array length integer width: 8-bit, 16-bit, 32-bit & Big Endian 8-bit, 16-bit, or 64-bit unsigned integer \\ \hline
Array & \texttt{0x90} to \texttt{0x9f} & 4-bit unsigned integer & In type byte \\ \hline
Array & \texttt{0xdc} to \texttt{0xdd} & Array length integer width: 16-bit, 32-bit & Big Endian 16-bit or 32-bit unsigned integer \\ \hline
Map & \texttt{0x80} to \texttt{0x8f} & 4-bit unsigned integer & In type byte \\ \hline
Map & \texttt{0xde} to \texttt{0xdf} & Map length integer width: 16-bit, 32-bit & Big Endian 16-bit or 32-bit unsigned integer \\
\bottomrule
\end{tabularx}
\end{table}

%%%%%%%%%%%%%%%%%%%%%%%%%%%%%%%%%%%%%%%%%%%%%%%%%
% NUMBERS
%%%%%%%%%%%%%%%%%%%%%%%%%%%%%%%%%%%%%%%%%%%%%%%%%

\textbf{Numbers.} MessagePack supports Big Endian IEEE 754 32-bit (type
definition \texttt{\textbf{0xca}}) and 64-bit (type definition
\texttt{\textbf{0xcb}}) floating-point numbers \cite{8766229}. In terms of
integers, MessagePack supports Big Endian 8-bit (type definition
\texttt{\textbf{0xcc}}), 16-bit (type definition \texttt{\textbf{0xcd}}),
32-bit (type definition \texttt{\textbf{0xce}}), and 64-bit (type definition
\texttt{\textbf{0xcf}}) unsigned integers and 8-bit (type definition
\texttt{\textbf{0xd0}}), 16-bit (type definition \texttt{\textbf{0xd1}}),
32-bit (type definition \texttt{\textbf{0xd2}}), and 64-bit (type definition
\texttt{\textbf{0xd3}}) signed Two's Complement \cite{twos-complement}
integers.

Unsigned integers less than $128$ and signed integers greater than or equal to
$-32$ are encoded as 8-bit integers without a preceding type definition
resulting in efficient use of space.  MessagePack can distinguish these
integers based on their constant most-significant bits (\texttt{\textbf{0}} for
the unsigned integers and \texttt{\textbf{111}} for the signed integers) as
they are not re-used in any other type definition.  MessagePack implementations
pick the smallest data type that can encode the given number.

%%%%%%%%%%%%%%%%%%%%%%%%%%%%%%%%%%%%%%%%%%%%%%%%%
% STRINGS
%%%%%%%%%%%%%%%%%%%%%%%%%%%%%%%%%%%%%%%%%%%%%%%%%

\textbf{Strings.} MessagePack produces UTF-8 \cite{UnicodeStandard} strings
that are not delimited with the \emph{NUL} character for both property names
and string values.  MessagePack supports four encoding variants of the type
definition depending on the byte-length of the string:

\begin{itemize}

\item If the byte-length of the string is less than $32$, then MessagePack
  prefixes the string with a type definition whose most-significant 3 bits
    equal \texttt{\textbf{101}} and whose remaining bits encode the string
    byte-length as an unsigned 5-bit integer.

  \item If the type definition is the byte \texttt{\textbf{0xd9}}, then
    MessagePack prefixes the string with the type definition followed by the
    byte-length of the string as an 8-bit unsigned integer.

  \item If the type definition is the byte \texttt{\textbf{0xda}}, then
    MessagePack prefixes the string with the type definition followed by the
    byte-length of the string as a Big Endian 16-bit unsigned integer.

  \item If the type definition is the byte \texttt{\textbf{0xdb}}, then
    MessagePack prefixes the string with the type definition followed by the
    byte-length of the string as a Big Endian 32-bit unsigned integer.

\end{itemize}

MessagePack does not attempt to deduplicate multiple occurrences of the same
property name or string value.

%%%%%%%%%%%%%%%%%%%%%%%%%%%%%%%%%%%%%%%%%%%%%%%%%
% BOOLEANS
%%%%%%%%%%%%%%%%%%%%%%%%%%%%%%%%%%%%%%%%%%%%%%%%%

\textbf{Booleans.} Booleans are encoded at the type level using the type
definition bytes \texttt{\textbf{0xc2}} (False) and \texttt{\textbf{0xc3}}
(True).

%%%%%%%%%%%%%%%%%%%%%%%%%%%%%%%%%%%%%%%%%%%%%%%%%
% ARRAYS
%%%%%%%%%%%%%%%%%%%%%%%%%%%%%%%%%%%%%%%%%%%%%%%%%

\textbf{Lists.} MessagePack encodes lists (arrays) as the concatenation of its
elements prefixed with a type definition that includes the logical size of the
array. MessagePack supports three encodings variants of the array type
definition depending on the size of the array:

\begin{itemize}

\item If the array contains less than $16$ elements, then MessagePack prefixes
  the array with a type definition whose most-significant 4 bits equal
    \texttt{\textbf{1001}} and whose remaining bits encode the array logical
    size as an unsigned 4-bit integer.

  \item If the type definition is the byte \texttt{\textbf{0xdc}}, then
    MessagePack prefixes the array with the type definition followed by the
    logical size of the array as a Big Endian 16-bit unsigned integer.

  \item If the type definition is the byte \texttt{\textbf{0xdd}}, then
    MessagePack prefixes the array with the type definition followed by the
    logical size of the array as a Big Endian 32-bit unsigned integer.

\end{itemize}

MessagePack does not attempt to deduplicate multiple occurrences of the same
element in an array.

\begin{figure*}[hb!]
  \frame{\includegraphics[width=\linewidth]{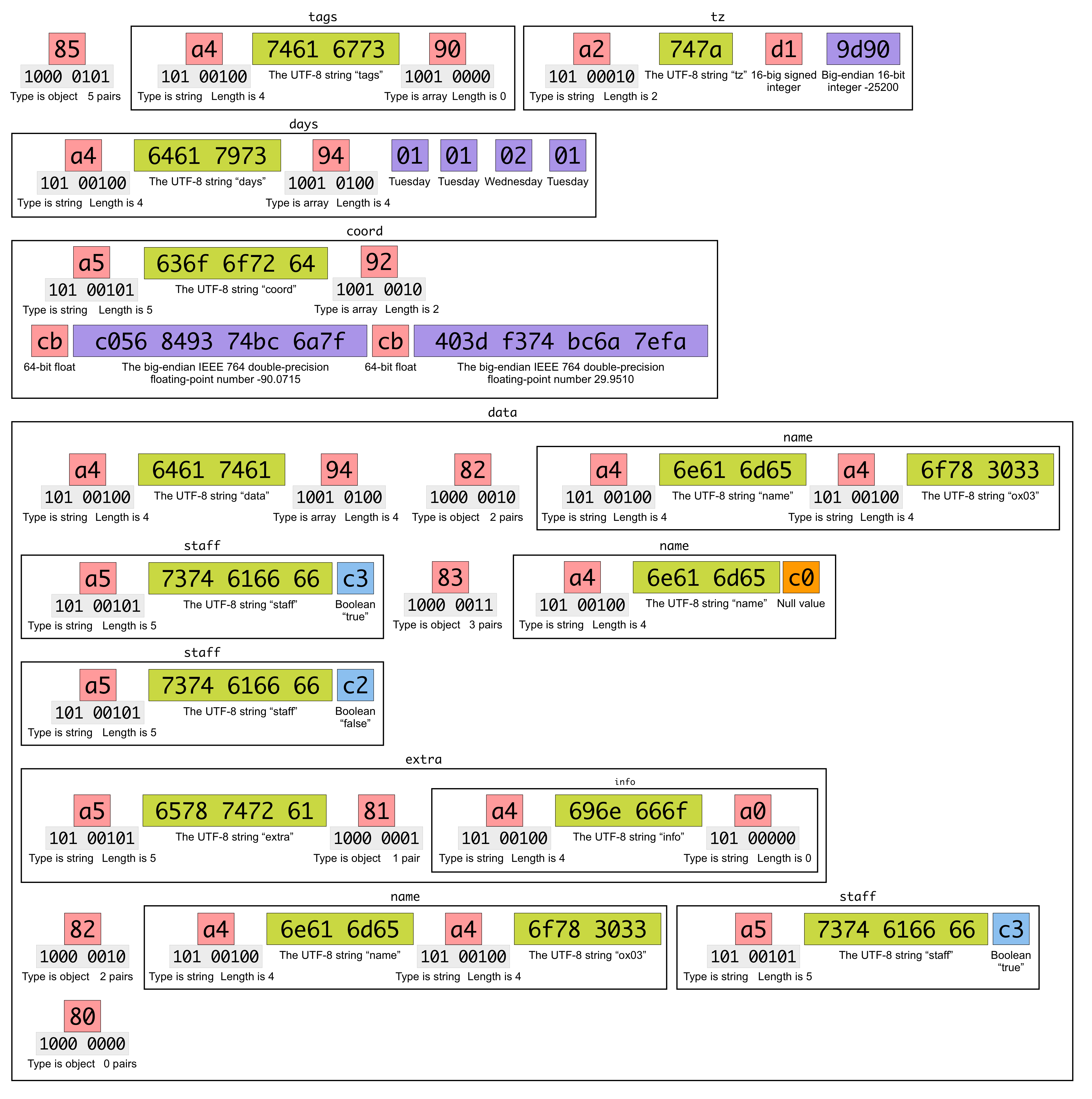}}
  \caption{Annotated hexadecimal output of serializing the
\autoref{lst:json-object-test} input data with MessagePack.}
\end{figure*}

\begin{table}[hb!]
\caption{A high-level summary of the MessagePack schema-less serialization specification.}
\label{table:messagepack}
\begin{tabularx}{\linewidth}{l|X}
  \toprule
  \textbf{Website} & \url{https://msgpack.org} \\ \hline
  \textbf{Company / Individual} & Sadayuki Furuhashi \\ \hline
  \textbf{Year} & 2009 \\ \hline
  \textbf{Specification} & \url{https://github.com/msgpack/msgpack/blob/master/spec.md} \\ \hline
  \textbf{License} & Apache License 2.0 \\ \hline
  \textbf{Layout} & Sequential \\ \hline
  \textbf{Languages} & ActionScript, C, C++, C\#, Clojure, Crystal, D, Dart, Delphi, Elixir, Erlang, F\#, Go, GNU Guile, Haskell, Haxe, HHVM, J, Java, JavaScript, Julia, Kotlin, Nim, MATLAB, OCaml, Objective-C, Pascal, PHP, Perl, Pony, Python, R, Racket, Ruby, Rust, Scala, Scheme, Smalltalk, SML, Swift \\ \hline
  \multicolumn{2}{c}{\textbf{Types}} \\ \hline

  \textbf{Numeric} &
  Big Endian 5-bit, 8-bit, 16-bit, 32-bit, and 64-bit Two's Complement \cite{twos-complement} signed integers \par\smallskip
  Big Endian 7-bit, 8-bit, 16-bit, 32-bit, and 64-bit unsigned integers \par\smallskip
  Big Endian 32-bit and 64-bit IEEE 764 floating-point numbers \cite{8766229} \\ \hline

  \textbf{String} & UTF-8 \cite{UnicodeStandard} \\ \hline
  \textbf{Composite} & Array, Map \\ \hline
  \textbf{Scalars} & Boolean, Nil \\ \hline
  \textbf{Other} &
  Bin (byte array) \par\smallskip
  32-bit, 64-bit, and 96-bit UNIX seconds and nanoseconds Epoch timestamps \cite{RFC8877} \\
  \bottomrule
\end{tabularx}
\end{table}

%% file: sections/formats/smile.tex
\label{sec:smile}

\begin{figure}[hb!]
  \frame{\includegraphics[width=\linewidth]{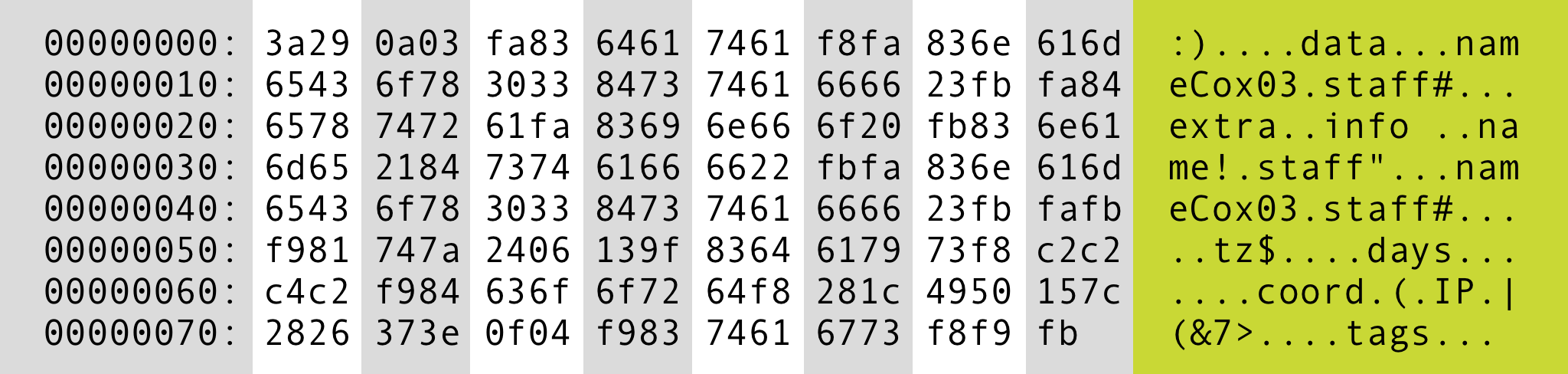}}
  \caption{Hexadecimal output (\texttt{xxd})
  of encoding \autoref{lst:json-object-test}
input data with Smile (127 bytes).}
\label{lst:hex-smile} \end{figure}

%%%%%%%%%%%%%%%%%%%%%%%%%%%%%%%%%%%%%%%%%%%%%%%%%
% HISTORY
%%%%%%%%%%%%%%%%%%%%%%%%%%%%%%%%%%%%%%%%%%%%%%%%%

\textbf{History.} Smile \cite{smile} is a schema-less binary serialization
specification developed by the team behind the popular Jackson JSON parser
\footnote{\url{https://github.com/FasterXML/jackson}}. Smile is an attempt to
create a standard JSON \cite{RFC8259} binary representation. Smile
development started on 2010 led by Tatu Saloranta
\footnote{\url{https://github.com/cowtowncoder}}, founder of FasterXML
\footnote{\url{http://fasterxml.com}} while also being a Principal Software
Engineer at Salesforce. Smile is released under the 2-clause BSD license
\footnote{\url{https://opensource.org/licenses/BSD-2-Clause}}.

%%%%%%%%%%%%%%%%%%%%%%%%%%%%%%%%%%%%%%%%%%%%%%%%%
% ADVANTAGES
%%%%%%%%%%%%%%%%%%%%%%%%%%%%%%%%%%%%%%%%%%%%%%%%%

\textbf{Characteristics.} \begin{itemize}

  \item \textbf{Performance Efficiency.} Smile's observation is that
    serialization specifications typically sacrifice write performance to speed-up
    read operations.  As a solution, Smile has been designed support equally
    runtime-efficient read and write operations.

  \item \textbf{Streaming.} Smile implementations can de-serialize bit-strings
    given a fixed amount of buffering.

\end{itemize}

%%%%%%%%%%%%%%%%%%%%%%%%%%%%%%%%%%%%%%%%%%%%%%%%%
% OVERALL STRUCTURE
%%%%%%%%%%%%%%%%%%%%%%%%%%%%%%%%%%%%%%%%%%%%%%%%%

\textbf{Layout.} A Smile bit-string is a self-delimited sequence of key-value
pairs, sometimes nested. The bit-strings produced by Smile are prefixed with a
header that consists of the ASCII string \texttt{\textbf{0x3a 0x29 0x0a}}, a
smiling face as the ASCII string ":)" plus a new-line character.  The header
string is followed by a byte that consists of the version number as a 4-bit
unsigned integer, followed by a reserved bit, followed by 3 bit flags:

\begin{itemize}
\item Whether the bit-string contains raw binary values.
\item Whether string values may be shared.
\item Whether property names may be shared.
\end{itemize}

Key-value pairs are encoded as the key element followed by the value element.
Smile objects are encoded as the concatenation of their key-value pairs
prefixed with the byte \texttt{\textbf{0xfa}} and suffixed with the byte
\texttt{\textbf{0xfb}}. Smile messages are guaranteed to not contain the
\texttt{\textbf{0xff}} byte as such byte is supported as an optional
end-of-message marker for framing purposes.

%%%%%%%%%%%%%%%%%%%%%%%%%%%%%%%%%%%%%%%%%%%%%%%%%
% TYPES
%%%%%%%%%%%%%%%%%%%%%%%%%%%%%%%%%%%%%%%%%%%%%%%%%

\textbf{Types.} Smile data types are prefixed with a 1-byte type definition.
The value might be embedded in the type definition if it is small enough
resulting in efficient use of space. There are three groups of type
definitions:

\begin{itemize}

\item Type bytes whose type is encoded in the 3 most-significant bits and the
  remaining 5-bits are type-dependent.

\item Constants such as \texttt{\textbf{0x21}} (null).

\item Structural markers such as \texttt{\textbf{0xf8}} (start of array).

\end{itemize}

%%%%%%%%%%%%%%%%%%%%%%%%%%%%%%%%%%%%%%%%%%%%%%%%%
% NUMBERS
%%%%%%%%%%%%%%%%%%%%%%%%%%%%%%%%%%%%%%%%%%%%%%%%%

\textbf{Numbers.} Smile supports Little Endian 5-bit, 32-bit, and 64-bit
ZigZag-encoded \ref{sec:zigzag-encoding} signed integers. 5-bit signed integers
are encoded as the least-significant bits of a type byte whose most-significant
3 bits equal \texttt{\textbf{110}}. Smile also supports a \emph{BigInteger}
type capable of encoding arbitrary-length signed integers. These type of
integers are encoded using the type byte \texttt{\textbf{0x30}}, followed by
the byte-length of the \emph{stringified} representation of the integer as an
unsigned Little Endian Base 128 (LEB128) (\ref{sec:varints}) variable-length
integer, followed by the \emph{stringified} UTF-8 \cite{UnicodeStandard}
representation of the integer encoded as a byte array.

\begin{figure}[hb!]
  \frame{\includegraphics[width=\linewidth]{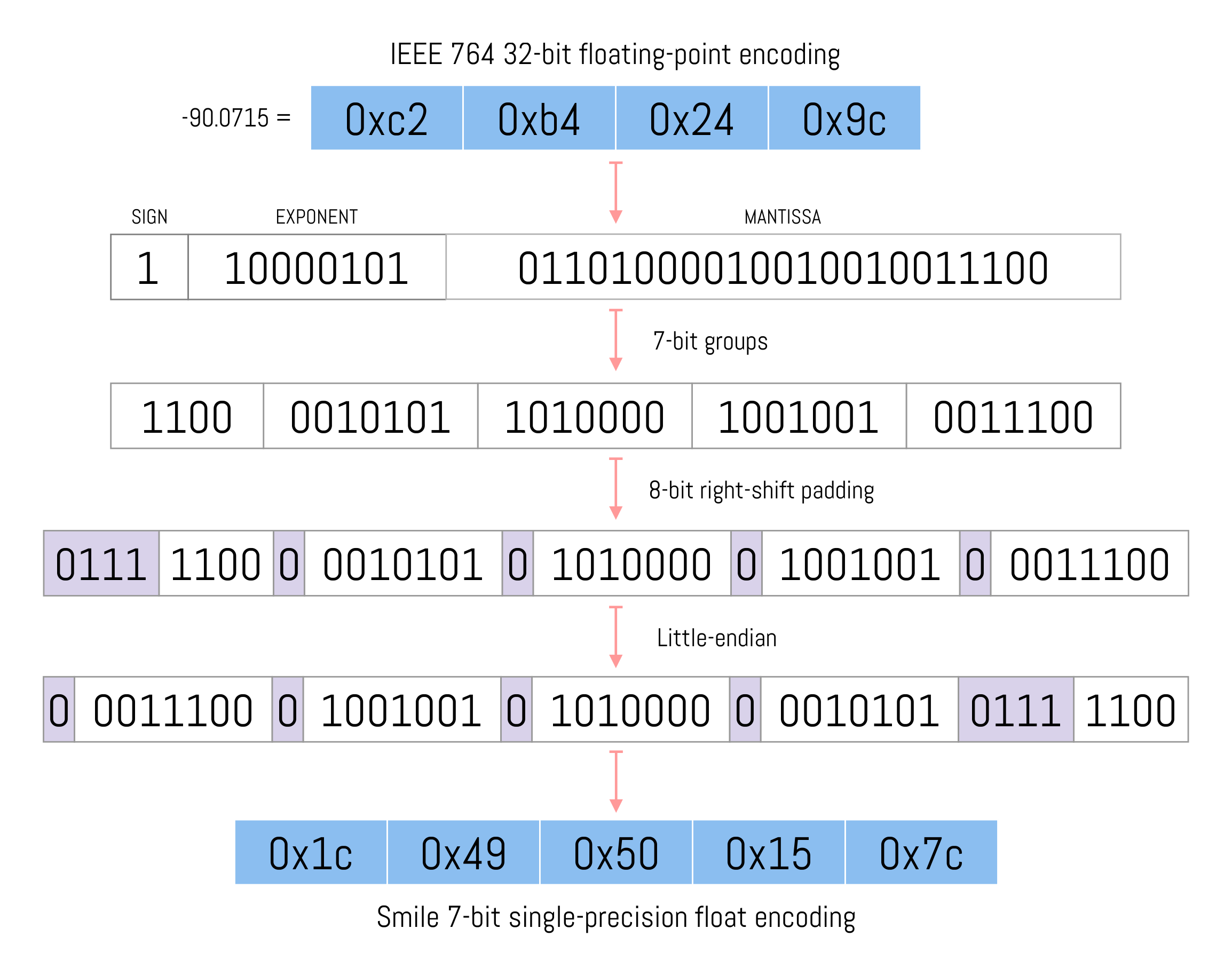}}
  \caption{A visual representation of converting the $-90.0715$ IEEE 764
  single-precision floating-point number \cite{8766229} to Smile's 7-bit
  float encoding.} \label{fig:smile-floats} \end{figure}

In terms of floats, Smile supports Little Endian 32-bit and 64-bit IEEE 764
\cite{8766229} floating-point numbers. However, Smile encodes
floating-point numbers using 7-bits to avoid including bytes that are reserved
for other type definitions. The encoding process consists in:

\begin{enumerate}

\item Obtaining the Big Endian IEEE 764 \cite{8766229} representation of
  the floating-point number.

\item Writing the least-significant 7 bits.

\item Right-shifting 7 bits.

\item Repeating the process until encoding the entire bit-string as shown in
  \autoref{fig:smile-floats}.

\end{enumerate}

Smile also supports a \emph{BigDecimal} type capable of encoding
arbitrary-precision decimal numbers.  These type of decimal numbers are encoded
using the type byte \texttt{\textbf{0x20}}, followed by the scale as a
ZigZag-encoded (\ref{sec:zigzag-encoding}) 32-bit signed integer, followed by
the \emph{stringified} UTF-8 \cite{UnicodeStandard} representation of the
integral part encoded as a byte array prefixed with its byte-length as an
unsigned Little Endian Base 128 (LEB128) (\ref{sec:varints}) variable-length
integer.

%%%%%%%%%%%%%%%%%%%%%%%%%%%%%%%%%%%%%%%%%%%%%%%%%
% STRINGS
%%%%%%%%%%%%%%%%%%%%%%%%%%%%%%%%%%%%%%%%%%%%%%%%%

\textbf{Strings.} Smile supports strings using ASCII \cite{STD80} and UTF-8
\cite{UnicodeStandard} encodings that are not \emph{NUL}-delimited. Based from
\our observations, property names are encoded using UTF-8 while string values
are encoded using ASCII unless the strings contain characters outside of the
ASCII range. Smile provides an empty string type and three type definition
variants for each of the supported encodings depending on the byte-length of
the string as shown in \autoref{table:smile-strings}.

\begin{table}[hb!]
\caption{The string encodings that Smile supports as specified in \cite{smile}.
  The \emph{From} and \emph{To} columns describe the string byte-lengths that each group can encode.}
\label{table:smile-strings}
\begin{tabularx}{\linewidth}{l|l|l|l|X|X}
  \toprule
  \textbf{Name} & \textbf{Type byte} & \textbf{From} & \textbf{To} & \textbf{String length} & \textbf{Suffix} \\ \midrule
  Empty string & \texttt{0x20} & 0 bytes & 0 bytes & 0 & None \\ \hline
  Tiny ASCII & \texttt{0x40} to \texttt{0x5f} & 1 byte & 32 bytes & Least-significant 5-bits as unsigned integer + 1 & None \\ \hline
  Tiny Unicode & \texttt{0x80} to \texttt{0x9f} & 2 bytes & 33 bytes & Least-significant 5-bits as unsigned integer + 2 & None \\ \hline
  Small ASCII & \texttt{0x60} to \texttt{0x7f} & 33 bytes & 64 bytes & Least-significant 5-bits as unsigned integer + 33 & None \\ \hline
  Small Unicode & \texttt{0xa0} to \texttt{0xbf} & 34 bytes & 65 bytes & Least-significant 5-bits as unsigned integer + 34 & None \\ \hline
  Long ASCII & \texttt{0xe0} & 1 byte & Any & Until suffix marker & \texttt{0xfc} end-of-string marker \\ \hline
  Long Unicode & \texttt{0xe4} & 2 bytes & Any & Until suffix marker & \texttt{0xfc} end-of-string marker \\
  \bottomrule
\end{tabularx}
\end{table}

For example, an ASCII \cite{STD80} string consisting of 4 characters can be
encoded as a \emph{Tiny ASCII} string. Therefore, it is prefixed with the type
byte \texttt{\textbf{0x43 = 01000011}}. The 5 least-significant bits are the
unsigned integer $3$ so the string length is $4 = 1 + 3$.

Alternatively, a UTF-8 \cite{UnicodeStandard} string consisting of 35
characters can be encoded as a \emph{Small Unicode} string. Therefore, it is
prefixed with the type byte \texttt{\textbf{0xa1 = 10100001}}. The 5
least-significant bits are the unsigned integer $1$ so the string length is $35
= 34 + 1$.

Note that the \emph{Tiny Unicode} string encoding group
cannot encode a 1-byte string. A 1-byte UTF-8 \cite{UnicodeStandard} string
must be a valid ASCII \cite{STD80}, therefore the \emph{Tiny ASCII}
encoding group is preferred.

%%%%%%%%%%%%%%%%%%%%%%%%%%%%%%%%%%%%%%%%%%%%%%%%%
% BOOLEANS
%%%%%%%%%%%%%%%%%%%%%%%%%%%%%%%%%%%%%%%%%%%%%%%%%

\textbf{Booleans.} Booleans are encoded as the type definition level using the
type bytes \texttt{\textbf{0x22}} (False) and \texttt{\textbf{0x23}} (True).

%%%%%%%%%%%%%%%%%%%%%%%%%%%%%%%%%%%%%%%%%%%%%%%%%
% ARRAYS
%%%%%%%%%%%%%%%%%%%%%%%%%%%%%%%%%%%%%%%%%%%%%%%%%

\textbf{Lists.} Smile encodes lists (arrays) as the concatenation of its
elements prefixed with the constant byte \texttt{\textbf{0xf8}} and suffixed
with the constant byte \texttt{\textbf{0xf9}}.

\begin{figure*}[hb!]
  \frame{\includegraphics[width=\linewidth]{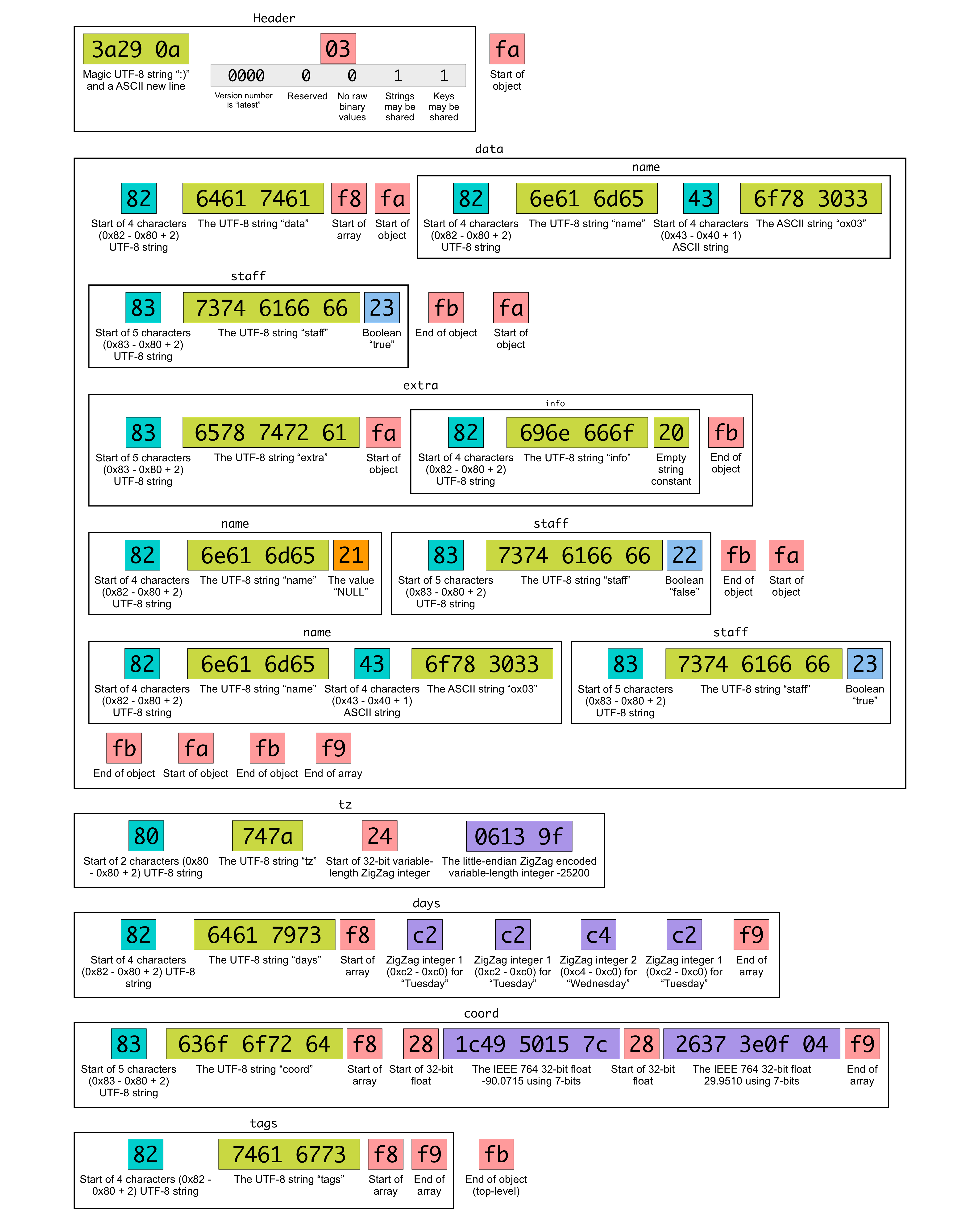}}
  \caption{Annotated hexadecimal output of serializing the
\autoref{lst:json-object-test} input data with Smile.}
\end{figure*}

\begin{table}[hb!]
\caption{A high-level summary of the Smile schema-less serialization specification.}
\label{table:smile}
\begin{tabularx}{\linewidth}{l|X}
  \toprule
  \textbf{Website} & \url{https://github.com/FasterXML/smile-format-specification} \\ \hline
  \textbf{Company / Individual} & FasterXML \\ \hline
  \textbf{Year} & 2010 \\ \hline
  \textbf{Specification} & \url{https://github.com/FasterXML/smile-format-specification/blob/master/smile-specification.md} \\ \hline
  \textbf{License} & 2-clause BSD \\ \hline
  \textbf{Layout} & Sequential \\ \hline
  \textbf{Languages} & C, Clojure, Go, Java, JavaScript, Python \\ \hline
  \multicolumn{2}{c}{\textbf{Types}} \\ \hline

  \textbf{Numeric} &
  5-bit, 32-bit, and 64-bit ZigZag-encoded (\ref{sec:zigzag-encoding}) signed Little Endian Base 128 (LEB128) (\ref{sec:varints}) variable-length integers \par\smallskip
  Little Endian 32-bit and 64-bit IEEE 764 floating-point numbers \cite{8766229} encoded using 7 bit groups \par\smallskip
  Arbitrary-length stringified signed integers \par\smallskip
  Arbitrary-precision decimals (with scale and stringified integral) \\ \hline

  \textbf{String} & ASCII \cite{STD80}, UTF-8 \cite{UnicodeStandard} \\ \hline
  \textbf{Composite} & Array, Object \\ \hline
  \textbf{Scalars} & Boolean, Null \\ \hline
  \textbf{Other} & Binary (byte array) \\
  \bottomrule
\end{tabularx}
\end{table}

%% file: sections/formats/ubjson.tex
\label{sec:ubjson}

\begin{figure}[hb!]
  \frame{\includegraphics[width=\linewidth]{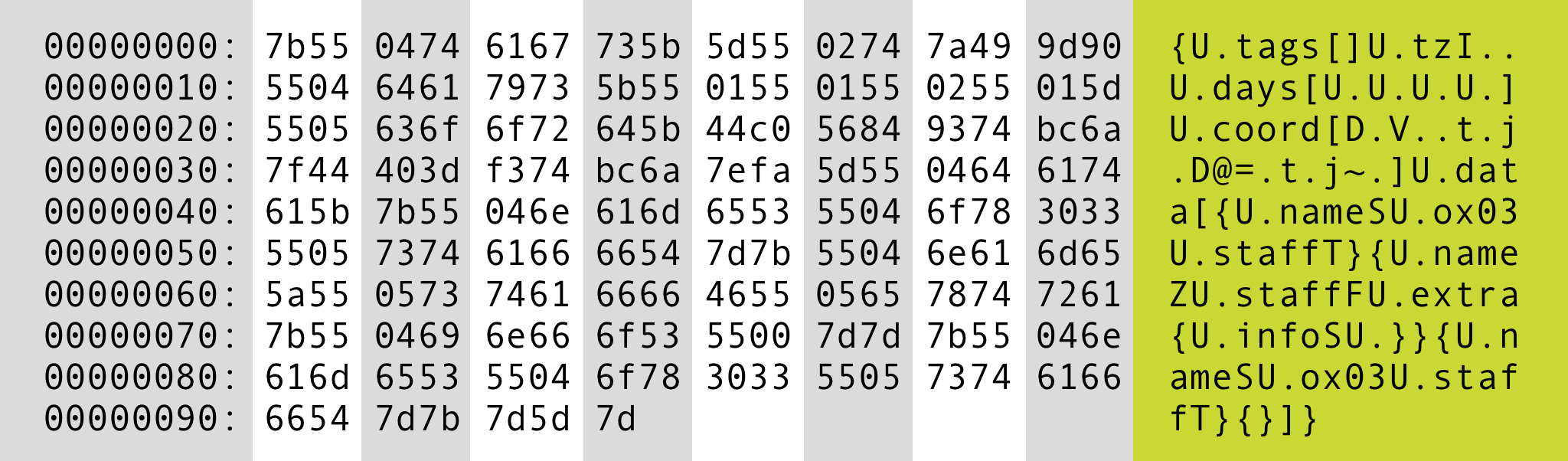}}
  \caption{Hexadecimal output (\texttt{xxd})
  of encoding \autoref{lst:json-object-test}
input data with UBJSON (151 bytes).}
\label{lst:hex-ubjson} \end{figure}

%%%%%%%%%%%%%%%%%%%%%%%%%%%%%%%%%%%%%%%%%%%%%%%%%
% HISTORY
%%%%%%%%%%%%%%%%%%%%%%%%%%%%%%%%%%%%%%%%%%%%%%%%%

\textbf{History.} UBJSON \cite{ubjson} is a schema-less binary serialization
specification that is a purely-compatible binary alternative to JSON \cite{RFC8259}.
Riyad Kalla \footnote{\url{https://github.com/rkalla}} started working on
UBJSON in 2012 while working as a Principal Lead at Genuitec
\footnote{\url{https://www.genuitec.com}}.  Many high-profile software
solutions include UBJSON support, such as Teradata
\footnote{\url{https://www.teradata.co.uk}} and Wolfram Mathematica
\footnote{\url{https://www.wolfram.com/mathematica/}}.  UBJSON is also natively
supported as part of the RIOT operating system for the Internet of Things
\cite{borgohain2015survey}. The UBJSON specification has been released under
the Apache License 2.0
\footnote{\url{http://www.apache.org/licenses/LICENSE-2.0.html}}.

%%%%%%%%%%%%%%%%%%%%%%%%%%%%%%%%%%%%%%%%%%%%%%%%%
% ADVANTAGES
%%%%%%%%%%%%%%%%%%%%%%%%%%%%%%%%%%%%%%%%%%%%%%%%%

\textbf{Characteristics.} \begin{itemize}

  \item \textbf{Readability.} Despite being a binary serialization specification,
    UBJSON is comparatively human-readable as it makes use of printable ASCII
    \cite{STD80} characters in field type definitions.

  \item \textbf{Simplicity.} The UBJSON specification is easy to understand and
    implement from a developer point of view as the specification is defined using a
    single core data structure throughout the entire specification

  \item \textbf{JSON Compatibility.} In comparison to serialization
    specifications such as BSON \cite{bson}, UBJSON is strictly compatible with
    the JSON specification \cite{RFC8259} and does not introduce additional
    data types.

\end{itemize}

%%%%%%%%%%%%%%%%%%%%%%%%%%%%%%%%%%%%%%%%%%%%%%%%%
% OVERALL STRUCTURE
%%%%%%%%%%%%%%%%%%%%%%%%%%%%%%%%%%%%%%%%%%%%%%%%%

\textbf{Layout.} A UBJSON bit-string is a sequence of key-value pairs,
sometimes nested. Every UBJSON element shares the same structure: a 1-byte type
definition, the content length if applicable, and the optional content data. A
key-value pair is the sequence of a string element and its corresponding value
element. UBJSON encodes objects as the concatenation of their key-value pairs
prefixed with the byte \texttt{\textbf{0x7b}} (the ASCII character
\texttt{\textbf{\{}} \ ) and suffixed with the byte \texttt{\textbf{0x7d}} (the
ASCII character \texttt{\textbf{\}}} \ ).

%%%%%%%%%%%%%%%%%%%%%%%%%%%%%%%%%%%%%%%%%%%%%%%%%
% TYPES
%%%%%%%%%%%%%%%%%%%%%%%%%%%%%%%%%%%%%%%%%%%%%%%%%

\begin{figure}[hb!]
  \frame{\includegraphics[width=\linewidth]{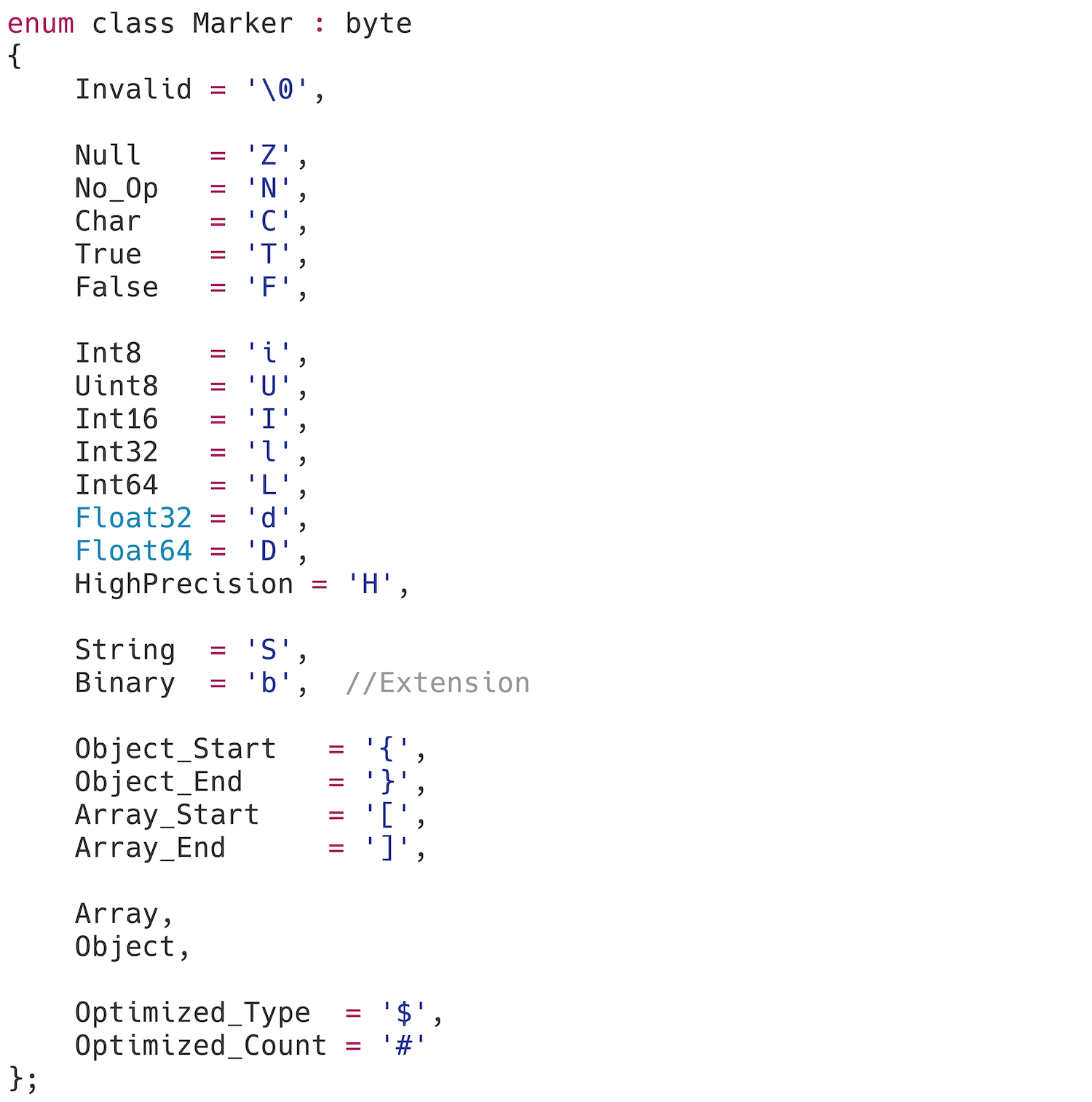}}
  \caption{UBJSON markers definition \protect\footnotemark.}
\label{lst:ubjson-markers} \end{figure}

\textbf{Types.} Each UBJSON element is prefixed with a 1-byte type definitions.
Type definitions are encoded using characters from the \emph{printable} ASCII
\cite{STD80} range. Each type definition character is a mnemonic of its
respective type. Refer to \autoref{lst:ubjson-markers} for a complete list. For
example, the \emph{string} type makes use of the character \texttt{\textbf{S}}.

\footnotetext{\url{https://github.com/WhiZTiM/UbjsonCpp/blob/7a7857f64247ce82b72072b04f87183b090fd554/include/types.hpp}}

%%%%%%%%%%%%%%%%%%%%%%%%%%%%%%%%%%%%%%%%%%%%%%%%%
% NUMBERS
%%%%%%%%%%%%%%%%%%%%%%%%%%%%%%%%%%%%%%%%%%%%%%%%%

\textbf{Numbers.} UBJSON supports Big Endian 8-bit (type definition
\texttt{\textbf{i}}), 16-bit (type definition \texttt{\textbf{I}}), 32-bit
(type definition \texttt{\textbf{l}}), and 64-bit (type definition
\texttt{\textbf{L}}) signed Two's Complement \cite{twos-complement}
integers and Big Endian 8-bit (type definition \texttt{\textbf{U}}) unsigned
integers. UBJSON also supports Big Endian IEEE 764 \cite{8766229} 32-bit
(type definition \texttt{\textbf{d}}) and 64-bit (type definition
\texttt{\textbf{D}}) floating-point numbers. Additionally, UBJSON supports
high-precision arbitrarily-large UTF-8 \cite{UnicodeStandard} stringified
numbers prefixed by the type definition \texttt{\textbf{H}} and the byte-length
of the string.  UBJSON implementations pick the smallest data type that can
encode the given number.

%%%%%%%%%%%%%%%%%%%%%%%%%%%%%%%%%%%%%%%%%%%%%%%%%
% STRINGS
%%%%%%%%%%%%%%%%%%%%%%%%%%%%%%%%%%%%%%%%%%%%%%%%%

\textbf{Strings.} UBJSON produces UTF-8 \cite{UnicodeStandard} strings that
are not delimited with the \emph{NUL} character for both property names and
string values. UBJSON strings are prefixed with the type definition
\texttt{\textbf{S}} and the byte-length of the string as a standalone integer
element with its own type definition. UBJSON does not attempt to deduplicate
multiple occurrences of the same property name or string value. UBJSON supports
a \emph{character} type to encode a single-character string without encoding
its byte-length resulting in efficient use of space.

%%%%%%%%%%%%%%%%%%%%%%%%%%%%%%%%%%%%%%%%%%%%%%%%%
% BOOLEANS
%%%%%%%%%%%%%%%%%%%%%%%%%%%%%%%%%%%%%%%%%%%%%%%%%

\textbf{Booleans.} UBJSON encodes booleans at the type definition level. UBJSON
supports two data types without content: \texttt{\textbf{0x54}} (True) and
\texttt{\textbf{0x46}} (False) which stand for the ASCII \cite{STD80}
characters \texttt{\textbf{T}} and \texttt{\textbf{F}}, respectively.

%%%%%%%%%%%%%%%%%%%%%%%%%%%%%%%%%%%%%%%%%%%%%%%%%
% ARRAYS
%%%%%%%%%%%%%%%%%%%%%%%%%%%%%%%%%%%%%%%%%%%%%%%%%

\textbf{Lists.} UBJSON encodes lists (arrays) as the concatenation of its
elements. UBJSON arrays are prefixed with the byte \texttt{\textbf{0x5b}} (the
ASCII character \texttt{\textbf{[}}) and suffixed with the byte
\texttt{\textbf{0x5d}} (the ASCII character \texttt{\textbf{]}}). UBJSON
arrays are not prefixed with their logical size nor their byte-length. UBJSON
does not attempt to deduplicate multiple occurrences of the same element in an
array.

\begin{figure*}[hb!]
  \frame{\includegraphics[width=\linewidth]{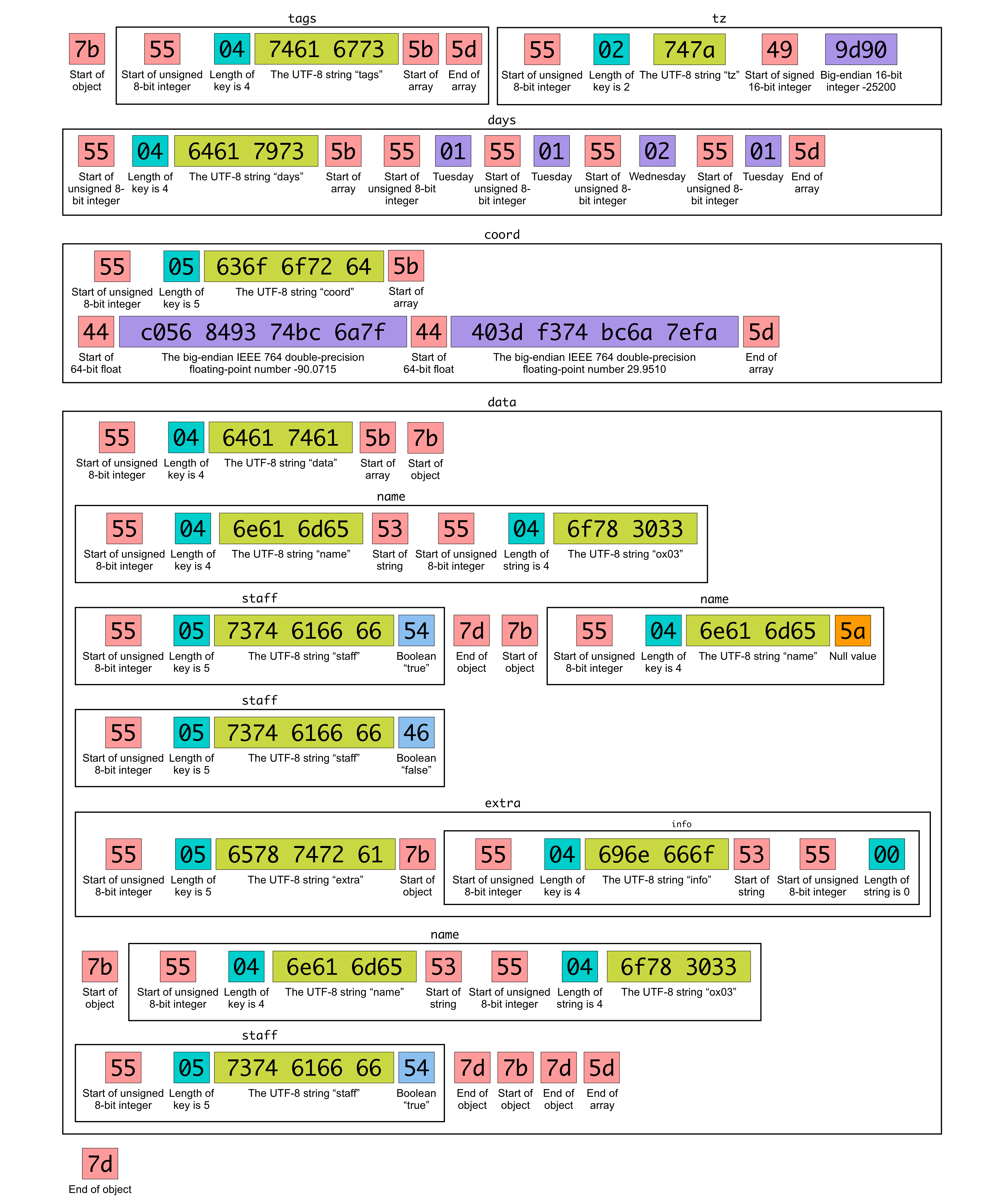}}
  \caption{Annotated hexadecimal output of serializing the
\autoref{lst:json-object-test} input data with UBJSON.}
\end{figure*}

\begin{table}[hb!]
\caption{A high-level summary of the UBJSON schema-less serialization specification.}
\label{table:ubjson}
  \begin{tabularx}{\linewidth}{l|X}
  \toprule
  \textbf{Website} & \url{https://ubjson.org} \\ \hline
  \textbf{Company / Individual} & Riyad Kalla \\ \hline
  \textbf{Year} & 2012 \\ \hline
  \textbf{Specification} & \url{https://github.com/ubjson/universal-binary-json} \\ \hline
  \textbf{License} & Apache License 2.0 \\ \hline
  \textbf{Layout} & Sequential \\ \hline
  \textbf{Languages} & C, C++, C\#, D, Go, Java, JavaScript, MATLAB, PHP, Python, Swift \\ \hline
  \multicolumn{2}{c}{\textbf{Types}} \\ \hline

  \textbf{Numeric} &
  Big Endian 8-bit, 16-bit, 32-bit, and 64-bit Two’s Complement \cite{twos-complement} signed integers \par\smallskip
  Big Endian 8-bit unsigned integers \par\smallskip
  Big Endian 32-bit and 64-bit IEEE 764 floating-point numbers \cite{8766229} \par\smallskip
  Arbitrary-precision ASCII-encoded numbers \\ \hline

  \textbf{String} & UTF-8 \cite{UnicodeStandard} \\ \hline
  \textbf{Composite} & Array, Object \\ \hline
  \textbf{Scalars} & Boolean, Null \\
  \bottomrule
\end{tabularx}
\end{table}

%% file: sections/evolution.tex
\label{sec:evolution}

%%%%%%%%%%%%%%%%%%%%%%%%%%%%%%%%%%%%%%%%%%%%%%%%%
% DEFINITION
%%%%%%%%%%%%%%%%%%%%%%%%%%%%%%%%%%%%%%%%%%%%%%%%%

Schema evolution is the problem of updating a schema definition while ensuring
that the programs relying on it can continue to operate. The study of schema
evolution originated in the context of relational databases to evolve database
schemas without disrupting client applications \cite{Roddick95asurvey}. In the
context of binary serialization specifications, schema evolution is concerned
with how bit-string producers and bit-string consumers can intercommunicate
despite future updates to the structure of the bit-strings they are concerned
with.

\begin{figure}[hb!]
  \frame{\includegraphics[width=\linewidth]{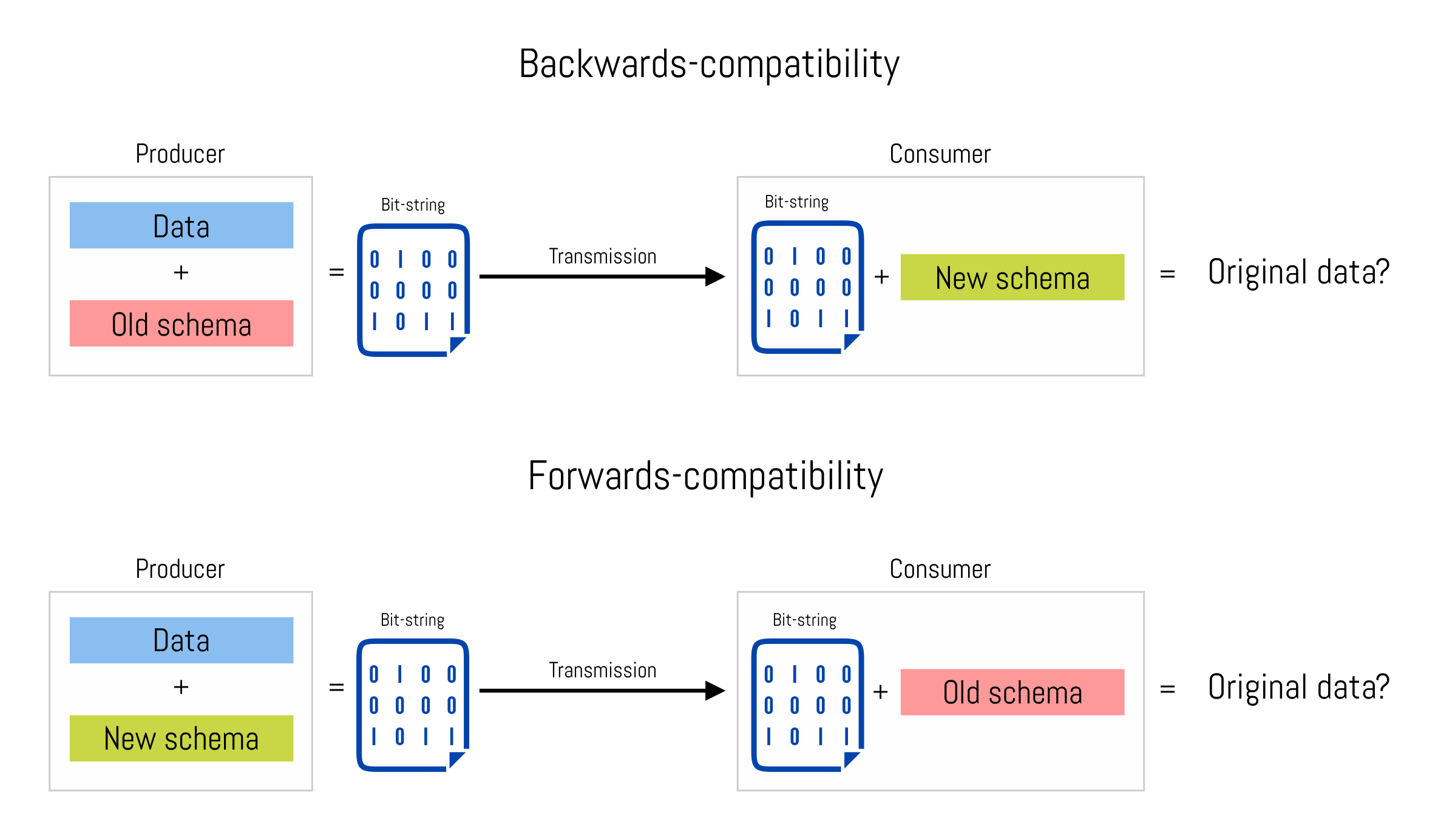}}
  \caption{In both of these cases, the producer serializes a data structure
  using one version of the schema and the consumer attempts to deserialize the
resulting bit-string using another version of the schema. Schema evolution is
concerned with whether the consumer will succeed in obtaining the original data
structure or not.} \label{fig:schema-evolution} \end{figure}

%%%%%%%%%%%%%%%%%%%%%%%%%%%%%%%%%%%%%%%%%%%%%%%%%
% RELEVANCE
%%%%%%%%%%%%%%%%%%%%%%%%%%%%%%%%%%%%%%%%%%%%%%%%%

As discussed in \autoref{sec:schema-less-driven}, programs using schema-driven
serialization specifications must know in advance the schema definitions of the
messages they are expecting to interchange. This problem is exacerbated by the
fact that schema definitions are typically updated in response to new or
changing requirements. \cite{5643642} state that software requirements
continuously change and as a result of these changes software projects tend to
fail. Schema evolution is an important topic in the context of schema-driven
serialization specifications as updating a schema definition may result in a
risky and expensive operation that requires re-compilation and coordinated
re-deployment of all the programs relying on such schema.

%%%%%%%%%%%%%%%%%%%%%%%%%%%%%%%%%%%%%%%%%%%%%%%%%
% INTRODUCTION TO SCHEMA COMPATIBILITY
%%%%%%%%%%%%%%%%%%%%%%%%%%%%%%%%%%%%%%%%%%%%%%%%%

Two schemas are \emph{compatible} if one schema can deserialize a bit-string
produced by the other schema and recover the original information. There are
three levels of schema compatibility:

\begin{itemize}

\item \textbf{Backwards.} The first schema is backwards-compatible with
  respect to the second schema if the first schema can deserialize data
    produced by the second schema.

\item \textbf{Forwards.} The first schema is forwards-compatible with respect
  to the second schema if the second schema can deserialize data produced by
    the first schema.

\item \textbf{Full.} The new schema is fully-compatible with respect to the old
  schema if it is both backwards and forwards compatible with respect to the
    old schema.

\end{itemize}

%% file: sections/evolution-theory.tex
We can think of a schema as a set of its valid instances where the following
rules apply:

\begin{itemize}

\item A backwards or forwards compatible transformation to the schema
  \emph{expands} or \emph{confines} the set of its valid instances,
    respectively. A fully-compatible transformation to the schema keeps the set
    of its valid instances intact.

\item A schema transformation results in an incompatible schema if neither of
  the sets is a subset of the other.

\item A schema is \emph{fully-compatible} with respect to itself.

\item The first schema is \emph{backwards-compatible} with respect to the
  second schema if and only if the second schema is \emph{forwards-compatible}
    with respect to the first schema.

\item Compatibility is a transitive property. A schema is transitive
  backwards, transitive forwards, or transitive fully-compatible with respect
  to a set of schemas if it is backwards, forwards, or fully compatible with
  respect to each of the schemas in the set, respectively.

\end{itemize}

%% file: sections/deploying-schema-transformations.tex
Consider a system involving a set of consumers and a set of producers that
exchange information using a schema-driven serialization specification. In such
a system, the rules for deploying \emph{compatible} schema transformations with
zero-downtime are as described in \autoref{table:compatible-deploys}. Deploying
\emph{incompatible} schema transformations typically involves re-deploying all
consumers and producers at the same time or including multiple incompatible
schemas in each of the programs and adding application-specific logic to decide
which schema to use when.

The same program in the system might be both a producer and a consumer. In this
case, consider the program to use one schema to produce data and another schema
to consume data where the two schemas may be equal. Therefore, each of the
schemas within the same program can be deployed separately using the rules
described in \autoref{table:compatible-deploys}.

\begin{table}[hb!]
\caption{These are the rules for deploying compatible schema transformations
  with zero-downtime.  For example, it is safe to deploy a forwards-compatible
  schema transformation to any producer. However, deploying a
  forwards-compatible schema transformation to any consumer requires first
  deploying the schema transformation to all producers.}

\label{table:compatible-deploys}
\begin{tabularx}{\linewidth}{l|X|X|X}
  \toprule
 & \textbf{Backwards-compatible schema transformation}
& \textbf{Forwards-compatible schema transformation}
& \textbf{Fully-compatible schema transformation} \\ \midrule
\textbf{Deploy to Producer} & Deploy transformation to consumers first & Safe & Safe \\ \hline
\textbf{Deploy to Consumer} & Safe & Deploy transformation to producers first & Safe \\
\bottomrule
\end{tabularx}
\end{table}

%% file: sections/evolution-analysis.tex
\label{sec:compatibility-analysis}

\We selected a set of structural and type conversion schema transformations and
tested if they result in compatible changes using the schema-driven
serialization implementations and encodings introduced in
\autoref{table:versions-schema-driven}. A different encoding of the same
schema-driven serialization specification may yield different results. The
results of the structural schema transformations are presented in
\autoref{table:evolution-structural} and the results of the type conversion
schema transformations are presented in \autoref{table:evolution-conversions}.
\We mark the test results as shown in \autoref{table:evolution-legend}.

\We found that sometimes the schema evolution features of a serialization
specification are subtly affected by the data types being used and by the
infinite possibilities of surrounding data. For this reason, \we recommend
schema-writers to use these results as a guide and to unit test the schema
transformations they plan to apply before deploying them. \We also encountered
various cases of undocumented compatible schema transformations. These
transformations may rely on accidental behaviour of either the serialization
specification design or the chosen implementation and may carry no future
guarantees.  \We encourage readers to consult the official schema evolution
documentation and check if their serialization specification of choice
satisfies the intended compatible transformation by design or by accident.

\begin{table}[hb!]
\caption{Descriptions of how \we will mark schema evolution transformation results.}
\label{table:evolution-legend}
  \begin{tabularx}{\linewidth}{l|X|X}
\toprule
\textbf{Symbol} & \textbf{Description} & \textbf{When} \\ \midrule
\texttt{\textbf{A}} & Fully-compatible & The schemas are fully-compatible for all tested instances \\ \hline
\texttt{\textbf{F}} & Forwards-compatible & The schemas are forwards-compatible for all tested instances, despite backwards-compatibility failures or exceptions \\ \hline
\texttt{\textbf{B}} & Backwards-compatible & The schemas are backwards-compatible for all tested instances, despite forwards-compatibility failures or exceptions \\ \hline
\texttt{\textbf{N}} & Silently-incompatible & The schemas are not forwards nor backwards-compatible in at least one tested instance but no exception is thrown \\ \hline
\texttt{\textbf{X}} & Runtime exception & The schemas are neither forwards nor backwards-compatible for all tested instances and at least one exception is thrown \\ \hline
 & Not-applicable & The schema transformation is not applicable to the serialization specification as it involves data types not supported by the serialization specification \\
 \bottomrule
\end{tabularx}
\end{table}

\begin{table}[hb!]
\caption{A schema transformation result is annotated as shown in
\autoref{table:evolution-legend}. The \emph{Type} column documents whether a
schema transformation confines (\texttt{\textbf{C}}), expands
(\texttt{\textbf{E}}), changes (\texttt{\textbf{!}}), or preserves
(\texttt{\textbf{=}}) the domain of the schema.}
\label{table:evolution-structural}
\begin{tabularx}{\linewidth}{|l|c|X|c|c|c|c|c|c|c|}

\toprule

% Header
\textbf{\rotatebox[origin=c]{90}{Category}} &
\textbf{\rotatebox[origin=c]{90}{Type}} &
\textbf{Schema Transformation} &
\textbf{\rotatebox[origin=c]{90}{ASN.1 PER Unaligned}} &
\textbf{\rotatebox[origin=c]{90}{Apache Avro Binary Encoding}} &
\textbf{\rotatebox[origin=c]{90}{Microsoft Bond Compact Binary v1}} &
\textbf{\rotatebox[origin=c]{90}{Cap'n Proto Packed Encoding}} &
\textbf{\rotatebox[origin=c]{90}{FlatBuffers Binary Wire Format}} &
\textbf{\rotatebox[origin=c]{90}{Protocol Buffers Binary Wire Format}} &
\textbf{\rotatebox[origin=c]{90}{Apache Thrift Compact Protocol}} \\ \midrule

\multirow{7}{*}{\rotatebox[origin=c]{90}{Structures / Tables}}
& E & Add an optional field to the end & A & A & A & A & A & A & A \\ \cline{2-10}
& C & Remove an optional field from the end & A & A & A & A & A & A & A \\ \cline{2-10}
& C & Add a required field & F & F & F &  & F &  & F \\ \cline{2-10}
& E & Remove a required field & B & B & B &  & B &  & B \\ \cline{2-10}
  & C & Optional to required & F & F & A {\tiny(1)} &  & F &  & F \\ \cline{2-10}
  & E & Required to optional & B & B & A {\tiny(1)} &  & B &  & B \\ \cline{2-10}
& ! & Change field default & N & N & N & N & N &  & N \\ \hline

\multirow{3}{*}{\rotatebox[origin=c]{90}{Lists}}
  & E & List of scalars to list of structures with scalar & A {\tiny(2)} & X & N & A {\tiny(2)} & X & X & X \\ \cline{2-10}
  & E & Scalar to list of scalars & X & X & X & N & N & B {\tiny(3)} & N \\ \cline{2-10}
& E & Composite to list of composites & X & X & X & X & X & B {\tiny(3)} & N \\ \hline

\multirow{8}{*}{\rotatebox[origin=c]{90}{Unions}}
  & ! & Move optional field to existing union & X & X &  & N & B {\tiny(4)} & F {\tiny(4)} & N \\ \cline{2-10}
& ! & Extract optional field from existing union & X & X &  & N & F {\tiny(4)} & B {\tiny(4)} & N \\ \cline{2-10}
& E & Move required field to existing union & X & X &  &  & B &  & N \\ \cline{2-10}
& C & Extract required field from existing union & X & X &  &  & F &  & N \\ \cline{2-10}
& ! & Move optional field to a new union & X & B {\tiny(5)} &  & B {\tiny(5)} & N & A {\tiny(5)} & N \\ \cline{2-10}
& E & Move required field to a new union & B & B &  &  & N &  & N \\ \cline{2-10}
& E & Add choice to existing union & B & B &  & B & B & B & B \\ \cline{2-10}
& C & Remove choice from existing union & F & F &  & F & F & F & F \\ \hline

\multirow{4}{*}{\rotatebox[origin=c]{90}{Enums}}
& C & Scalar to enumeration & F & X & F & F & F & F & F \\ \cline{2-10}
& E & Enumeration to scalar & B & X & B & B & B & B & B \\ \cline{2-10}
& E & Add enumeration constant & B & B & B & B & B & B & B \\ \cline{2-10}
& C & Remove enumeration constant & F & F & F & F & F & F & F \\
\bottomrule

\end{tabularx}
\end{table}

\textbf{Description from \autoref{table:evolution-structural}}.

(1) Microsoft Bond \cite{microsoft-bond} supports the concept of
\emph{required\_optional} fields that are required at serialization time but
optional at deserialization time. This concept enables schema-writers to make
an optional field requires and viceversa in a fully-compatible manner through a
two-step process: Changing an optional or required field to
\emph{required\_optional}, deploying the schema update to both producers and
consumers, and then changing the \emph{required\_optional} field to required or
optional.

\ifx\thesis\undefined
(2) ASN.1 PER Unaligned \cite{asn1-per} and Cap'n Proto \cite{capnproto}
support updating a list of scalars to a list of structures where the scalar is
the first and only element in a fully-compatible manner. In the case of ASN.1
PER Unaligned, this transformation is possible because structures are list of
values and a list of structures with a single required scalar is encoded in the
same manner as a list of such scalars. In the case of Cap'n Proto, a list
definition declares whether its element are scalars or composites as shown in
\autoref{table:capnproto-list-definitions}. If the elements are composite, the
list definition points to a 64-bit word that defines the composite elements,
allowing the deserializer to determine if following the pointer or not yields a
scalar of the same expected type. As an exception, Cap'n Proto does not support
this schema transformation on a list of booleans for runtime-efficiency reasons
\footnote{\url{https://capnproto.org/language.html\#evolving-your-protocol}}.
\else
(2) ASN.1 PER Unaligned \cite{asn1-per} and Cap'n Proto \cite{capnproto}
support updating a list of scalars to a list of structures where the scalar is
the first and only element in a fully-compatible manner. In the case of ASN.1
PER Unaligned, this transformation is possible because structures are list of
values and a list of structures with a single required scalar is encoded in the
same manner as a list of such scalars. In the case of Cap'n Proto, a list
definition declares whether its element are scalars or composites as shown in
\cite{viotti2022survey}. If the elements are composite, the
list definition points to a 64-bit word that defines the composite elements,
allowing the deserializer to determine if following the pointer or not yields a
scalar of the same expected type. As an exception, Cap'n Proto does not support
this schema transformation on a list of booleans for runtime-efficiency reasons
\footnote{\url{https://capnproto.org/language.html\#evolving-your-protocol}}.
\fi

(3) Protocol Buffers Binary Wire Format \cite{protocolbuffers} supports
transforming a field into a list of a compatible type in a backwards-compatible
manner. Protocol Buffers Binary Wire Format encodes lists as multiple
occurrences of the same field identifier or as a concatenation of the members
prefixed with a length-delimited type definition in the case of packed field
encoding. This design decision makes implementations using the new schema
interpret a standalone value as a list consisting of one value.

(4) Serialization specifications based on field identifiers that implement
unions without involving additional structures such as Protocol Buffers Binary
Wire Format \cite{protocolbuffers} support forwards-compatibility when moving
an optional field into an existing union. In this case, union choices and
fields outside of the union share the same field identifier context. This means
that an application using the older schema either leaves the union choices or
the optional field outside the union unset. In comparison, FlatBuffers
\cite{flatbuffers} requires creating a new data structure to hold the union
type. As a result, it supports backwards-compatibility when moving an optional
field into an existing union as an application using the newer schema will
ignore the optional field outside of the union. The converse is true when
extracting an optional field out of an existing union.

(5) Protocol Buffers \cite{protocolbuffers} implements unions based on field
identifiers on the current identifier context and supports unions of a single
choice. Therefore, an optional field and a union of the single field are
equivalent. A similar argument follows for Cap'n Proto \cite{capnproto},
however Cap'n Proto does not support unions of a single choice, making this
transformation only backwards-compatible. Apache Avro \cite{avro} supports
unions of a single choice, however its schema resolution rules throw an
exception on the forwards-compatible case.

\begin{table}[hb!]
\caption{A schema transformation result is annotated as shown in
\autoref{table:evolution-legend}. The \emph{Type} column documents whether a
schema transformation confines (\texttt{\textbf{C}}), expands
(\texttt{\textbf{E}}), changes (\texttt{\textbf{!}}), or preserves
(\texttt{\textbf{=}}) the domain of the schema.}
\label{table:evolution-conversions}
\begin{tabularx}{\linewidth}{|l|c|X|c|c|c|c|c|c|c|}

\toprule

% Header
\textbf{\rotatebox[origin=c]{90}{Category}} &
\textbf{\rotatebox[origin=c]{90}{Type}} &
\textbf{Schema Transformation} &
\textbf{\rotatebox[origin=c]{90}{ASN.1 PER Unaligned}} &
\textbf{\rotatebox[origin=c]{90}{Apache Avro Binary Encoding}} &
\textbf{\rotatebox[origin=c]{90}{Microsoft Bond Compact Binary v1}} &
\textbf{\rotatebox[origin=c]{90}{Cap'n Proto Packed Encoding}} &
\textbf{\rotatebox[origin=c]{90}{FlatBuffers Binary Wire Format}} &
\textbf{\rotatebox[origin=c]{90}{Protocol Buffers Binary Wire Format}} &
\textbf{\rotatebox[origin=c]{90}{Apache Thrift Compact Protocol}} \\ \midrule

\multirow{11}{*}{\rotatebox[origin=c]{90}{Type Conversions}}
& E & Increase integer width & N & B & B & N & B & B & N \\ \cline{2-10}
& C & Decrease integer width & N & F & F & N & F & F & N \\ \cline{2-10}
& E & Increase float precision & B & B & B & N & N & N &  \\ \cline{2-10}
& C & Decrease float precision & F & F & F & N & N & N &  \\ \cline{2-10}
& E & Unsigned to larger signed integer & N &  & X & N & B & B &  \\ \cline{2-10}
& C & Signed to smaller unsigned integer & N &  & X & N & F & F &  \\ \cline{2-10}
& E & Signed integer to float & X & B & X & N & N & N & N \\ \cline{2-10}
& C & Float to signed integer & X & F & X & N & N & N & N \\ \cline{2-10}
& E & String to byte-array & B & X & X & B & B & B & B \\ \cline{2-10}
& C & Byte-array to string & F & X & X & F & F & F & F \\ \cline{2-10}
& E & Boolean to integer & X & X & X & N & B & B & N \\ \cline{2-10}
& C & Integer to boolean & X & X & X & N & F & F & N \\ \cline{2-10}
  & = & Byte-array to array of 8-bit unsigned integers & A &  & A & A &  &  & \\
  \bottomrule

\end{tabularx}
\end{table}

%% file: sections/conclusions/use-cases.tex
In \autoref{table:use-cases}, \we identify a set of use-cases that binary
serialization specifications tend to optimize for and the characteristics that
typically enable those use-cases.

\begin{table}[hb!] \caption{Every serialization specification considered in
  this study supports a subset of these use-cases. The \emph{Enabling
  characteristics} column describes certain characteristics that \emph{tend} to
  result in a serialization specification that is a good fit for the respective
  use-case. The last column shows an example of a JSON-compatible binary
  serialization specification that has at least one of the respective enabling
  characteristics. However, the fact that a serialization specification has
  certain characteristics does not guarantee that its implementations make use
  of those characteristics to enable the respective use-cases, often for
  reasons other than technical.}

\label{table:use-cases}
  \begin{tabularx}{\linewidth}{p{3cm}|X|p{3cm}}
    \toprule
  \textbf{Use case} & \textbf{Enabling characteristics} & \textbf{Example} \\ \midrule
    Space-efficiency & The resulting bit-string embeds little metadata & ASN.1 PER Unaligned \cite{asn1-per} \\
  \cline{2-2} & Non-aligned data types & \\ \hline
  Runtime-efficient deserialization & Deserialization without additional memory allocations & Cap'n Proto \cite{capnproto} \\
  \cline{2-2} & Table of contents for the bit-string & \\
  \cline{2-2} & Aligned data types & \\ \hline
  Partial reads & Field byte-length serialized before content & FlatBuffers Binary Wire Format \cite{flatbuffers} \\
  \cline{2-2} & Table of contents of the bit-string & \\ \hline
  Streaming deserialization & Field byte-length serialized before content & Smile \cite{smile} \\
  \cline{2-2} & Sequential and standalone-encoded fields & \\ \hline
    Streaming serialization & No byte-length prefix metadata, mainly for nested structures & UBJSON \cite{ubjson} \\
  \cline{2-2} & Content serialized before structure & \\
  \cline{2-2} & Positional structural markers instead of length prefixes & \\ \hline
  In-place updates & Field spatial locality & BSON \cite{bson} \\
  \cline{2-2} & No byte-length field metadata & \\
  \cline{2-2} & Positional structural markers instead of length prefixes & \\ \hline
    Constrained devices & Simple specification and binary layout & CBOR \cite{RFC7049} \\
  \cline{2-2} & Small generated code and/or runtime library & \\ \hline
  Drop-in JSON replacement & The resulting bit-string embeds all metadata & MessagePack \cite{messagepack} \\
    \bottomrule
\end{tabularx}

\end{table}

None of the binary serialization specifications from this study support every
use-case listed in \autoref{table:use-cases} as some enabling characteristics
tend to conflict:

\begin{itemize}

\item The \emph{Space-efficiency} use-case typically involves a schema-driven
  serialization specification. However, JSON \cite{RFC8259} is a schema-less
    serialization specification. Therefore, the \emph{Drop-in JSON replacement}
    requires a schema-less serialization specification.

\item The \emph{Space-efficiency} use-case requires bit-strings to be as
  compact as possible. However, the \emph{Runtime-efficient deserialization}
    use-case may require aligned data types and alignment may involve
    significant padding. For example, Cap'n Proto \cite{capnproto} aligns data
    types to 64-bit words for runtime-performance reasons and supports a simple
    compression scheme to mitigate the additional space overhead.

\item The \emph{Space-efficiency} use-case typically requires bit-strings to
  contain minimal metadata. However, the \emph{Partial reads} use-case may
    require a table of contents for the bit-string, which may result in more
    encoded metadata. The extra overhead is amortized when encoding large
    amounts of data sharing the same structures. The input data JSON document
    from \autoref{lst:json-object-test} is a small data structure that consists
    of significant structure and relatively little data. In the case of the
    Cap'n Proto \cite{capnproto} and FlatBuffers \cite{flatbuffers}
    schema-driven serialization specifications, roughly half of the bit-strings
    produced by serializing the input data consists of pointers and structural
    information that represent a table of contents.

\item The \emph{Streaming serialization} use-case may involve serializing the
  scalar types before the composite types, like FlexBuffers
    \cite{flexbuffers}, given that an implementation may not know the size
    of a composite data type before its members are encoded. However, this
    approach tends to conflict with the \emph{Streaming deserialization}
    use-case as an implementation would have to wait until all scalar types are
    received before starting to understand how they interconnect.

\item The \emph{Runtime-efficient deserialization} and \emph{Partial reads}
  use-cases typically involve a pointer-based table of contents of the
    bit-string.  As a result, implementing \emph{In-place updates} is usually
    not runtime-efficient as some updates might involve adjusting pointer
    references in multiple parts of the bit-string as noted by
    \cite{10.1007/978-981-15-8697-2_23} when using the FlatBuffers
    \cite{flatbuffers} serialization specification. For example, adding a new
    field to a FlexBuffers \cite{flexbuffers} map may involve creating a
    new keys vector, updating the metadata and contents of the vector data
    section, and adjusting most of the pointers in the bit-string.

\end{itemize}

\We could not identify a fundamental conflict involving the \emph{Constrained
devices} use-case. \We believe that whether a binary serialization
specification is a good fit for constrained devices tends to be a consequence
of how it is implemented rather than a property of the serialization
specification. For example, the official Protocol Buffers
\cite{protocolbuffers} implementations are typically not suitable for
constrained devices as they tend to generate large amounts of code and incur
significant binary size and memory allocation overheads.  Kenton Varda, one of
Protocol Buffers former authors, argues that the official Protocol Buffers
implementations \say{\textit{were designed for use in Google's servers, where
binary size is mostly irrelevant, while speed and features (e.g. reflection)
are valued}} \footnote{\url{https://news.ycombinator.com/item?id=25586632}}.
However, \emph{nanopb} \footnote{\url{https://github.com/nanopb/nanopb}} is a
Protocol Buffers implementation targeted at 32-bit micro-controllers and other
constrained devices. Refer to \cite{10589/150617} for discussions and examples
of \emph{nanopb}.

%% file: sections/conclusions/sequential-vs-pointer.tex
\label{sec:sequential-pointer-based}

\begin{figure}[ht!]
\frame{\includegraphics[width=\linewidth]{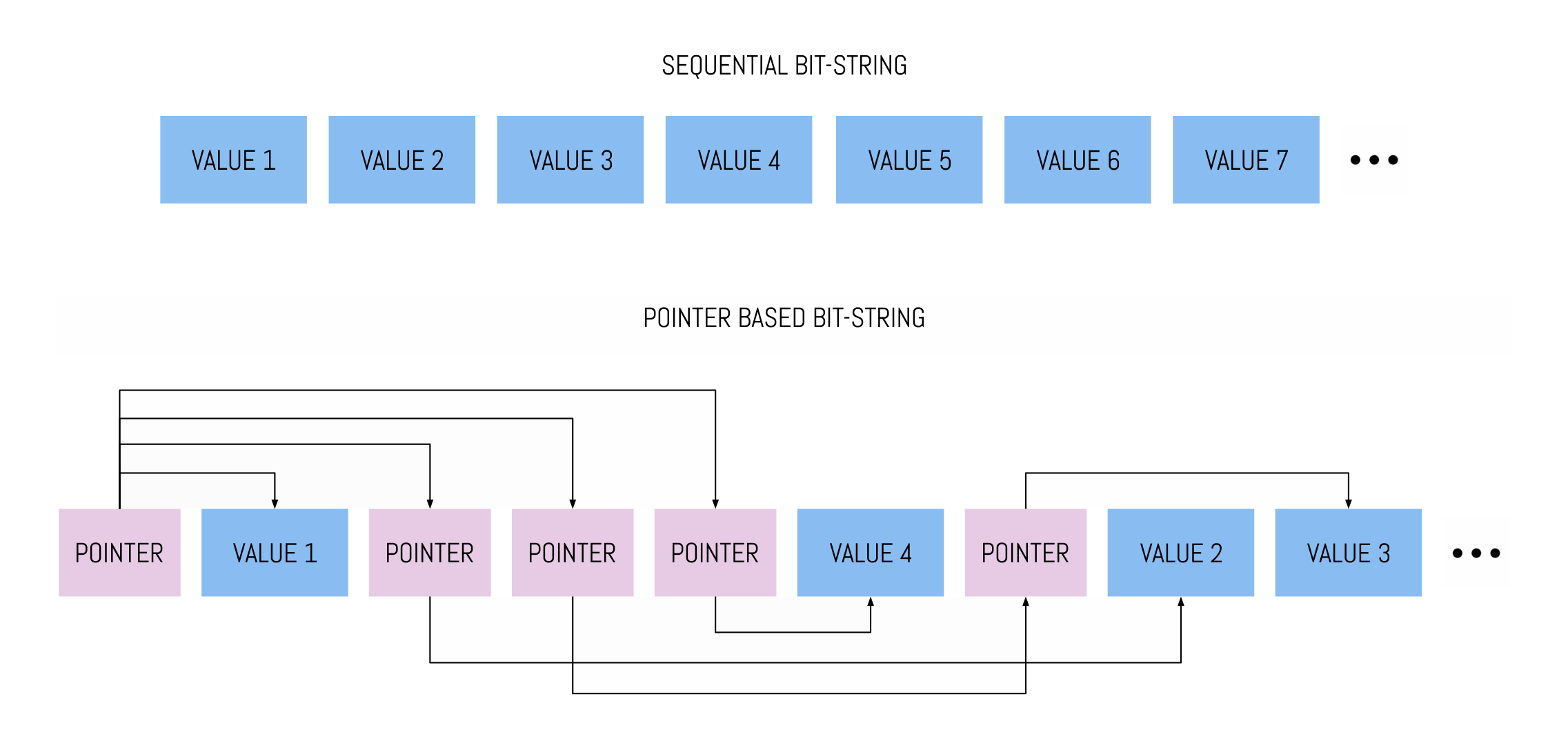}}
\caption{Visual representations of a sequential bit string (top) and a
pointer-based bit string (bottom).} \label{fig:sequential-pointer-based}
\end{figure}

\We found that serialization specifications can be categorized into that are
orthogonal to whether a serialization specification is schema-driven or
schema-less: whether the resulting bit-string is \emph{sequential} or
\emph{pointer-based} as shown in \autoref{fig:sequential-pointer-based}.

\textbf{Sequential.} Serialization specifications are \emph{sequential} if the
bit-strings they produce are concatenations of independently-encoded data
types. The majority of the serialization specifications discussed in this study
are sequential. As an example, Protocol Buffers \cite{protocolbuffers} is a
sequential serialization specification as its bit-strings consist of a
non-deterministic concatenation of fields
\footnote{\url{https://developers.google.com/protocol-buffers/docs/encoding\#implications}}
that are standalone with respect to the rest of the message.

\textbf{Pointer-based.} Serialization specifications are \emph{pointer-based}
if the bit-strings they produce are tree structures where each node is either a
scalar type or a composite value consisting of pointers to further nodes. In
comparison to sequential serialization specifications, this layout typically
results in larger bit-strings that are complicated to understand.  However,
pointer-based serialization specifications enable efficient streaming
deserialization, efficient random access reads and no additional memory
allocations during deserialization which translates to better deserialization
runtime performance. The pointer-based serialization specifications discussed
in this study are Cap'n Proto \cite{capnproto}, FlatBuffers \cite{flatbuffers},
and FlexBuffers \cite{flexbuffers}.

%% file: sections/conclusions/compatibility-resolution.tex
Every schema-driven serialization specifications discussed in this study allow
the schema-writer to perform certain schema transformations in compatible
manners. \We found that we can categorize schema-driven serialization
specifications into two groups based on how they approach schema compatibility
resolution: \emph{data-based resolution} or \emph{schema-based resolution}.

\textbf{Data-based resolution.} Every schema-driven serialization specification
discussed in this study except for Apache Avro \cite{avro} fall into this
category. In this type of schema compatibility resolution, the serialization
specification tries to understand the data by deserializing the bit-string as
if it was produced with the new schema and trying to accomodate to potential
mismatches at runtime.

\textbf{Schema-based resolution.} This approach is pioneered by Apache Avro
\cite{avro}, which refers to it as \emph{symbolic resolution}. In comparison to
the other schema-driven serialization specifications analyzed in this study, an
application deserializing an Apache Avro bit-string has to provide both the
\emph{exact schema} that was used to produce the bit-string and the new schema.
The implementation attempts to resolve the differences between the schemas
before deserializing the bit-string in order to determine how to adapt any
instance to the new representation. The bit-string is deserialized using the
old schema and transformed to match the new schema.
\cite{10.1145/3318464.3384417} briefly discusses the problem of integrating
heterogenous JSON datasets by resolving differences at the schema-level using a
similar approach. \cite{wilkinsdrys} propose a similar approach based on
version control systems where the codebase only maintains the latest schema
definition and code to upgrade older bit-strings to the latest version is
auto-generated based on the project commit history.

\We found that implementations of \emph{data-based resolution} schema-driven
serialization specifications, with some exceptions, tend to perform little
runtime checks to ensure data consistency, presumably for performance reasons.
For example, if the schema declares that the piece of data to follow is a
Little Endian 64-bit unsigned integer, then the deserialization specification
may blindly try to interpret the next 64-bits of the bit-string as such,
resulting in many cases in silently-incompatible unpredictable results rather
than informative runtime exceptions. In comparison to \emph{data-based
resolution} schema-driven serialization specifications, \we found that
\emph{schema-based resolution} tends to produce informative runtime exceptions
rather than unpredictable silently-incompatible results. However,
\emph{schema-based resolution} specifications require the consumer to know the
exact schema that was used to produce the data and have it available at the
deserialization process which may result in more complicated schema
transformation deployments.

\ifx\thesis\undefined
Based on the schema evolution experiments performed in
\autoref{sec:compatibility-analysis}, \we conclude that none of these
approaches produce specifications that are clearly more advantageous with
regards to compatible schema transformations: with some specification-specific
exceptions, most specifications tend to support the same compatible schema
transformations.
\else
Based on the schema evolution experiments performed in
\autoref{sec:thesis-appendix-schema-evolution-analysis}, \we conclude that none
of these approaches produce specifications that are clearly more advantageous
with regards to compatible schema transformations: with some
specification-specific exceptions, most specifications tend to support the same
compatible schema transformations.
\fi

%% file: sections/conclusions/similarities.tex
The bit-strings produced by the selection of binary serialization
specifications from \autoref{sec:serialization-format-selection} were more
similar than the \paperauthors expected. Each serialization specification has a
certain degree of unique characteristics and its tuned to particular use-cases.
However, most serialization specifications share the same underlying ideas and
approach to encoding.  The only notable exception to this pattern were the
sequential and pointer-based serialization specification groups discussed in
\autoref{sec:sequential-pointer-based}. Leaving that difference aside, \we
found that \we could largely infer the overall structure of the bit-strings
produced by a serialization specification without the need of a specification
after studying a handful of serialization specifications in depth.

%% file: sections/reproducibility.tex
\label{sec:reproducibility}

The hexadecimal bit-strings discussed in this study can be recreated by the
reader using the code files hosted on GitHub
\footnote{\url{https://github.com/jviotti/binary-json-survey}}.  This GitHub
repository contains a folder called \emph{analysis} including the input data
document from \autoref{lst:json-object-test} and the schema and code files for
each binary serialization specification implementation discussed in
\autoref{table:versions-schema-driven} and
\autoref{table:versions-schema-less}. The repository includes a \emph{Makefile}
for serializing the input data document with each of the selected serialization
specifications.

%% file: sections/future-work.tex
\label{sec:future-work}

\textbf{Space-efficient benchmarks.} \We plan to run space-efficiency benchmarks
involving the JSON-compatible binary serialization specifications discussed in
this paper using a range of JSON documents \cite{RFC8259} differing in content,
structure, and size. The goal of these space-efficiency benchmarks is dual.
First, \we want to understand what are the most space-efficient binary
serialization alternatives to JSON at the time of this writing. More
importantly, \we want to understand what serialization specification
characteristics and optimizations typically lead to more compact results and
what are the space-related bottlenecks that the new generation of
JSON-compatible binary serialization specifications need to solve to make a
breakthrough in the context of space-efficiency.

\textbf{Strict JSON-compatibility analysis.} As discussed in
\autoref{sec:serialization-format-selection}, \we discarded binary
serialization specifications that could not represent the \emph{input data}
JSON document from \autoref{lst:json-object-test} without changes. The fact
that a serialization specification can encode the \emph{input data} JSON
document provides a loose guarantee that such serialization specification is
JSON-compatible.  Given the relevance of JSON at the time of this writing, \we
believe that it is important to formally analyze whether the serialization
specifications discussed in this paper can represent \emph{any} valid JSON
document before claiming that they are JSON-compatible.

\textbf{Formal schema evolution compatibility analysis.} In
\autoref{sec:compatibility-analysis}, \we showcase a list of common schema
transformations and try to determine the level of schema compatibility
supported by the schema-driven serialization specifications discussed in this
paper through simple test cases. \We envision a formal analysis of the various
schema languages and their schema transformation compatibility levels that can
provide high-assurance and a more detailed view of what type of transformations
are compatible under what contexts.

\textbf{Schema semantic versioning.} To the best of \our knowledge, there is no
human readable versioning scheme that can distingush between backwards and
forwards compatible changes. Software libraries typically rely on
\emph{Semantic Versioning} \footnote{\url{https://semver.org}} to succintly
communicate whether a software library update is safe by distingushing between
incompatible changes, backwards-compatible new functionality, and
backwards-compatible bug fixes. \We envision a similar versioning convention
that is more applicable to schemas and distingushes between backwards,
forwards, and fully compatible changes.

\textbf{Schema-driven comparison metric.} As discussed in
\autoref{sec:schema-less-subset-schema-driven}, whether a serialization
specification is schema-driven is not a boolean characteristic. The
schema-driven serialization specifications that \we studied in this paper
leverage their respective schemas to different degrees during the
deserialization process. How much they leverage their schemas depends on the
expressiveness of their schema languages and on the amount of metadata they
embed into the bit-strings they produce. \We can envision a metric that can be
used to compare schema-driven serialization specifications in terms of
\emph{how much} schema-driven they are. \We believe that such metric can
formalize the understanding of why some schema-driven specifications are
generally more space-efficient than others. \We think that this metric is
analogous to Big O-notation \cite{danziger2010big} from the context of
algorithm analysis.